\documentclass[12pt]{book_phd}
\usepackage{epsfig,rotate,fancyhdr}
\newcommand{\hst}{\footnotesize \bfseries \MakeUppercase}
\newcommand{\beq}{\begin{equation}}
\newcommand{\enq}{\end{equation}}
\newcommand{\emp}{\end{minipage}}
\newcommand{\bmp}{\begin{minipage}}
\newcommand{\bfr}{\begin{flushright}}
\newcommand{\efr}{\end{flushright}}
\newcommand{\beqar}{\begin{eqnarray}}
\newcommand{\enqar}{\end{eqnarray}}
\newcommand{\bc}{\begin{center}}
\newcommand{\ec}{\end{center}}
\newcommand{\bef}{\begin{figure}}
\newcommand{\enf}{\end{figure}}
\newcommand{\mb}{\mathbf}
\newcommand{\tb}{\textbf}
\newcommand{\dst}{\displaystyle}
\newcommand{\tst}{\textstyle}
\newcommand{\tsc}{\textsuperscript}
\newcommand{\Tr}{\mathrm{Tr}\,}
\newcommand{\ms}{$\overline{\mathrm{MS}}$}
\newcommand{\mss}{\overline{\mathrm{MS}}}
\newcommand{\mrm}{\mathrm}
\newcommand{\scr}{\mathrm{screen}}
\newcommand{\tscs}{\textsuperscript}
\newcommand{\ordo}[1]{{\cal{O}}({#1})}
\newcommand{\dperd}[2]{\frac{d #1}{d #2}}
\newcommand{\6}{\partial}

\newcommand{\re}{{\mathrm{Re}}}

\newcommand{\ra}{$\rightarrow$\ }

\newcommand{\skk}{\qquad \quad \quad}
\newcommand{\sk}{\quad \quad}
\newcommand{\kk}{\quad}

\def\lsi{\raise0.3ex\hbox{$<$\kern-0.75em\raise-1.1ex\hbox{$\sim$}}}
\def\gsi{\raise0.3ex\hbox{$>$\kern-0.75em\raise-1.1ex\hbox{$\sim$}}}
\def\ha{H_1}
\def\hb{H_2}
\def\DD{{\cal D}}

\author{Pir\'oth Attila}
\title{Az elektrogyenge f\'azis\'atmenet}
\begin{document}
\pagestyle{empty}
\maketitle
\section*{Abstract}
The electroweak phase transition provides the most attractive framework to
account for the observed baryon asymmetry of the universe. 
In the literature it has been studied both perturbatively and
nonperturbatively, however, the comparison of perturbative and
nonperturbative results is not straightforward due to the different
coupling constant definitions. The
perturbative definition stems from the $\overline{\mathrm{MS}}$
subtraction scheme, while the nonperturbative one uses the static quark
potential. 

The momentum-space perturbative static potential is calculated in the
SU(2)--Higgs model, and is Fourier transformed into coordinate space
numerically.
Having established the connection between the coupling constants, two-loop
perturbative results are contrasted with 4-dimensional lattice simulation
results. The thermodynamically relevant parameters of the phase
transition
indicate that the perturbative results are reliable for low Higgs masses,
while for values around the endpoint ($m_H \approx 72$ GeV) the
perturbative approach breaks down. The previously known vale of the
endpoint can be refined to $72.1 \pm 1.4$ GeV, which excludes the standard
model baryogenesis scenario. The calculation also allows us to identify
the Higgs mass range for which dimensional reduction yields reliable
results.

As an extension of the standard model, the MSSM is also studied.
A simple one-loop perturbative approach is presented, which indicates some
useful trends: the baryogenesis requirements are more easily met if the
right-handed stop is lighter than the top; for certain parameters
colour-breaking phase transition is possible, etc. 

In order to perform 4-dimensional nonperturbative studies a special
purpose machine, PMS (Poor Man's Supercomputer) was built at
E\"otv\"os University between June 1998 and February 2000.

The results and the techniques of the simulations of the bosonic sector of
the MSSM performed on PMS are presented in some detail. A phase diagram is
presented, the bubble wall between the symmetric and the Higgs phase is
studied.  The cosmologically relevant part of the parameter space is
analysed. The results show that baryogenesis is possible within the MSSM
if $m_h \leq 103 \pm 4$ GeV, which can be tested in collider experiments
in the near future.

\pagestyle{fancy}
\tableofcontents
\fancyhead[CE,CO]{\hst{Tartalomjegyz\'ek}}
\chapter*{Be\-ve\-ze\-t\'es }
\fancyhead[CE]{\hst{Bevezet\'es}}
\addcontentsline{toc}{chapter}{Be\-ve\-ze\-t\'es }
A vi\-l\'a\-ge\-gye\-tem k\"o\-r\"u\-l\"ot\-t\"unk l\'e\-v\H o r\'e\-sze ba\-ri\-o\-nos, nem pe\-dig
an\-ti\-ba\-ri\-o\-nos a\-nyag\-b\'ol \'e\-p\"ul fel. Hon\-nan e\-red, m\'as sz\'o\-val a Big
Bang u\-t\'an h\H u\-l\H o vi\-l\'a\-ge\-gye\-tem mely f\'a\-zi\-s\'a\-ban ke\-let\-ke\-zett az a
je\-len\-t\'e\-keny ba\-ri\-on-t\"obb\-let, mely az an\-ti\-ba\-ri\-o\-no\-nok\-kal va\-l\'o
\"ut\-k\"o\-z\'e\-sek ha\-t\'a\-s\'a\-ra be\-k\"o\-vet\-ke\-z\H o an\-ni\-hi\-l\'a\-ci\-\'os fo\-lya\-ma\-tok
el\-le\-n\'e\-re fenn\-ma\-radt? Ad\-ha\-t\'o-e a fen\-ti k\'er\-d\'e\-sek\-re me\-ga\-la\-po\-zott
r\'e\-szecs\-ke\-fi\-zi\-ka\-i mo\-dell\-re \'e\-p\"u\-l\H o ma\-gya\-r\'a\-zat?

A ba\-ri\-o\-ge\-n\'e\-zis\-hez sz\"uk\-s\'e\-ges Sza\-ha\-rov-fel\-t\'e\-te\-lek \cite{sakharov} --
ba\-ri\-on\-sz\'am\-s\'er\-t\H o fo\-lya\-ma\-tok, C \'es CP-s\'er\-t\H o fo\-lya\-ma\-tok, ter\-mi\-kus
e\-gyen\-s\'uly\-t\'ol va\-l\'o el\-t\'e\-r\'es -- az a\-no\-m\'a\-lis szfa\-le\-ro\-n\'at\-me\-ne\-tek
\cite{thoft} r\'e\-v\'en elv\-ben ki\-e\-l\'e\-g\'{\i}t\-he\-t\H ok a stan\-dard mo\-dell
ke\-re\-t\'en be\-l\"ul. B\'ar a stan\-dard mo\-dell\-ben ta\-pasz\-tal\-ha\-t\'o CP-s\'er\-t\'es
m\'er\-t\'e\-ke t\'ul ki\-csi, \'{\i}gy leg\-fel\-jebb kva\-li\-ta\-t\'{\i}v ma\-gya\-r\'a\-zat
le\-het\-s\'e\-ges, a mo\-dell egy\-sze\-r\H u\-s\'e\-ge \'es k\'{\i}\-s\'er\-le\-ti
a\-l\'a\-t\'a\-masz\-tott\-s\'a\-ga mi\-att a k\'er\-d\'es vizs\-g\'a\-la\-t\'a\-nak je\-len\-t\H o\-s\'e\-ge
nem be\-cs\"ul\-he\-t\H o t\'ul. A ma\-gya\-r\'a\-zat\-hoz egy o\-lyan fo\-lya\-mat
sz\"uk\-s\'e\-ges, mely\-nek so\-r\'an a rend\-szer ki\-e\-sik a ter\-mi\-kus e\-gyen\-s\'uly
\'al\-la\-po\-t\'a\-b\'ol -- ez a vi\-l\'a\-ge\-gye\-tem h\H u\-l\'e\-se fo\-lya\-m\'an v\'eg\-be\-me\-n\H o
e\-lekt\-ro\-gyen\-ge f\'a\-zi\-s\'at\-me\-net r\'e\-v\'en le\-het\-s\'e\-ges.

A stan\-dard mo\-dell\-be\-li e\-lekt\-ro\-gyen\-ge f\'a\-zi\-s\'at\-me\-net \cite{rub1}
ta\-nul\-m\'a\-nyo\-z\'a\-sa e\-l\H o\-sz\"or per\-tur\-ba\-t\'{\i}v meg\-k\"o\-ze\-l\'{\i}\-t\'es\-ben t\"or\-t\'ent
\cite{arn,fod94/6,fod-heb}. Az egy- \'es k\'et\-hu\-rok\-ren\-d\H u sz\'a\-mo\-l\'a\-sok
a\-zon\-ban e\-r\H o\-sen k\'et\-s\'eg\-be von\-t\'ak a per\-tur\-ba\-t\'{\i}v meg\-k\"o\-ze\-l\'{\i}\-t\'es
al\-kal\-maz\-ha\-t\'o\-s\'a\-g\'at \cite{fod-heb}, \'{\i}gy (r\'esz\-ben) nem\-per\-tur\-ba\-t\'{\i}v
m\'od\-sze\-rek ki\-fej\-lesz\-t\'e\-se v\'alt sz\"uk\-s\'e\-ges\-s\'e. E\-zek k\"o\-z\"ul az
SU(2)--Higgs-mo\-dell n\'egy\-di\-men\-zi\-\'os r\'acsszi\-mu\-l\'a\-ci\-\'o\-ja \cite{fod94},
va\-la\-mint a di\-men\-zi\-\'os re\-duk\-ci\-\'o\-val ka\-pott h\'a\-rom\-di\-men\-zi\-\'os mo\-del\-lek
r\'acsszi\-mu\-l\'a\-ci\-\'o\-ja \cite{kaj, phil} bi\-zo\-nyult le\-gin\-k\'abb
gy\"u\-m\"ol\-cs\"o\-z\H o\-nek.

I\-gen fon\-tos a\-zon\-ban, hogy meg\-fe\-le\-l\H o\-k\'epp \"ossze tud\-juk vet\-ni a
per\-tur\-ba\-t\'{\i}v \'es nem\-per\-tur\-ba\-t\'{\i}v meg\-k\"o\-ze\-l\'{\i}\-t\'es\-sel ka\-pott
e\-red\-m\'e\-nye\-ket, hi\-szen egy i\-lyen kap\-cso\-lat a k\'e\-s\H ob\-bi\-ek\-ben
vizs\-g\'a\-lan\-d\'o bo\-nyo\-lul\-tabb mo\-del\-lek szem\-pont\-j\'a\-b\'ol na\-gyon hasz\-nos
le\-het. Az \"ossze\-ve\-t\'es a k\'et\-faj\-ta meg\-k\"o\-ze\-l\'{\i}\-t\'es el\-t\'e\-r\H o csa\-to\-l\'a\-si
\'al\-lan\-d\'o de\-fi\-n\'{\i}\-ci\-\'o\-ja mi\-att nem ma\-g\'a\-t\'ol \'er\-te\-t\H o\-d\H o; a
per\-tur\-ba\-t\'{\i}v meg\-k\"o\-ze\-l\'{\i}\-t\'es a j\'o\-lis\-mert \ms \ s\'e\-m\'a\-ban de\-fi\-ni\-\'al\-ja a
csa\-to\-l\'a\-si \'al\-lan\-d\'ot, m\'{\i}g a n\'egy\-di\-men\-zi\-\'os r\'acsszi\-mu\-l\'a\-ci\-\'ok a
sta\-ti\-kus kvark po\-ten\-ci\-\'al\-ra \cite{suss} \'e\-p\"ul\-nek. \'Igy a
nem\-per\-tur\-ba\-t\'{\i}v de\-fi\-n\'{\i}\-ci\-\'o\-hoz sz\"uk\-s\'e\-ges sta\-ti\-kus kvark po\-ten\-ci\-\'al
egy\-hu\-rok ren\-d\H u, per\-tur\-ba\-t\'{\i}v sz\'a\-mo\-l\'a\-sa r\'e\-v\'en a csa\-to\-l\'a\-si
\'al\-lan\-d\'ok k\"o\-z\"ott kap\-cso\-la\-tot te\-remt\-he\-t\"unk \cite{cikk1, la}. Ez
a prob\-l\'e\-ma k\'e\-pe\-zi a je\-len dok\-to\-ri \'er\-te\-ke\-z\'es el\-s\H o fe\-l\'e\-nek
t\'e\-m\'a\-j\'at.

Az el\-s\H o fe\-je\-zet\-ben a mo\-ti\-v\'a\-ci\-\'o\-ul szol\-g\'a\-l\'o ba\-ri\-o\-ge\-n\'e\-zis
t\'e\-m\'a\-j\'at is\-mer\-te\-tem, egy egy\-sze\-r\H u e\-lekt\-ro\-di\-na\-mi\-ka\-i a\-na\-l\'o\-gi\-\'a\-val
\'er\-z\'e\-kel\-tet\-ve a ba\-ri\-on\-sz\'am\-s\'er\-t\H o fo\-lya\-ma\-tok stan\-dard mo\-dell\-be\-li
meg\-va\-l\'o\-su\-l\'a\-s\'at. A fen\-ti\-ek\-n\'el r\'esz\-le\-te\-sebb in\-di\-k\'a\-ci\-\'o\-kat a\-dok
ar\-ra, mi\-\'ert e\-len\-ged\-he\-tet\-len a r\'acsszi\-mu\-l\'a\-ci\-\'ok al\-kal\-ma\-z\'a\-sa az
e\-lekt\-ro\-gyen\-ge f\'a\-zi\-s\'at\-me\-net vizs\-g\'a\-la\-t\'a\-ban.

A m\'a\-so\-dik fe\-je\-zet\-ben be\-ve\-ze\-tem a sta\-ti\-kus kvark po\-ten\-ci\-\'alt
\cite{suss}, me\-ga\-dom az im\-pul\-zus\-t\'er\-be\-li po\-ten\-ci\-\'al ki\-sz\'a\-m\'{\i}\-t\'a\-s\'a\-hoz
sz\"uk\-s\'e\-ges (Feyn\-man-m\'er\-t\'ek\-be\-li) gr\'af\-sza\-b\'a\-lyo\-kat \'es gr\'a\-fo\-kat,
majd a stan\-dard mo\-dell egy\-hu\-rok ren\-d\H u re\-nor\-m\'a\-l\'a\-sa so\-r\'an fel\-l\'e\-p\H o
gr\'a\-fok \cite{velt} ki\-v\'e\-te\-l\'e\-vel ki\-sz\'a\-m\'{\i}\-tom a gr\'a\-fok j\'a\-ru\-l\'e\-ka\-it.

A har\-ma\-dik fe\-je\-zet\-ben az e\-l\H ob\-bi\-ek\-ben meg\-ka\-pott im\-pul\-zus\-t\'er\-be\-li
po\-ten\-ci\-\'alt ko\-or\-di\-n\'a\-ta\-t\'er\-be transz\-for\-m\'a\-lom nu\-me\-ri\-kus m\'od\-szer\-rel.
A ko\-or\-di\-n\'a\-ta\-t\'er\-be\-li po\-ten\-ci\-\'al a\-lap\-j\'an oly m\'o\-don de\-fi\-ni\-\'a\-lom a
csa\-to\-l\'a\-si \'al\-lan\-d\'ot, hogy az a r\'acs\-t\'e\-rel\-m\'e\-let\-ben al\-kal\-ma\-zot\-tal a
le\-he\-t\H o leg\-szo\-ro\-sabb kap\-cso\-lat\-ban \'all\-jon. E\-z\'al\-tal me\-ga\-dom a
$g_{\overline{\mrm{MS}}}$ \'es a $g_{\mrm{pot}}$ csa\-to\-l\'a\-si \'al\-lan\-d\'ok
kap\-cso\-la\-t\'at.

A ne\-gye\-dik fe\-je\-zet\-ben a csa\-to\-l\'a\-si \'al\-lan\-d\'ok k\"u\-l\"on\-b\"o\-z\H o\-s\'e\-g\'e\-b\H ol
a\-d\'o\-d\'o ne\-h\'e\-zs\'e\-gek ki\-k\"u\-sz\"o\-b\"o\-l\'e\-se a\-lap\-j\'an meg\-vizs\-g\'a\-lom a
f\'a\-zi\-s\'at\-me\-ne\-tet jel\-lem\-z\H o ter\-mo\-di\-na\-mi\-ka\-i mennyi\-s\'e\-gek
\"ossze\-e\-gyez\-tet\-he\-t\H o\-s\'e\-g\'e\-nek k\'er\-d\'e\-s\'et a k\'et\-hu\-rok ren\-d\H u
per\-tur\-ba\-t\'{\i}v meg\-k\"o\-ze\-l\'{\i}\-t\'es, il\-let\-ve az SU(2)--Higgs-mo\-dell
n\'egy\-di\-men\-zi\-\'os r\'acsszi\-mu\-l\'a\-ci\-\'o\-j\'an a\-la\-pu\-l\'o m\'od\-sze\-rek k\"o\-z\"ott.
A vizs\-g\'a\-lat\-b\'ol le\-sz\H ur\-he\-t\H o, hogy a per\-tur\-ba\-t\'{\i}v meg\-k\"o\-ze\-l\'{\i}\-t\'es
csak bi\-zo\-nyos tar\-to\-m\'any\-ban m\H u\-k\"o\-dik he\-lye\-sen, az e\-lekt\-ro\-gyen\-ge
f\'a\-zi\-s\'at\-me\-net (nem\-per\-tur\-ba\-t\'{\i}v m\'od\-sze\-rek\-kel meg\-j\'o\-solt)
v\'eg\-pont\-j\'a\-t\'ol t\'a\-vol. A csa\-to\-l\'a\-si \'al\-lan\-d\'ok \"ossze\-ve\-t\'e\-s\'e\-b\H ol a
f\'a\-zi\-s\'at\-me\-net v\'eg\-pont\-j\'a\-ra a\-d\'o\-d\'o \'er\-t\'ek $72.0 \pm 1.4$ GeV
\cite{cikk1}, mely ki\-z\'ar\-ja az e\-lekt\-ro\-gyen\-ge f\'a\-zi\-s\'at\-me\-net
le\-he\-t\H o\-s\'e\-g\'et a stan\-dard mo\-dell\-ben, \'es ez\-zel e\-gy\"utt a ba\-ri\-o\-ge\-n\'e\-zis
k\'{\i}\-s\'er\-le\-ti\-leg i\-ga\-zolt fi\-zi\-ka\-i mo\-del\-len a\-la\-pu\-l\'o ma\-gya\-r\'a\-za\-t\'at.

A dol\-go\-zat m\'a\-sik fe\-l\'e\-ben a stan\-dard mo\-dell legp\-rag\-ma\-ti\-ku\-sabb
ki\-ter\-jesz\-t\'e\-s\'en, a mi\-ni\-m\'a\-lis szu\-per\-szim\-met\-ri\-kus ki\-ter\-jesz\-t\'e\-sen
be\-l\"ul vizs\-g\'a\-lom meg az e\-lekt\-ro\-gyen\-ge f\'a\-zi\-s\'at\-me\-net k\'er\-d\'e\-s\'et. A
sok\-di\-men\-zi\-\'os pa\-ra\-m\'e\-ter\-t\'er tel\-jes fel\-t\'er\-k\'e\-pe\-z\'e\-se re\-m\'eny\-te\-len\-nek
t\H u\-n\H o fe\-la\-dat, \'{\i}gy c\'e\-lom n\'e\-h\'any \'al\-ta\-l\'a\-nos ten\-den\-ci\-a
me\-g\'al\-la\-p\'{\i}\-t\'a\-sa, va\-la\-mint a pa\-ra\-m\'e\-ter\-t\'er ba\-ri\-o\-ge\-n\'e\-zis\-re al\-kal\-mas
(va\-la\-mely) sz\"og\-le\-t\'e\-nek dur\-va ki\-je\-l\"o\-l\'e\-se.

A stan\-dard mo\-dell\-be\-li vizs\-g\'a\-la\-tok a\-lap\-j\'an itt is nem\-per\-tur\-ba\-t\'{\i}v
m\'od\-sze\-rek\-re c\'el\-sze\-r\H u ha\-gyat\-koz\-ni. Az MSSM bo\-zo\-ni\-kus szek\-to\-r\'at
k\'{\i}\-v\'an\-juk majd n\'egy\-di\-men\-zi\-\'os szi\-mu\-l\'a\-ci\-\'ok r\'e\-v\'en vizs\-g\'a\-lat a\-l\'a
ven\-ni; ezt a vizs\-g\'a\-la\-tot meg\-k\"onny\'{\i}\-ti, ha e\-l\H o\-re ren\-del\-ke\-z\"unk
n\'e\-h\'any per\-tur\-ba\-t\'{\i}v m\'od\-sze\-rek\-kel ka\-pott j\'os\-lat\-tal. \'Igy az
\"o\-t\"o\-dik fe\-je\-zet\-ben egy i\-gen le\-egy\-sze\-r\H u\-s\'{\i}\-tett MSSM mo\-dellt
vizs\-g\'a\-lok \cite{car97}, az egy\-hu\-rok ren\-d\H u ef\-fek\-t\'{\i}v po\-ten\-ci\-\'al\-ban a
r\'acsszi\-mu\-l\'a\-ci\-\'ok\-n\'al hasz\-n\'alt pa\-ra\-m\'e\-te\-rek\-n\'el nu\-me\-ri\-kus \'u\-ton
ha\-t\'a\-ro\-zom meg a f\'a\-zi\-s\'at\-me\-ne\-ti pon\-to\-kat. Az MSSM-ben meg\-va\-l\'o\-su\-l\'o
h\'a\-rom f\'a\-zis (szim\-met\-ri\-kus, sz\'{\i}n\-s\'er\-t\H o \cite{bod}, Higgs-f\'a\-zis)
je\-len\-l\'e\-t\'et ez\-zel az egy\-sze\-r\H u\-s\'{\i}\-tett mo\-del\-lel is de\-monst\-r\'a\-lom,
meg\-ha\-t\'a\-ro\-zom a\-dott pa\-ra\-m\'e\-te\-r\'e\-t\'e\-kek\-n\'el a f\'a\-zis\-di\-ag\-ram e\-gyes
\'a\-ga\-it. Meg\-mu\-ta\-tom, ho\-gyan han\-gol\-ha\-t\'ok a pa\-ra\-m\'e\-te\-rek \'ugy, hogy a
fi\-zi\-ka\-i mennyi\-s\'e\-gek (pl.\ Higgs-W t\"o\-me\-ga\-r\'any) \'er\-t\'e\-ke ne
v\'al\-toz\-zon. A mo\-dell e\-r\H o\-sen le\-egy\-sze\-r\H u\-s\'{\i}\-tett jel\-le\-ge mi\-att ezt
kva\-li\-ta\-t\'{\i}\-ven hasz\-n\'a\-lom a to\-v\'ab\-bi vizs\-g\'a\-la\-tok so\-r\'an.
\fancyhead[CO]{\hst{Bevezet\'es}}

Az MSSM n\'egy\-di\-men\-zi\-\'os szi\-mu\-l\'a\-ci\-\'o\-ja \'o\-ri\-\'a\-si sz\'a\-m\'{\i}\-t\'o\-g\'e\-pes
i\-g\'e\-nye\-ket t\'a\-maszt. En\-nek ki\-e\-l\'e\-g\'{\i}\-t\'e\-s\'e\-re az El\-m\'e\-le\-ti Fi\-zi\-ka\-i
Tan\-sz\'e\-ken 1998.\ nya\-r\'an el\-kez\-d\H o\-d\"ott a 32 PC e\-lem\-b\H ol \'al\-l\'o
PMS (Po\-or Man's Su\-per\-com\-pu\-ter) szu\-per\-sz\'a\-m\'{\i}\-t\'o\-g\'ep \'e\-p\'{\i}\-t\'e\-se.
N\'e\-h\'any h\'o\-na\-pos mun\-k\'a\-val a sz\'a\-m\'{\i}\-t\'o\-g\'ep m\H u\-k\"o\-d\H o\-k\'e\-pes
\'al\-la\-pot\-ba ju\-tott, b\'ar a szu\-per\-sz\'a\-m\'{\i}\-t\'o\-g\'ep i\-ga\-zi kva\-li\-t\'a\-s\'at a\-d\'o
n\'o\-du\-sok k\"o\-z\"ot\-ti kom\-mu\-ni\-k\'a\-ci\-\'o csak 2000.\ feb\-ru\-\'ar\-j\'a\-ban va\-l\'o\-sult
meg \cite{pmscikk}. A ha\-to\-dik fe\-je\-zet\-ben a szu\-per\-sz\'a\-m\'{\i}\-t\'o\-g\'ep
\'e\-p\'{\i}\-t\'e\-s\'et, fe\-l\'e\-p\'{\i}\-t\'e\-s\'et \'es tel\-je\-s\'{\i}t\-m\'e\-ny\'et (egy\-sze\-res
pon\-tos\-s\'a\-g\'u m\H u\-ve\-le\-tek e\-se\-t\'en kb.\ 27 Gflop, 0.45 \$/Mflop
\'ar--tel\-je\-s\'{\i}t\-m\'eny h\'a\-nya\-dos) fog\-la\-lom r\"o\-vi\-den \"ossze. A
szu\-per\-sz\'a\-m\'{\i}\-t\'o\-g\'ep mel\-lett 1999.\ nya\-r\'a\-t\'ol egy to\-v\'ab\-bi 64
PC-e\-lem\-b\H ol \'al\-l\'o clus\-ter (PMS2) is ren\-del\-ke\-z\'es\-re \'allt a
szi\-mu\-l\'a\-ci\-\'ok so\-r\'an.

A n\'egy\-di\-men\-zi\-\'os r\'acsszi\-mu\-l\'a\-ci\-\'ok\-kal a he\-te\-dik fe\-je\-zet fog\-lal\-ko\-zik.
E\-l\H o\-sz\"or is\-mer\-te\-tem a r\'acs\-ra he\-lye\-zett Lag\-ran\-ge-f\"ugg\-v\'enyt, majd a
(v\'e\-ges il\-let\-ve z\'e\-rus h\H o\-m\'er\-s\'ek\-le\-t\H u) r\'acsszi\-mu\-l\'a\-ci\-\'ok\-ban
m\'e\-ren\-d\H o mennyi\-s\'e\-ge\-ket. A f\'a\-i\-zs\'at\-me\-ne\-ti pon\-to\-kat
Le\-e--Yang-m\'od\-szer\-rel ha\-t\'a\-ro\-zom meg, mi\-\'al\-tal a per\-tur\-ba\-t\'{\i}v
meg\-k\"o\-ze\-l\'{\i}\-t\'es\-ben ka\-pot\-tal kva\-li\-ta\-t\'{\i}v e\-gye\-z\'es\-ben \'al\-l\'o
f\'a\-zis\-di\-ag\-ra\-mot ka\-pok. Vizs\-g\'a\-lom e\-zen k\'{\i}\-v\"ul a ba\-ri\-o\-ge\-n\'e\-zis
szem\-pont\-j\'a\-b\'ol i\-gen fon\-tos \cite{brhlik, clin-gdm99}
bu\-bo\-r\'ek\-fal-vas\-tag\-s\'a\-got. \\
A r\'acsszi\-mu\-l\'a\-ci\-\'ok\-ban v\'eg\-re\-haj\-tan\-d\'o kon\-ti\-nu\-um-li\-mesz k\'ep\-z\'e\-se
nagy ne\-h\'e\-zs\'e\-gek\-be \"ut\-k\"o\-zik: min\-ded\-dig nem si\-ke\-r\"ult a\-zo\-nos fi\-zi\-ka\-i
pa\-ra\-m\'e\-te\-re\-ket biz\-to\-s\'{\i}\-ta\-ni egy\-re ki\-sebb r\'a\-cs\'al\-lan\-d\'o mel\-lett. \'Igy
a n\'egy\-di\-men\-zi\-\'os szi\-mu\-l\'a\-ci\-\'ok\-b\'ol sz\'ar\-maz\-tat\-ha\-t\'o mennyi\-s\'e\-gek
k\"o\-re kor\-l\'a\-to\-zott, pl.\ a koz\-mo\-l\'o\-gi\-a\-i je\-len\-t\H o\-s\'e\-g\H u $v/T_c$
h\'a\-nya\-dost a r\'acsszi\-mu\-l\'a\-ci\-\'ok a\-lap\-j\'an kel\-l\H o pon\-tos\-s\'ag\-gal nem
le\-het k\"oz\-vet\-le\-n\"ul meg\-be\-cs\"ul\-ni. A r\'a\-cse\-red\-m\'e\-nyek\-hez al\-kal\-maz\-ko\-d\'o
egy\-hu\-rok\-ren\-d\H u per\-tur\-b\'a\-ci\-\'o\-sz\'a\-m\'{\i}\-t\'as \cite{jak2} se\-g\'{\i}t\-s\'e\-g\'e\-vel
fel\-raj\-zol\-ha\-t\'o a leg\-ki\-sebb t\"o\-me\-g\H u szu\-per\-szim\-met\-ri\-kus Higgs t\"o\-meg
\'es a jobb\-ke\-zes stop-t\"o\-meg s\'{\i}k\-j\'an a koz\-mo\-l\'o\-gi\-a\-i\-lag re\-le\-v\'ans
tar\-to\-m\'any; ez a\-lap\-j\'an a ba\-ri\-o\-ge\-n\'e\-zis MSSM-be\-li meg\-va\-l\'o\-su\-l\'a\-s\'a\-nak
(4-di\-men\-zi\-\'os r\'acsszi\-mu\-l\'a\-ci\-\'ok\-ra a\-la\-po\-zott) fel\-t\'e\-te\-le\-k\'ent $m_h
\leq 103 \pm 4$ GeV a\-d\'o\-dik, a per\-tur\-ba\-t\'{\i}v \cite{los} \'es a di\-men\-zi\-\'os
re\-duk\-ci\-\'os m\'od\-sze\-ren a\-la\-pu\-l\'o e\-red\-m\'e\-nyek\-kel \cite{la98} \"ossz\-hang\-ban.
Ez az \'er\-t\'ek a k\'{\i}\-s\'er\-le\-ti\-leg m\'eg le\-het\-s\'e\-ges tar\-to\-m\'any\-ba e\-sik,
me\-lyet a CERN-ben \'es a Te\-vat\-ron\-ban v\'eg\-re\-haj\-tott k\'{\i}\-s\'er\-le\-tek
ha\-ma\-ro\-san el\-le\-n\H o\-riz\-ni fog\-nak.

K\'et f\"ug\-ge\-l\'e\-ket mel\-l\'e\-ke\-lek, me\-lyek a szta\-ti\-kus kvark po\-ten\-ci\-\'al
sz\'a\-m\'{\i}\-t\'a\-s\'a\-hoz kap\-cso\-l\'od\-nak. Az el\-s\H o\-ben egy hu\-ro\-kin\-teg\-r\'alt
sz\'a\-m\'{\i}\-tok ki, mely az egy\-hu\-rok-ren\-d\H u k\"o\-ze\-l\'{\i}\-t\'es\-ben nem l\'ep fel
k\"oz\-vet\-le\-n\"ul; a m\'a\-so\-dik\-ban r\'esz\-le\-te\-sen ki\-sz\'a\-m\'{\i}\-tom a
kvan\-tum\-sz\'{\i}n\-di\-na\-mi\-ka r\'e\-g\'o\-ta is\-mert po\-ten\-ci\-\'al\-j\'at \cite{suss, fis}.
Az i\-ro\-da\-lom\-ban k\"o\-z\"olt \'er\-t\'e\-kek rep\-ro\-du\-k\'a\-l\'a\-sa in\-di\-rekt
bi\-zo\-ny\'{\i}\-t\'ek\-k\'ent szol\-g\'al a bo\-nyo\-lul\-tabb SU(2)--Higgs-mo\-dell\-be\-li
po\-ten\-ci\-\'al he\-lyes\-s\'e\-g\'et il\-le\-t\H o\-en.

A szta\-ti\-kus kvark po\-ten\-ci\-\'al\-lal kap\-cso\-la\-tos sz\'a\-mo\-l\'a\-so\-kat Csi\-kor
Fe\-renc\-cel, Fo\-dor Zol\-t\'an\-nal \'es He\-ge\-d\"us P\'al\-lal v\'e\-gez\-tem. A PMS
szu\-per\-sz\'a\-m\'{\i}\-t\'o\-g\'ep \'e\-p\'{\i}\-t\'e\-s\'e\-ben Csi\-kor Fe\-renc\-cel, Fo\-dor
Zol\-t\'an\-nal, He\-ge\-d\"us P\'al\-lal, Hor\-v\'ath Vik\-tor\-ral \'es Katz S\'an\-dor\-ral
e\-gy\"utt vet\-tem r\'eszt. Az MSSM-be\-li e\-lekt\-ro\-gyen\-ge f\'a\-zi\-s\'at\-me\-ne\-tet
Csi\-kor Fe\-renc\-cel, Fo\-dor Zol\-t\'an\-nal, He\-ge\-d\"us P\'al\-lal, Ja\-ko\-v\'ac
An\-tal\-lal \'es Katz S\'an\-dor\-ral e\-gy\"utt vizs\-g\'al\-tam. A dol\-go\-zat\-ban
k\"o\-z\"olt e\-red\-m\'e\-nyek k\"o\-z\"ul az a\-l\'ab\-bi\-ak a sa\-j\'at\-ja\-im:
\begin{itemize}
\item
Meg\-ha\-t\'a\-roz\-tam a szta\-ti\-kus kvark po\-ten\-ci\-\'alt im\-pul\-zus\-t\'er\-ben; az
i\-ro\-da\-lom\-b\'ol is\-mert egy\-hu\-rok-ren\-d\H u bo\-zon-pro\-pa\-g\'a\-tor ki\-v\'e\-te\-l\'e\-vel
Feyn\-man-m\'er\-t\'ek\-ben me\-gad\-tam a po\-ten\-ci\-\'al\-ban sze\-rep\-l\H o gr\'a\-fok
j\'a\-ru\-l\'e\-ka\-it. El\-le\-n\H or\-z\'es\-k\'ent rep\-ro\-du\-k\'al\-tam a QCD i\-ro\-da\-lom\-b\'ol
is\-mert egy\-hu\-rok ren\-d\H u szta\-ti\-kus po\-ten\-ci\-\'al\-j\'at.
\item
A \emph{Maple} prog\-ram se\-g\'{\i}t\-s\'e\-g\'e\-vel ko\-or\-di\-n\'a\-ta\-t\'er\-be
transz\-for\-m\'al\-tam a po\-ten\-ci\-\'alt, majd ezt nu\-me\-ri\-ku\-san dif\-fe\-ren\-ci\-\'al\-tam;
A $dV/dx$ mennyi\-s\'eg\-b\H ol meg\-ha\-t\'a\-roz\-tam a k\'et\-f\'e\-le csa\-to\-l\'a\-si \'al\-lan\-d\'o
kap\-cso\-la\-t\'at.
\item
A \emph{Maple} prog\-ram\-mal nu\-me\-ri\-ku\-san vizs\-g\'al\-tam az MSSM bo\-zo\-ni\-kus
szek\-to\-r\'at egy\-hu\-rok-ren\-d\H u per\-tur\-ba\-t\'{\i}v k\"o\-ze\-l\'{\i}\-t\'es\-ben, a
\cite{car97} cikk\-ben k\"o\-z\"olt ef\-fek\-t\'{\i}v po\-ten\-ci\-\'al a\-lap\-j\'an. Ki\-mu\-tat\-tam
a h\'a\-rom f\'a\-zis je\-len\-l\'e\-t\'et. Be\-mu\-tat\-tam a m\'od\-szer
to\-v\'abb\-fej\-lesz\-t\'e\-s\'e\-nek le\-he\-t\H o\-s\'e\-g\'et \'es sz\"uk\-s\'e\-ges\-s\'e\-g\'et.
\item
1998.\ \'es 1999.\ nya\-r\'an r\'eszt vet\-tem a PMS szu\-per\-sz\'a\-m\'{\i}\-t\'o\-g\'ep
\'e\-p\'{\i}\-t\'e\-s\'e\-ben.
\item
V\'e\-ges h\H o\-m\'er\-s\'ek\-le\-t\H u Mon\-te Car\-lo szi\-mu\-l\'a\-ci\-\'o\-kat haj\-tot\-tam v\'eg\-re
a szu\-per\-sz\'a\-m\'{\i}\-t\'o\-g\'e\-pen, a ka\-pott a\-da\-tok\-b\'ol t\"obb e\-set\-ben
meg\-ha\-t\'a\-roz\-tam a v\'eg\-te\-len t\'er\-fo\-ga\-t\'u ha\-t\'a\-r\'er\-t\'e\-ket. A
f\'a\-zi\-s\'at\-me\-ne\-ti pon\-tok meg\-ha\-t\'a\-ro\-z\'a\-s\'a\-ra a Le\-e--Yang-m\'od\-sze\-ren
k\'{\i}\-v\"ul a k\'et\-cs\'u\-cs\'u hisz\-tog\-ram m\'od\-szert \'es a hisz\-te\-r\'e\-zis
m\'od\-szert is hasz\-n\'al\-tam. Vizs\-g\'al\-tam a f\'a\-zi\-s\'at\-me\-net le\-he\-t\H o\-s\'e\-g\'et
mind\-h\'a\-rom f\'a\-zis k\"o\-z\"ott. V\'e\-ges t\'er\-fo\-ga\-ton meg\-ha\-t\'a\-roz\-tam a
f\'a\-zis\-di\-ag\-ram sz\'{\i}n\-s\'er\-t\H o \'es Higgs-f\'a\-zis k\"oz\-ti \'a\-g\'at, majd
v\'eg\-te\-len t\'er\-fo\-ga\-t\'u ha\-t\'a\-r\'at\-me\-ne\-tet haj\-tot\-tam v\'eg\-re.
\end{itemize}
A dol\-go\-zat\-hoz kap\-cso\-l\'o\-d\'o, re\-fe\-r\'alt fo\-ly\'o\-i\-rat\-ban meg\-je\-lent
pub\-li\-k\'a\-ci\-\'ok:
\begin{itemize}
\item
F.~Csi\-kor, Z.~Fo\-dor, P.~He\-ged\"us, A.~Pir\'oth, \emph{Sta\-tic
po\-ten\-ti\-al in the SU(2)--Higgs mo\-del and co\-up\-ling cons\-tant de\-fi\-ni\-ti\-ons
in lat\-ti\-ce and con\-ti\-nu\-um mo\-dels}, Physi\-cal Re\-vi\-ew \textbf{D60}, 114511
(1999)
\item
F.~Csi\-kor, Z.~Fo\-dor, P.~He\-ged\"us, A.~Ja\-kov\'ac, S.~D.~Katz,
A.~Pir\'oth, \emph{E\-lect\-ro\-we\-ak Pha\-se Tran\-si\-ti\-ons in the MSSM:
4-di\-men\-si\-o\-nal Lat\-ti\-ce Si\-mu\-la\-ti\-ons}, hep-ph/0001087, k\"oz\-l\'es\-re
el\-fo\-gad\-va a \emph{Physi\-cal Re\-vi\-ew Let\-ters} c.\ fo\-ly\'o\-i\-rat\-n\'al
\item
F.~Csi\-kor, Z.~Fo\-dor, P.~He\-ged\"us, V.~K.~Horv\'ath, S.~D.~Katz,
A.~Pir\'oth, \emph{The PMS Pro\-ject -- Po\-or Man's Su\-per\-com\-pu\-ter},
hep-lat/9912059 -- k\"oz\-l\'es\-re el\-fo\-gad\-va a \emph{Com\-pu\-ter Physics
Com\-mu\-ni\-ca\-ti\-ons} c.\ fo\-ly\'o\-i\-rat\-n\'al
\end{itemize}
A dol\-go\-zat\-hoz kap\-cso\-l\'o\-d\'o to\-v\'ab\-bi pub\-li\-k\'a\-ci\-\'ok:
\begin{itemize}
\item
A.~Pir\'oth, \emph{The Sta\-tic Po\-ten\-ti\-al in the SU(2)--Higgs mo\-del and
the E\-lect\-ro\-we\-ak Pha\-se Tran\-si\-ti\-on} hep-ph/9909552.
\end{itemize}

\chapter{Ba\-ri\-o\-ge\-n\'e\-zis }
\pagestyle{fancy}
\fancyhead[CE]{\hst{\thechapter{}.\ fe\-je\-zet \quad Ba\-ri\-o\-ge\-n\'e\-zis}}
\bfr
\emph{
Well be\-yond the tro\-post\-ra\-ta\\
The\-re is a re\-gi\-on stark and stel\-lar \\
Whe\-re, on a pi\-e\-ce of an\-ti-mat\-ter \\
Li\-ved Dr.\ Ed\-ward An\-ti-Tel\-ler. \medskip \\
}
H.~P.~Fruth \cite{physspeak}
\efr
Wer\-ner He\-i\-sen\-berg az an\-ti\-a\-nyag fel\-fe\-de\-z\'e\-s\'et tar\-tot\-ta a hu\-sza\-dik
sz\'a\-za\-di fi\-zi\-ka ta\-l\'an leg\-fon\-to\-sabb e\-l\H o\-re\-l\'e\-p\'e\-s\'e\-nek
\cite{physspeak}. A fel\-fe\-de\-z\'es ha\-t\'a\-s\'a\-ra a
\bc
{\emph{\fbox{Mi\-\'ert van va\-la\-mi a sem\-mi he\-lyett --
a\-vagy ho\-gyan zaj\-lott le a vi\-l\'a\-ge\-gye\-tem te\-rem\-t\'e\-se?}}}
\ec
\noindent k\'er\-d\'es \'ev\-sz\'a\-za\-dos--\'e\-vez\-re\-des t\"or\-t\'e\-ne\-t\'e\-ben \'uj
fe\-je\-zet -- de le\-ga\-l\'ab\-bis \'uj l\'ab\-jegy\-zet -- ny\'{\i}lt meg: a te\-o\-l\'o\-gi\-a
\'es a fi\-lo\-z\'o\-fi\-a u\-t\'an a r\'e\-szecs\-ke\-fi\-zi\-ka is r\'eszt k\"o\-ve\-telt a
k\'er\-d\'es\-b\H ol.

A r\'e\-szecs\-ke\-fi\-zi\-ka\-i v\'a\-lasz\-ke\-re\-s\'es konk\-r\'e\-tab\-ban a k\'er\-d\'es
k\"o\-vet\-ke\-z\H o as\-pek\-tu\-s\'a\-ra \"ossz\-pon\-to\-s\'{\i}t:
\bc
{\emph{\fbox{Mi\-\'ert csak ba\-ri\-o\-nos a\-nya\-got l\'a\-tunk ma\-gunk
k\"o\-r\"ul?}}}
\ec

A k\'er\-d\'es a\-lap\-ja k\"oz\-vet\-len ta\-pasz\-ta\-la\-tunk: an\-ti\-a\-nyag a F\"ol\-d\"on
csak mik\-rosz\-ko\-pi\-kus mennyi\-s\'eg\-ben for\-dul e\-l\H o, \'es a Nap\-rend\-szer
t\'a\-vo\-lab\-bi r\'e\-sze\-i\-r\H ol vissza\-t\'e\-r\H o \H ur\-szon\-d\'ak bi\-zo\-ny\'{\i}\-t\'e\-ka\-i is
meggy\H o\-z\H o\-ek. T\'av\-cs\"o\-ve\-ink min\-ded\-dig nem ta\-l\'al\-tak a\-nyag \'es
an\-ti\-a\-nyag cso\-m\'ok ha\-t\'a\-r\'an le\-zaj\-l\'o, l\'at\-v\'a\-nyos sz\'et\-su\-g\'ar\-z\'as\-sal
j\'a\-r\'o na\-gye\-ner\-gi\-\'a\-j\'u fo\-lya\-ma\-tok\-ra u\-ta\-l\'o je\-le\-ket. Becs\-l\'e\-se\-ink
sze\-rint min\-tegy $10^{13}$ nap\-t\"o\-meg\-nyi k\"or\-nye\-ze\-t\"unk pusz\-t\'an
a\-nyag\-b\'ol \'all \cite{coh}.

A r\'e\-szecs\-ke\-fi\-zi\-ka kel\-l\H o s\'u\-ly\'u \'erv hi\-\'a\-ny\'a\-ban el\-ve\-ti an\-nak
le\-he\-t\H o\-s\'e\-g\'et, hogy a vi\-l\'a\-ge\-gye\-tem i\-gen nagy sk\'a\-l\'as szer\-ke\-ze\-te
egy\-m\'as\-sal v\'al\-ta\-ko\-z\'o, $10^{13}$ nap\-t\"o\-meg\-n\'el na\-gyobb m\'e\-re\-t\H u
a\-nyag-, il\-let\-ve an\-ti\-a\-nyag\-hal\-ma\-zok\-b\'ol \'all\-na, mint\-hogy min\-ded\-dig nem
is\-mert e\-gyet\-len o\-lyan me\-cha\-niz\-mus sem, mely ek\-ko\-ra sk\'a\-l\'an k\'e\-pes
len\-ne az a\-nya\-got \'es az an\-ti\-a\-nya\-got sz\'et\-v\'a\-lasz\-ta\-ni. \'Igy,
r\'e\-szecs\-ke\-fiz\-ka\-i ta\-pasz\-ta\-la\-tunk a\-lap\-j\'an fel\-t\'e\-te\-lez\-z\"uk, hogy a
vi\-l\'a\-ge\-gye\-tem m\'a\-sutt is a\-nyag\-b\'ol \'all -- ezt a hi\-po\-t\'e\-zist
ne\-vez\-z\"uk a \emph{vi\-l\'a\-ge\-gye\-tem ba\-ri\-on-a\-szim\-met\-ri\-\'a\-j\'a}nak.

A ba\-ri\-on-a\-szim\-met\-ri\-a e\-gyik le\-het\-s\'e\-ges ma\-gya\-r\'a\-za\-ta a kez\-de\-ti
fel\-t\'e\-te\-lek meg\-fe\-le\-l\H o meg\-v\'a\-lasz\-t\'a\-sa len\-ne. Ez a ma\-gya\-r\'a\-zat
t\'ul\-s\'a\-go\-san konk\-lu\-z\'{\i}v, \'{\i}gy meg kell vizs\-g\'al\-nunk a m\'a\-sik
le\-he\-t\H o\-s\'e\-get; mi\-vel u\-t\'ob\-bi sok\-kal szer\-te\-\'a\-ga\-z\'obb \'es sz\'{\i}\-ne\-sebb
ma\-gya\-r\'a\-zat ke\-re\-s\'e\-s\'et je\-len\-ti, ezt az u\-tat fog\-juk v\'a\-lasz\-ta\-ni -- a
kez\-de\-ti fel\-t\'e\-te\-lek\-re va\-l\'o t\'a\-masz\-ko\-d\'ast pe\-dig meg\-fe\-le\-l\H o \'er\-vek
hi\-\'a\-ny\'a\-ban \'es esz\-t\'e\-ti\-ka\-i szem\-pon\-tok a\-lap\-j\'an el\-vet\-j\"uk.

O\-lyan fi\-zi\-ka\-i fo\-lya\-ma\-tot kell ke\-res\-n\"unk, mely\-ben di\-na\-mi\-ka\-i
ma\-gya\-r\'a\-za\-tot ad\-ha\-tunk a ba\-ri\-on-a\-szim\-met\-ri\-\'a\-ra. A k\"o\-vet\-ke\-z\H o mo\-za\-i\-kot
k\'{\i}\-v\'an\-juk te\-h\'at fi\-zi\-ka\-i\-lag (mi\-n\'el in\-k\'abb) a\-l\'a\-t\'a\-masz\-tott
fo\-lya\-ma\-tok\-b\'ol fe\-l\'e\-p\'{\i}\-te\-ni: kez\-det\-ben az u\-ni\-ver\-zum \"ossz-ba\-ri\-on\-sz\'a\-ma
0, majd a nagy bumm u\-t\'a\-ni el\-s\H o m\'a\-sod\-perc ki\-csiny t\"o\-re\-d\'e\-k\'e\-ben
o\-lyan fo\-lya\-ma\-tok j\'at\-sz\'od\-nak le, me\-lyek\-ben a ba\-ri\-on--an\-ti\-ba\-ri\-on
szim\-met\-ri\-a eny\-h\'en meg\-bom\-lik, \'{\i}gy k\"o\-zel (de nem tel\-je\-sen) a\-zo\-nos
sz\'a\-m\'u ba\-ri\-on \'es an\-ti\-ba\-ri\-on j\"on l\'et\-re. E\-zek egy\-m\'as\-sal
k\"ol\-cs\"on\-hat\-va sz\'et\-su\-g\'ar\-z\'od\-nak -- a\-mi \'al\-tal ren\-ge\-teg fo\-ton
ke\-let\-ke\-zik. A ki\-csiny meg\-ma\-radt ba\-ri\-on\-t\"obb\-let\-b\H ol pe\-dig v\'e\-g\"ul
l\'et\-re\-j\"o\-het az u\-ni\-ver\-zum ma\-i szer\-ke\-ze\-te.

A vi\-l\'a\-ge\-gye\-tem\-ben ta\-l\'al\-ha\-t\'o k\"onny\H u e\-le\-mek re\-la\-t\'{\i}v gya\-ko\-ri\-s\'a\-ga
az e\-l\H o\-z\H o\-ek\-ben t\'ar\-gyalt fo\-lya\-ma\-tok\-b\'ol vissza\-ma\-radt ba\-ri\-o\-nok \'es
fo\-to\-nok a\-r\'a\-ny\'at a
\beq
3 \times 10^{-10} < \eta \equiv \frac{n_B}{n_\gamma} < 10^{-9}
\enq
ha\-t\'a\-rok k\"o\-z\"ott \'al\-la\-p\'{\i}t\-ja meg \cite{ol}.

Eh\-hez a sz\'am\-hoz k\'{\i}\-v\'a\-nunk te\-h\'at egy r\'e\-szecs\-ke\-fi\-zi\-ka\-i mo\-dellt
al\-kot\-ni. Nyil\-v\'an\-va\-l\'o\-an o\-lyan mo\-dell\-re van sz\"uk\-s\'e\-g\"unk, mely
ki\-e\-l\'e\-g\'{\i}\-ti a h\'a\-rom Sza\-ha\-rov-fel\-t\'e\-telt \cite{sakharov}:
\begin{itemize}
\item \emph{Ba\-ri\-on\-sz\'am s\'er\-t\'es}
\item \emph{C \'es CP s\'er\-t\'es}
\item \emph{El\-t\'e\-r\'es a h\H o\-m\'er\-s\'ek\-le\-ti e\-gyen\-s\'uly\-t\'ol}
\end{itemize}
Mi\-vel e\-gyet\-len sz\'a\-mon \'all vagy bu\-kik az e\-gyes mo\-del\-lek
\'e\-let\-k\'e\-pes\-s\'e\-ge, a fen\-ti sz\'am\-mal va\-l\'o egy\-be\-csen\-g\'es
kri\-t\'e\-ri\-u\-m\'a\-val gya\-kor\-la\-ti\-lag csak \emph{kiz\'arhatunk} bi\-zo\-nyos
mo\-del\-le\-ket, \emph{igazolni} sem\-mi e\-set\-re sem i\-ga\-zol\-hat\-juk \H o\-ket.

\section{Ba\-ri\-on\-sz\'am\-s\'er\-t\'es a stan\-dard mo\-dell\-ben}
\fancyhead[CO]{\hst{\thesection \quad Ba\-ri\-on\-sz\'am\-s\'er\-t\'es a stan\-dard
mo\-dell\-ben}}
A stan\-dard mo\-dell\-ben nem raj\-zol\-ha\-t\'o fel e\-gyet\-len ba\-ri\-on\-sz\'am\-s\'er\-t\H o
Feyn\-man-gr\'af sem. E\-mi\-att so\-k\'a\-ig \'ugy v\'el\-t\'ek \cite{nanopoulos}, hogy
a ba\-ri\-o\-ge\-n\'e\-zis fel\-t\'e\-te\-le\-it csak egy k\'{\i}\-s\'er\-le\-ti\-leg m\'eg hossz\'u
i\-de\-ig nem tesz\-tel\-he\-t\H o nagy e\-gye\-s\'{\i}\-tett el\-m\'e\-let ke\-re\-te\-in be\-l\"ul le\-het
meg\-te\-rem\-te\-ni. A nagy e\-gye\-s\'{\i}\-tett el\-m\'e\-le\-tek\-re a\-la\-po\-zott k\"oz\-vet\-len
ba\-ri\-o\-ge\-n\'e\-zis mo\-del\-lek mel\-lett lep\-to\-ge\-n\'e\-zis mo\-del\-lek is
l\'e\-tez\-nek \cite{fuk-yan}, me\-lyek\-ben a ne\-h\'ez ste\-ril ne\-ut\-r\'{\i}\-n\'ok
bom\-l\'a\-sa ter\-mi\-kus e\-gyen\-s\'uly hi\-\'a\-ny\'a\-ban lep\-to\-na\-szim\-met\-ri\-\'a\-ra ve\-zet;
ez szfa\-le\-ro\-n\'at\-me\-ne\-tek r\'e\-v\'en ba\-ri\-o\-na\-szim\-met\-ri\-\'at hoz l\'et\-re.
(To\-v\'ab\-bi al\-ter\-na\-t\'{\i}v ba\-ri\-o\-ge\-n\'e\-zis mo\-del\-le\-ket il\-le\-t\H o\-en l\'asd a
\cite{clin00} cikk hi\-vat\-ko\-z\'a\-sa\-it.)

1976-ban a\-zon\-ban ki\-de\-r\"ult, hogy a stan\-dard mo\-del\-le\-ben is
l\'e\-tez\-nek o\-lyan, \emph{nemperturbat\'{\i}v} fo\-lya\-ma\-tok, me\-lyek a
ba\-ri\-on- \'es lep\-ton\-sz\'a\-mot egy\-szer\-re v\'al\-toz\-tat\-j\'ak \cite{thoft}. A
vi\-l\'a\-ge\-gye\-tem ba\-ri\-on-a\-szim\-met\-ri\-\'a\-j\'a\-nak egy k\'{\i}\-s\'er\-le\-ti\-leg szin\-te
tel\-je\-sen i\-ga\-zolt fi\-zi\-ka\-i mo\-dell ke\-re\-t\'e\-ben t\"or\-t\'e\-n\H o
ma\-gya\-r\'a\-za\-t\'a\-nak le\-he\-t\H o\-s\'e\-g\'et fog\-juk az a\-l\'ab\-bi\-ak\-ban meg\-vizs\-g\'al\-ni
\cite{rub1, rub2}.

A C \'es CP-s\'er\-t\'es kva\-li\-ta\-t\'{\i}\-ve je\-len van a stan\-dard mo\-dell\-ben, \'{\i}gy
fi\-gyel\-m\"un\-ket a m\'a\-sik k\'et Sza\-ha\-rov-fel\-t\'e\-tel\-re for\-d\'{\i}t\-juk.
A ba\-ri\-on\-sz\'am-s\'er\-t\H o fo\-lya\-ma\-tok fi\-zi\-ka\-i a\-lap\-ja\-it eb\-ben a sza\-kasz\-ban
te\-kin\-tem \'at, m\'{\i}g a ter\-mo\-di\-na\-mi\-ka\-i e\-gyen\-s\'uly\-t\'ol va\-l\'o el\-t\'e\-r\'est,
az e\-lekt\-ro\-gyen\-ge f\'a\-zi\-s\'at\-me\-net fo\-lya\-ma\-t\'at a k\"o\-vet\-ke\-z\H o sza\-kasz\-ban
kez\-dem vizs\-g\'al\-ni. \bigskip

A stan\-dard mo\-dell\-be\-li ba\-ri\-on\-sz\'am\-s\'er\-t\H o fo\-lya\-ma\-to\-k\'ert min\-de\-nek
e\-l\H ott az al\-kal\-mas to\-po\-l\'o\-gi\-a \'es a ki\-ra\-li\-t\'as a fe\-le\-l\H os. Hogy
mi\-k\'ent, azt vizs\-g\'al\-juk meg egy egy\-sze\-r\H ubb mo\-del\-len, az 1+1
di\-men\-zi\-\'os
\beq
{\cal{L}} = - \frac14 F^2 + \bar \psi i \gamma^\mu {\cal{D}}_\mu \psi - h
\phi \bar \psi \psi + |{\cal{D}} \varphi|^2 - V(\varphi)
\enq
Lag\-ran\-ge-f\"ugg\-v\'ennyel le\-\'{\i}rt a\-be\-li Higgs-mo\-del\-len, mely\-ben az
U(1)-szim\-met\-ri\-\'at mu\-ta\-t\'o szim\-met\-ri\-a\-s\'er\-t\H o po\-ten\-ci\-\'al mi\-ni\-mu\-ma egy
$S^1$ so\-ka\-s\'ag \cite{turok}.

Az egy\-sze\-r\H u\-s\'eg ked\-v\'e\-\'ert t\'e\-te\-lez\-z\"uk to\-v\'ab\-b\'a fel, hogy e\-gyet\-len
t\'er\-di\-men\-zi\-\'on\-kat egy $L$ su\-ga\-r\'u k\"or men\-t\'en kom\-pak\-ti\-fi\-k\'al\-tuk oly
m\'o\-don, hogy a Higgs-t\'er\-re \'es a m\'er\-t\'ek\-te\-rek\-re pe\-ri\-o\-di\-kus
ha\-t\'ar\-fel\-t\'e\-te\-le\-ket szab\-tunk ki. Ek\-kor k\'et csa\-va\-ro\-d\'a\-si sz\'am
de\-fi\-ni\-\'al\-ha\-t\'o:
\beq
{\cal{N}}_H = \frac{1}{2 \pi} \int dx \, \6_x\alpha, \qquad
{\cal{N}}_{CS} = \frac{g}{2 \pi} \int dx \, A_x,
\enq
a\-hol $\alpha$ a Higgs-t\'er f\'a\-zi\-sa ($\varphi = \phi e^{i \alpha}$). A
Higgs-t\'er csa\-va\-ro\-d\'a\-si sz\'a\-ma te\-h\'at azt mu\-tat\-ja meg, h\'any\-szor
for\-dul k\"or\-be a Higgs-t\'er, mi\-k\"oz\-ben v\'e\-gig\-me\-gy\"unk az $L$-su\-ga\-r\'u
gy\H u\-r\H un. Nyil\-v\'an\-va\-l\'o a\-zon\-ban, hogy ha a Higgs-t\'er va\-la\-me\-lyik
pont\-ban el\-t\H u\-nik, ak\-kor az ${\cal{N}}_H$ csa\-va\-ro\-d\'a\-si sz\'am nem
meg\-ha\-t\'a\-ro\-zott. \'Igy a Higgs- \'es m\'er\-t\'ek\-te\-rek si\-ma i\-d\H o\-fej\-l\H o\-d\'e\-se
e\-se\-t\'en is le\-he\-t\H o\-s\'e\-g\"unk van az ${\cal{N}}_H$ csa\-va\-ro\-d\'a\-si sz\'am
e\-g\'esz \'er\-t\'e\-kek\-kel va\-l\'o v\'al\-toz\-ta\-t\'a\-s\'a\-ra.

Ez\-zel szem\-ben a m\'er\-t\'ek\-te\-rek csa\-va\-ro\-d\'a\-si sz\'a\-ma, az ${\cal{N}}_{CS}$
Chern--Si\-mons-sz\'am min\-den\-k\'ep\-pen j\'ol de\-fi\-ni\-\'alt. A\-mennyi\-ben tisz\-ta
m\'er\-t\'ek\-te\-rek\-kel dol\-go\-zunk \'es az $i g A_x = \6_x U(x) U^{-1}(x)$
k\'ep\-let\-ben sze\-rep\-l\H o $U(x)$ U(1)-be\-li e\-lem $x$-nek e\-gy\'er\-t\'e\-k\H u
f\"ugg\-v\'e\-nye, pl.\ $U(x) = e^{i \theta(x)}$, \'ugy a $\theta(x + L) =
\theta(x) + 2 \pi N$ ($N$ e\-g\'esz) pe\-ri\-o\-di\-kus ha\-t\'ar\-fel\-t\'e\-tel
k\"o\-vet\-kez\-t\'e\-ben ${\cal{N}}_{CS}=N$.

Az el\-m\'e\-let klasszi\-kus v\'a\-ku\-u\-ma\-it e\-gyet\-len sz\'am\-mal in\-de\-xel\-het\-j\"uk:
h\'any\-szor csa\-va\-ro\-dik k\"o\-r\"ul a Higgs-t\'er a po\-ten\-ci\-\'al mi\-ni\-mu\-m\'a\-ban. A
v\'a\-ku\-u\-m\'al\-la\-po\-tok\-ban a Higgs-t\'er gra\-di\-en\-se el kell t\H un\-j\"on, te\-h\'at
$i g A_x = - \6_x \alpha$, a\-hon\-nan
\beq
{\cal{N}}_{CS}= - {\cal{N}}_{H}= N = \mathrm{eg\acute{e}sz}
\enq
k\"o\-vet\-ke\-zik. A sz\'o\-ban\-for\-g\'o \'al\-la\-po\-tot az el\-m\'e\-let $N.$
v\'a\-ku\-u\-m\'al\-la\-po\-t\'a\-nak ne\-vez\-z\"uk.

A Higgs-t\'er csa\-va\-ro\-d\'a\-si sz\'a\-m\'at v\'al\-toz\-ta\-t\'o fo\-lya\-ma\-tok r\'e\-v\'en
az e\-gyik klasszi\-kus v\'a\-ku\-u\-m\'al\-la\-pot\-b\'ol a m\'a\-sik\-ba jut\-ha\-tunk. Ek\-kor a
Chern--Si\-mon-sz\'am meg\-v\'al\-to\-z\'a\-s\'at az ${\cal{E}}$ e\-lekt\-ro\-mos t\'er
ha\-t\'a\-roz\-za meg:
\beq
\6_t {\cal{N}}_{CS} \propto \int dx \, {\cal{E}}
\enq

K\"onnyen ki\-sz\'a\-m\'{\i}t\-ha\-t\'o az az e\-ner\-gi\-a, a\-mely $N$ egy\-s\'eg\-nyi
meg\-v\'al\-to\-z\'a\-s\'a\-hoz sz\"uk\-s\'e\-ges: a ska\-l\'ar Higgs-e\-gyen\-le\-tek
pa\-ri\-t\'a\-sin\-va\-ri\-\'an\-sak, \'{\i}gy az e\-ner\-gi\-a\-f\"ugg\-v\'eny az
${\cal{N}}_{CS} \rightarrow {\cal{N}}_{CS}+1$ transz\-for\-m\'a\-ci\-\'o
mel\-lett az ${\cal{N}}_{CS} \rightarrow -{\cal{N}}_{CS}$
transz\-for\-m\'a\-ci\-\'o\-ra is in\-va\-ri\-\'ans.
Az e\-ner\-gi\-a\-f\"ugg\-v\'eny (e\-gyik) ma\-xi\-mu\-ma te\-h\'at ${\cal{N}}_{CS}=
\frac12$-ben van. Az e\-lekt\-ro\-gyen\-ge el\-m\'e\-let\-ben ez a szfa\-le\-ron-kor\-l\'at
10 TeV k\"o\-r\"u\-li\-nek a\-d\'o\-dik \cite{turok}.

K\"o\-vet\-ke\-z\H o l\'e\-p\'es\-k\'ent ve\-gy\"unk fi\-gye\-lem\-be fer\-mi\-o\-no\-kat is. Az
egy\-sze\-r\H u\-s\'eg ked\-v\'e\-\'ert t\"o\-me\-g\"uk le\-gyen 0, m\'er\-t\'ek\-csa\-to\-l\'a\-su\-kat
jel\-le\-mez\-ze a $g_F$ csa\-to\-l\'a\-si \'al\-lan\-d\'o, \'{\i}gy a Lag\-ran\-ge-s\H u\-r\H u\-s\'eg\-ben
fel\-l\'e\-p\H o de\-ri\-v\'alt ${\cal{D}}_\mu = \6_\mu + i g_F A_\mu$. Ek\-kor a
$\psi$ fer\-mi\-on\-te\-rek $\psi' = e^{i \theta \gamma_5} \psi$
transz\-for\-m\'a\-ci\-\'o\-j\'a\-nak glo\-b\'a\-lis a\-xi\-\'al\-szim\-met\-ri\-\'a\-ja csak
l\'at\-sz\'o\-la\-gos, u\-gya\-nis a hoz\-z\'a tar\-to\-z\'o $j_5^\mu = \bar \psi
\gamma^\mu \gamma^5 \psi$ No\-et\-her-\'a\-ra\-mot a\-no\-m\'a\-li\-a-tag s\'er\-ti:
\beq
\6_\mu j^\mu_5 = \frac{g}{2 \pi} \epsilon^{\mu \nu} F_{\mu \nu},
\label{anomalia}
\enq
\'es en\-nek k\"o\-vet\-kez\-t\'e\-ben a bal- il\-let\-ve jobb\-ke\-zes r\'e\-szecs\-k\'ek
sz\'a\-ma nem ma\-rad meg.

Hogy a (\ref{anomalia}) e\-gyen\-let k\"o\-vet\-kez\-m\'e\-nye\-it k\"onnyen
\'at\-l\'at\-has\-suk, vizs\-g\'al\-juk az 1+1 di\-men\-zi\-\'os Di\-rac-e\-gyen\-le\-tet az $A_0
= 0$ m\'er\-t\'ek\-v\'a\-lasz\-t\'as mel\-lett:
\beq
i \6_0 \psi = - i \gamma^0 \gamma^1 {\cal{D}}_x \psi,
\enq
mely\-ben $\psi$ egy k\'et\-kom\-po\-nen\-s\H u Di\-rac-spi\-nor, to\-v\'ab\-b\'a
$\gamma^5 \psi_{L,R} = \pm \psi_{L,R}$.

Kelt\-s\"unk most e\-gyen\-le\-tes e\-lekt\-ro\-mos t\'e\-re\-r\H os\-s\'e\-get 1 di\-men\-zi\-\'os
(gy\H u\-r\H u) men\-ti vi\-l\'a\-gunk\-ban -- ez p\'el\-d\'a\-ul po\-zi\-t\'{\i}v \'es ne\-ga\-t\'{\i}v
t\"ol\-t\'e\-sek p\'ar\-kel\-t\'e\-se, sz\'et\-v\'a\-lasz\-t\'a\-sa, majd sz\'et\-su\-g\'ar\-z\'a\-sa
r\'e\-v\'en me\-gold\-ha\-t\'o; a vizs\-g\'alt fi\-zi\-ka\-i prob\-l\'e\-m\'a\-ban ezt a
k\"u\-l\"on\-b\"o\-z\H o $N$-v\'a\-ku\-u\-mok k\"oz\-ti \'at\-me\-ne\-tek\-hez tar\-to\-z\'o
Higgs-t\'er-\'a\-ram fog\-ja biz\-to\-s\'{\i}\-ta\-ni.

Mi\-vel a $\6_t A_x$ e\-lekt\-ro\-mos me\-z\H o ho\-mo\-g\'en, \'{\i}gy $A_x$ kons\-tans.
A Di\-rac-e\-gyen\-let me\-gol\-d\'a\-s\'a\-hoz te\-kint\-s\"uk a $\psi \sim e^{i
\left( Et - px \right)}$ pr\'o\-ba\-f\"ugg\-v\'enyt, mely\-ben $E$ \'es $p$
kons\-tans. A jobb\-ra (R) il\-let\-ve bal\-ra (L) moz\-g\'o fer\-mi\-o\-nok\-ra a\-d\'o\-d\'o
disz\-per\-zi\-\'os re\-l\'a\-ci\-\'o ek\-kor
\beq
E_{R,L} = \pm \left(p + g_F A_x \right).
\enq

Mi\-lyen ha\-t\'a\-sa van az e\-lekt\-ro\-mos t\'er n\"o\-ve\-l\'e\-s\'e\-nek? Az e\-gyes
\'al\-la\-po\-tok ka\-no\-ni\-kus im\-pul\-zu\-s\'at a pe\-ri\-o\-di\-kus ha\-t\'ar\-fel\-t\'e\-tel szab\-ja
meg, $p = 2 \pi n /L$, a\-hol $L$ a vi\-l\'ag su\-ga\-ra, $n$ pe\-dig egy e\-g\'esz.
Ez a mennyi\-s\'eg kvan\-t\'alt, \'{\i}gy $A_x$ n\"o\-ve\-l\'e\-se\-kor nem v\'al\-to\-zik --
az \'al\-la\-pot\-hoz tar\-to\-z\'o e\-ner\-gi\-a a\-zon\-ban v\'al\-to\-zik. A
Chern--Si\-mons-sz\'am egy\-s\'eg\-nyi meg\-v\'al\-to\-z\'a\-sa\-kor az e\-ner\-gi\-a\-szin\-tek
el\-to\-l\'o\-d\'a\-sa $\delta E_{L,R} = \mp 2 \pi / L$, te\-h\'at a bal\-ra ha\-la\-d\'o
\'al\-la\-po\-tok egy l\'ep\-cs\H o\-vel lej\-jebb, a jobb\-ra ha\-la\-d\'ok egy l\'ep\-cs\H o\-vel
fel\-jebb ke\-r\"ul\-nek. Ha te\-h\'at a Di\-rac-ten\-ger fel\-sz\'{\i}\-n\'e\-r\H ol in\-du\-lunk --
te\-h\'at e\-re\-de\-ti\-leg az \"osszes ne\-ga\-t\'{\i}v e\-ner\-gi\-\'a\-j\'u \'al\-la\-pot be van
t\"olt\-ve, de e\-gyet\-len po\-zi\-t\'{\i}v e\-ner\-gi\-\'a\-j\'u sem --, ak\-kor az e\-lekt\-ro\-mos
t\'er nagy\-s\'a\-g\'at \'ugy v\'al\-toz\-tat\-va, hogy a Chern--Si\-mins-sz\'am
egy\-s\'eg\-nyi\-vel n\H o\-j\"on, egy jobb\-ra ha\-la\-d\'o r\'e\-szecs\-ke \'es egy bal\-ra
ha\-la\-d\'o lyuk fog ke\-let\-kez\-ni. A jobb\-ra il\-let\-ve bal\-ra ha\-la\-d\'o
r\'e\-szecs\-k\'ek sz\'a\-m\'a\-nak k\"u\-l\"onb\-s\'e\-ge te\-h\'at
\beq
\Delta (N_R - N_L) = 2 \Delta {\cal{N}}_{CS}
\enq
sze\-rint v\'al\-to\-zik.

A fen\-ti k\'ep ma\-te\-ma\-ti\-ka\-i\-lag kis\-s\'e in\-go\-v\'a\-nyos a\-la\-pon \'all: a
v\'a\-ku\-um\-b\'ol kel\-tett, gya\-kor\-la\-ti\-lag 0 im\-pul\-zu\-s\'u r\'e\-szecs\-k\'e\-ket egy
o\-lyan Di\-rac-ten\-ger\-b\H ol h\'uz\-za e\-l\H o -- il\-let\-ve az el\-t\"un\-te\-ten\-d\H o\-ket
egy o\-lyan Di\-rac-ten\-ger\-be rej\-ti -- mely v\'eg\-te\-len m\'ely; ezt a
v\'eg\-te\-len m\'ely\-s\'e\-get a\-zon\-ban a re\-gu\-la\-ri\-z\'a\-l\'as\-n\'al el\-tom\-p\'{\i}t\-juk. A
na\-iv No\-et\-her-t\'e\-tel\-re a\-la\-po\-zott meg\-ma\-ra\-d\'a\-si t\'e\-telt s\'er\-t\H o
a\-no\-m\'a\-li\-\'ak te\-h\'at i\-gen tr\"uk\-k\"os m\'o\-don hoz\-z\'ak kap\-cso\-lat\-ba a na\-gyon
nagy \'es a na\-gyon kis e\-ner\-gi\-\'a\-kat.

Az $\psi$ fer\-mi\-on\-t\'er $L$ bal, il\-let\-ve $R$ jobb kom\-po\-nen\-s\'e\-nek $g_L$
il\-let\-ve $g_R$ m\'er\-t\'ek\-t\"ol\-t\'est ad\-va a m\'er\-t\'ek\-t\'er \'a\-ra\-ma
\beq
j^\mu = g_L \bar \psi_L \gamma^\mu \psi_L + g_R \bar \psi_R
\gamma^\mu \psi_R,
\enq
mely\-ben a $g_L = -g_R$ v\'a\-lasz\-t\'as a t\"ol\-t\'es\-meg\-ma\-ra\-d\'ast biz\-to\-s\'{\i}t\-ja,
de a r\'e\-szecs\-k\'ek p\'ar\-kel\-t\'e\-s\'et is le\-he\-t\H o\-v\'e te\-szi.

A 3+1 di\-men\-zi\-\'os e\-lekt\-ro\-gyen\-ge el\-m\'e\-let\-ben ha\-son\-l\'o a\-no\-m\'a\-li\-a\-ta\-gok
se\-g\'{\i}t\-s\'e\-g\'e\-vel ke\-r\"ul\-he\-t\H o meg a r\'e\-szecs\-ke\-sz\'am-meg\-ma\-ra\-d\'as. Az ott
fel\-l\'e\-p\H o
\beq
j^\mu_B = \frac13 \sum_{g, c} g_R \bar q_R \gamma^\mu q_R + g_L \bar
q_L \gamma^\mu q_L
\enq
ba\-ri\-on-\'a\-ram\-ban ($g$ a ge\-ne\-r\'a\-ci\-\'o-in\-dex, $c$ a sz\'{\i}n-in\-dex) csak a
bal ke\-zes kvar\-kok csa\-to\-l\'od\-nak az SU(2)-m\'er\-t\'ek\-te\-rek\-hez, \'{\i}gy a
ba\-ri\-on\-sz\'am (\'es ha\-son\-l\'o\-k\'ep\-pen a lep\-ton\-sz\'am) s\'e\-r\"ul:
\beq
\6_\mu j^\mu_B = \6_\mu j^\mu_L = - \frac{g^2}{32 \pi^2} N_g
F^{a*}_{\mu \nu} F^{a, \mu \nu}, \label{barsz}
\enq
a\-hol $N_g$ a ge\-ne\-r\'a\-ci\-\'ok sz\'a\-ma, $F^{a}_{\mu \nu}$ pe\-dig az SU(2)
t\'e\-re\-r\H os\-s\'eg ten\-zor. \bigskip

A fen\-ti\-ek\-ben in\-di\-k\'a\-ci\-\'ot ad\-tunk ar\-ra, mi\-k\'ent s\'e\-r\"ul a ba\-ri\-on\-sz\'am a
stan\-dard mo\-dell\-ben; a (\ref{barsz}) k\'ep\-let a\-lap\-j\'an azt is l\'at\-juk,
hogy a ba\-ri\-on\-sz\'am mel\-lett a lep\-ton\-sz\'am is s\'e\-r\"ul, a\-zon\-ban a $B-L$
kvan\-tum\-sz\'am meg\-ma\-rad.
\section{Az e\-lekt\-ro\-gyen\-ge f\'a\-zi\-s\'at\-me\-net}
\fancyhead[CO]{\hst{\thesection \quad Az e\-lekt\-ro\-gyen\-ge
f\'a\-zis\'at\-me\-net}}
Eb\-ben a sza\-kasz\-ban azt a fi\-zi\-ka\-i fo\-lya\-ma\-tot vizs\-g\'a\-lom meg, a\-mely a
stan\-dard mo\-dell ke\-re\-t\'e\-ben le\-he\-t\H o\-s\'e\-get ad a har\-ma\-dik
Sza\-ha\-rov-fel\-t\'e\-tel ki\-e\-l\'e\-g\'{\i}\-t\'e\-s\'e\-re.

A kis e\-ner\-gi\-\'a\-kon s\'e\-r\"u\-l\H o e\-lekt\-ro\-gyen\-ge szim\-met\-ri\-a ma\-gas
h\H o\-m\'er\-s\'ek\-le\-ten hely\-re\-\'all \cite{rub2}. Az \H os\-rob\-ba\-n\'as u\-t\'an
t\'a\-gu\-l\'o \'es h\H u\-l\H o vi\-l\'a\-gye\-tem \'{\i}gy \'a\-te\-sett az e\-lekt\-ro\-gyen\-ge
f\'a\-zi\-s\'at\-me\-ne\-ten, el\-k\'ep\-zel\-he\-t\H o te\-h\'at, hogy a f\'a\-zi\-s\'at\-me\-ne\-tek\-n\'el
meg\-szo\-kott m\'o\-don a ba\-ri\-o\-ge\-n\'e\-zis\-hez sz\"uk\-s\'e\-ges e\-gyen\-s\'u\-lyi
\'al\-la\-pot\-t\'ol va\-l\'o el\-t\'e\-r\'es is meg\-va\-l\'o\-sult.

Az e\-lekt\-ro\-gyen\-ge f\'a\-zi\-s\'at\-me\-net kvan\-ti\-ta\-t\'{\i}v le\-\'{\i}\-r\'a\-s\'a\-hoz az
ef\-fek\-t\'{\i}v po\-ten\-ci\-\'alt hasz\-n\'al\-juk.

\bef
\bc
\epsfig{file=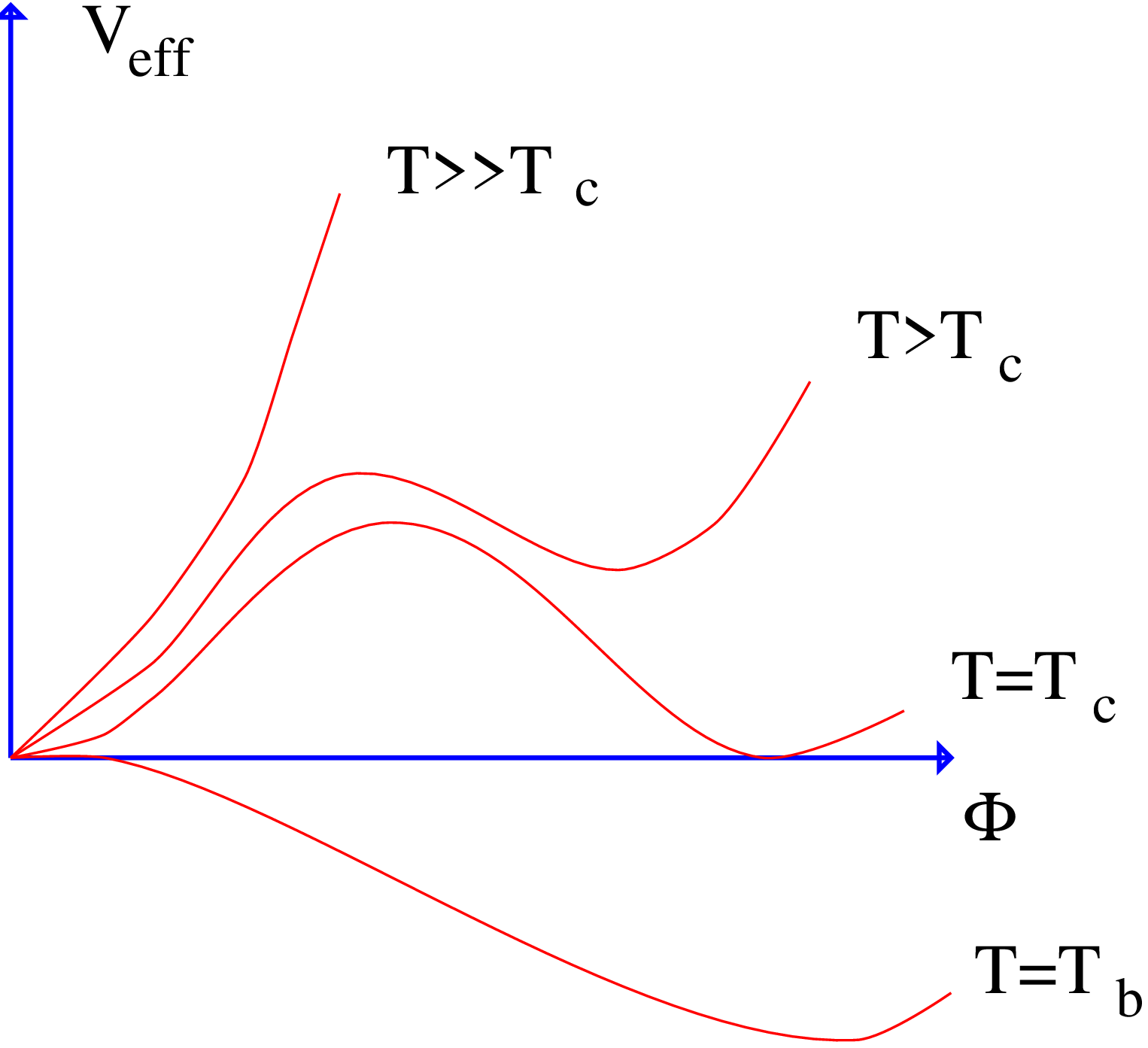,angle=270,width=8cm} \label{veff}
\vspace{1.8cm}
\caption{Az e\-lekt\-ro\-gyen\-ge f\'a\-zi\-s\'at\-me\-net le\-\'{\i}\-r\'a\-s\'a\-hoz hasz\-n\'alt
ef\-fek\-t\'{\i}v po\-ten\-ci\-\'al. A g\"or\-b\'ek k\"u\-l\"on\-b\"o\-z\H o h\H o\-m\'er\-s\'ek\-le\-tek\-nek
fe\-lel\-nek meg.}
\ec
\enf

Ma\-gas h\H o\-m\'er\-s\'ek\-le\-ten az ef\-fek\-t\'{\i}v po\-ten\-ci\-\'al szim\-met\-ri\-kus,
mi\-ni\-mum\-he\-lye a $\Phi = 0$ pont\-ban van. Z\'e\-rus h\H o\-m\'er\-s\'ek\-le\-ten a
po\-ten\-ci\-\'al\-nak $\Phi = 0$ lo\-k\'a\-lis ma\-xi\-mu\-ma, a mi\-ni\-mum va\-la\-mely
szim\-met\-ri\-a\-s\'er\-t\H o $\Phi$ \'er\-t\'ek mel\-lett va\-l\'o\-sul meg. En\-n\'el
va\-la\-mi\-vel ma\-ga\-sabb h\H o\-m\'er\-s\'ek\-len mind az o\-ri\-go, mind egy
szim\-met\-ri\-a\-s\'er\-t\H o $\Phi$ \'er\-t\'ek mel\-lett a po\-ten\-ci\-\'al\-nak mi\-ni\-mu\-ma van.
Ez a k\'et  mi\-ni\-mum egy j\'ol meg\-ha\-t\'a\-ro\-zott h\H o\-m\'er\-s\'ek\-le\-t\'er\-t\'ek\-n\'el
egy\-be\-e\-sik: ezt a pon\-tot ne\-vez\-z\"uk f\'a\-zi\-s\'at\-me\-ne\-ti pont\-nak. A
f\'a\-zi\-s\'at\-me\-ne\-tet le\-\'{\i}\-r\'o rend\-pa\-ra\-m\'e\-ter a $\Phi$ Higgs-t\'er
v\'a\-ku\-um v\'ar\-ha\-t\'o \'er\-t\'e\-ke.

A ba\-ri\-o\-ge\-n\'e\-zis fo\-lya\-ma\-ta ek\-kor a k\"o\-vet\-ke\-z\H o\-k\'ep\-pen j\'at\-sz\'od\-hat le.
A szim\-met\-ri\-kus f\'a\-zis\-ban le\-v\H o rend\-szer h\H u\-l\'e\-se\-kor ki\-a\-la\-kul egy
m\'a\-sik, nemt\-ri\-vi\-\'a\-lis mi\-ni\-mum; a to\-v\'ab\-bi h\H u\-l\'es fo\-lya\-m\'an az \'uj
mi\-ni\-mum\-ba va\-l\'o \'at\-ju\-t\'as va\-l\'o\-sz\'{\i}\-n\H u\-s\'e\-ge egy\-re na\-gyobb lesz. Ha a
vi\-l\'a\-ge\-gye\-tem va\-la\-mely pont\-j\'a\-ban az \'uj, s\'er\-tett f\'a\-zis va\-l\'o\-sul
meg, egy s\'er\-tett f\'a\-zi\-s\'u bu\-bo\-r\'ek a\-la\-kul ki, mely t\'a\-gul\-ni kezd. A
fal k\"o\-ze\-l\'e\-ben, m\'eg s\'er\-tet\-len f\'a\-zis\-ban le\-v\H o r\'e\-szecs\-k\'ek
k\"ol\-cs\"on\-hat\-nak a fal v\'al\-to\-z\'o pro\-fi\-l\'u Higgs-te\-r\'e\-vel; a
k\"ol\-cs\"on\-ha\-t\'as CP-s\'er\-t\H o jel\-le\-ge mi\-att a k\"u\-l\"on\-b\"o\-z\H o ki\-ra\-li\-t\'a\-s\'u
r\'e\-szecs\-k\'ek fa\-lon va\-l\'o \'at\-ha\-la\-d\'a\-sa il\-let\-ve ar\-r\'ol t\"or\-t\'e\-n\H o
vissza\-ve\-r\H o\-d\'e\-se nem u\-gya\-no\-lyan va\-l\'o\-sz\'{\i}\-n\H u\-s\'e\-g\H u lesz. A m\'er\-t\'ek-,
Yu\-ka\-wa- \'es e\-r\H os szfa\-le\-ron \'at\-me\-ne\-tek ha\-t\'a\-s\'a\-ra a CP-s\'er\-t\H o
fo\-lya\-ma\-tok\-ban ke\-let\-ke\-z\H o bal\-ke\-zes kvar\-kok s\H u\-r\H u\-s\'e\-ge u\-gya\-no\-lyan
m\'er\-t\'ek\-ben ha\-lad\-ja meg a bal\-ke\-zes an\-tik\-var\-ko\-k\'et, mint a jobb\-ke\-zes
an\-tik\-var\-kok s\H u\-r\H u\-s\'e\-ge a jobb\-ke\-zes kvar\-ko\-k\'et -- eb\-ben a pil\-la\-nat\-ban
te\-h\'at az \"ossz\-ba\-ri\-on\-sz\'am m\'eg nul\-la. A\-zon\-ban a CP-a\-szim\-met\-ri\-a
gyen\-ge szfa\-le\-ron fo\-lya\-ma\-tok r\'e\-v\'en ba\-ri\-on--an\-ti\-ba\-ri\-on
a\-szim\-met\-ri\-\'a\-v\'a a\-la\-kul\-hat \'at \cite{brhlik, clin-gdm99}. En\-nek egy
r\'e\-sze be\-dif\-fun\-d\'al a s\'er\-tett f\'a\-zi\-s\'u bu\-bo\-r\'ek bel\-se\-j\'e\-be, a\-hol az
a\-no\-m\'a\-lis fo\-lya\-ma\-tok ex\-po\-nen\-ci\-\'a\-li\-san el van\-nak nyom\-va \cite{thoft}.

A ba\-ri\-on\-sz\'am-s\'er\-t\'es fel\-t\'e\-te\-le te\-h\'at az, hogy egy e\-set\-le\-ge\-sen
ke\-let\-ke\-z\H o $B+L$ a\-szim\-met\-ri\-\'a\-nak ne le\-gyen i\-de\-je ki\-mo\-s\'od\-ni, te\-h\'at
a szim\-met\-ri\-a\-s\'er\-tett f\'a\-zis\-ban a k\"u\-l\"on\-b\"o\-z\H o v\'a\-ku\-u\-m\'al\-la\-po\-tok
k\"oz\-ti \'at\-me\-net el le\-gyen nyom\-va. Mi\-vel $T<T_C$ e\-se\-t\'en az
i\-d\H o\-egy\-s\'eg a\-lat\-ti \'at\-me\-ne\-tek sz\'a\-ma
\beq
\Gamma \propto \exp \left( - \frac{E_{szf}(T)}{T} \right),
\enq
a\-hol a szfa\-le\-ron sza\-ba\-de\-ner\-gi\-a-kor\-l\'at\-ja
\beq
E_{szf}(T) = \frac{2 m_W(T)}{\alpha_W} B\left(\frac{m_H}{m_W}\right).
\enq
$T>T_C$ e\-se\-t\'en a r\'esz\-le\-tes sz\'a\-m\'{\i}\-t\'a\-sok ta\-n\'u\-s\'a\-ga sze\-rint a
szfa\-le\-ron-\'at\-me\-ne\-tek gya\-ko\-ri\-s\'a\-ga $\Gamma \propto \alpha^5 T^4 \times
\left( \log (m/g^2 T) +\delta \right)$, l\'asd \cite{asy97, bod99, gdm99,
smit99}. An\-nak fel\-t\'e\-te\-le, hogy a f\'a\-zi\-s\'at\-me\-net\-re jel\-lem\-z\H o $T_C$
h\H o\-m\'er\-s\'ek\-let a\-latt a szfa\-le\-ron-fo\-lya\-ma\-tok hir\-te\-len be\-fagy\-ja\-nak
\cite{rub2}
\beq
\frac{E_{szf}(T_C)}{T_C} > 45,
\enq
mely e\-set\-ben fen\-n\'all az egy\-sze\-r\H u
\beq
{\langle \Phi \rangle_{T_C} > T_C} \label{vtpert}
\enq
e\-gyen\-l\H ot\-len\-s\'eg. Ez azt je\-len\-ti, hogy a ba\-ri\-o\-ge\-n\'e\-zis ma\-gya\-r\'a\-za\-ta
e\-r\H os el\-s\H o\-ren\-d\H u f\'a\-zi\-s\'at\-me\-ne\-tet i\-g\'e\-nyel.

M\'as\-r\'eszt a CP-s\'er\-t\H o fo\-lya\-ma\-tok
\beq
\delta_{ms} = \left[ \sigma(\mrm{in} \to \mrm{out}) -
\sigma(\overline{\mrm{in}} \to \overline{\mrm{out}}) \right] / \left[
\sigma(\mrm{in} \to \mrm{out}) + \sigma(\overline{\mrm{in}} \to
\overline{\mrm{out}}) \right]
\enq
mik\-rosz\-ko\-pi\-kus a\-szim\-met\-ri\-a\-pa\-ra\-m\'e\-te\-r\'e\-vel fe\-l\'{\i}r\-ha\-t\'o az
a\-nyag--an\-ti\-a\-nyag sz\'et\-su\-g\'ar\-z\'as u\-t\'an meg\-ma\-ra\-d\'o ba\-ri\-on--fo\-ton
h\'a\-nya\-dos \cite{shap87}:
\beq
\frac{n_B}{n_\gamma} \sim \frac{\mu_B}{T} \sim \frac{\tau_0}{\tau_U}
\delta_{ms},
\enq
a\-hol $\mu_B$ a ba\-ri\-o\-nok k\'e\-mi\-a\-i po\-ten\-ci\-\'al\-ja, $\tau_0$ a
ba\-ri\-on\-sz\'am\-s\'er\-t\H o fo\-lya\-ma\-tok ka\-rak\-te\-risz\-ti\-kus i\-de\-je, $\tau_U$ pe\-dig
az u\-ni\-ver\-zum t\'a\-gu\-l\'a\-s\'a\-\'e. A stan\-dard mo\-dell Ko\-ba\-yas\-hi--Mas\-ka\-wa
m\'at\-ri\-x\'a\-ban sze\-rep\-l\H o CP s\'er\-t\H o $\delta$ pa\-ra\-m\'e\-ter \'{\i}gy d\"on\-t\H o
sze\-re\-pe\-hez jut. A CP-s\'er\-t\'es m\'er\-t\'e\-k\'e\-nek vizs\-g\'a\-la\-ta a\-zon\-ban egy
m\'a\-sik dok\-to\-ri \'er\-te\-ke\-z\'es t\'e\-m\'a\-ja le\-het\-ne -- je\-len dol\-go\-zat\-ban
az e\-gyen\-s\'uly\-t\'ol va\-l\'o el\-t\'e\-r\'es vizs\-g\'a\-la\-t\'at t\'ar\-gya\-lom.

A stan\-dard mo\-dell\-ben ma az e\-gyet\-len is\-me\-ret\-len pa\-ra\-m\'e\-ter a
Higgs-bo\-zon t\"o\-me\-ge. Az e\-lekt\-ro\-gyen\-ge f\'a\-zi\-s\'at\-me\-net ter\-mo\-di\-na\-mi\-ka\-i
jel\-lem\-z\H o\-i i\-gen \'er\-z\'e\-ke\-nyen f\"ugg\-nek et\-t\H ol a pa\-ra\-m\'e\-ter\-t\H ol. A
k\"o\-vet\-ke\-z\H o sza\-kasz\-ban azt vizs\-g\'a\-lom meg, mi\-lyen
\emph{m\'od\-sze\-rek\-kel }ta\-nul\-m\'a\-nyoz\-ha\-t\'o ez a f\"ug\-g\'es.

\section{Az e\-lekt\-ro\-gyen\-ge f\'a\-zi\-s\'at\-me\-net per\-tur\-ba\-t\'{\i}v \'es
nem\-per\-tur\-ba\-t\'{\i}v ta\-nul\-m\'a\-nyo\-z\'a\-sa \label{pnp}}
\fancyhead[CO]{\hst{\thesection \quad Per\-tur\-bat\'{\i}v \'es
nem\-per\-tur\-bat\'{\i}v megk\"o\-zel\'{\i}t\'es}}
Az e\-lekt\-ro\-gyen\-ge f\'a\-zi\-s\'at\-me\-net vizs\-g\'a\-la\-t\'a\-nak ta\-l\'an
leg\-k\'e\-zen\-fek\-v\H obb m\'od\-sze\-re a per\-tur\-b\'a\-ci\-\'o\-sz\'a\-m\'{\i}\-t\'as. Mind a
stan\-dard mo\-dell\-ben, mind en\-nek mi\-ni\-m\'a\-lis szu\-per\-szim\-met\-ri\-kus
ki\-ter\-jesz\-t\'e\-s\'e\-ben sz\'a\-mos per\-tur\-ba\-t\'{\i}v dol\-go\-zat t\'ar\-gyal\-ja a
f\'a\-zi\-s\'at\-me\-net jel\-lem\-z\H o\-it \cite{arn,fod94/6,fod-heb,fod94/5}.
A kez\-de\-ti l\'at\-sz\'o\-la\-gos si\-ke\-rek u\-t\'an a\-zon\-ban a k\'et\-hu\-rok-ren\-d\H u
sz\'a\-mo\-l\'a\-sok i\-gen ko\-moly prob\-l\'e\-m\'a\-ra h\'{\i}v\-t\'ak fel a
fi\-gyel\-met \cite{fod-heb}: a per\-tur\-b\'a\-ci\-\'o\-sz\'a\-m\'{\i}\-t\'as m\'a\-sod\-rend\-je az
el\-s\H o\-ren\-d\H u e\-red\-m\'e\-nyek\-hez $\ordo{100\%}$-os kor\-rek\-ci\-\'o\-kat ad, a\-mi
a\-lap\-ja\-i\-ban k\'er\-d\H o\-je\-le\-zi meg a per\-tur\-ba\-t\'{\i}v meg\-k\"o\-ze\-l\'{\i}\-t\'es
l\'et\-jo\-go\-sult\-s\'a\-g\'at.

En\-nek a je\-len\-s\'eg\-nek az o\-k\'at r\'esz\-le\-te\-sen a \ref{pertalk} sza\-kasz\-ban
vizs\-g\'a\-lom meg. E\-gye\-l\H o\-re e\-l\'eg annyit hang\-s\'u\-lyoz\-ni, hogy a ma\-gas
h\H o\-m\'er\-s\'ek\-le\-t\H u szim\-met\-ri\-kus f\'a\-zis\-ban a bo\-zo\-ni\-kus szek\-tor\-ban s\'u\-lyos
inf\-ra\-v\"o\-r\"os prob\-l\'e\-m\'ak buk\-kan\-nak fel, me\-lyek\-nek na\-iv
per\-tur\-b\'a\-ci\-\'o\-sz\'a\-m\'{\i}\-t\'as\-sal va\-l\'o ke\-ze\-l\'e\-se meg\-b\'{\i}z\-ha\-tat\-lan
e\-red\-m\'e\-nyek\-re ve\-zet.

A m\'a\-sik le\-he\-t\H o\-s\'eg a tel\-jes stan\-dard mo\-dell n\'egy\-di\-men\-zi\-\'os Mon\-te
Car\-lo szi\-mu\-l\'a\-ci\-\'o\-ja len\-ne. A fer\-mi\-o\-nok -- k\"u\-l\"o\-n\"os\-k\'ep\-pen a
ki\-r\'a\-lis fer\-mi\-o\-nok -- r\'a\-cson t\"or\-t\'e\-n\H o ke\-ze\-l\'e\-se a\-zon\-ban o\-lyan i\-gen
ko\-moly ne\-h\'e\-zs\'e\-ge\-ket vet fel.

\'Igy f\'e\-lig per\-tur\-ba\-t\'{\i}v, f\'e\-lig nem\-per\-tur\-ba\-t\'{\i}v m\'od\-sze\-rek\-hez kell
fo\-lya\-mod\-nunk. Az e\-gyik leg\-be\-vet\-tebb el\-j\'a\-r\'as a di\-men\-zi\-\'os re\-duk\-ci\-\'o
m\'od\-sze\-r\'en a\-la\-pul: egy ma\-gas h\H o\-m\'er\-s\'ek\-le\-ten j\'ol m\H u\-k\"o\-d\H o
h\'a\-rom\-di\-men\-zi\-\'os ef\-fek\-t\'{\i}v el\-m\'e\-let Mon\-te Car\-lo szi\-mu\-l\'a\-ci\-\'o\-ja \'ut\-j\'an
kap\-juk meg az e\-red\-m\'e\-nye\-ket \cite{kaj, phil, kaj2, kar, gur}.

Az a\-l\'ab\-bi\-ak\-ban egy m\'a\-sik m\'od\-szert fo\-gok t\'ar\-gyal\-ni, mely\-ben az
U(1) cso\-por\-tot \'es a fer\-mi\-o\-no\-kat per\-tur\-ba\-t\'{\i}\-ve ke\-zel\-j\"uk, a
fenn\-ma\-ra\-d\'o bo\-zo\-ni\-kus el\-m\'e\-le\-tet pe\-dig n\'egy\-di\-men\-zi\-\'os Mon\-te Car\-lo
szi\-mu\-l\'a\-ci\-\'ok\-kal vizs\-g\'al\-juk. (Mi\-vel az e\-l\H ob\-bi\-ek\-ben em\-l\'{\i}\-tett
inf\-ra\-v\"o\-r\"os prob\-l\'e\-m\'ak a bo\-zo\-ni\-kus szek\-tor\-ban buk\-kan\-nak csak fel, a
r\'a\-cson ne\-he\-zen ke\-zel\-he\-t\H o fer\-mi\-o\-nok per\-tur\-ba\-t\'{\i}v fi\-gye\-lem\-be\-v\'e\-te\-le
in\-do\-kolt.) Az SU(2)--Higgs mo\-dell -- vagy az MSSM-b\H ol ki\-in\-dul\-va
a\-d\'o\-d\'o bo\-zo\-ni\-kus mo\-dell -- f\'a\-zi\-s\'at\-me\-ne\-t\'e\-nek jel\-lem\-z\H o
pa\-ra\-m\'e\-te\-re\-i\-b\H ol per\-tur\-ba\-t\'{\i}v kor\-rek\-ci\-\'ok se\-g\'{\i}t\-s\'e\-g\'e\-vel
ha\-t\'a\-roz\-ha\-t\'ok meg a tel\-jes el\-m\'e\-let\-re jel\-lem\-z\H o pa\-ra\-m\'e\-te\-rek.

A n\'egy\-di\-men\-zi\-\'o mo\-dell szi\-mu\-l\'a\-ci\-\'o\-ja sok\-kal i\-d\H o\-i\-g\'e\-nye\-sebb, mint
a h\'a\-rom\-di\-men\-zi\-\'os re\-du\-k\'alt mo\-del\-l\'e \cite{far, jak}. A k\'et\-faj\-ta
meg\-k\"o\-ze\-l\'{\i}\-t\'es e\-red\-m\'e\-ny\'e\-nek kom\-pa\-ti\-bi\-li\-t\'a\-sa a\-zon\-ban me\-ge\-r\H o\-s\'{\i}t
ben\-n\"un\-ket ab\-ban a hit\-ben, hogy az \'{\i}gy ka\-pott e\-red\-m\'e\-nyek j\'ok --
az i\-lyes\-faj\-ta el\-le\-n\H or\-z\'e\-si m\'od\-sze\-rek sz\"uk\-s\'e\-ges\-s\'e\-g\'e\-re \'ep\-pen a
per\-tur\-ba\-t\'{\i}v e\-red\-m\'e\-nyek sor\-sa h\'{\i}v\-ja fel fi\-gyel\-m\"un\-ket.

A h\'a\-rom- \'es n\'egy\-di\-men\-zi\-\'os szi\-mu\-l\'a\-ci\-\'ok e\-red\-m\'e\-nye\-i e\-gya\-r\'ant azt
mu\-tat\-j\'ak, hogy a stan\-dard mo\-dell ke\-re\-t\'e\-ben a k\'{\i}\-s\'er\-le\-tek \'al\-tal
le\-het\-s\'e\-ges\-nek mi\-n\H o\-s\'{\i}\-tett Higgs-t\"o\-meg tar\-to\-m\'any\-ban nincs
e\-lekt\-ro\-gyen\-ge f\'a\-zi\-s\'at\-me\-net; a kis Higgs-t\"o\-meg e\-se\-t\'en m\'eg
el\-s\H o\-ren\-d\H u f\'a\-zi\-s\'at\-me\-net a Higgs-t\"o\-meg n\"o\-ve\-ked\-t\'e\-vel egy\-re
gyen\-g\"ul, majd 72 GeV k\"or\-ny\'e\-k\'en a f\'a\-zi\-s\'at\-me\-net m\'a\-sod\-ren\-d\H u\-v\'e
v\'a\-lik -- e v\'eg\-pont f\"o\-l\"ott pe\-dig si\-ma cross-o\-ver van a k\'et f\'a\-zis
k\"o\-z\"ott. Nincs te\-h\'at e\-r\H os e\-lekt\-ro\-gyen\-ge f\'a\-zi\-s\'at\-me\-net, m\'as
sz\'o\-val a ba\-ri\-o\-ge\-n\'e\-zis nem ma\-gya\-r\'az\-ha\-t\'o a stan\-dard mo\-dell
ke\-re\-t\'e\-ben. Po\-zi\-t\'{\i}\-vab\-ban fo\-gal\-maz\-va: a r\'acsszi\-mu\-l\'a\-ci\-\'ok ar\-ra
u\-tal\-nak, hogy van fi\-zi\-ka a stan\-dard mo\-del\-len t\'ul.

A dol\-go\-zat el\-s\H o r\'e\-sze m\'e\-gis a stan\-dard mo\-dell ke\-re\-t\'en be\-l\"ul
ma\-rad. A r\'acsszi\-mu\-l\'a\-ci\-\'ok azt is meg\-mu\-tat\-t\'ak, hogy kis Higgs-t\"o\-meg
e\-se\-t\'en a per\-tur\-b\'a\-ci\-\'o\-sz\'a\-m\'{\i}\-t\'as \'es a r\'acsszi\-mu\-l\'a\-ci\-\'ok
e\-red\-m\'e\-nye\-i \"ossze\-e\-gyez\-tet\-he\-t\H o\-ek. Az e\-red\-m\'e\-nyek \"o\-sze\-ve\-t\'e\-se a\-zon\-ban
nem tri\-vi\-\'a\-lis: a k\"oz\-pon\-ti sze\-rep\-pel b\'{\i}\-r\'o csa\-to\-l\'a\-si \'al\-lan\-d\'o
el\-t\'e\-r\H o m\'o\-don van de\-fi\-ni\-\'al\-va a per\-tur\-b\'a\-ci\-\'o\-sz\'a\-m\'{\i}\-t\'as\-ban,
il\-let\-ve a nem\-per\-tur\-ba\-t\'{\i}v r\'acsszi\-mu\-l\'a\-ci\-\'ok\-ban. A per\-tur\-ba\-t\'{\i}v
meg\-k\"o\-ze\-l\'{\i}\-t\'es a\-dott $\mu$ t\"o\-megs\-k\'a\-l\'an az \ms \ le\-vo\-n\'a\-si el\-j\'a\-r\'as
ke\-re\-t\'e\-ben a meg\-szo\-kott m\'o\-don de\-fi\-ni\-\'al\-ja a csa\-to\-l\'a\-si \'al\-lan\-d\'ot. A
r\'acsszi\-mu\-l\'a\-ci\-\'ok a szta\-ti\-kus kvark po\-ten\-ci\-\'al\-ra \'e\-p\"u\-l\H o csa\-to\-l\'a\-si
\'al\-lan\-d\'o de\-fi\-n\'{\i}\-ci\-\'ot hasz\-n\'al\-nak.

A k\'et\-f\'e\-le meg\-k\"o\-ze\-l\'{\i}\-t\'es \"ossze\-ve\-t\'e\-s\'e\-hez te\-h\'at a szta\-ti\-kus kvark
po\-ten\-ci\-\'al per\-tur\-ba\-t\'{\i}v ki\-sz\'a\-m\'{\i}\-t\'a\-sa sz\"uk\-s\'e\-ges. A csa\-to\-l\'a\-si
\'al\-lan\-d\'ok k\"o\-z\"ot\-ti kap\-cso\-lat meg\-te\-rem\-t\'e\-s\'e\-vel a n\'egy\-di\-men\-zi\-\'os
SU(2)--Higgs-mo\-dell\-be\-li e\-red\-m\'e\-nyek tel\-jes stan\-dard mo\-dell\-be\-li
e\-red\-m\'e\-nyek\-k\'e va\-l\'o kon\-ver\-t\'a\-l\'a\-sa so\-r\'an fel\-l\'e\-p\H o, a k\'et\-f\'e\-le
(per\-tur\-ba\-t\'{\i}v \'es nem\-per\-tur\-ba\-t\'{\i}v) l\'e\-p\'es al\-kal\-ma\-z\'a\-s\'a\-b\'ol fa\-ka\-d\'o
hi\-ba is cs\"ok\-kent\-he\-t\H o. A dol\-go\-zat el\-s\H o fe\-l\'e\-ben te\-h\'at
ki\-sz\'a\-m\'{\i}\-tom a szta\-ti\-kus kvark po\-ten\-ci\-\'alt im\-pul\-zus\-t\'er\-ben, majd az
\'{\i}gy ka\-pott ki\-fe\-je\-z\'est Fo\-u\-ri\-er-transz\-for\-m\'al\-va meg\-te\-rem\-tem a
k\'et\-f\'e\-le csa\-to\-l\'a\-si \'al\-lan\-d\'o k\"oz\-ti kap\-cso\-la\-tot.
Ez a kap\-cso\-lat a n\'egy\-di\-men\-zi\-i\-\'os, \'es a di\-men\-zi\-\'os re\-duk\-ci\-\'o\-val
ka\-pott h\'a\-rom\-di\-men\-zi\-\'os e\-red\-m\'e\-nyek \"ossze\-ve\-t\'e\-s\'et te\-szi le\-he\-t\H o\-v\'e,
a\-mi \'al\-tal nem\-per\-tur\-ba\-t\'{\i}v e\-red\-m\'e\-nyek\-kel t\'a\-maszt\-ja a\-l\'a a
di\-men\-zi\-\'os re\-duk\-ci\-\'o el\-ter\-jedt m\'od\-sze\-r\'e\-nek al\-kal\-maz\-ha\-t\'o\-s\'a\-g\'at.

A per\-tur\-ba\-t\'{\i}v \'es nem\-per\-tur\-ba\-t\'{\i}v e\-red\-m\'e\-nyek \"ossze\-ve\-t\'e\-se egy\-r\'eszt
meg\-mu\-tat\-ja, mi\-lyen pa\-ra\-m\'e\-ter\-tar\-to\-m\'any\-ban szol\-g\'al\-tat meg\-b\'{\i}z\-ha\-t\'o
e\-red\-m\'e\-nye\-ket a per\-tur\-b\'a\-ci\-\'o\-sz\'a\-m\'{\i}\-t\'as, m\'as\-r\'eszt in\-di\-k\'a\-ci\-\'ot
ad\-hat ar\-ra, hogy a per\-tur\-ba\-t\'{\i}\-ve ke\-zel\-he\-tet\-len tar\-to\-m\'any\-ban mi\-\'ert
nem m\H u\-k\"o\-dik a per\-tur\-b\'a\-ci\-\'o\-sz\'a\-m\'{\i}\-t\'as -- \'{\i}gy al\-kal\-mas
m\'od\-sze\-rek\-kel (pl.\ meg\-fe\-le\-l\H o fe\-l\"osszeg\-z\'e\-sek stb.) ez e\-set\-leg
or\-vo\-sol\-ha\-t\'o le\-het. Er\-re a\-z\'ert van sz\"uk\-s\'eg, mi\-vel a bo\-nyo\-lul\-tabb
mo\-del\-lek -- min\-de\-nek e\-l\H ott a dol\-go\-zat m\'a\-so\-dik r\'e\-sz\'e\-nek a\-nya\-g\'at
k\'e\-pe\-z\H o mi\-ni\-m\'a\-lis szu\-per\-szim\-met\-ri\-kus stan\-dard mo\-dell -- e\-se\-t\'e\-ben
sz\'a\-mos per\-tur\-ba\-t\'{\i}v dol\-go\-zat vizs\-g\'al\-ja az e\-lekt\-ro\-gyen\-ge
f\'a\-zi\-s\'at\-me\-ne\-tet -- k\"u\-l\"o\-n\"os te\-kin\-tet\-tel a\-zo\-kat a
pa\-ra\-m\'e\-ter\-tar\-to\-m\'a\-nyo\-kat, a\-hol a ba\-ri\-o\-ge\-n\'e\-zis sz\"uk\-s\'e\-ges fel\-t\'e\-te\-le\-i
meg\-va\-l\'o\-sul\-hat\-nak --, va\-la\-mint j\'o\-n\'e\-h\'any di\-men\-zi\-\'os re\-duk\-ci\-\'on
a\-la\-pu\-l\'o e\-red\-m\'eny is l\'e\-te\-zik, a\-zon\-ban ke\-v\'es n\'egy\-di\-men\-zi\-\'os
szi\-mu\-l\'a\-ci\-\'o\-b\'ol ka\-pott e\-red\-m\'eny \'all m\'eg ren\-del\-ke\-z\'e\-s\"unk\-re. A
n\'egy\-di\-men\-zi\-\'os szi\-mu\-l\'a\-ci\-\'ok\-ban vizs\-g\'alt pa\-ra\-m\'e\-ter\-tar\-to\-m\'a\-nyok
meg\-v\'a\-lasz\-t\'a\-s\'a\-hoz j\'o a\-la\-pot ny\'ujt\-hat\-nak a per\-tur\-ba\-t\'{\i}v e\-red\-m\'e\-nyek
-- me\-lyek\-nek meg\-b\'{\i}z\-ha\-t\'o\-s\'a\-ga mel\-lett ko\-moly \'er\-vek hoz\-ha\-t\'o\-ak fel,
a\-zon\-ban ezt (a stan\-dard mo\-dell\-ben l\'a\-tot\-tak k\"o\-vet\-kez\-t\'e\-ben) csak a
meg\-fe\-le\-l\H o nem\-per\-tur\-ba\-t\'{\i}v e\-red\-m\'e\-nyek\-kel va\-l\'o e\-gye\-z\'es t\'a\-maszt\-hat\-ja
a\-l\'a  kel\-l\H o\-k\'ep\-pen.

\chapter{Az im\-pul\-zus\-t\'er\-be\-li po\-ten\-ci\-\'al}
\section{A szta\-ti\-kus kvark po\-ten\-ci\-\'al be\-ve\-ze\-t\'e\-se \label{p_bev}}
\fancyhead[CE]{\hst{\thechapter{}.\ fe\-je\-zet \quad Az
im\-pul\-zust\'er\-be\-li po\-ten\-ci\'al}}
A szta\-ti\-kus kvark po\-ten\-ci\-\'al fo\-gal\-m\'at a Yang--Mills-el\-m\'e\-le\-tek\-ben
fel\-l\'e\-p\H o a\-szimp\-to\-ti\-kus sza\-bad\-s\'ag per\-tur\-ba\-t\'{\i}v vizs\-g\'a\-la\-ta
c\'el\-j\'a\-b\'ol a k\"o\-vet\-ke\-z\H o gon\-do\-lat\-k\'{\i}\-s\'er\-let\-tel ve\-zet\-te be
L.~Suss\-kind \cite{suss} 1976-os les ho\-u\-ches-i e\-l\H o\-a\-d\'a\-s\'a\-ban.

Kelt\-s\"unk a $-T/2$ pil\-la\-nat\-ban egy na\-gyon ne\-h\'ez kvark--an\-tik\-vark
(for\-r\'as--an\-ti\-for\-r\'as) p\'art a v\'a\-ku\-um\-b\'ol, kv\'a\-zisz\-ta\-ti\-kus
(a\-di\-a\-ba\-ti\-kus) k\"o\-r\"ul\-m\'e\-nyek k\"o\-z\"ott t\'a\-vo\-l\'{\i}t\-suk el \H o\-ket
egy\-m\'as\-t\'ol $R$ t\'a\-vol\-s\'ag\-ra; $T$ i\-de\-ig ne v\'al\-toz\-tas\-sunk a
kon\-fi\-gu\-r\'a\-ci\-\'on, majd $+T/2$-ben k\"o\-ze\-l\'{\i}t\-s\"uk egy\-m\'as\-hoz a k\'et
for\-r\'ast, \'es hagy\-juk, hogy an\-ni\-hi\-l\'a\-l\'od\-ja\-nak \cite{suss, kog}. A
fo\-lya\-ma\-tot az a\-l\'ab\-bi \'ab\-ra szem\-l\'el\-te\-ti. \\
\begin{minipage}{0.9cm}
{\ }
\end{minipage}
\begin{minipage}{5.9cm}
\begin{picture}(200, 120)(-40,-20)
\put(0,40){\oval (40, 100)}
\put(0,40){\oval (43, 103)}
\put(-5,-25) {\vector(-1,0){15}}
\put(5,-25) {\vector(1,0){15}}
\put(0,-25){\makebox (0,0){R}}
\put(45,30) {\vector(0,-1){40}}
\put(45,50) {\vector(0,1){40}}
\put(45,40){\makebox (0,0){T}}
\end{picture}
\medskip \\
Neh\'ez kvark hu\-rok
\end{minipage}
\begin{minipage}{10cm}
\vspace{1.4cm}
A fo\-lya\-mat e\-uk\-li\-de\-szi amp\-li\-tu\-d\'o\-ja az $\exp(-HT)$ i\-d\H o\-fej\-lesz\-t\H o
o\-pe\-r\'a\-tor $| i \rangle$ kez\-de\-ti- \'es $| f \rangle$ v\'e\-g\'al\-la\-pot
k\"o\-z\"ot\-ti m\'at\-ri\-xe\-le\-me:
\beq
\left< i | e^{-HT} | f \right>. \label{matrixelem}
\enq
A $T \to \infty$ ha\-t\'a\-re\-set\-ben $| i \rangle$ \'es $| f \rangle$
e\-gya\-r\'ant az egy\-m\'as\-t\'ol $R$ t\'a\-vol\-s\'ag\-ra le\-v\H o kvark--an\-tik\-vark
\'al\-la\-pot\-nak fe\-lel meg; $H$ a vizs\-g\'alt el\-m\'e\-let Ha\-mil\-ton-f\"ugg\-v\'e\-nye.
\vspace{1.4cm}
\end{minipage}

A (\ref{matrixelem}) k\'ep\-le\-tet tisz\-ta SU(3) (SU(N)) m\'er\-t\'e\-kel\-m\'e\-let
e\-se\-t\'en az a\-l\'ab\-bi m\'o\-don \'{\i}r\-hat\-juk \'at p\'a\-lya\-in\-teg\-r\'al a\-lak\-j\'a\-ba:
\beq
\left< i | e^{-HT} | f \right> = \frac{Z(J)}{Z(0)}= \frac{\dst \int \left[
D A^a_{\mu} \right] \left[D c_a \right] \left[ D c_a^* \right] \exp
\left[ - S + i g \int A^a_\mu J^a_\mu d^4 x \right]} {\dst \int \left[D A^a_{\mu}
\right] \left[D c_a \right] \left[ D c_a^* \right] \exp \left[- S \right] }.
\enq
$S$ a ha\-t\'as, $J^a_\mu (a=1, \ldots 8)$ a ne\-h\'ez kvar\-kok
vi\-l\'ag\-vo\-na\-l\'a\-b\'ol ka\-pott k\"ul\-s\H o \'a\-ram, $A^a_\mu$ a m\'er\-t\'ek\-t\'er
(glu\-on\-t\'er), $c^a$ pe\-dig a Fa\-gye\-jev--Po\-pov-f\'e\-le szel\-lem\-t\'er. A
fen\-ti \'ab\-r\'an l\'a\-tott vi\-l\'ag\-vo\-na\-lak\-ra az in\-teg\-ran\-dus
sz\'am\-l\'a\-l\'o\-j\'a\-ban le\-v\H o \'a\-ram\-tag\-ban
\beq
\int A^a_\mu J^a_\mu d^4 x =
\oint A^a_\mu \textstyle{\frac{1}{2}} \lambda^a dx_\mu
\enq \'{\i}r\-ha\-t\'o. Mi\-vel a fo\-lya\-mat szta\-ti\-kus \'es az $|i \rangle$ kez\-de\-ti
\'es az $| f \rangle$ v\'e\-g\'al\-la\-pot me\-ge\-gye\-zik,
\beq
\left< i | e^{-HT} | f \right> = e^{-V(R)T} \left< i | f \right>,
\enq
a\-hol $V(R)$ a szta\-ti\-kus kvark po\-ten\-ci\-\'al. Az e\-l\H ob\-bi k\'ep\-le\-tek
egy\-m\'as\-ba he\-lyet\-te\-s\'{\i}\-t\'e\-s\'e\-vel a szta\-ti\-kus kvark po\-ten\-ci\-\'al\-ra
\beq
V(R) = - \lim_{T \to \infty} \frac1T \frac{\dst \ln \left< \Tr P \exp
\left[ i g \oint A^a_\mu \textstyle{\frac12} \lambda^a dx_\mu \right]
\right>} {\dst \left< \Tr 1 \right>}
\enq
a\-d\'o\-dik. A k\"o\-vet\-ke\-z\H o sza\-ka\-szok\-ban c\'e\-lunk a fen\-ti kon\-t\'u\-rin\-teg\-r\'al
ki\-sz\'a\-m\'{\i}\-t\'a\-sa lesz. Ezt le\-gegy\-sze\-r\H ub\-ben az im\-pul\-zus\-t\'er\-ben
fel\-raj\-zolt Feyn\-man-gr\'a\-fok ki\-\'er\-t\'e\-ke\-l\'e\-s\'e\-vel fog\-juk tud\-ni
meg\-va\-l\'o\-s\'{\i}\-ta\-ni.
\fancyhead[CO]{\hst{\thesection \quad A szta\-ti\-kus kvark po\-ten\-ci\'al
be\-ve\-zet\'e\-se}}

A szta\-ti\-kus kvark po\-ten\-ci\-\'al\-nak m\'as de\-fi\-n\'{\i}\-ci\-\'o\-ja is le\-het\-s\'e\-ges: a
kvark--an\-tik\-vark hur\-kot e\-gyik, vagy mind\-k\'et v\'e\-g\'en be\-z\'a\-rat\-la\-nul
hagy\-hat\-juk \cite{montvay}. Ez kvark--an\-tik\-vark p\'ar kel\-t\'e\-s\'e\-nek,
l\'e\-te\-z\H o kvark--an\-tik\-vark\-p\'ar sz\'et\-su\-g\'ar\-z\'a\-s\'a\-nak, vagy \"o\-r\"ok\-k\'e
l\'e\-te\-z\H o kvark--an\-tik\-vark p\'ar\-nak fe\-lel meg. N\'e\-h\'any
r\'acs\-t\'e\-rel\-m\'e\-le\-ti mun\-k\'a\-ban a fen\-ti (z\'art hur\-kos) de\-fi\-n\'{\i}\-ci\-\'o
he\-lyett m\'a\-sik de\-fi\-n\'{\i}\-ci\-\'o sz\"uk\-s\'e\-ges (pl.\ a h\'ur\-sza\-ka\-d\'as
le\-\'{\i}\-r\'a\-s\'a\-ra, v\"o.\ \cite{montvay} \'es \cite{hursz}). Az itt vizs\-g\'alt
prob\-l\'e\-ma\-k\"or\-ben a\-zon\-ban a z\'art hur\-kos de\-fi\-n\'{\i}\-ci\-\'o t\"o\-k\'e\-le\-te\-sen
al\-kal\-maz\-ha\-t\'o.

A szta\-ti\-kus kvark po\-ten\-ci\-\'al k\"onnyen de\-fi\-ni\-\'al\-ha\-t\'o r\'a\-cson, mint a
meg\-fe\-le\-l\H o t\'er\-be\-li m\'e\-re\-t\H u Wil\-son-hur\-kok \cite{wilson} $T \to
\infty$ li\-mesz\-ben vett ha\-t\'a\-r\'er\-t\'e\-ke. A Wil\-son-hur\-kok a r\'a\-cson i\-gen
k\"onnyen m\'er\-he\-t\H o\-ek (er\-re b\H o\-veb\-ben a \ref{mssm_wilson} sza\-kasz\-ban
t\'e\-rek ki), \'{\i}gy a szta\-ti\-kus kvark po\-ten\-ci\-\'al\-b\'ol sz\'ar\-maz\-tat\-ha\-t\'o
csa\-to\-l\'a\-si \'al\-lan\-d\'o a\-lap\-ve\-t\H o fon\-tos\-s\'a\-g\'u a r\'acs\-t\'e\-rel\-m\'e\-let\-ben.

A kvan\-tum\-sz\'{\i}n\-di\-na\-mi\-ka\-i e\-set\-ben a po\-ten\-ci\-\'al\-b\'ol l\'e\-nye\-g\'e\-ben
e\-gy\'er\-tel\-m\H u a csa\-to\-l\'a\-si \'al\-lan\-d\'o sz\'ar\-maz\-ta\-t\'a\-sa; v\'a\-ra\-ko\-z\'a\-sunk\-nak
meg\-fe\-le\-l\H o\-en a po\-ten\-ci\-\'al im\-pul\-zus\-t\'er\-be\-li a\-lak\-ja
\beq
V(q^2) = -C_F \frac{4 \pi \alpha_V(\mathbf{q}^2)}{\mathbf{q}^2}
\enq
lesz, a\-hol ${\mathbf{q}}$ a ki\-cse\-r\'elt h\'ar\-ma\-sim\-pul\-zus. Eh\-hez ha\-son\-l\'o
ki\-fe\-je\-z\'es a\-d\'o\-dik az \'al\-ta\-lunk vizs\-g\'alt SU(2)--Higgs-mo\-dell
e\-se\-t\'e\-ben, az\-zal a l\'e\-nye\-ges k\"u\-l\"onb\-s\'eg\-gel, hogy ott a po\-ten\-ci\-\'al
nem mu\-tat inf\-ra\-v\"o\-r\"os di\-ver\-gen\-ci\-\'at: a ne\-ve\-z\H o\-ben ${\mathbf{q}^2}
+ m^2$ \'all, a\-hol $m$ a m\'er\-t\'ek\-bo\-zon t\"o\-me\-g\'e\-vel \'all kap\-cso\-lat\-ban.
Az $m$ t\"o\-meg\-pa\-ra\-m\'e\-tert a\-zon\-ban t\"obb\-f\'e\-le\-k\'epp is meg\-v\'a\-laszt\-hat\-juk:
a fag\-r\'af-szin\-t\H u W-bo\-zon t\"o\-meg ($M_W^0$) el\-vi\-leg \'ep\-p\'ugy meg\-fe\-lel,
mint az egy\-hu\-rok-szin\-t\H u ($M_W^1$); de v\'a\-laszt\-ha\-tunk va\-la\-mely
r\'acs\-t\'e\-rel\-m\'e\-le\-ti ke\-ret\-ben de\-fi\-ni\-\'alt t\"o\-meg\-pa\-ra\-m\'e\-tert is, mint
p\'el\-d\'a\-ul a kor\-re\-l\'a\-ci\-\'os f\"ugg\-v\'e\-nyek (t\"ob\-b\'e-ke\-v\'es\-b\'e)
ex\-po\-nen\-ci\-\'a\-lis le\-csen\-g\'e\-s\'e\-b\H ol a\-d\'o\-d\'o \'ar\-ny\'e\-ko\-l\'a\-si
t\"o\-meg \cite{fod94}. (Kvan\-tum\-sz\'{\i}n\-di\-na\-mi\-k\'a\-ban ez
a prob\-l\'e\-ma nem
l\'ep fel, hi\-szen a Ward--Ta\-ka\-has\-hi-a\-zo\-nos\-s\'a\-gok \'er\-tel\-m\'e\-ben a
glu\-on\-t\"o\-meg a per\-tur\-b\'a\-ci\-\'o\-sz\'a\-m\'{\i}\-t\'as tet\-sz\H o\-le\-ges rend\-j\'e\-ben 0.)
Er\-re a fon\-tos pont\-ra r\'esz\-le\-te\-sen a \ref{csatall} sza\-kasz\-ban t\'e\-rek
ki.

\section{Gr\'af\-sza\-b\'a\-lyok }
\fancyhead[CO]{\hst{\thesection \quad Gr\'af\-szab\'a\-lyok}}
Eb\-ben a pa\-rag\-ra\-fus\-ban a szta\-ti\-kus kvark po\-ten\-ci\-\'al egy\-hu\-rok-ren\-d\H u
ki\-sz\'a\-m\'{\i}\-t\'a\-s\'a\-hoz sz\"uk\-s\'e\-ges gr\'af\-sza\-b\'a\-lyo\-kat a\-dom meg. Eh\-hez
elv\-ben nincs sz\"uk\-s\'eg m\'er\-t\'ek\-r\"og\-z\'{\i}\-t\'es\-re, hi\-szen a fi\-zi\-ka\-i\-lag
m\'er\-he\-t\H o po\-ten\-ci\-\'al\-ra a\-d\'o\-d\'o e\-red\-m\'eny m\'er\-t\'ek\-f\"ug\-get\-len kell
le\-gyen. Je\-len\-t\H o\-sen egy\-sze\-r\H u\-s\'{\i}\-ti a\-zon\-ban a sz\'a\-mo\-l\'ast, ha
le\-mon\-dunk er\-r\H ol az \'al\-ta\-l\'a\-nos ke\-ret\-r\H ol, \'es a Feyn\-man-m\'er\-t\'e\-ket
v\'a\-laszt\-juk.\footnote{
Je\-len\-t\H o\-sen r\"o\-vi\-d\"ul a sz\'a\-mo\-l\'as ak\-kor is, ha m\'as\-k\'ent r\"og\-z\'{\i}t\-j\"uk
a m\'er\-t\'e\-ket; hogy mi\-\'ert leg\-c\'el\-sze\-r\H ubb m\'e\-gis Feyn\-man-m\'er\-t\'e\-ket
hasz\-n\'al\-ni, az a \ref{grafok} pa\-rag\-ra\-fus\-ban fog ki\-de\-r\"ul\-ni.}
Ez ter\-m\'e\-sze\-te\-sen azt je\-len\-ti, hogy az e\-red\-m\'eny\-nek m\'ar nem kell
\'at\-men\-ni\-e a m\'er\-t\'ek\-f\"ug\-get\-len\-s\'e\-gi tesz\-ten -- me\-lyen egy e\-set\-le\-ges
hi\-b\'as sz\'a\-mo\-l\'as k\"onnyen fen\-na\-kad\-ha\-tott vol\-na. El\-le\-n\H or\-z\'es\-k\'ep\-pen
te\-h\'at v\'eg\-re\-haj\-tot\-tam az i\-ro\-da\-lom\-b\'ol j\'ol is\-mert
kvan\-tum\-sz\'{\i}n\-di\-na\-mi\-ka\-i (tisz\-ta SU(3) m\'er\-t\'e\-kel\-m\'e\-let\-be\-li)
sz\'a\-mo\-l\'a\-so\-kat; e\-ze\-ket a \ref{qcderedmeny} sza\-kasz\-ban \'es a \ref{qcd}
f\"ug\-ge\-l\'ek\-ben te\-kin\-tem \'at. E\-red\-m\'e\-nyem me\-ge\-gye\-zik M.~La\-i\-ne
ha\-son\-l\'o, m\'er\-t\'ek\-r\"og\-z\'{\i}\-t\'est nem hasz\-n\'a\-l\'o sz\'a\-mo\-l\'a\-s\'a\-val
\cite{la}.
\subsection{A szta\-ti\-kus kvark pro\-pa\-g\'a\-tor}
A szta\-ti\-kus, vagy v\'eg\-te\-len ne\-h\'ez kvark de\-fi\-n\'{\i}\-ci\-\'o\-j\'a\-n\'al fog\-va
csu\-p\'an i\-d\H o\-i\-r\'any\-ba k\'e\-pes pro\-pa\-g\'al\-ni; a szo\-k\'a\-sos kon\-ven\-ci\-\'o
\'er\-tel\-m\'e\-ben a szta\-ti\-kus kvark i\-d\H o\-ben e\-l\H o\-re, az an\-tik\-vark pe\-dig
i\-d\H o\-ben vissza\-fe\-l\'e. A ko\-or\-di\-n\'a\-ta\-t\'er\-be\-li pro\-pa\-g\'a\-to\-rok te\-h\'at
\\
\bmp{9cm}
\beqar
&& i S_Q^{ab}(y,x) = \delta^{ab} \delta (\mb{x-y}) \theta(y_0-x_0), \\
&& i S_A^{ab}(y,x) = \delta^{ab} \delta (\mb{x-y}) \theta(-y_0+x_0),
\label{propag}
\enqar
a\-hol $Q$ a kvark\-ra, $A$ az an\-tik\-vark\-ra u\-tal, \'es a pro\-pa\-g\'a\-tor
ar\-gu\-men\-tu\-m\'a\-ban a szo\-k\'a\-sos \emph{v\'e\-g\'al\-la\-pot, kez\-d\H o \'al\-la\-pot}
sor\-ren\-det v\'a\-lasz\-tot\-tuk.
\emp
\hspace{1cm}
\bmp{6.9cm}
\bc
\epsfig{file=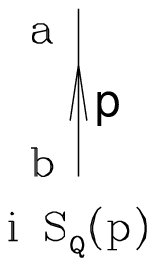,width=1.4cm}\
\epsfig{file=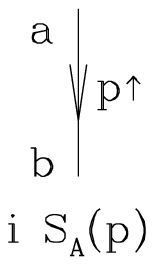,width=1.4cm}
\\
Szta\-ti\-kus kvark- \'es \\
an\-tik\-vark pro\-pa\-g\'a\-tor \label{prop}
\ec
\emp

Az im\-pul\-zus\-t\'er\-be\-li pro\-pa\-g\'a\-to\-ro\-kat tri\-vi\-\'a\-lis
Fo\-u\-ri\-er-transz\-for\-m\'a\-ci\-\'o\-val kap\-hat\-juk meg, mely\-nek e\-red\-m\'e\-nye
\beqar
i S_Q^{ab}(p) = \frac{i \delta^{ab}}{v p + i \epsilon}, &&
i S_A^{ab}(p) = \frac{i \delta^{ab}}{v p + i \epsilon}.
\enqar
$p$ a ki\-cse\-r\'elt n\'e\-gye\-sim\-pul\-zus, $v$ pe\-dig a szta\-ti\-kus for\-r\'as
n\'e\-gyes\-se\-bes\-s\'e\-ge, el\-s\H o k\"o\-ze\-l\'{\i}\-t\'es\-ben $v^\mu = (1, \mb{0})$.
\footnote{Az e\-gyik le\-het\-s\'e\-ges gr\'af\-n\'al fel\-buk\-ka\-n\'o di\-ver\-gen\-ci\-\'ak
le\-v\'a\-lasz\-t\'a\-sa in\-do\-kol\-ja a pro\-pa\-g\'a\-tor fen\-ti a\-lak\-j\'a\-nak meg\-tar\-t\'a\-s\'at
az egy\-sze\-r\H ubb $i/(p_0 + i \epsilon)$ le\-he\-t\H o\-s\'eg\-gel szem\-ben.}
\subsection{A szta\-ti\-kus kvark--m\'er\-t\'ek\-bo\-zon ver\-tex}
\bmp{5cm}
\bc
\epsfig{file=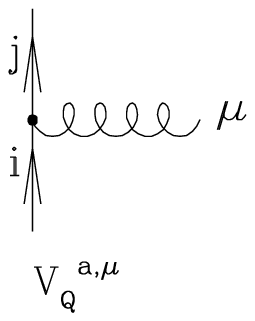,width=2.2cm}\
\epsfig{file=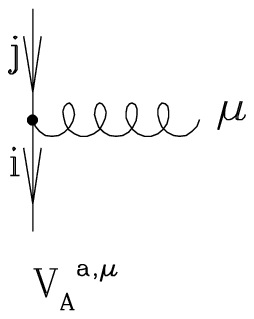,width=2.2cm}
\ec
A szta\-ti\-kus kvark--m\'er\-t\'ek\-bo\-zon ver\-tex
\emp
\hspace{1cm}
\bmp{10.9cm}
A m\'er\-t\'ek\-bo\-zon\-hoz va\-l\'o csa\-to\-l\'as k\"o\-ve\-tel\-m\'e\-nye az, hogy a
fer\-mi\-on n\'el\-k\"u\-li kvan\-tum\-e\-lekt\-ro\-di\-na\-mi\-ka\-i (U(1)) e\-set\-ben a szo\-k\'a\-sos
Co\-u\-lomb-po\-ten\-ci\-\'alt kap\-juk vissza. A szo\-k\'a\-sos
kvan\-tum\-e\-lekt\-ro\-di\-na\-mi\-ka\-i fer\-mi\-on--fo\-ton csa\-to\-l\'as\-hoz ha\-son\-l\'o
ki\-fe\-je\-z\'es l\'ep fel itt is, az\-zal a k\"u\-l\"onb\-s\'eg\-gel, hogy a
szta\-ti\-kus kvar\-kok v\'eg\-te\-len t\"o\-me\-g\'e\-b\H ol k\"o\-vet\-ke\-z\H o\-en az \'a\-ta\-dott
im\-pul\-zus t\'er\-sze\-r\H u kell le\-gyen, te\-h\'at egy ext\-ra $\delta^{\mu, 0}$
is fel\-buk\-kan. A ver\-tex j\'a\-ru\-l\'e\-ka te\-h\'at
\emp
\beqar
V_Q^{a, \mu} = i g T^a_{i,j} \delta^{\mu, 0}, &&
V_A^{a, \mu} = - i g T^a_{i,j} \delta^{\mu, 0}
\enqar
a\-hol a bal ol\-dal\-r\'ol le\-hagy\-tuk a tri\-vi\-\'a\-li\-san ke\-ze\-len\-d\H o $i,j$
in\-de\-xe\-ket.
A k\"u\-l\"on\-b\"o\-z\H o ren\-d\H u j\'a\-ru\-l\'e\-kok fe\-l\"osszeg\-z\'e\-se\-kor fel\-l\'e\-p\H o
ex\-po\-nen\-ci\-\'a\-li\-z\'a\-l\'o\-d\'as\-r\'ol sz\'o\-l\'o pa\-rag\-ra\-fus\-ban meg fo\-gom mu\-tat\-ni,
hogy ez a QED e\-se\-t\'e\-ben va\-l\'o\-ban a Co\-u\-lomb-po\-ten\-ci\-\'al\-ra ve\-zet.
\subsection{To\-v\'ab\-bi gr\'af\-sza\-b\'a\-lyok}
\bmp{11.9cm}
A to\-v\'ab\-bi gr\'af\-sza\-b\'a\-lyok mind a QED, a QCD \'es az
SU(2)--Higgs-mo\-dell e\-se\-t\'en j\'ol is\-mer\-tek; a sz\'a\-mo\-l\'as so\-r\'an a
\cite{hol} cikk kon\-ven\-ci\-\'o\-it vet\-tem \'at.
E\-gye\-d\"ul a h\'a\-rom m\'er\-t\'ek\-bo\-zon csa\-to\-l\'ast \'{\i}r\-juk itt fel, mi\-vel er\-re
a k\"o\-vet\-ke\-z\H o fe\-je\-zet\-ben exp\-li\-ci\-ten hi\-vat\-koz\-ni fo\-gok:
\emp
\bmp{5cm}
\bc
\epsfig{file=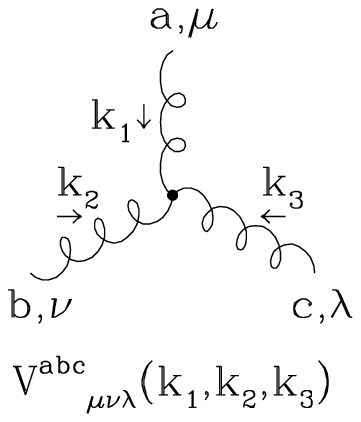,width=2.5cm}
\\ H\'a\-rom\-bo\-zon-ver\-tex
\ec
\emp
\beq
V_{\mu \nu \lambda}^{abc}(k_1,k_2,k_3) = igC^{abc} \left[ (k_1-
k_2)_\lambda g_{\mu \nu} + (k_2-k_3)_\mu  g_{\nu \lambda} + (k_3-k_1)_\nu
g_{\lambda \mu} \right]. \label{3boz}
\enq
\section{A j\'a\-ru\-l\'e\-kot a\-d\'o gr\'a\-fok \label{grafok}}
\fancyhead[CO]{\hst{\thesection \quad A j\'a\-rul\'e\-kot ad\'o gr\'a\-fok}}
Az e\-l\H o\-z\H o sza\-kasz\-ban fe\-l\'{\i}rt gr\'af\-sza\-b\'a\-lyok a\-lap\-j\'an
fel\-raj\-zol\-hat\-juk a j\'a\-ru\-l\'e\-kot a\-d\'o gr\'a\-fo\-kat.
\\
\bmp{4cm}
\bc
\epsfig{file=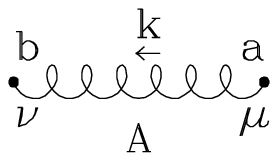,width=2cm}
\vspace{0.5cm}
\ec
\emp
\bmp{12.9cm}
A po\-ten\-ci\-\'al $g^2$ ren\-d\H u j\'a\-ru\-l\'e\-ka az $A$ gr\'af\-b\'ol a\-d\'o\-dik:
\beq
V_{Q}^{a, \mu} \cdot i D_{\nu \mu}^{ba}(k) \cdot V_{A}^{b, \nu} =
-i g^2 T^aT^a {1 \over {k^2 - M_W^2 + i \epsilon}}. \label{tree}
\enq
\emp

A $g^4$ ren\-d\H u j\'a\-ru\-l\'e\-kot a\-d\'o gr\'a\-fok k\"o\-z\"ul e\-l\H o\-sz\"or a
\ref{graph_be} \'ab\-r\'an sze\-rep\-l\H o 4 k\'et\-bo\-zon\-cse\-r\'es gr\'a\-fot kell
ki\-sz\'a\-m\'{\i}\-ta\-nunk; az e\-zek\-ben fel\-l\'e\-p\H o hu\-ro\-kin\-teg\-r\'a\-lok l\'e\-nye\-ge\-sen
k\"u\-l\"on\-b\"oz\-nek pl.\ az e\-lekt\-ro\-gyen\-ge el\-m\'e\-let re\-nor\-m\'a\-l\'a\-sa so\-r\'an
fel\-l\'e\-p\H ok\-t\H ol. A hu\-ro\-kin\-teg\-r\'a\-lok ki\-sz\'a\-m\'{\i}\-t\'a\-s\'at a k\"o\-vet\-ke\-z\H o
sza\-kasz\-ban v\'eg\-zem el.
\bef[ht]
\bc
\epsfig{file=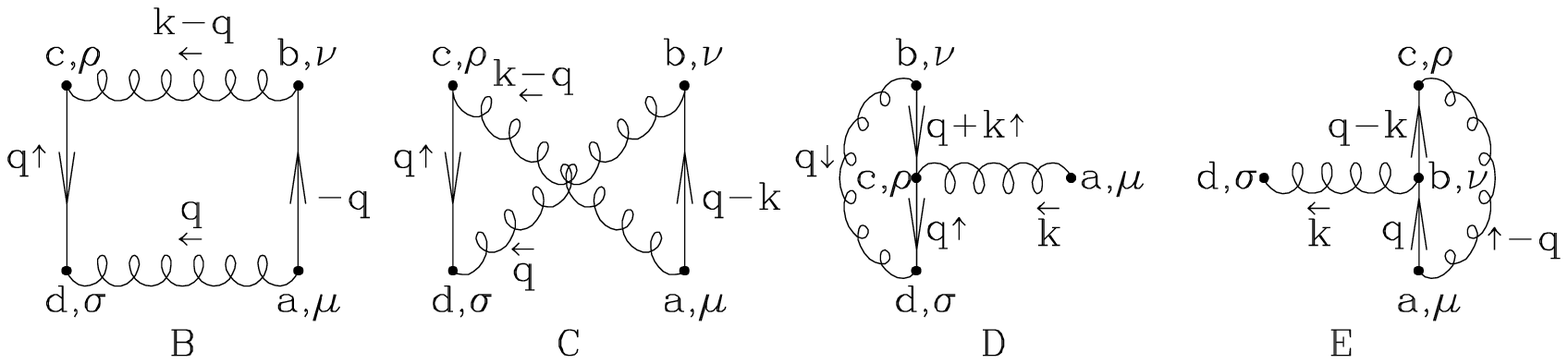,width=13cm}
\caption{A k\'et\-bo\-zon cse\-r\'es gr\'a\-fok \label{graph_be}}
\ec
\enf
\bigskip \\
{\underline {B gr\'af}}
\beqar
&& \int {{d^D q} \over {(2 \pi)^D}} V_{Q}^{a, \mu} \cdot i D_{\sigma \mu}^{d
a}(q) \cdot i S_Q^{ba}(-q) \cdot V_{Q}^{b, \nu} \cdot i D_{\rho \nu}^{cb}
(k-q) \cdot V_{A}^{c, \rho} \cdot i S_A^{cd}(q) \cdot V_{A}^{d, \sigma} =
\nonumber \\*
&& -g^4 T^a T^b T^b T^a \int {{d^D q} \over {(2 \pi)^D}} {1 \over {-q_0+i
\epsilon}} {1 \over {-q_0+i \epsilon}} {1 \over {(k-q)^2-M_W^2+i \epsilon}}
{1 \over {q^2 - M_W^2 + i \epsilon}}.
\end{eqnarray}
\bigskip \\
{\underline {C gr\'af}}
\beqar
&& \int {{d^D q} \over {(2 \pi)^D}} V_{Q}^{a, \mu} \cdot i
S_Q^{ba}(q-k) \cdot i D_{\rho \mu}^{ca}(k-q) \cdot V_{Q}^{b, \nu}
\cdot i D_{\sigma \nu}^{ db}(q) \cdot V_{A}^{c, \rho} \cdot i
S_A^{cd}(q) \cdot V_{A}^{d, \sigma} = \nonumber \\*
&& g^4 T^a T^b T^a T^b \int {{d^D q} \over {(2 \pi)^D}}
{1 \over {q_0+i \epsilon}} {1 \over {-q_0+i \epsilon}} {1 \over
{(k-q)^2-M_W^2 +i \epsilon}} {1 \over {q^2 - M_W^2 + i \epsilon}}.
\enqar
\bigskip \\
{\underline {D gr\'af}}
\beqar
&& \int {{d^D q} \over {(2 \pi)^D}} V_{Q}^{a, \mu} \cdot i D_{\rho \mu}^{ca}
(k) \cdot V_{A}^{b, \nu} \cdot i S_A^{bc}(k-q) \cdot i D_{\nu \sigma}^{bd}(
q) \cdot V_{A}^{c, \rho} \cdot i S_A^{cd}(-q) \cdot V_{A}^{d, \sigma} =
\nonumber \\
&& -g^4 T^a T^b T^a T^b {1 \over {k^2 - M_W^2 + i \epsilon}} \int
{{d^D q} \over {(2 \pi)^D}} {1 \over {q_0+i \epsilon}} {1 \over {q_0+
i \epsilon}} {1 \over {q^2 - M_W^2 + i \epsilon}}.
\enqar
\bigskip \\
{\underline {E gr\'af}}
\beqar
&& \int {{d^D q} \over {(2 \pi)^D}} V_{Q}^{a, \mu} \cdot i D_{\rho
\mu}^{ca} (-q) \cdot i S_Q^{ba}(q) \cdot V_{Q}^{b, \nu} \cdot i
D_{\sigma \nu}^{db}(k) \cdot i S_Q^{cb}(q-k) \cdot V_{Q}^{c, \rho}
\cdot V_{A}^{d, \sigma} = \nonumber \\
&& -g^4 T^a T^b T^a T^b {1 \over {k^2 - M_W^2 + i \epsilon}} \int
{{d^D q} \over {(2 \pi)^D}} {1 \over {q_0+i \epsilon}} {1 \over {q_0+
i \epsilon}} {1 \over {q^2 - M_W^2 + i \epsilon}}.
\enqar
\bigskip \\
Az e\-l\H o\-z\H o k\'ep\-le\-tek\-ben fel\-buk\-ka\-n\'o cso\-por\-tel\-m\'e\-le\-ti fak\-to\-rok\-ra az
a\-l\'ab\-bi je\-l\"o\-l\'est ve\-zet\-j\"uk be:
\beqar
\Tr T^a T^a / \Tr 1 & = & C(R), \\
C^{acd} C^{bcd}     & = & C(G) \delta^{ab},
\enqar
\'{\i}gy
\beqar
\Tr T^a T^b T^b T^a / \Tr 1 & = & C^2(R), \\
\Tr T^a T^b T^a T^b / \Tr 1 & = & C^2(R) - \frac12 C(R) C(G).
\enqar
\subsection{Kvan\-tu\-me\-lekt\-ro\-di\-na\-mi\-ka\-i ki\-t\'e\-r\H o}
A fen\-ti gr\'a\-fok se\-g\'{\i}t\-s\'e\-g\'e\-vel k\"onnyen ki\-sz\'a\-m\'{\i}t\-hat\-juk a
fer\-mi\-on-men\-tes QED szta\-ti\-kus kvark po\-ten\-ci\-\'al\-j\'at. B\'ar a fe\-la\-dat
l\'e\-nye\-ge\-sen egy\-sze\-r\H ubb, mint a QCD, vagy az SU(2)--Higgs e\-set
vizs\-g\'a\-la\-ta, k\'et ok\-b\'ol is \'er\-de\-mes v\'eg\-re\-haj\-ta\-ni. Egy\-r\'eszt a
Co\-u\-lomb-po\-ten\-ci\-\'al le\-ve\-ze\-t\'e\-se biz\-to\-s\'{\i}t min\-ket a\-fe\-l\H ol, hogy j\'o
\'u\-ton j\'a\-runk. M\'as\-r\'eszt a QED e\-set k\"onnyen fe\-l\"ossze\-gez\-he\-t\H o a
per\-tur\-b\'a\-ci\-\'o\-sz\'a\-m\'{\i}\-t\'as \"osszes rend\-j\'e\-re; mi\-vel pe\-dig a k\'et m\'a\-sik
mo\-dell\-ben fel\-l\'e\-p\H o gr\'a\-fok a\-be\-li j\'a\-ru\-l\'e\-ka\-i me\-ge\-gyez\-nek a QED
e\-set\-ben fel\-l\'e\-p\H o gr\'a\-fok j\'a\-ru\-l\'e\-ka\-i\-val, e\-z\'ert a QCD il\-let\-ve
SU(2)--Higgs e\-set vizs\-g\'a\-la\-t\'a\-n\'al e\-le\-gen\-d\H o lesz a gr\'a\-fok ne\-ma\-be\-li
j\'a\-ru\-l\'e\-ka\-it ki\-sz\'a\-m\'{\i}\-ta\-nunk. \'Igy p\'el\-d\'a\-ul a $B$ gr\'af j\'a\-ru\-l\'e\-k\'at
nem kell v\'e\-gig\-sz\'a\-mol\-nunk; hogy l\'as\-suk, hogy ez mek\-ko\-ra
k\"onnyebb\-s\'e\-get je\-lent, a sz\'a\-mo\-l\'ast m\'e\-gis el\-v\'eg\-zem a
\ref{bgraf} f\"ug\-ge\-l\'ek\-ben.

\'Al\-l\'{\i}\-t\'a\-sunk a k\"o\-vet\-ke\-z\H o: \bigskip \\
\emph{A per\-tur\-b\'a\-ci\-\'o\-sz\'a\-m\'{\i}\-t\'as \"osszes rend\-j\'e\-re fe\-l\"ossze\-gez\-ve a
kvan\-tu\-me\-lekt\-ro\-di\-na\-mi\-ka\-i szta\-ti\-kus kvark po\-ten\-ci\-\'al\-ra a\-d\'o\-d\'o
j\'a\-ru\-l\'e\-ko\-kat \'ep\-pen az egy-bo\-zon-cse\-r\'es
(egy-fo\-ton-cse\-r\'es)\footnote{A to\-v\'ab\-bi\-ak\-ban az \'al\-ta\-l\'a\-no\-sabb
'bo\-zon' sz\'ot fog\-juk hasz\-n\'al\-ni} gr\'af j\'a\-ru\-l\'e\-k\'a\-nak
ex\-po\-nen\-ci\-a\-li\-z\'alt\-j\'at kap\-juk.} \cite{fis} \bigskip \\
M\'as sz\'o\-val: az $N$ m\'er\-t\'ek\-bo\-zon-cse\-r\'es (fo\-ton-cse\-r\'es) gr\'a\-fok
\"ossze\-ge = $1/N! \cdot$ (1-bo\-zon-cse\-r\'es gr\'af j\'a\-ru\-l\'e\-ka)\tscs{N}.

Az \'al\-l\'{\i}\-t\'ast ko\-or\-di\-n\'a\-ta-t\'er\-ben c\'el\-sze\-r\H u be\-l\'at\-ni.
\bef[ht]
\bc
\epsfig{file=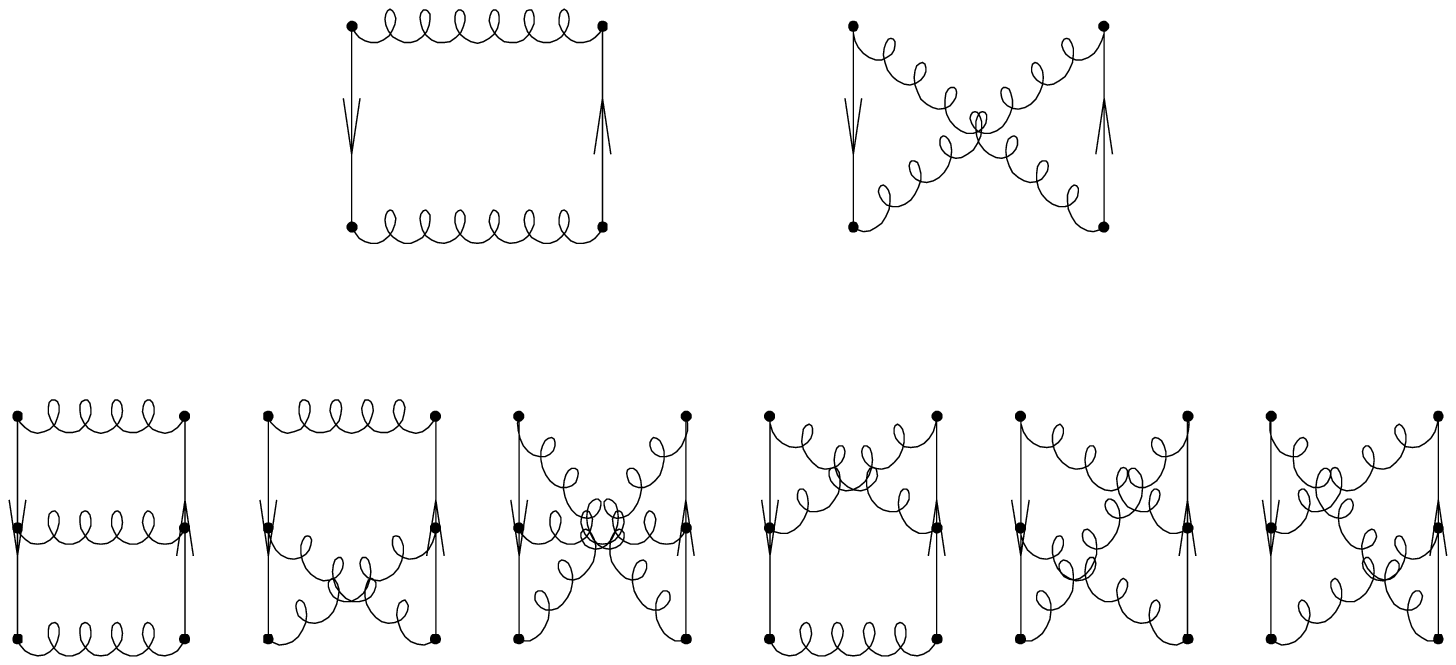,width=10cm}
\caption{Az a\-be\-li j\'a\-ru\-l\'e\-kok ex\-po\-nen\-ci\-a\-li\-z\'a\-l\'o\-d\'a\-sa \label{expo}}
\ec
\enf
Te\-kint\-s\"uk e\-l\H o\-sz\"or a \ref{expo} \'ab\-ra k\'et-bo\-zon-cse\-r\'es gr\'af\-ja\-it.
A k\'et bo\-zon-pro\-pa\-g\'a\-to\-ron k\'{\i}\-v\"ul nyil\-v\'an fel\-buk\-kan k\'et
i\-d\H o\-ren\-de\-z\'es \'es a n\'egy ver\-tex\-pont i\-d\H o\-ko\-or\-di\-n\'a\-t\'a\-i\-ra vett
in\-teg\-r\'a\-l\'as. A k\'et gr\'a\-fot \"ossze\-ad\-va az e\-gyik szta\-ti\-kus kvark
pro\-pa\-g\'a\-to\-ron je\-len\-le\-v\H o k\'et\-f\'e\-le $\theta$ f\"ugg\-v\'eny 1-re
e\-g\'e\-sz\'{\i}\-ti ki egy\-m\'ast. Ha mind\-k\'et gr\'a\-fot k\'et\-szer vessz\"uk, ak\-kor a
$\theta$ f\"ugg\-v\'e\-nyek mind\-k\'et vi\-l\'ag\-vo\-na\-lon 1-re e\-g\'e\-sz\'{\i}\-tik ki
egy\-m\'ast, \'{\i}gy
\beq
2 * (B + C) = (\mathrm{bozon-propag\acute{a}tor})^2.
\enq
Ha\-son\-l\'o\-k\'epp, az $N!$ da\-rab $N$-bo\-zon-cse\-r\'es gr\'a\-fot \"ossze\-ad\-va
az e\-gyik szta\-ti\-kus for\-r\'as vi\-l\'ag\-vo\-na\-l\'an a $\theta$ f\"ugg\-v\'e\-nyek 1-re
e\-g\'e\-sz\'{\i}\-tik ki egy\-m\'ast, ha te\-h\'at min\-den e\-gyes gr\'a\-fot $N!$-szor
te\-kin\-t\"unk, mind\-k\'et vi\-l\'ag\-vo\-nal men\-t\'en v\'eg\-re\-hajt\-ha\-t\'o a $\theta$
f\"ugg\-v\'e\-nyek ki\-ej\-t\'e\-se, \'{\i}gy
\beq
N! * \sum \mathrm{N\!\!-\!\! bo\-zon\!\!-\!\! cser\acute{e}s\
gr\acute{a}fok} = (\mathrm{bo\-zon\!\! -\!\! pro\-pag\acute{a}tor})^N.
\enq
\subsection{To\-v\'ab\-bi gr\'a\-fok}
\bmp{8cm}
\bc
\epsfig{file=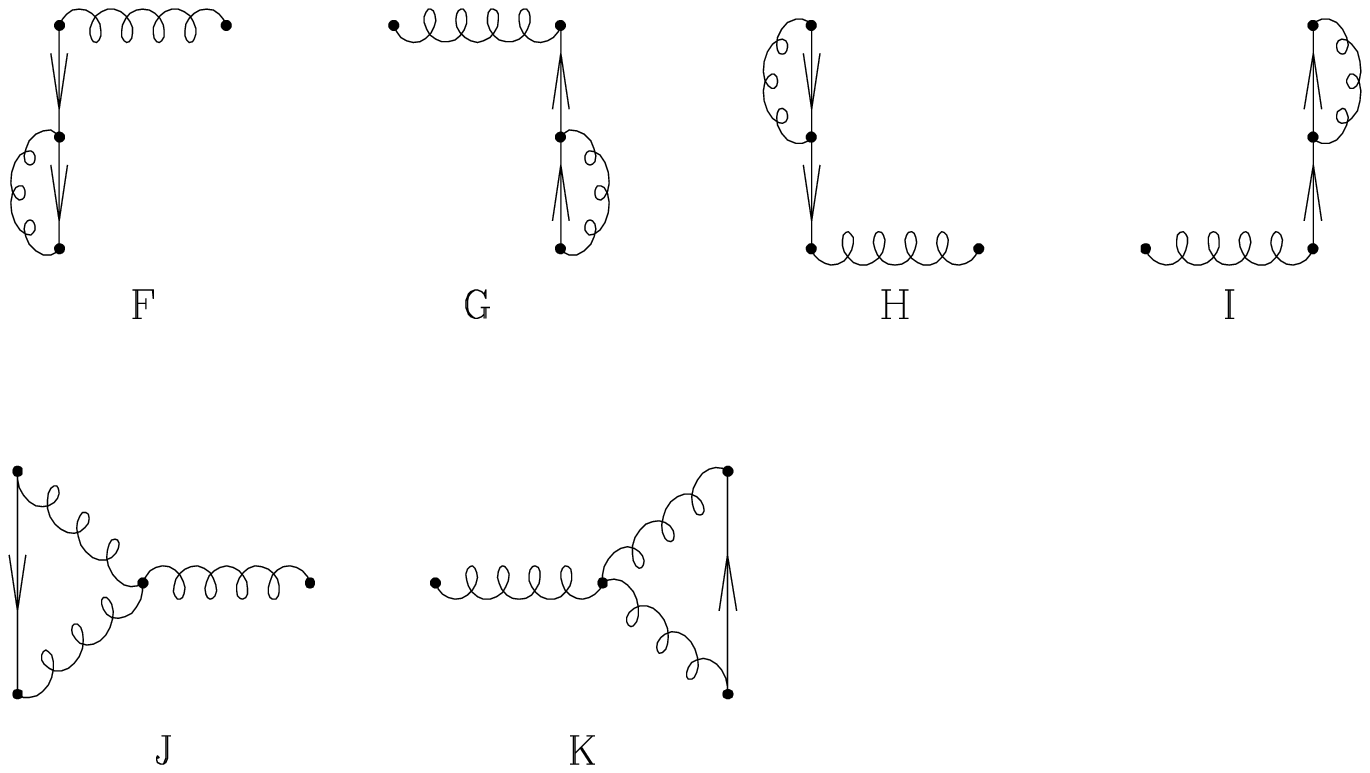,width=7cm} \\
Feyn\-man-m\'er\-t\'ek\-ben z\'e\-rus j\'a\-ru\-l\'e\-kot \\
a\-d\'o $g^4$ ren\-d\H u gr\'a\-fok
\ec
\emp
\bmp{8.9cm}
Nem ad j\'a\-ru\-l\'e\-kot a jobb ol\-da\-li gr\'af\-t\'{\i}\-pus, hi\-szen a szta\-ti\-kus kvark
pro\-pa\-g\'a\-tor\-ra ra\-k\'o\-d\'o hu\-rok csu\-p\'an t\"o\-meg\-re\-nor\-m\'a\-l\'ast o\-koz\-hat\-na; a
for\-r\'as v\'eg\-te\-len t\"o\-me\-ge mi\-att ezt nyil\-v\'an nem kell fi\-gye\-lem\-be
ven\-n\"unk.

U\-gyan\-csak z\'e\-rus j\'a\-ru\-l\'e\-kot ad a $J$ \'es a $K$ gr\'af
mint\-hogy a (\ref{3boz}) 3-bo\-zon-ver\-tex a Lo\-rentz-in\-de\-xek\-ben tel\-je\-sen
an\-ti\-szim\-met\-ri\-kus, m\'{\i}g a ne\-h\'ez (an\-ti)kvark--m\'er\-t\'ek\-bo\-zon ver\-tex
mind\-h\'a\-rom e\-set\-ben a Lo\-rentz-in\-dex 0.\ kom\-po\-nen\-s\'e\-vel a\-r\'a\-nyos.
\vspace{0.3cm}
\emp

E\-ze\-ken a gr\'a\-fo\-kon k\'{\i}\-v\"ul az egy\-hu\-rok-ren\-d\H u
m\'er\-t\'ek\-bo\-zon-pro\-pa\-g\'a\-tor, va\-la\-mint n\'e\-h\'any tad\-po\-le-gr\'af ad
po\-ten\-ci\-\'al-j\'a\-ru\-l\'e\-kot. E\-ze\-ket k\"u\-l\"on al\-sza\-ka\-szok\-ban te\-kin\-tem \'at.
\subsection{Az egy\-hu\-rok-ren\-d\H u m\'er\-t\'ek\-bo\-zon pro\-pa\-g\'a\-tor}
Az egy\-hu\-rok-ren\-d\H u m\'er\-t\'ek\-bo\-zon pro\-pa\-g\'a\-tor mind a QCD (pl.\
\cite{muta}), mind az e\-lekt\-ro\-gyen\-ge el\-m\'e\-let (pl.\ \cite{velt, jeg})
e\-se\-t\'e\-ben j\'ol is\-mert; u\-t\'ob\-bi\-b\'ol egy\-sze\-r\H u\-en sz\'ar\-maz\-tat\-ha\-t\'o az
itt vizs\-g\'alt fer\-mi\-on-men\-tes SU(2)--Higgs-mo\-dell e\-se\-te is. \'Igy
az a\-l\'ab\-bi\-ak\-ban csak a fel\-l\'e\-p\H o gr\'a\-fo\-kat te\-kin\-tem \'at. A
pon\-to\-zott vo\-nal Fa\-gye\-jev--Po\-pov-szel\-le\-met, a szag\-ga\-tott
ska\-l\'ar-r\'e\-szecs\-k\'e\-ket je\-l\"ol, me\-lyek k\"o\-z\"ul $\Phi^1$ a
Higgs-r\'e\-szecs\-ke.
\bef[h]
\bc
\epsfig{file=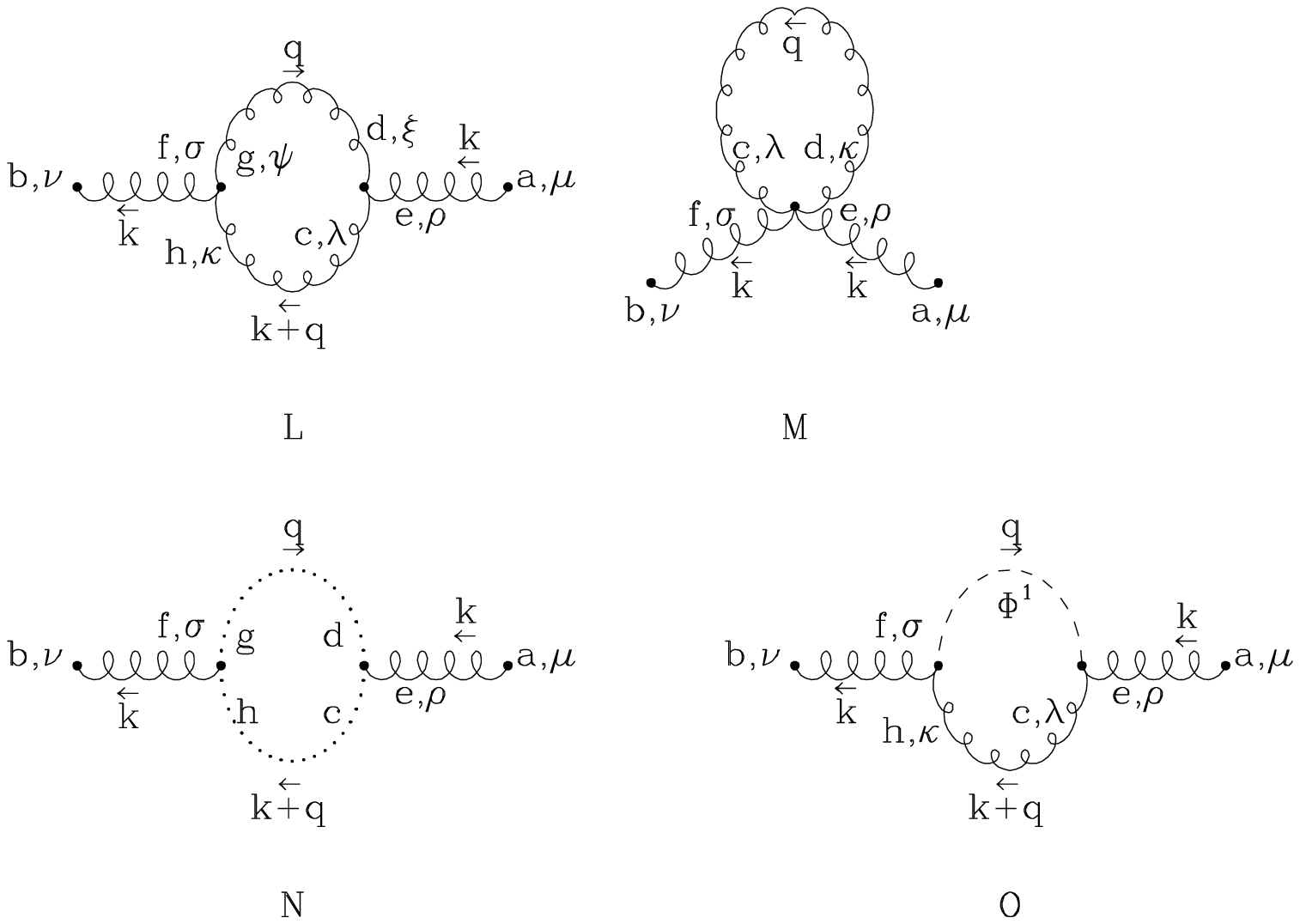,height=6cm}\
\epsfig{file=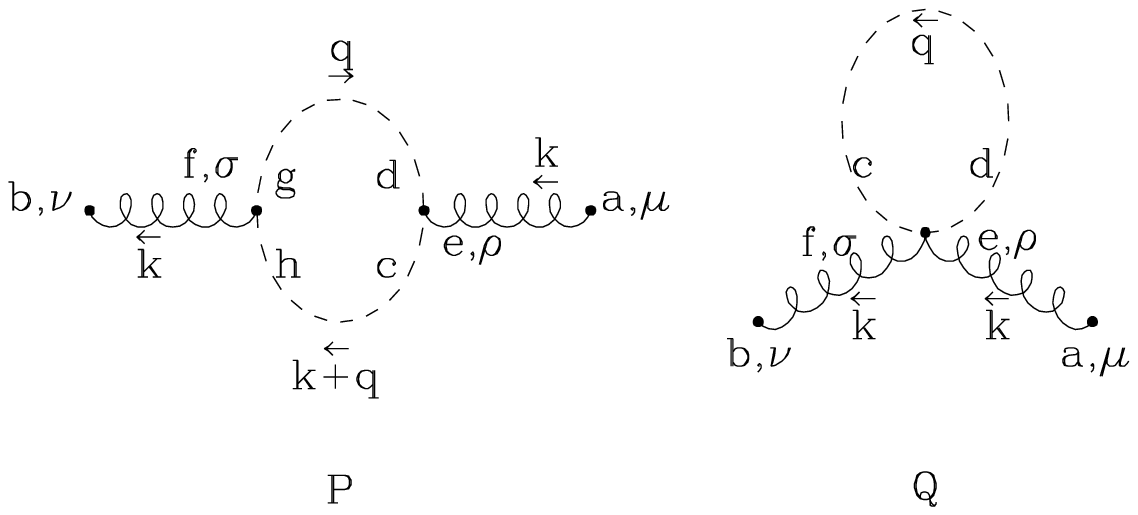,height=6cm}\
\label{graph_lq}
\caption{A m\'er\-t\'ek\-bo\-zon pro\-pa\-g\'a\-tor egy\-hu\-rok-ren\-d\H u kor\-rek\-ci\-\'o\-ja}
\ec
\enf
A kvan\-tum\-sz\'{\i}n\-di\-na\-mi\-ka\-i e\-set\-ben csak az $L$, $M$, $N$ gr\'a\-fok l\'ep\-nek
fel.
\subsection{A tad\-po\-le-gr\'a\-fok}
\'Al\-ta\-l\'a\-nos e\-set\-ben to\-v\'ab\-bi 6 tad\-po\-le-gr\'af is fel\-l\'ep;
\bef[ht]
\bc
\epsfig{file=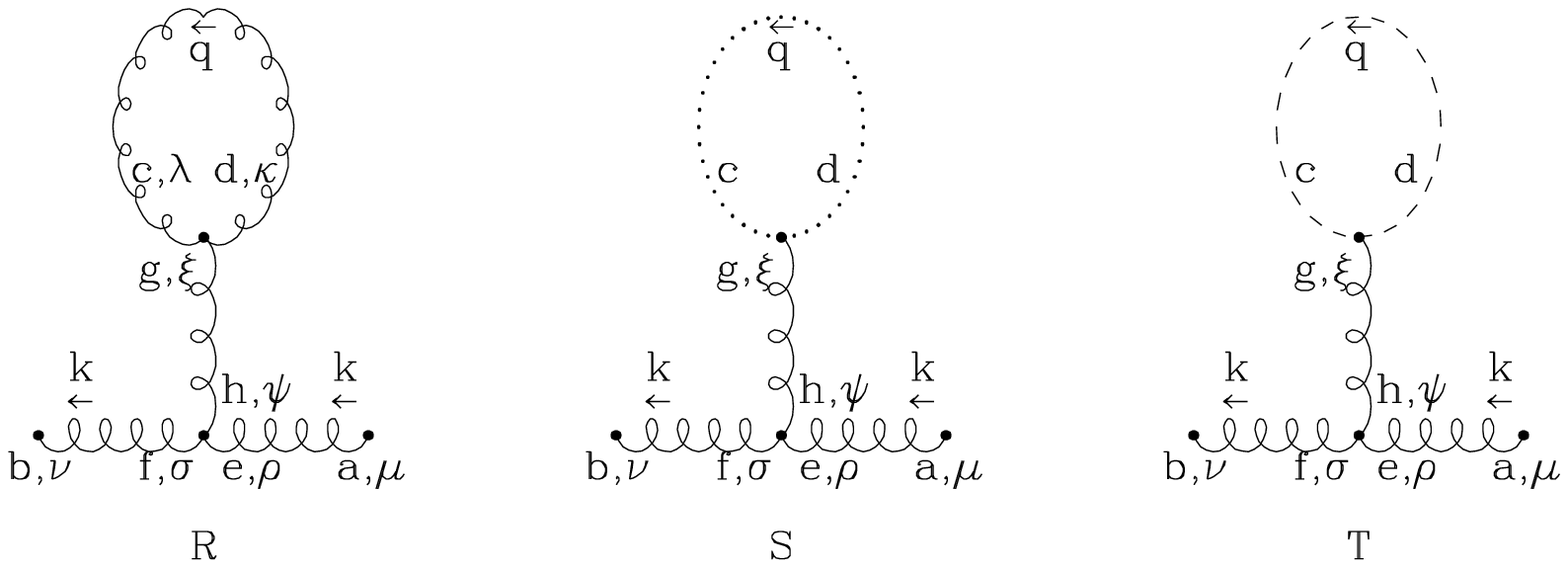,width=8cm} \quad
\epsfig{file=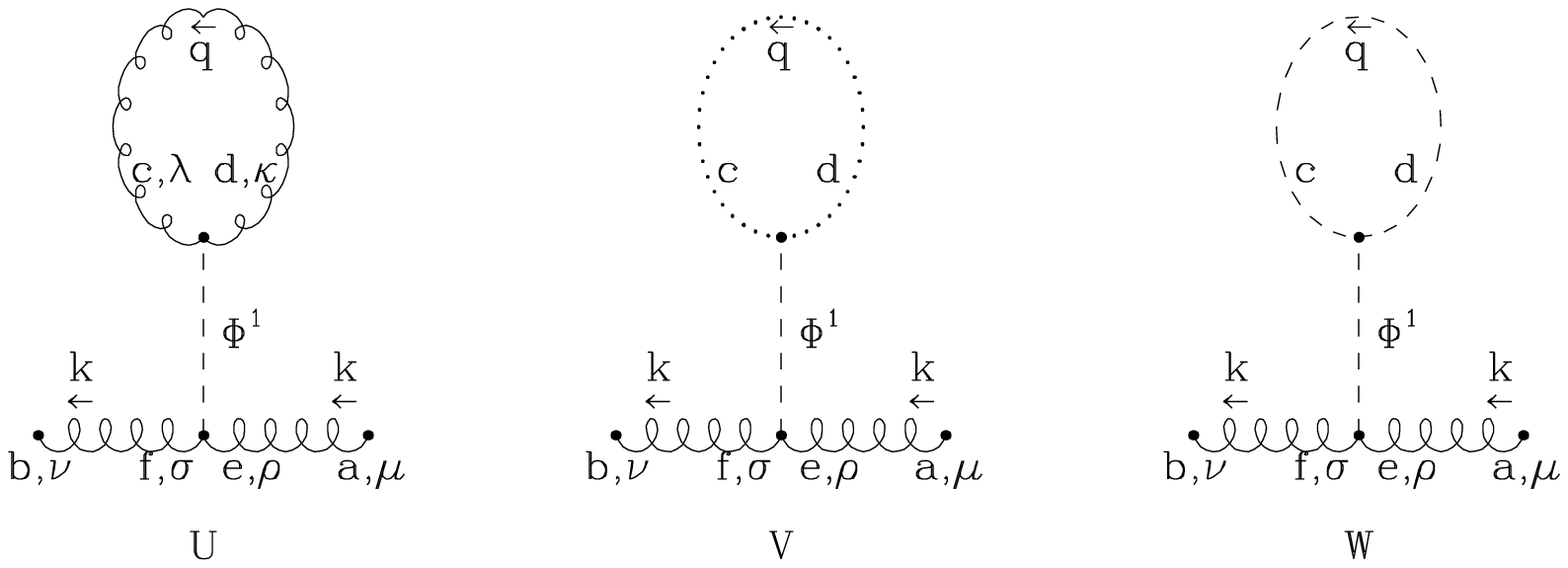,width=8cm}
\caption{A tad\-po\-le-gr\'a\-fok}
\ec
\enf
Feyn\-man-m\'er\-t\'ek\-ben a\-zon\-ban a Lo\-rentz-in\-de\-xek\-ben tel\-je\-sen
an\-ti\-szim\-met\-ri\-kus (\ref{3boz}) 3-bo\-zon-ver\-tex mi\-att az $R$, $S$,
$T$ gr\'a\-fok (te\-h\'at me\-lyek\-nek `m\'er\-t\'ek\-bo\-zon\-b\'ol van a nya\-ka')
nem ad\-nak j\'a\-ru\-l\'e\-kot. \'Igy tad\-po\-le-j\'a\-ru\-l\'ek csak az
SU(2)--Higgs-mo\-dell\-ben l\'ep fel \cite{hol, jeg}, a
kvan\-tum\-sz\'{\i}n\-di\-na\-mi\-ka\-i e\-set\-ben nem.

\section{Hu\-ro\-kin\-teg\-r\'a\-lok }
\fancyhead[CO]{\hst{\thesection \quad Hu\-ro\-kin\-tegr\'a\-lok}}
Eb\-ben a sza\-kasz\-ban az e\-l\H o\-z\H o\-ek\-ben fe\-l\'{\i}rt gr\'a\-fok ki\-sz\'a\-m\'{\i}\-t\'a\-sa
so\-r\'an fel\-l\'e\-p\H o hu\-ro\-kin\-teg\-r\'a\-lo\-kat \'er\-t\'e\-ke\-lem ki. Az
egy\-hu\-rok-ren\-d\H u m\'er\-t\'ek\-bo\-zon-pro\-pa\-g\'a\-tor sz\'a\-m\'{\i}\-t\'a\-sa so\-r\'an
fel\-l\'e\-p\H o in\-teg\-r\'a\-lok az i\-ro\-da\-lom\-b\'ol j\'ol is\-mer\-tek \cite{velt}.
A k\'et-bo\-zon-cse\-r\'es gr\'a\-fok\-n\'al h\'a\-rom \'uj hu\-ro\-kin\-teg\-r\'al buk\-kan fel:
\beqar
K (M,k) & = & \mu_0^{4-D} \int \frac{d^D q}{(2 \pi)^D}\frac{1}
{q^2-M^2+i \epsilon} \frac{1}{(k-q)^2-M^2+i \epsilon} \left(\frac{1}
{q_0+i \epsilon} \right)^2, \\
K' (M,k) & = & \mu_0^{4-D} \int \frac{d^D q}{(2 \pi)^D}\frac{1}
{q^2-M^2+i \epsilon} \frac{1}{(k-q)^2-M^2+i \epsilon} \frac{1}
{q_0+i \epsilon} \frac{1}{-q_0+i \epsilon}, \label{k'} \\
L (M,k) & = & \mu_0^{4-D} \int \frac{d^D q}{(2 \pi)^D} \frac{1}
{q^2-M^2+i \epsilon} \left( \frac{1}{q_0+i \epsilon} \right)^2.
\enqar
Az in\-teg\-r\'a\-lok\-ban ult\-ra\-i\-bo\-lya \'es (az $M=0$ e\-set\-ben) inf\-ra\-v\"o\-r\"os
di\-ver\-gen\-ci\-\'ak l\'ep\-nek fel. E\-zek ke\-ze\-l\'e\-s\'e\-re be\-ve\-zet\-j\"uk
$\epsilon_{UV} = {{4-D} \over 2} $-t \'es $\epsilon_I = {{D-4} \over
2}$-t.

A hu\-ro\-kin\-teg\-r\'a\-lok k\"o\-z\"ul $K'$ a \pageref{graph_be}.\ ol\-da\-lon le\-v\H o
$B$ gr\'af\-ban l\'ep fel; mint\-hogy e gr\'af\-nak nincs ne\-ma\-be\-li j\'a\-ru\-l\'e\-ka,
$K'$ ki\-\'er\-t\'e\-ke\-l\'e\-se k\"oz\-vet\-le\-n\"ul nem sz\"uk\-s\'e\-ges az egy\-hu\-rok-ren\-d\H u
po\-ten\-ci\-\'al meg\-a\-d\'a\-s\'a\-hoz, \'{\i}gy $K'$ vizs\-g\'a\-la\-t\'at a \ref{bgraf}
f\"ug\-ge\-l\'ek\-re hagy\-juk.
\subsection{$K$ \label{ksec}}
\beq
K (M,k) = \mu_0^{4-D} \int \frac{d^D q}{(2 \pi)^D}\frac{1}
{q^2-M^2+i \epsilon} \frac{1}{(k-q)^2-M^2+i \epsilon} \left(\frac{1}
{q_0+i \epsilon} \right)^2 \label{k}
\enq
ki\-\'er\-t\'e\-ke\-l\'e\-s\'e\-hez az a\-l\'ab\-bi k\'ep\-le\-tet fog\-juk
fel\-hasz\-n\'al\-ni \cite{mark}:
\beq
{1 \over {a^n b^m }} = {{\Gamma(n+m)} \over {\Gamma(n) \Gamma(m)}}
\int_0^\infty \alpha^{m-1} d \alpha {1 \over {(a+ \alpha b)^{n+m}}},
\label{trukk}
\enq
Eh\-hez e\-l\H o\-sz\"or $K$ ne\-ve\-z\H o\-j\'e\-ben v\'eg\-re\-hajt\-juk a szo\-k\'a\-sos
\begin{eqnarray}
&&{1 \over {q^2 - M^2 + i \epsilon}} {1 \over {(q-k)^2 - M^2 + i \epsilon}} =
\label{fey} \nonumber \\
&& \qquad \int_0^1 {{d \alpha} \over {[\alpha (q^2-M^2+ i \epsilon) +
(1- \alpha) ((k-q)^2-M^2+i \epsilon)]^2}} = \nonumber \\
&& \qquad \int_0^1 {{d \alpha} \over {(q^2 - 2(1- \alpha)kq +
(1-\alpha) k^2 - M^2 + i \epsilon)^2}}
\end{eqnarray}
he\-lyet\-te\-s\'{\i}\-t\'est, majd a (\ref{trukk}) k\'ep\-let\-ben a k\"o\-vet\-ke\-z\H o
v\'a\-lasz\-t\'a\-sok\-kal \'e\-l\"unk:
\begin{eqnarray}
a=q^2-2(1-\alpha)kq+(1-\alpha)k^2-M^2+i \epsilon, \hskip 0.25truecm
n=2, \hskip 0.5truecm && b=q_0+i \epsilon, \hskip 0.25truecm m=2.
\enqar
Ek\-kor
\beqar
K = && 6 \mu_0^{4-D} \int_0^1 d \alpha \int_0^\infty d \beta \int {{d^D q}
\over {(2 \pi)^D}} \cdot \nonumber \\
&& \qquad {\beta \over {[q_0^2 - {\bf q}^2 + 2(1- \alpha) {\bf qk} -
(1-\alpha) {\bf k}^2 -M^2 + \beta q_0 + i \epsilon ]^4}}.
\enqar
A ki\-je\-l\"olt in\-teg\-r\'a\-lok k\"o\-z\"ul e\-l\H o\-sz\"or a $D$-di\-men\-zi\-\'os t\'er\-re vett
in\-teg\-r\'a\-l\'ast hajt\-juk v\'eg\-re egy Wick-for\-ga\-t\'as se\-g\'{\i}t\-s\'e\-g\'e\-vel. Eh\-hez
\'{\i}r\-juk a t\'er sze\-rin\-ti in\-teg\-r\'alt a k\"o\-vet\-ke\-z\H o a\-lak\-ba:
\beqar
&& \int {{d^D q}\over {(2 \pi)^D}} {1 \over {[q_0^2 + \beta q_0 - A + i
\epsilon]^4}}, \label{intk}
\enqar
a\-hol
\beq
A = {\bf q}^2 -2(1-\alpha) {\bf qk} + (1-\alpha ){\bf k}^2 + M^2.
\enq
Az in\-teg\-r\'a\-l\'a\-si v\'al\-to\-z\'o meg\-fe\-le\-l\H o el\-to\-l\'a\-s\'a\-val \'es a $q_4'=-i
q_0'$ \'uj v\'al\-to\-z\'o be\-ve\-ze\-t\'e\-s\'e\-vel (Wick-for\-ga\-t\'as), a k\"o\-vet\-ke\-z\H o
tel\-jes n\'egy\-zet\-hez ju\-tunk:
\beq
\int {{d^D q_E'}\over {(2 \pi)^D}} {i \over {({q_E'}^2+B^2)^4}},
\enq
mely\-ben $E$ az e\-uk\-li\-de\-szi t\'er hasz\-n\'a\-la\-t\'a\-ra u\-tal, \'es
\beq
B^2=\alpha (1- \alpha ) {\bf k}^2 + {\beta^2 \over 4} + M^2.
\enq
En\-nek ki\-\'er\-t\'e\-ke\-l\'e\-se tri\-vi\-\'a\-lis:
\beq
\int {{d^D q_E'}\over {(2 \pi)^D}} {i \over {({q_E'}^2+B^2)^4}} =
{i \over {(4 \pi)^{D/2}}} {{\Gamma(4-{D\over 2})}\over {\Gamma(4)}}
(B^2)^{{D \over 2} -4} \label{ddim},
\enq
\'es a k\"o\-vet\-ke\-z\H o ki\-fe\-je\-z\'es\-re ve\-zet:
\beq
K = {{i \mu_0^{4-D}\Gamma(4-{D\over 2}) }\over{(4 \pi)^{D/2}}} \int_0^1 d
\alpha \int_0^\infty \beta \left[ \alpha (1-\alpha) {\bf k}^2 + {\beta^2
\over 4} + M^2 \right]^{{D \over 2}-4} d\beta. \label{kint}
\enq
Az $M=0$ e\-set\-ben (\ref{kint}) szin\-gu\-l\'a\-ris; en\-nek ke\-ze\-l\'e\-s\'e\-re
ve\-zes\-s\"uk be a $\epsilon_I = {{D-4} \over 2}$ se\-g\'ed\-v\'al\-to\-z\'ot.

E\-l\H o\-sz\"or a $\beta$ sze\-rin\-ti in\-teg\-r\'a\-l\'ast hajt\-juk v\'eg\-re:
\beq
\int_0^\infty {\beta \over {\left[ \alpha (1-\alpha) {\bf k}^2 + {\beta^2
\over 4} + M^2 \right]^{2-\epsilon_I}}} d\beta = {2 \over {1 - \epsilon_I}}
{1 \over {(\alpha (1-\alpha) {\bf k}^2 + M^2)^{1-\epsilon_I }}}
\enq

Az $\alpha$ sze\-rin\-ti in\-teg\-r\'a\-l\'as k\"u\-l\"on\-b\"o\-z\H o\-k\'ep\-pen t\"or\-t\'e\-nik az
$M=0$, il\-let\-ve $M \neq 0$ e\-set\-ben.

\paragraph{Ha $M=0$, }ak\-kor
\beqar
\int_0^1 {{d \alpha}\over {(-\alpha^2 {\bf k}^2+\alpha {\bf
k}^2)^{1 - \epsilon_I}}} & = & ({\bf k}^2)^{-1+\epsilon_I } \int_{-0.
5}^{0.5} {{d \alpha} \over {({1 \over 4} - \alpha^2)^{1-
\epsilon_I}}} \nonumber \\
& = & ({\bf k}^2)^{-1+\epsilon_I} 4^{1- \epsilon_I} {1 \over 2} \int_0^1
{{{1 \over \sqrt x} dx} \over {(1-x)^{1- \epsilon_I}}} = {2^{1-
2\epsilon_I} \over ({\bf k}^2)^{1-\epsilon_I}} {{\Gamma(\epsilon_I)
\Gamma({1 \over 2})} \over {\Gamma({1 \over 2}+ \epsilon_I)}},
\label{alphaint1}
\enqar
te\-h\'at
\beq
K = {i \over {4 \pi^2 {\bf k}^2}} \Gamma(\epsilon_I) {{\Gamma({1 \over
2})} \over {\Gamma({1 \over 2} + \epsilon_I)}} \Gamma(2- \epsilon_I) {1 \over
{1- \epsilon_I}} \left( 16 \pi {{\mu_0^2} \over {{\bf k}^2} } \right)^{-
\epsilon_I}.
\enq
Mi\-vel el\-s\H o rend\-ben $ {1 \over {1 - \epsilon_I}} \Gamma(2-
\epsilon_I) = \Gamma (1- \epsilon_I)$, \'es
\beqar
&
\Gamma \left( {1 \over 2} + \epsilon_I \right) = \Gamma \left( {1
\over 2} \right) + \epsilon_I \Gamma'\left({1 \over 2} \right) =
\Gamma \left( {1 \over 2} \right) + \epsilon_I \Gamma \left( {1 \over
2} \right) (- \gamma - 2\ln 2), &
\enqar
a\-hol $\gamma$ az E\-u\-ler--Mas\-che\-ro\-ni-f\'e\-le \'al\-lan\-d\'o,
\beq
K = {i \over {4 \pi^2 {\bf k}^2}} \Gamma (\epsilon_I) (1 + 2(\gamma +
\ln 2) \epsilon_I) \left( {{16 \pi \mu_0^2} \over {\bf k}^2}
\right)^{- \epsilon_I}.
\enq
$K$ di\-ver\-gens r\'e\-sz\'et a
\beqar
a^{- \epsilon_I} = 1 - \epsilon_I \ln a + \ordo{\epsilon_I^2} &, &
\Gamma (1 + \epsilon_I) = 1 - \gamma \epsilon_I + \ordo{\epsilon_I^2}
\label {divpart}
\nonumber
\enqar
sor\-fej\-t\'es se\-g\'{\i}t\-s\'e\-g\'e\-vel v\'a\-laszt\-hat\-juk le; v\'e\-ge\-red\-m\'e\-ny\"unk te\-h\'at
\beq
K = {i \over {4 \pi^2 {\bf k}^2}} \left[ {1 \over \epsilon_I} +
\left( \gamma - \ln {{4 \pi \mu_0^2} \over {\bf k}^2} \right) +
\ordo{\epsilon_I} \right].
\enq
\paragraph{Ha $M\neq 0$, }ak\-kor az in\-teg\-r\'al $D=4$ e\-se\-t\'en re\-gu\-l\'a\-ris,
\beq
\int_0^1 {{d \alpha}\over {-\alpha^2 {\bf k}^2+\alpha {\bf k}^2 +
M^2}} = {1 \over \sqrt{{\bf k}^4 + 4M^2{\bf k}^2}} \ln \left( {{\bf
k}^2 + \sqrt {{\bf k}^4 + 4M^2{\bf k}^2}} \over {{\bf k}^2 -
\sqrt{{\bf k}^4 + 4M^2{\bf k}^2}} \right)^2 \label{alphaint2},
\enq
\'{\i}gy
\beq
K= {i \over {8 \pi^2}} {1\over {\sqrt{{\bf k}^4 + 4M^2{\bf k}^2}}} \ln
\left( {{\bf k}^2 + \sqrt {{\bf k}^4 + 4M^2{\bf k}^2}} \over {{\bf k}^2 -
\sqrt{{\bf k}^4 + 4M^2{\bf k}^2}} \right)^2.
\enq
\subsection{$L$}
\begin{eqnarray}
L & = & 2 \mu_0^{4-D} \int_0^\infty \alpha d \alpha \int {{d^D q} \over
{(2 \pi)^D}} {1 \over {[(q^2-M^2) + \alpha q_0 + i\epsilon ]^3}} \\
& = & 2 \mu_0^{4-D} \int_0^\infty \alpha d \alpha \int {{d^D q} \over
{(2 \pi)^D}} {1 \over {[(q_0^2-{\bf q}^2-M^2) + \alpha q_0 +
i \epsilon ]^3}}.
\nonumber
\end{eqnarray}
Ha az in\-teg\-r\'a\-l\'a\-si v\'al\-to\-z\'ot
$ d q_0 \rightarrow d ( q_0 + \alpha /2)$ sze\-rint el\-tol\-juk, \'es a
$q_4'=-i q_0'$ \'uj v\'al\-to\-z\'o be\-ve\-ze\-t\'e\-s\'e\-vel Wick-for\-ga\-t\'ast haj\-tunk
v\'eg\-re, a (D-di\-men\-zi\-\'os) e\-uk\-li\-de\-szi t\'er sze\-rin\-ti in\-teg\-r\'al a
\beq \int {{d^D q_E'}\over {(2 \pi)^D}} {-i \over {({q_E'}^2+
{{\alpha^2}\over 4} + M^2)^3}} \enq
a\-la\-kot \"ol\-ti. (\ref{ddim}) a\-lap\-j\'an ez
\begin{eqnarray}
&& \int {{d^D q_E'}\over {(2 \pi)^D}}{-i \over {({q_E'}^2+
{{\alpha^2}\over 4} + M^2)^3}} ={-i \over{(4 \pi)^{D/2}}}
{{\Gamma(3-{D\over 2})}\over {\Gamma(3)}} \left( {{\alpha^2}\over 4} +
M^2 \right)^{{D \over 2}-3}.
\end{eqnarray}
A ki\-fe\-je\-z\'es ult\-ra\-i\-bo\-lya-di\-ver\-gens; az $M=0$ e\-set\-ben pe\-dig
inf\-ra\-v\"o\-r\"os di\-ver\-gen\-ci\-\'at is tar\-tal\-maz. Ezk ke\-ze\-l\'e\-s\'e\-re ve\-zes\-s\"uk be
a $\epsilon_{UV} = {{4-D} \over 2}$, $\epsilon_I={{D-4} \over 2}$
v\'al\-to\-z\'o\-kat.

\paragraph{Ha $M\neq0$, }ak\-kor
\begin{eqnarray}
L & = & {{-i \mu_0^{4-D} \Gamma \left(3- {D \over 2} \right)} \over
{(4 \pi)^ {D/2}}} \int_0^\infty \alpha \left( {{\alpha^2}\over 4} +
M^2 \right)^{{D \over 2}-3} d \alpha \nonumber \\
& = & {{-i \mu_0^{4-D} \Gamma \left(3- {D \over 2} \right)} \over {(4 \pi)^{D/2}}
} \int_0^\infty 2^{5-D} {{d(\alpha^2)}\over {( \alpha^2 + 4 M^2 )^{3 -
{D \over 2}}}} = {{-i \mu_0^{4-D} \Gamma \left(3- {D \over 2}
\right)} \over {(4 \pi)^{D/2} }} 2^{5-D} {2 \over {4-D}}
(4M^2)^{{D-4}\over 2} \nonumber \\
& = & {{-i} \over {8 \pi^2}} {2 \over {4-D}}
{\tst{\Gamma\left(3-{D\over 2} \right)}} \left( 4 \pi {\mu_0^2 \over
M^2} \right)^{\epsilon_{UV}}.
\end{eqnarray}
Ek\-kor $L$ di\-ver\-gens r\'e\-sze (\ref{divpart}) sze\-rint v\'a\-laszt\-ha\-t\'o le:
\beq
L = {-i \over {8 \pi^2}} \left[ {1 \over \epsilon_{UV}} + \left( \ln
\left( {{4\pi \mu_0^2} \over M^2} \right) - \gamma \right) +
\ordo{\epsilon_{UV}} \right].
\enq
\paragraph{Ha $M=0$, } ak\-kor
\beq
L = {{-i \mu_0^{4-D} \Gamma \left( 3- {D \over 2} \right) 4^{3 - {D
\over 2}} } \over {2 (4 \pi)^{D/2}}} \int_0^\infty {{d \alpha^2}
\over {(\alpha^2)^{3- {D \over 2}}}}
\enq
Az in\-teg\-r\'a\-l\'a\-si tar\-to\-m\'anyt ek\-kor k\'et r\'esz\-re v\'ag\-juk:
$\int_0^\infty = \int_0^1 + \int _1^\infty$, \'{\i}gy az inf\-ra\-v\"o\-r\"os \'es
az ult\-ra\-i\-bo\-lya di\-ver\-gen\-ci\-\'ak sz\'et\-v\'al\-nak:
\begin{eqnarray}
&& \int_0^1 (\alpha^2)^{{D \over 2} - 3} d \alpha^2 = \int_0^1
(\alpha^2)^{ \epsilon_I - 1} = \left[ {1 \over \epsilon_I}
(\alpha^2)^\epsilon_I \right]^1 _0 = {1 \over \epsilon_I}, \\
&& \int_1^\infty (\alpha^2)^{{D \over 2} - 3} d \alpha^2 =
\int_1^\infty ( \alpha^2)^{-\epsilon_{UV} - 1} = \left[ {-1 \over
\epsilon_{UV}} (\alpha^2)^{-\epsilon_{UV}} \right]^\infty_1 = {1 \over
\epsilon_{UV}},
\end{eqnarray}
A ket\-t\H o \"ossze\-ge\-k\'ent
\beq
L= {{-i} \over {8 \pi^2}} \left[ {1 \over \epsilon_I} + {1 \over
\epsilon_{UV}} \right].
\enq
a\-d\'o\-dik, mi\-vel a k\'et sor\-fej\-t\'es v\'e\-ges tag\-ja\-i ki\-ej\-tik egy\-m\'ast.

\section{A szta\-ti\-kus kvark po\-ten\-ci\-\'al im\-pul\-zus\-t\'er\-ben}
\fancyhead[CO]{\hst{\thesection \quad A szta\-ti\-kus kvark po\-ten\-ci\'al
im\-pul\-zust\'er\-ben}}
\subsection{Re\-nor\-m\'a\-l\'as }
Az e\-l\H o\-z\H o\-ek\-ben ki\-sz\'a\-m\'{\i}\-tot\-tam a szta\-ti\-kus kvark po\-ten\-ci\-\'al\-hoz
j\'a\-ru\-l\'e\-kot a\-d\'o gr\'a\-fo\-kat, az i\-ro\-da\-lom\-b\'ol j\'ol is\-mert
egy\-hu\-rok-ren\-d\H u bo\-zonp\-ro\-pa\-g\'a\-tor \'es a tad\-po\-le-gr\'a\-fok
ki\-v\'e\-te\-l\'e\-vel \cite{velt, jeg}. N\'e\-h\'any gr\'af e\-se\-t\'e\-ben
di\-ver\-gen\-ci\-\'ak l\'ep\-tek fel; e\-zek\-t\H ol re\-nor\-m\'a\-l\'as r\'e\-v\'en le\-het
meg\-sza\-ba\-dul\-ni. A sok\-f\'e\-le le\-he\-t\H o\-s\'eg k\"o\-z\"ul a
leggyak\-rab\-ban hasz\-n\'a\-la\-tos \ms \ prog\-ra\-mot k\"oz\-ve\-tem, \'{\i}gy a
k\'e\-s\H ob\-bi\-ek\-ben a szta\-ti\-kus kvark po\-ten\-ci\-\'al\-b\'ol a\-d\'o\-d\'o csa\-to\-l\'a\-si
\'al\-lan\-d\'ot a $g_{\overline{\mrm{MS}}}$ csa\-to\-l\'a\-si \'al\-lan\-d\'o\-val ho\-zom
e\-l\H o\-sz\"or kap\-cso\-lat\-ba.
\subsubsection*{\underline{$A$ gr\'af}}
A (\ref{tree}) k\'ep\-let\-ben a Wick-for\-ga\-t\'ast nem is kell v\'eg\-re\-haj\-ta\-nunk,
hi\-szen a k\'et szta\-ti\-kus for\-r\'as k\"o\-z\"ot\-ti im\-pul\-zus\-cse\-re i\-d\H o\-sze\-r\H u
kom\-po\-nen\-se 0, \'{\i}gy a gr\'af j\'a\-ru\-l\'e\-ka
\beq
G_A = i g^2 C(R) \frac{1}{\mb{k}^2 + M_W^2} = R_A.
\enq
Ez ter\-m\'e\-sze\-te\-sen v\'e\-ges mennyi\-s\'eg, \'{\i}gy re\-nor\-m\'a\-l\'as\-ra nincs
sz\"uk\-s\'eg.
\subsubsection*{\underline{$B, C, D, E$ gr\'af}}
A $B$ gr\'af j\'a\-ru\-l\'e\-ka tisz\-t\'an a\-be\-li, \'{\i}gy ezt fi\-gyel\-men k\'{\i}\-v\"ul
hagy\-hat\-juk, a m\'a\-sik h\'a\-rom gr\'af j\'a\-ru\-l\'e\-k\'a\-nak pe\-dig csak a ne\-ma\-be\-li
r\'e\-sz\'et kell fi\-gye\-lem\-be ven\-n\"unk.
\paragraph{$M_W \neq0$}
\beqar
G_{B+C+D+E}^{nemabeli} & = & \frac{i g^4}{16 \pi^2} C(R) C(G)
\left[ \frac{1}{{\bf k}^2 + M_W^2} \left( \frac{1}{\epsilon} + \ln
(4 \pi) - \gamma + \ln \left( \frac{\mu_0^2}{M_W^2} \right) \right)
\nonumber \right. \\*
&& \left. - \frac{1}{\sqrt{{\bf k}^4 + 4M_W^2 {\bf k}^2}} \ln \left(
\frac{{\bf k}^2 + \sqrt{{\bf k}^4 + 4 M_W^2 {\bf k}^2}}{{\bf k}^2 -
\sqrt{{\bf k}^4 + 4M_W^2 {\bf k}^2}} \right)^2 \right].
\label{G_bcde}
\enqar
A di\-ver\-gen\-ci\-\'ak (va\-gyis az $\left[ {1 \over \epsilon} - \gamma + \ln
(4 \pi) \right]$-vel a\-r\'a\-nyos ta\-gok) el\-t\'a\-vo\-l\'{\i}\-t\'a\-sa u\-t\'an
\beqar
R_{B+C+D+E}^{nemabeli} & = & \frac{i g^4}{16 \pi^2} C(R) C(G) \left[
\frac{1}{{\bf k}^2 + M_W^2} \ln \left( \frac{\mu_0^2}{M_W^2} \right) -
\right. \nonumber \\*
&& \left. \frac{1}{\sqrt{{\bf k}^4 + 4M_W^2 {\bf k}^2}} \ln \left(
\frac{{\bf k}^2 + \sqrt{{\bf k}^4 + 4 M_W^2 {\bf k}^2}}{{\bf k}^2 -
\sqrt{{\bf k}^4 + 4M_W^2 {\bf k}^2}} \right)^2 \right]
\label{rbcde}
\end{eqnarray}
ma\-rad. Mi\-e\-l\H ott eh\-hez hoz\-z\'a\-ad\-n\'ank a(z \ms) re\-nor\-m\'alt egy-hu\-rok ren\-d\H u
m\'er\-t\'ek\-bo\-zon-pro\-pa\-g\'a\-tor\-b\'ol, il\-let\-ve a tad\-po\-le-gr\'a\-fok\-b\'ol a\-d\'o\-d\'o
j\'a\-ru\-l\'e\-kot, hogy (v\'eg\-re) meg\-kap\-juk a szta\-ti\-kus kvark po\-ten\-ci\-\'alt,
\'{\i}r\-juk fel a 0 t\"o\-me\-g\H u e\-set\-re is a k\'et-bo\-zon-cse\-r\'es gr\'a\-fok
re\-nor\-m\'alt j\'a\-ru\-l\'e\-k\'at.
\paragraph{$M_W = 0$}
\beqar
G_{B+C+D+E}^{nemabeli} & = & \frac{i}{8 \pi^2 {\bf k}^2} g^4 C(R) C(G)
\left[ \frac{1}{\epsilon} - \gamma + \ln (4 \pi) + \ln \left(
\frac{\mu_0^2}{{\bf k}^2} \right) \right].
\enqar
A di\-ver\-gen\-ci\-\'ak \ms \ ki\-k\"u\-sz\"o\-b\"o\-l\'e\-se a k\"o\-vet\-ke\-z\H o egy\-sze\-r\H u
e\-red\-m\'eny\-re ve\-zet:
\beqar
R_{B+C+D+E}^{nemabeli} & = & \frac{i}{8 \pi^2 {\bf k}^2} g^4 C(R) C(G) \ln
\left( \frac{\mu_0^2}{{\bf k}^2} \right). \label{qcdrenbos}
\end{eqnarray}
\subsection{Az im\-pul\-zus\-t\'er\-be\-li po\-ten\-ci\-\'al}
\subsubsection*{$M=0$, QCD \label{qcderedmeny}}
Az egy-hu\-rok-szin\-t\H u re\-nor\-m\'alt m\'er\-t\'ek\-bo\-zon-pro\-pa\-g\'a\-tor \ref{qcd}
f\"ug\-ge\-l\'ek\-ben me\-ga\-dott k\'ep\-le\-te a\-lap\-j\'an a tisz\-ta SU(3)
m\'er\-t\'e\-kel\-m\'e\-let\-be\-li szta\-ti\-kus kvark po\-ten\-ci\-\'al
\beqar
R^{QCD} & = & ig^2 C(R) \frac{1}{{\bf k}^2} + \frac{i}{8 \pi^2
{\bf k}^2} g^4 C(R) C(G) \ln \left( \frac{\mu_0^2}{{\bf k}^2}
\right) \nonumber \\
&& + \frac{i}{(4 \pi)^2} g^4 C(G) C(R) \frac{1}{{\bf k}^2} \left(
\frac53 \ln \frac{\mu_0^2}{{\bf k}^2} + \frac{31}{9} \right)
\nonumber \\
& = & {\frac{ig^2 C(R)}{{\bf k}^2} \left[ 1 + \frac{g^2 C(G)}{16
\pi^ 2} \left( \frac{11}{3} \ln \frac{\mu_0^2}{{\bf k}^2} +
\frac{31}{9} \right) \right],}
\end{eqnarray}
a\-mi\-vel si\-ke\-r\"ult rep\-ro\-du\-k\'al\-ni az i\-ro\-da\-lom\-b\'ol is\-mert egy-hu\-rok
szin\-t\H u e\-red\-m\'enyt \cite{fis, mark, app, schr}.
\subsubsection*{$M \neq 0$, SU(2)--Higgs-mo\-dell}
Az egy\-hu\-rok-ren\-d\H u (re\-nor\-m\'alt) m\'er\-t\'ek\-bo\-zon-pro\-pa\-g\'a\-tor\-ra \'es a
tad\-po\-le-gr\'a\-fok j\'a\-ru\-l\'e\-k\'a\-ra vo\-nat\-ko\-z\'o k\'ep\-le\-tek \'es a (\ref{rbcde})
k\'ep\-let a\-lap\-j\'an az im\-pul\-zus\-t\'er\-ben fe\-l\'{\i}rt SU(2)--Higgs-mo\-dell\-be\-li
szta\-ti\-kus kvark po\-ten\-ci\-\'al:
\beqar \label{mom_pot}
&\dst{V_{{\mathrm{1-loop}}}(k)=- \frac{3g^4}{32\pi^2}\frac{1}{k^2+
M_W^2}} \nonumber \\*
&\dst{\left\{\frac{k^2+M_W^2}{k}\frac{2}{\sqrt{k^2+4M_W^2}}
\log \frac{\sqrt{k^2+4M_W^2}-k}{\sqrt{k^2+4M_W^2}+k} + \right.}
\nonumber \\*
&\dst{\frac{1}{k^2+M_W^2} \left[\frac{1}{24R_{HW}^2}\left(86R_{HW}^2
k^2 -9(6-3 R_{HW}^2 +R_{HW}^4 )M_W^2\right)\log\frac{\mu_0^2}{M_W^2}
\right.} \nonumber \\*
&\dst{+ \frac{1}{8} (13 k^2-20 M_W^2 ) F(k^2;M_W^2,M_W^2)}  \nonumber
\\
&\dst{-\frac{1}{24}\left( (R_{HW}^2-1)^2 \frac{M_W^4}{k^2} +k^2 +
2(R_{HW}^2-5) M_W^2\right) F(k^2;M_W^2,M_H^2)} \nonumber \\
&\dst{+\frac{R_{HW}^2 \cdot \log R_{HW}}{12(R_{HW}^2-1)} \left( k^2+
(9R_{HW}^2-17)M_W^2\right)} \nonumber \\
&\dst{\left. \left. +\frac{1}{72R_{HW}^2}\left( R_{HW}^2 k^2+3 (-18+
R_{HW}^2-11 R_{HW}^4)M_W^2\right) \right] \right\} },
\enqar
a\-hol $F(k^2;m_1^2,m_2^2) $ a t\"o\-me\-ges el\-m\'e\-le\-tek
hu\-ro\-kin\-teg\-r\'al\-ja\-i\-b\'ol j\'ol is\-mert
\beqar \label{fv}
F(k^2;m_1^2,m_2^2)=1+\frac{m_1^2 +m_2^2 }{m_1^2 -m_2^2 } \log
\frac{m_1}{m_2} + \frac{m_1^2 -m_2^2 }{k^2} \log \frac{m_1}{m_2}
\nonumber \\
+\frac{1}{k^2} \sqrt{ (m_1 +m_2 )^2 +k^2 )((m_1 -m_2 )^2 +k^2 )} \log
\dst{\frac{1-\sqrt{\frac{(m_1 -m_2 )^2 +k^2}{(m_1 +m_2 )^2 +k^2 }}}
{1+\sqrt{\frac{(m_1 -m_2 )^2 +k^2}{(m_1 +m_2 )^2 +k^2 }}}}.
\enqar
\begin{itemize}
\item E\-red\-m\'e\-nyem me\-ge\-gye\-zik M.~La\-i\-ne \'al\-ta\-l\'a\-nos $R_\xi$
m\'er\-t\'ek\-ben v\'eg\-re\-haj\-tott sz\'a\-mo\-l\'a\-s\'a\-nak e\-red\-m\'e\-ny\'e\-vel.
\item A $k \to 0$ inf\-ra\-v\"o\-r\"os ha\-t\'a\-re\-set\-ben a k\'ep\-let\-ben t\"obb
di\-ver\-gens tag is sze\-re\-pel, e\-zek a\-zon\-ban v\'a\-ra\-ko\-z\'a\-sunk\-nak meg\-fe\-le\-l\H o\-en
ki\-ej\-tik egy\-m\'ast.
\end{itemize}
\chapter{A ko\-or\-di\-n\'a\-ta\-t\'er\-be\-li po\-ten\-ci\-\'al}
\fancyhead[CE]{\hst{\thechapter{}.\ fe\-je\-zet \quad A
ko\-or\-din\'a\-tat\'er\-be\-li po\-ten\-ci\'al}}
\section{Fo\-u\-ri\-er-transz\-for\-m\'a\-ci\-\'o }
Az SU(2)--Higgs szta\-ti\-kus po\-ten\-ci\-\'al ki\-sz\'a\-m\'{\i}\-t\'a\-s\'a\-nak el\-s\H od\-le\-ges
mo\-ti\-v\'a\-ci\-\'o\-ja a kon\-ti\-nu\-um-t\'e\-rel\-m\'e\-let\-ben \'es a r\'acs\-t\'e\-rel\-m\'e\-let\-ben
hasz\-n\'a\-la\-tos csa\-to\-l\'a\-si \'al\-lan\-d\'ok \"ossze\-ve\-t\'e\-se volt. Eh\-hez a\-zon\-ban
ko\-or\-di\-n\'a\-ta\-t\'er\-be\-li ki\-fe\-je\-z\'es\-re van sz\"uk\-s\'eg, hogy ar\-ra a
Som\-mer-f\'e\-le de\-fi\-n\'{\i}\-ci\-\'ot \cite{som94},
\beq
g^2 \propto - x^2 \left. \dperd{V}{x} \right|_{x=\mathrm{const.}}
\enq
vagy en\-nek r\'acs\-t\'e\-rel\-m\'e\-le\-ti meg\-fe\-le\-l\H o\-j\'et \cite{fod94}
al\-kal\-maz\-has\-suk.

A Fo\-u\-ri\-er-transz\-for\-m\'a\-ci\-\'on k\'{\i}\-v\"ul egy t\'a\-vol\-s\'ag sze\-rin\-ti
dif\-fe\-ren\-ci\-\'a\-l\'ast is el kell v\'e\-gez\-ni. A ko\-or\-di\-n\'a\-ta\-ten\-ge\-lyek
al\-kal\-mas meg\-v\'a\-lasz\-t\'a\-s\'a\-val a h\'a\-rom\-di\-men\-zi\-\'os
Fo\-u\-ri\-er-transz\-for\-m\'a\-ci\-\'o r\"o\-vid \'u\-ton vissza\-ve\-zet\-he\-t\H o
egy\-di\-men\-zi\-\'os\-ra. K\'e\-zen\-fek\-v\H o\-nek t\H u\-nik az a meg\-k\"o\-ze\-l\'{\i}\-t\'es, hogy
e\-zu\-t\'an (m\'eg im\-pul\-zus\-t\'er\-ben) a dif\-fe\-ren\-ci\-\'a\-l\'ast hajt\-juk v\'eg\-re,
majd a ka\-pott ki\-fe\-je\-z\'est Fo\-u\-ri\-er-transz\-for\-m\'al\-juk: ek\-kor a
$\dperd{}{z}$ dif\-fe\-ren\-ci\-\'a\-lo\-pe\-r\'a\-tor e\-gye\-d\"ul az $e^{ipz}$ tag\-ra hat,
a\-mi egy $i p$ szor\-z\'ot e\-red\-m\'e\-nyez. En\-nek e\-red\-m\'e\-nye\-k\'ep\-pen a\-zon\-ban az
$\left(\frac{1}{p^2 + M^2}\right)^2$-tel a\-r\'a\-nyos tag a
Fo\-u\-ri\-er-transz\-for\-m\'a\-ci\-\'o so\-r\'an m\'ar nem $p^2 dp$-vel, ha\-nem $p^3
dp$-vel szor\-z\'o\-dik, \'{\i}gy a ki\-fe\-je\-z\'es bo\-nyo\-lult\-s\'a\-ga mi\-att
el\-ke\-r\"ul\-he\-tet\-len nu\-me\-ri\-kus in\-teg\-r\'a\-l\'as $\infty$-be\-li fel\-s\H o
ha\-t\'a\-r\'a\-nak va\-la\-mely e\-l\'eg nagy v\'e\-ges \'er\-t\'ek\-kel t\"or\-t\'e\-n\H o
he\-lyet\-te\-s\'{\i}\-t\'e\-se nem le\-het\-s\'e\-ges.

\'Igy te\-h\'at e\-l\H o\-sz\"or a (h\'a\-rom\-r\'ol egy\-di\-men\-zi\-\'os\-ra re\-du\-k\'alt)
Fo\-u\-ri\-er-transz\-for\-m\'a\-ci\-\'ot kell v\'eg\-re\-haj\-ta\-ni, majd a ka\-pott
ki\-fe\-je\-z\'est (bo\-nyo\-lult\-s\'a\-ga mi\-att) nu\-me\-ri\-ku\-san kell dif\-fe\-ren\-ci\-\'al\-ni
a t\'a\-vol\-s\'ag sze\-rint.
\subsection{3 di\-men\-zi\-\'o \ra 1 di\-men\-zi\-\'o}
A 3 di\-men\-zi\-\'os Fo\-u\-ri\-er-transz\-for\-m\'a\-ci\-\'o he\-lyett e\-le\-gen\-d\H o 1
di\-men\-zi\-\'o\-sat v\'eg\-re\-haj\-ta\-ni, mi\-vel a transz\-for\-m\'a\-lan\-d\'o ki\-fe\-je\-z\'es\-ben
a h\'ar\-ma\-sim\-pul\-zus csak n\'egy\-ze\-tes a\-lak\-ban, te\-h\'at ska\-l\'a\-ris
kom\-bi\-n\'a\-ci\-\'o\-ban buk\-kan fel. E\-l\H o\-sz\"or a
\beq
\int\! \int\! \int_{-\infty}^{\infty} \frac{1}{(2 \pi)^3} e^{i
\mathbf{k \cdot r}} \cdot f(k^2)\ dk_x\, dk_y\, dk_z \label{polarba}
\enq
ki\-fe\-je\-z\'est kell po\-l\'ar\-ko\-or\-di\-n\'a\-ta\-rend\-szer\-be \'a\-t\'{\i}r\-ni. Eh\-hez a
de\-r\'ek\-sz\"o\-g\H u ko\-or\-di\-n\'a\-ta\-ten\-ge\-lye\-ket meg\-v\'a\-laszt\-hat\-juk \'ugy, hogy a
szta\-ti\-kus kvark\-b\'ol az an\-tik\-vark\-ba mu\-ta\-t\'o $\mathbf r$ hely\-vek\-tor\-nak
csak $x$-kom\-po\-nen\-se le\-gyen. Ek\-kor a
\beq
k_x = k \cos \vartheta, \quad k_y = k \sin \vartheta \sin \varphi,
\quad k_z = k \sin \vartheta \cos \varphi,
\enq
v\'a\-lasz\-t\'as\-sal
\beq
(\ref{polarba}) = \int_0^{2 \pi} d \varphi \int_0^\pi d \vartheta \,
\sin \vartheta \int_0^\infty dk\, k^2\, e^{i k r \cos \vartheta}\,
f(k^2).
\enq

A $\varphi$ sze\-rin\-ti in\-teg\-r\'a\-l\'as egy tri\-vi\-\'a\-lis $2 \pi$
szor\-z\'o\-fak\-tort ad.

A $\vartheta$ sze\-rin\-ti in\-teg\-r\'a\-l\'as is k\"onnyen v\'eg\-re\-hajt\-ha\-t\'o:
\beq
\int_0^\pi d \vartheta \, \sin \vartheta \rightarrow \int_1^{-1} d(\cos
\vartheta),
\enq
\'{\i}gy
\begin{eqnarray}
(\ref{polarba}) & = & 2 \pi \int_0^\infty dk\, k^2 \int_1^{-1} dw\,
e^{ikx\cdot w} f(k^2) = 2 \pi \int_0^\infty dk \, k^2 \frac{e^{-ikx}
- e^{ikx}}{ikx}\, f(k^2) \nonumber \\*
& = & \frac{-4 \pi}{x} \int_0^\infty dk \, k\, \sin(kx)\, f(k^2).
\end{eqnarray}
A h\'a\-rom\-di\-men\-zi\-\'os in\-teg\-r\'al egy\-di\-men\-zi\-\'os\-ra va\-l\'o vissza\-ve\-ze\-t\'e\-se
te\-h\'at (a $(2 \pi)^{-3}$ fak\-tor be\-ol\-vasz\-t\'a\-s\'a\-val e\-gy\"utt) a
\newcommand{\jac}{{\mathrm{Jac}}}
\beq
\jac = - \frac{\pi\, k \sin(kx)}{2 \pi^3 x}
\enq
Ja\-co\-bi-de\-ter\-mi\-n\'ans\-sal va\-l\'o szor\-z\'as\-sal \'{\i}r\-ha\-t\'o le -- e\-zu\-t\'an m\'ar
csak a $k$ sze\-rin\-ti in\-teg\-r\'a\-l\'ast kell el\-v\'e\-gez\-ni.

\section{A fag\-r\'af szin\-t\H u j\'a\-ru\-l\'ek}
\fancyhead[CO]{\hst{\thesection \quad A fagr\'af szint\H{u}
j\'a\-rul\'ek}}
Az
\beq
R_1 = i \frac{3}{4} \frac{g^2}{k^2 + M_W^2}.
\enq
fag\-r\'af szin\-t\H u po\-ten\-ci\-\'al\-j\'a\-ru\-l\'ek a\-na\-li\-ti\-ku\-san is k\"onnyen
ki\-\'er\-t\'e\-kel\-he\-t\H o (pl.\ a re\-zi\-du\-um-t\'e\-tel se\-g\'{\i}t\-s\'e\-g\'e\-vel).
\beq
\int_0^\infty dk \, (R_1) \cdot \jac = - \frac{3 \pi\, g^2}{8\pi^3 x}
\cdot \re \int_0^\infty dk \, \frac{k \, e^{(ikx)}}{k^2 + M_W^2} =
-\frac{3}{16 \pi x} g^2 e^{-M_W x}, \label{fagr}
\enq
a\-hol $\re$ va\-la\-mely komp\-lex mennyi\-s\'eg va\-l\'os r\'e\-sz\'et je\-l\"o\-li.

A ka\-pott ki\-fe\-je\-z\'es $-1$-sze\-re\-s\'et $x$ sze\-rint dif\-fe\-ren\-ci\-\'al\-va,
majd a t\'a\-vol\-s\'a\-got $x=1$ sze\-rint r\"og\-z\'{\i}t\-ve fag\-r\'af szin\-ten
\beq
\mathrm{Potenci\acute{a}l} = - \frac{3}{8 \pi e} \cdot g^2_{\mss} = -
0.043912 \cdot g^2_{\mss}
\enq
a\-d\'o\-dik.

A fen\-ti v\'a\-lasz\-t\'as a t\"o\-meg\-di\-men\-zi\-\'o\-j\'u mennyi\-s\'e\-gek va\-la\-mely $M^0$
t\"o\-meg\-pa\-ra\-m\'e\-ter\-rel t\"or\-t\'e\-n\H o di\-men\-zi\-\'ot\-la\-n\'{\i}\-t\'a\-s\'a\-nak fe\-lel meg.

Az egy\-hu\-rok\-ren\-d\H u kor\-rek\-ci\-\'o ezt
\beq
\mathrm{Potenci\acute{a}l} = -0.043912 \cdot g^2_{\mss} + ... \cdot
g^4_{\mss} \label{fa}
\enq
a\-lak\-ban m\'o\-do\-s\'{\i}t\-ja, a\-hon\-nan m\'ar csak egy l\'e\-p\'es a po\-ten\-ci\-\'al\-b\'ol
sz\'ar\-maz\-tat\-ha\-t\'o, il\-let\-ve az \ms \ csa\-to\-l\'a\-si \'al\-lan\-d\'o k\"o\-z\"ot\-ti
kap\-cso\-lat meg\-ha\-t\'a\-ro\-z\'a\-sa.

\section{Az egy\-hu\-rok-ren\-d\H u j\'a\-ru\-l\'ek}
\fancyhead[CO]{\hst{\thesection \quad Az egy\-hu\-rok-rend\H{u}
j\'a\-rul\'ek}}
\subsection{Nu\-me\-ri\-kus in\-teg\-r\'a\-l\'as}
Az im\-pul\-zus\-t\'er\-be\-li po\-ten\-ci\-\'al egy\-hu\-rok-ren\-d\H u tag\-j\'a\-nak bi\-zo\-nyos
r\'e\-sze\-i a\-na\-li\-ti\-ku\-san is Fo\-u\-ri\-er-transz\-for\-m\'al\-ha\-t\'o\-ak. I\-lyen t\"ob\-bek
k\"o\-z\"ott a di\-men\-zi\-\'os re\-gu\-la\-ri\-z\'a\-ci\-\'o so\-r\'an be\-ve\-ze\-tett $\mu_0$
t\"o\-meg\-pa\-ra\-m\'e\-tert tar\-tal\-ma\-z\'o tag.

B\'ar a $\mu_0$-f\"ug\-get\-len r\'esz\-b\H ol is le\-v\'a\-laszt\-ha\-t\'o n\'e\-h\'any,
a\-na\-li\-ti\-ku\-san ke\-zel\-he\-t\H o tag, ezt a sz\'et\-v\'a\-lasz\-t\'ast el\-ve\-tet\-tem,
u\-gya\-nis a \emph{Maple} prog\-ram\-mal t\"or\-t\'e\-n\H o nu\-me\-ri\-kus in\-teg\-r\'a\-l\'as
so\-r\'an a n\'e\-h\'any egy\-sze\-r\H u tag ki\-ik\-ta\-t\'a\-s\'a\-b\'ol fa\-ka\-d\'o
i\-d\H o\-nye\-re\-s\'eg i\-gen cse\-k\'ely. Nem \'{\i}gy j\'art el M.~La\-i\-ne \cite{la},
\'{\i}gy az \H o r\'esz\-ben a\-na\-li\-ti\-kus e\-red\-m\'e\-nye\-i\-vel va\-l\'o \"ossze\-ha\-son\-l\'{\i}\-t\'as
e\-red\-m\'e\-nyem he\-lyes\-s\'e\-g\'e\-nek \'u\-jabb el\-le\-n\H or\-z\'e\-s\'e\-re a\-dott
le\-he\-t\H o\-s\'e\-get.

V\'e\-g\"ul a $\mu_0$-f\"ug\-g\H o r\'eszt nu\-me\-ri\-ku\-san is in\-teg\-r\'al\-tam; en\-nek
e\-red\-m\'e\-nye vissza\-ad\-ta az a\-na\-li\-ti\-kus sz\'a\-mo\-l\'a\-s\'et, a\-mi a hasz\-n\'alt
nu\-me\-ri\-kus m\'od\-szer he\-lyes\-s\'e\-g\'et t\'a\-maszt\-ja a\-l\'a.
\bigskip

A nu\-me\-ri\-kus in\-teg\-r\'a\-l\'as so\-r\'an e\-l\H o\-sz\"or di\-men\-zi\-\'ot\-la\-n\'{\i}\-ta\-ni kell a
t\"o\-meg\-di\-men\-zi\-\'o\-j\'u mennyi\-s\'e\-get. Er\-re t\"obb le\-he\-t\H o\-s\'eg is van,
p\'el\-d\'a\-ul a fag\-r\'af-szin\-t\H u $W$-bo\-zon-t\"o\-meg ($M_W^0$), az
egy\-hu\-rok\-szin\-t\H u $W$-bo\-zon-t\"o\-meg ($M_W^1$), vagy a
r\'acs\-szi\-mu\-l\'a\-ci\-\'ok\-ban hasz\-n\'alt $M_{\mathrm{screen}}$ \'ar\-ny\'e\-ko\-l\'a\-si
t\"o\-meg. A k\"u\-l\"on\-b\"o\-z\H o t\"o\-meg\-pa\-ra\-m\'e\-te\-rek csak a csa\-to\-l\'a\-si
\'al\-lan\-d\'o ma\-ga\-sabb rend\-je\-i\-ben t\'er\-nek el egy\-m\'as\-t\'ol, \'{\i}gy
b\'ar\-me\-lyik\-kel hajt\-juk is v\'eg\-re a Fo\-u\-ri\-er-transz\-for\-m\'a\-ci\-\'ot, o\-lyan
e\-red\-m\'enyt ka\-punk, mely k\"onnyen \"ossze\-vet\-he\-t\H o egy m\'a\-sik
t\"o\-meg\-pa\-ra\-m\'etr\-rel v\'eg\-re\-haj\-tott sz\'a\-mo\-l\'as e\-red\-m\'e\-ny\'e\-vel.
Az el\-j\'a\-r\'ast e\-z\'ert a k\'e\-zen\-fek\-v\H o $M_W^0$ v\'a\-lasz\-t\'as mel\-lett
mu\-ta\-tom be r\'esz\-le\-te\-sen, b\'ar a fi\-zi\-ka\-i al\-kal\-ma\-z\'a\-sok\-hoz egy et\-t\H ol
el\-t\'e\-r\H o (de az itt be\-mu\-ta\-tan\-d\'o m\'od\-szer\-rel u\-gya\-n\'ugy ke\-zel\-he\-t\H o)
t\"o\-meg\-pa\-ra\-m\'e\-tert, az \'ar\-ny\'e\-ko\-l\'a\-si t\"o\-me\-get v\'a\-lasz\-tot\-tam (l\'asd a
\ref{csatall} sza\-kaszt).

A $V(x)$ po\-ten\-ci\-\'alt $x=1$ k\"o\-r\"ul t\"obb pont\-ban kell meg\-ha\-t\'a\-roz\-ni,
hogy az $x$ sze\-rin\-ti nu\-me\-ri\-kus dif\-fe\-ren\-ci\-\'a\-l\'as meg\-b\'{\i}z\-ha\-t\'o
e\-red\-m\'enyt ad\-jon. A fe\-la\-dat nem ma\-g\'a\-t\'ol \'er\-te\-t\H o\-d\H o: $x$ nem le\-het
t\'ul\-s\'a\-go\-san t\'a\-vol az 1-t\H ol, hi\-szen li\-ne\-\'a\-ris k\"o\-ze\-l\'{\i}\-t\'est
k\'{\i}\-v\'a\-nunk al\-kal\-maz\-ni. M\'as\-r\'eszt $x$ nem le\-het t\'ul\-s\'a\-go\-san k\"o\-zel
sem az 1-hez, hi\-szen az e\-gyes pon\-tok\-ban v\'eg\-re\-haj\-tott nu\-me\-ri\-kus
in\-teg\-r\'a\-l\'a\-sok hi\-b\'a\-i j\'o\-val ki\-seb\-bek kell le\-gye\-nek, mint a
k\"u\-l\"on\-b\"o\-z\H o pon\-tok\-ban ta\-l\'alt \'er\-t\'e\-kek k\"u\-l\"onb\-s\'e\-ge\-i. Az $x=0.96$,
$x=0.98$, $x=1.00$, $x=1.02$, $x=1.04$ v\'a\-lasz\-t\'as mind\-k\'et
fel\-t\'e\-tel\-nek meg\-fe\-lelt. A ko\-or\-di\-n\'a\-ta\-t\'er\-be\-li po\-ten\-ci\-\'alt az
in\-teg\-ran\-dus\-ban sze\-rep\-l\H o $R_{HW} = M_H/M_W$ h\'a\-nya\-dos ki\-lenc
k\"u\-l\"on\-f\'e\-le \'er\-t\'e\-ke e\-se\-t\'en k\'{\i}\-v\'an\-tuk meg\-ha\-t\'a\-roz\-ni.
\bigskip

Az e\-gyes nu\-me\-ri\-kus in\-teg\-r\'a\-l\'a\-sok so\-r\'an fel\-me\-r\"ul a k\'er\-d\'es: hol
le\-het le\-v\'ag\-ni az el\-vi\-leg $+ \infty$-ig fu\-t\'o in\-teg\-r\'alt? M\'as
sz\'o\-val: ho\-gyan hajt\-suk v\'eg\-re a nu\-me\-ri\-kus in\-teg\-r\'a\-l\'ast, hogy hi\-b\'a\-ja
ki\-csi le\-gyen \'es j\'ol ke\-zel\-he\-t\H o.

Azt ta\-l\'al\-tam leg\-c\'el\-sze\-r\H ubb\-nek, ha az osz\-cil\-l\'a\-l\'o in\-teg\-r\'alt
\'ugy bont\-juk t\"obb r\'esz\-re, hogy az egy\-m\'as u\-t\'a\-ni
r\'e\-szek el\-len\-ke\-z\H o e\-l\H o\-je\-l\H u j\'a\-ru\-l\'e\-kot ad\-ja\-nak, \'es e\-zen
j\'a\-ru\-l\'e\-kok ab\-szo\-l\'ut \'er\-t\'e\-ke a le\-he\-t\H o leg\-ki\-sebb.
Ab\-ban a tar\-to\-m\'any\-ban, a\-hol a f\"ugg\-v\'eny m\'ar las\-san le\-cseng, a
Ja\-co\-bi-de\-ter\-mi\-n\'ans\-b\'ol a\-d\'o\-d\'o szi\-nusz-f\"ugg\-v\'eny ha\-t\'a\-roz\-za meg az
in\-teg\-ran\-dus jel\-le\-g\'et. Ez annyit tesz, hogy m\'{\i}g a f\"ugg\-v\'eny
z\'e\-rus\-he\-lye\-i eg\-zak\-tan me\-ge\-gyez\-nek a szi\-nusz-f\"ugg\-v\'eny gy\"o\-ke\-i\-vel, a
ma\-xi\-mu\-mok \'es a mi\-ni\-mu\-mok is a meg\-k\"o\-ve\-telt pon\-tos\-s\'a\-gi
ha\-t\'a\-ron be\-l\"ul a szi\-nusz-f\"ugg\-v\'eny sz\'el\-s\H o\-\'er\-t\'e\-ke\-i\-vel e\-gyez\-nek
meg. K\'et\-f\'e\-le lo\-gi\-kus v\'a\-lasz\-t\'as le\-het\-s\'e\-ges. Egy\-r\'eszt
in\-teg\-r\'al\-ha\-tunk null\-hely\-t\H ol null\-he\-lyig, \'ugy, hogy a
szi\-nusz\-f\"ugg\-v\'eny f\'e\-le\-g\'esz sz\'a\-m\'u pe\-ri\-\'o\-dust ha\-lad e\-l\H o\-re egy
in\-teg\-r\'a\-l\'a\-si tar\-to\-m\'a\-nyon be\-l\"ul -- ek\-kor az egy\-m\'as u\-t\'a\-ni
in\-ter\-val\-lu\-mok j\'a\-ru\-l\'e\-ka nyil\-v\'an\-va\-l\'o\-an el\-len\-t\'e\-tes e\-l\H o\-je\-l\H u.
M\'as\-r\'eszt in\-teg\-r\'al\-ha\-tunk ma\-xi\-mum\-t\'ol mi\-ni\-mu\-mig. K\"onnyen
be\-l\'at\-ha\-t\'o, hogy a m\'a\-so\-dik le\-he\-t\H o\-s\'e\-get c\'el\-sze\-r\H u k\"o\-vet\-ni; az
egy\-m\'as u\-t\'a\-ni el\-len\-t\'e\-tes e\-l\H o\-je\-l\H u ne\-gyed\-pe\-ri\-\'o\-du\-sok \'{\i}gy majd\-nem
tel\-je\-sen ki\-ej\-tik egy\-m\'ast. (Az is j\'ol l\'at\-szik, hogy a m\'a\-sik
``lo\-gi\-kus v\'a\-lasz\-t\'as'' a leg\-rosszabb a f\'e\-le\-g\'esz-pe\-ri\-\'o\-du\-s\'u
in\-teg\-r\'a\-l\'a\-si in\-ter\-val\-lu\-mok k\"o\-z\"ott.)

Mi\-lyen nu\-me\-ri\-kus in\-teg\-r\'a\-l\'a\-si for\-mu\-l\'at al\-kal\-maz\-zunk? H\'any \'es
mi\-lyen hossz\'u in\-ter\-val\-lu\-mot kell fel\-ven\-ni? H\'any osz\-t\'o\-pon\-tot kell
fel\-ven\-ni az e\-gyes in\-ter\-val\-lu\-mok\-ban?

A t\'eg\-la\-lap-, vagy a tra\-p\'ez\-sza\-b\'aly se\-g\'{\i}t\-s\'e\-g\'e\-vel is kel\-l\H o
pon\-tos\-s\'ag \'er\-he\-t\H o el, a\-zon\-ban ek\-kor sok osz\-t\'o\-pont fel\-v\'e\-te\-le
sz\"uk\-s\'e\-ges: a fen\-ti nu\-me\-ri\-kus in\-teg\-r\'a\-l\'a\-si m\'od\-sze\-rek\-kel $h^2$
pon\-tos\-s\'ag \'er\-he\-t\H o el (a\-hol $h$ a szom\-sz\'e\-dos osz\-t\'o\-pon\-tok k\"oz\-ti
t\'a\-vol\-s\'ag). Bo\-nyo\-lul\-tabb for\-mu\-l\'ak e\-se\-t\'en az in\-teg\-r\'a\-l\'a\-si
in\-ter\-val\-lum sz\'e\-le\-in le\-v\H o n\'e\-h\'any pon\-tot k\"u\-l\"on\-b\"o\-z\H o
s\'uly\-fak\-to\-rok\-kal kell fi\-gye\-lem\-be ven\-ni; nyil\-v\'an\-va\-l\'o\-an n\'e\-h\'any\-sz\'az
osz\-t\'o\-pont e\-se\-t\'en min\-tegy t\'{\i}z ``sz\'el\-s\H o'' pont i\-lyen fi\-gye\-lem\-be
v\'e\-te\-le el\-ha\-nya\-gol\-ha\-t\'o g\'e\-pi\-d\H o-n\"o\-ve\-ke\-d\'es\-sel j\'ar. \'Igy a
Nu\-me\-ri\-cal Re\-ci\-pes \cite{numrec} c.\ k\"onyv\-ben ta\-l\'al\-ha\-t\'o $h^5$
pon\-tos\-s\'a\-g\'u k\'ep\-le\-tet v\'a\-lasz\-tot\-tam (l\'asd a\-l\'abb).

Egy 40 Mbyte me\-m\'o\-ri\-\'a\-val ren\-del\-ke\-z\H o sze\-m\'e\-lyi sz\'a\-m\'{\i}\-t\'o\-g\'e\-pen \'ugy
ta\-l\'al\-tam, hogy nagy\-s\'ag\-ren\-di\-leg 1000 osz\-t\'o\-pont ve\-he\-t\H o fel
a\-n\'el\-k\"ul, hogy a \emph{Maple}-nek me\-m\'o\-ri\-a\-ke\-ze\-l\'e\-si ne\-h\'e\-zs\'e\-ge\-i
len\-n\'e\-nek. A di\-men\-zi\-\'ot\-la\-n\'{\i}\-tott ar\-gu\-men\-tu\-m\'u in\-teg\-ran\-dust $15 \pi$
hossz\'u\-s\'a\-g\'u in\-ter\-val\-lu\-mok\-ra osz\-tot\-tam fel -- a\-zaz az e\-gyes
in\-ter\-val\-lu\-mok fel\-s\H o ha\-t\'a\-r\'a\-nak $\left(15*N + \frac12 \right) \cdot
\pi$-t v\'a\-lasz\-tot\-tuk -- ek\-kor a nu\-me\-ri\-kus in\-teg\-r\'a\-l\'as hi\-b\'a\-ja
e\-le\-gen\-d\H o\-en ki\-csi volt. Er\-r\H ol \'ugy gy\H o\-z\H od\-tem meg, hogy 180, 360,
540 \'es 720 osz\-t\'o\-pont fel\-v\'e\-te\-l\'e\-vel haj\-tot\-tuk v\'eg\-re a nu\-me\-ri\-kus
in\-teg\-r\'a\-l\'ast; az \'{\i}gy ka\-pott e\-red\-m\'e\-nyek ki\-e\-l\'e\-g\'{\i}\-t\H o gyor\-sa\-s\'a\-g\'u
kon\-ver\-gen\-ci\-\'a\-ja a\-lap\-j\'an ar\-ra a k\"o\-vet\-kez\-te\-t\'es\-re ju\-tot\-tam, hogy
nincs sz\"uk\-s\'eg 720 osz\-t\'o\-pont\-n\'al t\"obb\-re.

4 in\-ter\-val\-lum fel\-v\'e\-te\-le gya\-kor\-la\-ti\-lag e\-le\-gen\-d\H o\-nek bi\-zo\-nyult. Ez az
\'al\-l\'{\i}\-t\'as a k\"o\-vet\-ke\-z\H o\-k\'ep\-pen \'er\-ten\-d\H o. Az in\-teg\-ran\-dus\-ban k\'et
v\'al\-toz\-tat\-ha\-t\'o pa\-ra\-m\'e\-ter sze\-re\-pel: az $x$ t\'a\-vol\-s\'ag \'es az
$R_{HW}$ t\"o\-me\-ga\-r\'any. K\'{\i}\-s\'er\-le\-tez\-ge\-t\'e\-sek so\-r\'an ki\-de\-r\"ult,
hogy a 4.\ in\-ter\-val\-lum u\-t\'a\-ni j\'a\-ru\-l\'e\-kok nem\-csak e\-l\'eg gyor\-san
csen\-ge\-nek le, ha\-nem a meg\-k\"o\-ve\-telt pon\-tos\-s\'a\-gon be\-l\"ul f\"ug\-get\-le\-nek
$R_{HW}$-t\H ol. \'Igy az 5.--20.\ in\-ter\-val\-lu\-mok j\'a\-ru\-l\'e\-k\'at az \"ot
k\"u\-l\"on\-b\"o\-z\H o $x$ \'er\-t\'ek e\-se\-t\'en e\-l\'eg volt egy\-szer--egy\-szer
ki\-sz\'a\-m\'{\i}\-ta\-ni; az \'{\i}gy ka\-pott \'er\-t\'ek se\-g\'{\i}t\-s\'e\-g\'e\-vel az el\-s\H o n\'egy
in\-ter\-val\-lum\-be\-li j\'a\-ru\-l\'e\-kok \"ossze\-g\'et kor\-ri\-g\'al\-ni tud\-tam.
V\'e\-ge\-ze\-t\"ul azt is meg\-fi\-gyel\-tem, hogy az egy\-m\'ast k\"o\-ve\-t\H o
in\-ter\-val\-lu\-mok\-ban egy\-re ke\-ve\-sebb pont is e\-l\'eg a meg\-k\"o\-ve\-telt
pon\-tosd\-s\'ag e\-l\'e\-r\'e\-s\'e\-hez -- a\-mi ter\-m\'e\-sze\-te\-sen az in\-teg\-ran\-dus
le\-csen\-g\H o jel\-le\-g\'e\-b\H ol fa\-kad. \'Igy az el\-s\H o in\-ter\-val\-lu\-mot 720, a
m\'a\-so\-di\-kat 540, a har\-ma\-di\-kat 360, a ne\-gye\-di\-ket 180 osz\-t\'o\-pont\-tal
in\-teg\-r\'al\-va a vizs\-g\'alt fi\-zi\-ka\-i pon\-tok\-ban meg tud\-tam ha\-t\'a\-roz\-ni a
ko\-or\-di\-n\'a\-ta\-t\'er\-be\-li po\-ten\-ci\-\'alt. Eh\-hez a k\"o\-vet\-ke\-z\H o u\-ta\-s\'{\i}\-t\'ast kap\-ta
a \emph{Maple}: \bigskip \\
\ttfamily
sup[0]:=0.001; f:='f': for f from 1 to 4 do; \\
sup[f]:=(15*f+0.5)*Pi/x; \\
inf[f]:= sup[f-1]; \\
q[f]:=sup[f]-inf[f]: \\
for h from 1 to (5-f) do: N[h]:= 180*h: \\
e\-red\-meny[f][h]:= \= e\-valf((q[f]/N[h])*((3/8)* e\-valf(subs(k=inf[f],
S2))+  \\
(7/6)* e\-valf(subs(k=inf[f]+q[f]/N[h], S2)) + \\
(23/24)* e\-valf(subs(k=inf[f]+2*q[f]/N[h], S2)) + \\
sum(e\-valf(subs(k=inf[f]+ j*q[f]/N[h], S2)),j=3..N[h]-3) +  \\
(23/24)* e\-valf(subs(k=inf[f]+ (N[h]-2)*q[f]/N[h], S2)) +  \\
(7/6)* e\-valf(subs(k = inf[f]+ (N[h]-1)*q[f]/N[h], S2)) +  \\
(3/8) * e\-valf(subs(k=inf[f]+ N[h]*q[f]/N[h], S2)))); \\
print(e\-red\-meny[f][h]); od; od;
\bigskip \\
\rmfamily
a\-hol {\texttt{S2}} az in\-teg\-ran\-dus.

A m\'a\-so\-dik t\'ab\-l\'a\-zat azt mu\-tat\-ja meg, hogy az el\-s\H o n\'egy
in\-terr\-val\-lum e\-se\-t\'e\-ben egy\-re ke\-ve\-sebb osz\-t\'o\-pont fi\-gye\-lem\-be\-v\'e\-te\-le is
e\-l\'eg a meg\-k\"o\-ve\-telt pon\-tos\-s\'ag\-hoz. A har\-ma\-dik t\'ab\-l\'a\-zat a
ko\-or\-di\-n\'a\-ta\-t\'er\-be\-li po\-ten\-ci\-\'al k\"u\-l\"on\-b\"o\-z\H o \'er\-t\'e\-ke\-it tar\-tal\-maz\-za
$R_{HW}$ \'es $x$ k\"u\-l\"on\-b\"o\-z\H o \'er\-t\'e\-ke\-i mel\-lett.

Az e\-gy\"utt\-ha\-t\'o\-kat \'ugy nor\-m\'al\-tam, hogy a po\-ten\-ci\-\'al a\-lak\-ja $x=1$-ben
\begin{equation}
V = \mathrm{const.} \cdot g^2 \left( 1 + g^2 * \ldots \right)
\end{equation}
le\-gyen; a fen\-ti e\-gy\"utt\-ha\-t\'ok \'{\i}\-ran\-d\'ok \ldots he\-ly\'e\-be.

\begin{table}
\begin{center}
\small
\begin{tabular}{|| r || r | r | r | r | r ||}
\hline
    & $x = 0.96$\sk & $x = 0.98$\sk & $x = 1.00$\sk & $x = 1.02$\sk &
$x = 1.04$ \sk	 \\
\hline
\hline
5.  & +.00001458006 & +.00001420744 & +.00001385120 & +.00001351032 &
+.00001318388 \\
6.  & --.00001018534 & --.00000992769 & --.00000968133 & --.00000944556 &
--.00000921974 \\
7.  & +.00000755733 & +.00000736764 & +.00000718624 & +.00000701261 &
+.00000684628 \\
8.  & --.00000585143 & --.00000570546 & --.00000556585 & --.00000543222 &
--.00000530419 \\
9.  & +.00000467690 & +.00000456082 & +.00000444979 & +.00000434350 &
+.00000424166 \\
10. & --.00000383144 & --.00000373675 & --.00000364617 & --.00000355946 &
--.00000347636 \\
11. & +.00000320127 & +.00000312245 & +.00000304704 & +.00000297485 &
+.00000290566 \\
12. & --.00000271822 & --.00000265150 & --.00000258768 & --.00000252657 &
--.00000246800 \\
13. & +.00000233927 & +.00000228202 & +.00000222725 & +.00000217480 &
+.00000212454 \\
14. & --.00000203619 & --.00000198648 & --.00000193892 & --.00000189338 &
--.00000184974 \\
15. & +.00000178975 & +.00000174616 & +.00000170446 & +.00000166452 &
+.00000162624 \\
16. & --.00000158652 & --.00000154796 & --.00000151106 & --.00000147573 &
--.00000144186 \\
17. & +.00000141682 & +.00000138245 & +.00000134957 & +.00000131807 &
+.00000128788 \\
18. & --.00000127359 & --.00000124275 & --.00000121324 & --.00000118498 &
--.00000115789 \\
19. & +.00000115153 & +.00000112369 & +.00000109705 & +.00000107154 &
+.00000104708 \\
20. & --.00000104661 & --.00000102135 & --.00000099717 & --.00000097402 &
--.00000095182 \\
\hline
\hline
$\sum_0$ & +.818359 e-5   & +.797273 e-5   & +.777118 e-5   & +.757829 e-5
&  +.739362 e-5   \\
\hline
\hline
$\sum_1$ & +.873312 e-5   & +.850899 e-5   & +.8294735 e-5  & +.808968 e-5
& +.7893345 e-5  \\
$\sum_2$ & +.862392 e-5   & +.840247 e-5   & +.8190778 e-5  & +.798818 e-5
& +.7794198 e-5  \\
\hline
\hline
$\sum$	 & +.8679 e-5	  & +.8456 e-5	   & +.8243 e-5     & +.8039 e-5
& +.7844 e-5	 \\
\hline
\end{tabular}
\caption{\label{tail1}
Az el\-s\H o n\'egy in\-teg\-r\'a\-l\'a\-si in\-ter\-val\-lum u\-t\'a\-ni j\'a\-ru\-l\'e\-kok}
\end{center}
\end{table}
A \ref{tail1} t\'ab\-l\'a\-zat\-ban $\sum_0$ ad\-ja az 5.--20.\ in\-ter\-val\-lu\-mok
j\'a\-ru\-l\'e\-ka\-i\-nak \"ossze\-g\'et. Mi\-vel ez egy t\'eg\-la\-lap-\"osszeg, k\"onnyen
fi\-no\-m\'{\i}t\-ha\-t\'o: $\sum_1$ \'es $\sum_2$ tra\-p\'ez-sza\-b\'a\-lyon a\-la\-pu\-l\'o
fel\-s\H o il\-let\-ve al\-s\'o becs\-l\'es. A to\-v\'ab\-bi\-ak\-ban az e\-zek\-b\H ol ka\-pott
$\sum$ becs\-l\'est al\-kal\-maz\-zuk a nu\-me\-ri\-kus in\-teg\-r\'a\-l\'as so\-r\'an az el\-s\H o
n\'egy in\-teg\-r\'a\-l\'a\-si in\-ter\-val\-lum ki\-z\'a\-r\'o\-la\-gos fi\-gye\-lem\-be\-v\'e\-te\-l\'e\-b\H ol
fa\-ka\-d\'o hi\-ba kor\-ri\-g\'a\-l\'a\-s\'a\-ra. $\sum$ hi\-b\'a\-ja a ki\-\'{\i}rt u\-tol\-s\'o
ti\-ze\-des\-jegy\-ben leg\-fel\-jebb 5.

Az $R_{HW}=49/80$, $x=0.96$ \'er\-t\'e\-kek mel\-lett fel\-vett \ref{tail2}
t\'ab\-l\'a\-zat azt mu\-tat\-ja meg, hogy az el\-s\H o n\'egy in\-terr\-val\-lum
e\-se\-t\'e\-ben egy\-re ke\-ve\-sebb osz\-t\'o\-pont fi\-gye\-lem\-be\-v\'e\-te\-le is e\-l\'eg a
meg\-k\"o\-ve\-telt pon\-tos\-s\'ag\-hoz. Az e\-red\-m\'e\-nyek szem\-mel l\'at\-ha\-t\'o\-an
pon\-tos\-sab\-bak, mint a \ref{tail1} t\'ab\-l\'a\-zat\-be\-li \'er\-t\'e\-kek; en\-nek
meg\-fe\-le\-l\H o\-en a k\"u\-l\"on\-b\"o\-z\H o $R_{HW}$ t\"o\-me\-ga\-r\'a\-nyok\-ra \'es $x$
t\'a\-vol\-s\'a\-gok\-ra a\-d\'o\-d\'o v\'e\-ge\-red\-m\'enyt tar\-tal\-ma\-z\'o \ref{koopot}
t\'ab\-l\'a\-zat a\-da\-ta\-it is 7 ti\-ze\-des\-jegy\-re ad\-tam meg.

\begin{table}
\begin{center}
\small
\begin{tabular}{||r||r|r|r|r||}
\hline
 & 180 \skk & 360 \skk & 540 \skk & 720 \skk \\
\hline
\hline
1.    &  +.0514051901709534 &  +.0512451778906175 &  +.0512385118271583
& +.0512376695413435 \\
2.    &  --.0001132248658153 &	--.0001133186132436 &  --.0001133222837346
& \\
3.    &  +.0000422640841745 &  +.0000423012037429 &
& \\
4.    &  --.0000228928299680 & & & \\
\hline
5.--  & .8679 e-5 & & & \\
\hline
$\sum$& .05115253204  & & &\\
\hline
\end{tabular}
\caption{\label{tail2}
Az osz\-t\'o\-pon\-tok sz\'a\-m\'a\-nak n\"o\-ve\-l\'e\-s\'e\-nek ha\-t\'a\-sa az el\-s\H o
n\'egy in\-ter\-val\-lum j\'a\-ru\-l\'e\-k\'a\-ra}
\end{center}
\end{table}

A ko\-or\-di\-n\'a\-ta\-t\'er\-be\-li SU(2)--Higgs-po\-ten\-ci\-\'al k\"u\-l\"on\-b\"o\-z\H o $R_{HW}$
\'es $x$ \'er\-t\'e\-kek\-n\'el fel\-vett \'er\-t\'e\-ke\-it a \ref{koopot} t\'ab\-l\'a\-zat
fog\-lal\-ja \"ossze.
\begin{table}
\begin{center}
\begin{tabular}{||c||r|r|r|r|r||}
\hline
$R_{HW}$ & $x=0.96$\kk	& $x=0.98$\kk  & $x=1.00$\kk  & $x=1.02$\kk  &
$x=1.04$\kk  \\
\hline
$19/80$  & .0511525 & .0496946 & .0483044 & .0469767 &
.0457072 \\
$35/80$  & .0207696 & .0199157 & .0191173 & .0183698 &
.0176687 \\
$49/80$  & .0149091 & .0141733 & .0134907 & .0128565 &
.0122665 \\
$64/80$  & .0126144 & .0119262 & .0112902 & .0107017 &
.0101563 \\
$1$	 & .0115523 & .0108874 & .0102740 & .0097076 &
.0091840 \\
$1.2$	 & .0109418 & .0102910 & .0096915 & .0091385 &
.0086279 \\
$1.5$	 & .0101157 & .0094841 & .0089032 & .0083684 &
.0078755 \\
$2$	 & .0079148 & .0073306 & .0067960 & .0063064 &
.0058577 \\
$3$	 &-.0022005 &-.0025787 &-.0029117 &-.0032040 &
-.0034596 \\
\hline
\end{tabular}
\caption{\label{koopot} A ko\-or\-di\-n\'a\-ta\-t\'er\-be\-li po\-ten\-ci\-\'al}
\end{center}
\end{table}

Az in\-teg\-ran\-dus (\ref{mom_pot}) l\'at\-sz\'o\-lag szin\-gu\-l\'a\-ris $R_{HW}=1$-re;
vol\-ta\-k\'ep\-pen t\"obb di\-ver\-gens tag is fel\-l\'ep, me\-lyek\-nek \"ossze\-ge
v\'e\-ges lesz, a\-zon\-ban a \emph{Maple} e\-ze\-ket a di\-ver\-gen\-ci\-\'a\-kat ne\-he\-zen
tud\-ja ke\-zel\-ni. E\-z\'ert az $R_{HW}=1$-hez tar\-to\-z\'o pon\-tot az $R_{HW}=0.
9999$ \'es az $R_{HW}=1.0001$ pon\-tok\-ra ka\-pott e\-red\-m\'eny sz\'am\-ta\-ni
k\"o\-ze\-pe\-k\'ent sz\'a\-m\'{\i}\-tot\-tam ki.

\subsubsection*{A $\mu$-f\"ug\-g\H o tag}
Az e\-l\H o\-z\H o\-ek\-hez tel\-je\-sen ha\-son\-l\'o m\'od\-szer\-rel a po\-ten\-ci\-\'al
$\mu$-f\"ug\-g\H o tag\-j\'at is Fo\-u\-ri\-er-transz\-for\-m\'al\-tam. Vi\-l\'a\-gos a\-zon\-ban,
hogy ez a tag sz\'a\-munk\-ra sok\-kal ke\-v\'es\-b\'e b\'{\i}r k\"oz\-vet\-len fi\-zi\-ka\-i
je\-len\-t\'es\-sel, mint a $\mu$-f\"ug\-get\-len tag: mi\-vel nem k\'{\i}\-v\'a\-nunk
re\-nor\-m\'a\-l\'a\-si-cso\-port vizs\-g\'a\-la\-tot v\'e\-gez\-ni, nyu\-god\-tan \'el\-het\-n\'enk a
$\mu = M_W$ v\'a\-lasz\-t\'as\-sal -- ek\-kor a $\mu$-f\"ug\-g\H o tag j\'a\-ru\-l\'e\-ka 0.

A $\mu$-f\"ug\-g\H o tag a\-zon\-ban na\-gyon egy\-sze\-r\H u\-en v\'e\-gig\-sz\'a\-mol\-ha\-t\'o: a
fag\-r\'af-szin\-ten ki\-sz\'a\-molt (\ref{fagr}) mel\-lett e\-gyet\-len \'u\-jabb
in\-teg\-r\'al buk\-kan fel:
\beq
\re \int_0^\infty dk \, \frac{k \, e^{(ikx)}}{(k^2 + M_W^2)^2}
\label{rezi}
\enq
mely\-nek ki\-\'er\-t\'e\-ke\-l\'e\-s\'e\-hez a
\beq
\int_0^\infty \frac{k \cos k}{(k^2 + 1)^2} dk = \frac{\pi^2}{e}
\frac{2189}{450} = 0.1766194388
\enq
\"ossze\-f\"ug\-g\'est hasz\-n\'al\-juk fel. A ko\-or\-di\-n\'a\-ta\-t\'er\-be\-li po\-ten\-ci\-\'al\-ban
\beqar
\label{pot}
\frac{V(r)}{M_W}= - \frac{3g^2}{16\pi} \frac{\exp(-M_W^0 r)}{M_W r} +
\frac{g^4}{16\pi^2} \left(A+B\log(\mu^2/M_W^2)\right)
\enqar
sze\-rep\-l\H o $A$ \'es $B$ f\"ugg\-v\'e\-nye\-ket e\-z\'al\-tal me\-gad\-tuk. $M_W^0
=M_W-\delta M_W$; $\delta M_W$ az egy\-hu\-rok-ren\-d\H u t\"o\-meg\-kor\-rek\-ci\-\'o.
Mi\-vel $\delta M_W$ sk\'a\-la-f\"ug\-g\H o, $M_W^0$ is az.

Az e\-red\-m\'e\-nye\-in\-ket a \ref{pot_fig} \'es a \ref{pot_R_fig} \'ab\-ra
t\"un\-te\-ti fel. Itt a k\'et\-v\'al\-to\-z\'os f\"ugg\-v\'eny\-nek egy--egy
v\'al\-to\-z\'o\-j\'at r\"og\-z\'{\i}\-tett: az el\-s\H o e\-set\-ben az $R_{HW}$
t\"o\-me\-ga\-r\'any az e\-lekt\-ro\-gyen\-ge f\'a\-zi\-s\'at\-me\-net v\'eg\-pont\-j\'at jel\-lem\-z\H o
\'er\-t\'ek\-kel e\-gyen\-l\H o \cite{csik99}, a m\'a\-sik e\-set\-ben a
di\-men\-zi\-\'ot\-la\-n\'{\i}\-tott t\'a\-vol\-s\'a\-got egy\-s\'eg\-nyi.
\bef
\bc
\epsfig{file=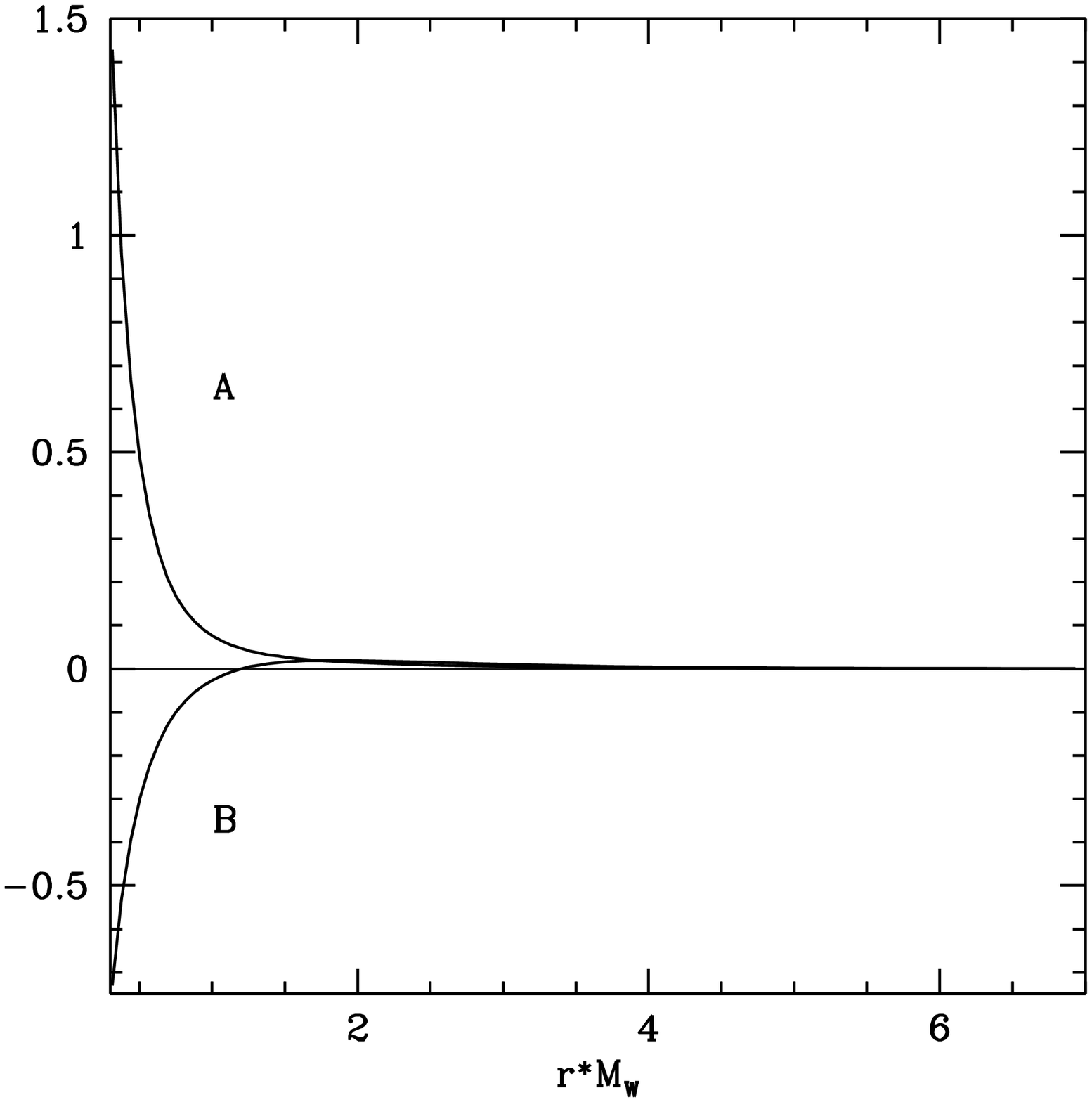,width=8.0cm}
\caption{\label{pot_fig}
{A $g^4/(16 \pi^2)$ tag szor\-z\'o\-f\"ugg\-v\'e\-nye -- A g\"or\-be --, il\-let\-ve a
$g^4/(16\pi^2) \log(\mu^2/M_W^2)$ ta\-g\'e -- B g\"or\-be -- mint a W
t\"o\-meg\-gel szor\-zott t\'a\-vol\-s\'ag f\"ugg\-v\'e\-nye. $R_{HW}$=0.8314.}}
\ec
\enf

\bef
\bc
\epsfig{file=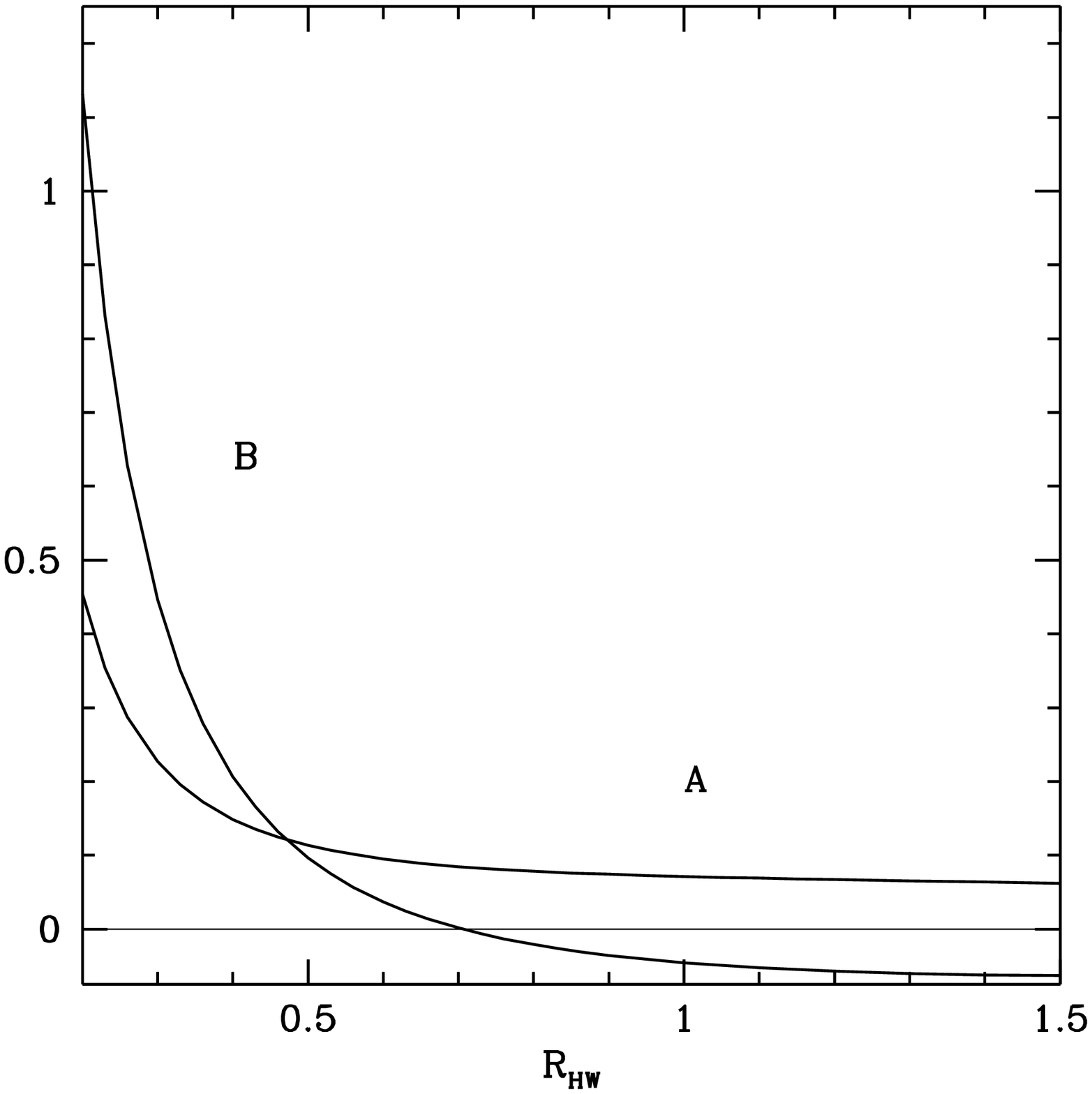,width=8.0cm}
\caption{\label{pot_R_fig}
{A $g^4/(16 \pi^2)$ tag szor\-z\'o\-f\"ugg\-v\'e\-nye -- A g\"or\-be --, il\-let\-ve a
$g^4/(16\pi^2) \log(\mu^2/M_W^2)$ ta\-g\'e -- B g\"or\-be -- mint $R_{HW} =
M_H / M_W$ f\"ugg\-v\'e\-nye, $x = M_W^{-1}$ t\'a\-vol\-s\'a\-g\'er\-t\'ek mel\-lett.}}
\ec
\enf
\subsection{Nu\-me\-ri\-kus dif\-fe\-ren\-ci\-\'a\-l\'as}
A csa\-to\-l\'a\-si \'al\-lan\-d\'o de\-fi\-n\'{\i}\-ci\-\'o\-j\'a\-ban a po\-ten\-ci\-\'al hely sze\-rin\-ti
de\-ri\-v\'alt\-ja fog sze\-re\-pel\-ni. Eh\-hez a \ref{koopot} t\'ab\-l\'a\-zat
e\-red\-m\'e\-nye\-in kell nu\-me\-ri\-kus dif\-fe\-ren\-ci\-\'a\-l\'ast v\'eg\-re\-haj\-ta\-ni. Eh\-hez a
k\"u\-l\"on\-b\"o\-z\H o $R_{HW}$ \'er\-t\'e\-kek\-hez tar\-to\-z\'o pont-\"o\-t\"o\-s\"ok\-re
m\'a\-sod\-fo\-k\'u po\-li\-no\-mo\-kat il\-lesz\-tet\-tem, me\-lyek\-nek $x=1$-be\-li
me\-re\-dek\-s\'e\-ge ad\-ta meg a de\-ri\-v\'al\-tat. Mint a nu\-me\-ri\-kus
dif\-fe\-ren\-ci\-\'a\-l\'as\-n\'al \'al\-ta\-l\'a\-ban, az igy ka\-pott \'er\-t\'ek j\'o\-val
ke\-v\'es\-b\'e pon\-tos, mint a po\-ten\-ci\-\'al\-ra ka\-pott \'er\-t\'e\-kek, a\-zon\-ban m\'eg
\'{\i}gy is gya\-kor\-la\-ti c\'el\-ja\-ink\-hoz meg\-fe\-le\-l\H o\-en pon\-tos \'er\-t\'e\-ke\-ket
kap\-tunk. Az e\-red\-m\'e\-nye\-ket a \ref{diff} t\'ab\-l\'a\-zat fog\-lal\-ja \"ossze.
\begin{table}
\begin{center}
\begin{tabular}{||c||r||}
\hline
$R_{HW}$ & $\dst{\left(-\dperd{V}{x}\right)_{x=1}}$ \\
\hline
\hline
$19/80$  &  0.06804(3)	\\
$35/80$  &  0.03874(3)	\\
$49/80$  &  0.03301(3)	\\
$64/80$  &  0.03070(3)	\\
$1$	 &  0.02958(3)	\\
$1.2$	 &  0.02890(3)	\\
$1.5$	 &  0.02798(3)	\\
$2$	 &  0.02569(3)	\\
$3$	 &  0.01572(3)	\\
\hline
\end{tabular}
\caption{\label{diff}
A po\-ten\-ci\-\'al $\mu$-f\"ug\-get\-len r\'e\-sz\'e\-nek hely sze\-rin\-ti
nu\-me\-ri\-kus de\-ri\-v\'alt\-ja.}
\end{center}
\end{table}

\section{A po\-ten\-ci\-\'al a\-lap\-j\'an de\-fi\-ni\-\'alt csa\-to\-l\'a\-si \'al\-lan\-d\'o
\label{csatall}}
\fancyhead[CO]{\hst{\thesection \quad A po\-ten\-ci\-\'al a\-lap\-j\'an de\-fi\-ni\-\'alt
csa\-to\-l\'a\-si \'al\-lan\-d\'o}}
A po\-ten\-ci\-\'al\-ra ka\-pott (\ref{fa}) t\'{\i}\-pu\-s\'u ki\-fe\-je\-z\'es a\-lap\-j\'an
de\-fi\-ni\-\'al\-ni k\'{\i}\-v\'a\-nunk egy $g_R(r)$ csa\-to\-l\'a\-si \'al\-lan\-d\'ot, mely a
sz\'a\-mo\-l\'as\-ban hasz\-n\'alt $g_{\mss}$ csa\-to\-l\'a\-si \'al\-lan\-d\'o\-val
\beq
g^2_R(r) = g_{\mss}^2(\mu) \left( 1 + g_{\mss}^2(\mu) \cdot \ldots
\right)
\enq
vi\-szony\-ban \'all. A Co\-u\-lomb-po\-ten\-ci\-\'al (vagy az en\-n\'el va\-la\-mi\-vel
\'al\-ta\-l\'a\-no\-sabb QCD-po\-ten\-ci\-\'al) \'es a be\-l\H o\-le sz\'ar\-maz\-tat\-ha\-t\'o
e\-lekt\-ro\-mos t\"ol\-t\'es k\"oz\-ti kap\-cso\-lat l\'e\-nye\-g\'e\-ben e\-gy\'er\-tel\-m\H u
\beq \label{csat}
g_R^2(r) = \frac{1}{C_F} \frac{\dst{- \dperd{V}{r}}}{\dst{\dperd{}{r}
\int \frac{d^3k}{(2 \pi)^3} \frac{\exp({i \strut{\vec k} \strut{\vec
r}})}{k^2}}}
\enq
kap\-cso\-la\-ta l\'at\-sz\'o\-lag a t\"o\-me\-ges el\-m\'e\-let\-re is k\"onnyen \'at\-vi\-he\-t\H o: a
ne\-ve\-z\H o in\-teg\-ran\-du\-s\'a\-ba a pro\-pa\-g\'a\-tort kell be\-\'{\i}r\-ni. A $k^2 + m^2$
tag\-ban sze\-rep\-l\H o t\"o\-meg a\-zon\-ban t\"obb\-f\'e\-le\-k\'epp is meg\-v\'a\-laszt\-ha\-t\'o:
a fag\-r\'af- il\-let\-ve az egy\-hu\-rok-szin\-t\H u W-t\"o\-meg e\-gya\-r\'ant be\-\'{\i}r\-ha\-t\'o
i\-de.

Az a\-l\'ab\-bi\-ak\-ban egy har\-ma\-dik pa\-ra\-m\'e\-ter v\'a\-lasz\-tunk, a
r\'acs\-t\'e\-rel\-m\'e\-let a\-lap\-j\'an de\-fi\-ni\-\'alt \emph{\'ar\-ny\'e\-ko\-l\'a\-si
t\"o\-me\-get} \cite{fod94}, mint\-hogy a fen\-ti sz\'a\-mo\-l\'as f\H o mo\-ti\-v\'a\-ci\-\'o\-ja
a per\-tur\-ba\-t\'{\i}v \'es nem\-per\-tur\-ba\-t\'{\i}v e\-red\-m\'e\-nyek \"ossze\-ve\-t\'e\-se.

A r\'a\-cson k\"onnyen m\'er\-he\-t\H o $r,t$ ki\-ter\-je\-d\'e\-s\H u Wil\-son-hur\-kok\-b\'ol $t
\to \infty$ ext\-ra\-po\-l\'a\-l\'as\-sal kap\-hat\-juk meg a szta\-ti\-kus po\-ten\-ci\-\'alt,
$r$ k\"u\-l\"on\-b\"o\-z\H o \'er\-t\'e\-ke\-i mel\-lett. Az \'{\i}gy ka\-pott
$V_{\mrm{latt}}(r)$ f\"ugg\-v\'enyt egy n\'e\-h\'any-pa\-ra\-m\'er\-te\-res
ki\-fe\-je\-z\'es\-sel k\'{\i}\-v\'an\-juk le\-\'{\i}r\-ni; leg\-c\'el\-sze\-r\H ubb v\'a\-lasz\-t\'as a
Yu\-ka\-wa-po\-ten\-ci\-\'al n\'egy\-pa\-ra\-m\'e\-te\-res r\'acs\-v\'al\-to\-za\-ta \cite{fod94}. A
t\'a\-vol\-s\'ag sze\-rin\-ti ex\-po\-nen\-ci\-\'a\-lis le\-csen\-g\'est jel\-lem\-z\H o pa\-ra\-m\'e\-ter az
\'ar\-ny\'e\-ko\-l\'a\-si t\"o\-meg, me\-lyet $M_{\mrm{latt}}$-tal je\-l\"o\-l\"unk.
A k\"u\-l\"on\-b\"o\-z\H o $r$ \'er\-t\'e\-kek mel\-lett (r\'a\-cson) m\'ert szta\-ti\-kus
po\-ten\-ci\-\'al \'er\-t\'e\-kek diszk\-r\'et $r$ sze\-rin\-ti de\-ri\-v\'alt\-ja a\-lap\-j\'an,
a (\ref{csat}) k\'ep\-let\-hez ha\-son\-l\'o\-an de\-fi\-ni\-\'al\-ha\-t\'o egy csa\-to\-l\'a\-si
\'al\-lan\-d\'ot, me\-lyet $g_{\mrm{latt}}$ fog je\-l\"ol\-ni.

A fen\-ti de\-fi\-n\'{\i}\-ci\-\'o k\"onnyen \'at\-vi\-he\-t\H o a per\-tur\-ba\-t\'{\i}v sz\'a\-mo\-l\'as\-ra:
az egy\-hu\-rok-szin\-t\H u po\-ten\-ci\-\'al\-ra n\'egy\-pa\-ra\-m\'e\-te\-res Yu\-ka\-wa-po\-ten\-ci\-\'alt
il\-leszt\-he\-t\"unk, mely\-ben az ex\-po\-nen\-ci\-\'a\-lis le\-csen\-g\'est egy per\-tur\-ba\-t\'{\i}v
``\'ar\-ny\'e\-ko\-l\'a\-si t\"o\-meg'' ha\-t\'a\-roz\-za meg, $M_{\scr}$. Ek\-kor a
csa\-to\-l\'a\-si \'al\-lan\-d\'ot
\beq
g_R^2(r) = \frac{1}{C_F} \frac{\dst{- \dperd{V}{r}}}{\dst{\dperd{}{r}
\int \frac{d^3k}{(2 \pi)^3} \frac{\exp({i \strut{\vec k} \strut{\vec
r}})}{k^2+M_{\scr}^2}}}
\enq
de\-fi\-ni\-\'al\-ja, mely az \ms\ csa\-to\-l\'a\-si \'al\-lan\-d\'o\-val a k\"o\-vet\-ke\-z\H o
kap\-cso\-lat\-ban \'all:
\beq \label{cd}
\dst{g_R^2(r) = g_{\mss}^2(\mu) \left[1 + \frac12 \left( 1 -
\frac{M_W^0}{M_{\scr}} \right) \right] + \frac{g_{\mss}^4(\mu)}{16
\pi^2} \left(C + D \log\frac{\mu^2}{M_W^2} \right)},
\enq
a\-hol az u\-tol\-s\'o tag\-ban sze\-rep\-l\H o $M_W$-t a\-k\'ar fag\-r\'af szin\-t\H u, a\-k\'ar
egy\-hu\-rok-szin\-t\H u W-t\"o\-meg\-nek v\'a\-laszt\-hat\-juk, hi\-szen a kor\-rek\-ci\-\'o csak
$g^6$ ren\-d\H u. A $C$ \'es $D$ f\"ugg\-v\'e\-nyek $R_{HW}$-t\H ol \'es az
\'ar\-ny\'e\-ko\-l\'a\-si t\"o\-meg\-t\H ol f\"ug\-ge\-nek.

Az egy\-hu\-rok-ren\-d\H u W-t\"o\-meg v\'a\-lasz\-t\'a\-s\'a\-nak szin\-t\'en ko\-moly e\-l\H o\-nye\-i
van\-nak \cite{la}. En\-nek o\-ka az, hogy a po\-ten\-ci\-\'al-ki\-fe\-je\-z\'es\-ben
sze\-rep\-l\H o $\mu$-f\"ug\-g\H o ta\-gok ki\-z\'a\-r\'o\-lag az egy-hu\-rok-szin\-t\H u
m\'er\-t\'ek\-bo\-zon-pro\-pa\-g\'a\-tor\-b\'ol a\-d\'od\-nak, me\-lyet egy
t\"o\-meg\-re\-nor\-m\'a\-l\'as\-sal is fi\-gye\-lem\-be ve\-he\-t\"unk. \'Igy a $\mu$-f\"ug\-g\'es
tel\-je\-sen ki\-k\"u\-sz\"o\-b\"ol\-he\-t\H o, \'es a k\'et csa\-to\-l\'a\-si \'al\-lan\-d\'o k\"o\-z\"ott
a
\beq
\dst{g_{\mathrm{Laine}}^2(M^{-1}) = g_{\mss}^2(M_W^1) \left[1 +
\frac12 \left( 1 - \frac{M}{M_W^1} \right) \right] +
\frac{g_{\mss}^4(M_W^1)}{16 \pi^2} f(R_{HW}) }
\enq
kap\-cso\-la\-tot kap\-juk.

A fen\-ti k\'et meg\-k\"o\-ze\-l\'{\i}\-t\'es ter\-m\'e\-sze\-te\-sen ek\-vi\-va\-lens; a vizs\-g\'alt
fi\-zi\-ka\-i pon\-tok\-ban a meg\-fe\-le\-l\H o f\"ugg\-v\'e\-nyek nu\-me\-ri\-kus el\-t\'e\-r\'e\-se
ki\-csi. Ezt j\'ol szem\-l\'e\-le\-te\-ti a \ref{couplings} t\'ab\-l\'a\-zat, mely a
r\'acsszi\-mu\-l\'a\-ci\-\'ok so\-r\'an hasz\-n\'alt k\"u\-l\"on\-b\"o\-z\H o Higgs-t\"o\-me\-gek\-hez
tar\-to\-z\'o \'ar\-ny\'e\-ko\-l\'a\-si t\"o\-me\-ge\-ket \'es csa\-to\-l\'a\-si \'al\-lan\-d\'o\-kat
fog\-lal\-ja \"ossze. $R_{HW}=0.8314$ a f\'a\-zi\-s\'at\-me\-ne\-ti v\'eg\-pont\-nak fe\-lel
meg. $T_c$ a f\'a\-zi\-s\'at\-me\-net kri\-ti\-kus h\H o\-m\'er\-s\'ek\-le\-te.

\begin{table}[htb]
\begin{center}
\begin{tabular}{||c||c|c|c|c||}
\hline \hline
 $R_{HW}$ & .2049 & .4220 & .595  & .8314 \\
 \hline
 $T_c$	(GeV)  & 38.3 & 72.6 & 100.0 & 128.4\\
 \hline
 $M_{\mathrm{latt}}$ (GeV)& 84.3(12) & 78.6(2) & 80.0(4) & 76.7(24) \\
 \hline
 $g^2_{\mathrm{latt}} (M^{-1} )$ & .5630(60) & .5788(16) & .5782(25) & .569(4) \\
 \hline
 $M_{\rm screen}$ (GeV) &74.97 & 80.44 & 80.70 & 81.77 \\
 \hline
 $g^2_{\overline {\rm {MS}}} (T_c)$ &0.540 & 0.592 & 0.585 & 0.570 \\
 \hline
 $g_{\rm {Laine}}^{2} (T_c)$ & 0.589 & 0.589 & 0.579 & 0.562
 \\
 \hline
 \end{tabular}
 \caption{\label{couplings}
A k\"u\-l\"on\-b\"o\-z\H o Higgs-t\"o\-me\-gek\-hez tar\-to\-z\'o t\"o\-meg\-pa\-ra\-m\'e\-te\-rek \'es
csa\-to\-l\'a\-si \'al\-lan\-d\'ok.
}
 \end{center}
 \end{table}

A (\ref{cd}) e\-gyen\-let\-ben sze\-rep\-l\H o $C$ \'es $D$ f\"ugg\-v\'e\-nyek
k\"u\-l\"on\-b\"o\-z\H o $R_{HW}$ t\"o\-me\-ga\-r\'a\-nyok e\-se\-t\'en fel\-vett \'er\-t\'e\-ke\-it a
\ref{cdtable} t\'ab\-l\'a\-zat tar\-tal\-maz\-za. Az \'ar\-ny\'e\-ko\-l\'a\-si t\"o\-me\-get itt
$M_{\scr} = M_W = 80$ GeV-nek v\'a\-lasz\-tot\-tuk.

\begin{table}[htb]
\begin{center}
\begin{tabular}{||c||c|c||}
\hline
 $R_{HW}$ & $C$ & $D$ \\
 \hline \hline
 0.2 &-41.54 & -22.19\\
 \hline
 0.3 &-8.26 & -6.58\\
 \hline
 0.4 &-6.47 & -1.12\\
 \hline
 0.5 & -5.66& 1.39\\
 \hline
 0.6 & -5.23& 2.74\\
 \hline
 0.7 &-4.98 & 3.55\\
 \hline
 0.8 & -4.83& 4.06\\
 \hline
 0.9 &-4.72 & 4.39\\
 \hline
 1.0 & -4.65& 4.62\\
 \hline
 1.1 & -4.59& 4.78\\
 \hline
 1.2 & -4.54& 4.89\\
 \hline
 1.3 & -4.50& 4.98\\
 \hline
 1.4 & -4.45& 4.98\\
 \hline
 1.5 &-4.40 & 5.01\\
 \hline
 \end{tabular}
 \caption{\label{cdtable}
$C$ \'es $D$ \'er\-t\'e\-ke k\"u\-l\"on\-b\"o\-z\H o $R_{HW}$ t\"o\-me\-ga\-r\'a\-nyok e\-se\-t\'en.
 }
 \end{center}
 \end{table}

\chapter{Per\-tur\-ba\-t\'{\i}v \'es nem\-per\-tur\-ba\-t\'{\i}v mennyi\-s\'e\-gek \"ossze\-ve\-t\'e\-se}
\fancyhead[CE]{\hst{\thechapter{}.\ fe\-je\-zet \quad Per\-tur\-bat\'{\i}v
\'es nem\-per\-tur\-bat\'{\i}v e\-redm\'e\-nyek \"ossze\-vet\'e\-se}}
A (\ref{cd}) e\-gyen\-let a per\-tur\-b\'a\-ci\-\'o\-sz\'a\-m\'{\i}\-t\'as il\-let\-ve a
r\'acs\-t\'e\-rel\-m\'e\-let ke\-re\-t\'en be\-l\"ul de\-fi\-ni\-\'alt csa\-to\-l\'a\-si \'al\-lan\-d\'ok
k\"o\-z\"ott te\-remt kap\-cso\-la\-tot. Az e\-lekt\-ro\-gyen\-ge f\'a\-zi\-s\'at\-me\-net
per\-tur\-ba\-t\'{\i}v \'es nem\-per\-tur\-ba\-t\'{\i}v vizs\-g\'a\-la\-t\'a\-nak \"ossze\-ve\-t\'e\-se
so\-r\'an ne\-h\'e\-zs\'e\-get, de le\-ga\-l\'ab\-bis \'u\-jabb hi\-ba\-for\-r\'ast je\-len\-tett a
csa\-to\-l\'a\-si \'al\-lan\-d\'ok el\-t\'e\-r\H o de\-fi\-n\'{\i}\-ci\-\'o\-ja -- a szta\-ti\-kus po\-ten\-ci\-\'al
ki\-sz\'a\-m\'{\i}\-t\'a\-s\'a\-nak ez volt az e\-gyik leg\-f\H obb mo\-ti\-v\'a\-ci\-\'o\-ja. Eb\-ben a
fe\-je\-zet\-ben ezt a hi\-ba\-for\-r\'ast ki\-k\"u\-sz\"o\-b\"ol\-ve ha\-son\-l\'{\i}t\-juk \"ossze a
v\'e\-ges h\H o\-m\'er\-s\'ek\-le\-t\H u f\'a\-zi\-s\'at\-me\-ne\-tet jel\-lem\-z\H o ter\-mo\-di\-na\-mi\-ka\-i
mennyi\-s\'e\-ge\-ket.
\section{Mi\-\'ert kell a per\-tur\-b\'a\-ci\-\'o\-sz\'a\-m\'{\i}\-t\'as?}
\subsection{A csa\-va\-ros e\-sz\H u Ark\-hi\-m\'e\-d\'esz}
A k\"o\-ze\-l\'{\i}\-t\H o\-m\'od\-sze\-rek ki\-fej\-lesz\-t\'e\-se Ark\-hi\-m\'e\-d\'esz ne\-v\'e\-hez
f\H u\-z\H o\-dik \cite{arkhi}. B\'ar m\'ar E\-uk\-li\-d\'esz is em\-l\'{\i}\-ti a k\'e\-tol\-da\-li
k\"o\-ze\-l\'{\i}\-t\'es m\'od\-sze\-r\'e\-nek le\-he\-t\H o\-s\'e\-g\'et, Ark\-hi\-m\'e\-d\'esz volt az, a\-ki
en\-nek je\-len\-t\H o\-s\'e\-g\'et fe\-lis\-mer\-te: a k\"or te\-r\"u\-le\-t\'et be\-\'{\i}rt \'es
k\"o\-r\"u\-l\'{\i}rt sok\-sz\"o\-gek\-kel k\"o\-ze\-l\'{\i}\-tet\-te, \'es 96-sz\"o\-gek
al\-kal\-ma\-z\'a\-s\'a\-val h\'a\-rom ti\-ze\-des\-jegy\-re pon\-to\-san meg\-ha\-t\'a\-roz\-ta a $\pi$
sz\'a\-mot. ``Me\-cha\-ni\-ka\-i m\'od\-sze\-re'' -- mely gya\-kor\-la\-ti\-lag a
dif\-fe\-ren\-ci\-\'al- \'es in\-teg\-r\'al\-sz\'a\-m\'{\i}\-t\'as kez\-det\-le\-ges for\-m\'a\-ja -- is
e\-zen a\-la\-pult, \'es se\-g\'{\i}t\-s\'e\-g\'e\-vel o\-lyan bo\-nyo\-lult ma\-te\-ma\-ti\-ka\-i
t\'e\-te\-le\-ket tu\-dott be\-bi\-zo\-ny\'{\i}\-ta\-ni, me\-lyek kor\-t\'ar\-sa\-i fe\-j\'e\-ben meg sem
for\-dul\-tak.

Me\-cha\-ni\-ka\-i m\'od\-sze\-r\'e\-vel ka\-pott t\'e\-te\-le\-it a\-zon\-ban Ark\-hi\-m\'e\-d\'esz m\'as
m\'od\-sze\-rek\-kel is be\-bi\-zo\-ny\'{\i}\-tot\-ta, mi\-vel az e\-l\H ob\-bi meg\-k\"o\-ze\-l\'{\i}\-t\'est a
szi\-go\-r\'u g\"o\-r\"og ge\-o\-met\-ri\-a-szem\-l\'e\-let nem te\-kin\-tet\-te tel\-jes
\'er\-t\'e\-k\H u\-nek. A g\"o\-r\"og tu\-d\'os ma\-ga is ko\-moly fenn\-tar\-t\'a\-sok\-kal
te\-kin\-tett m\'od\-sze\-r\'e\-re -- b\'ar je\-len\-t\H o\-s\'e\-g\'et \'es hasz\-nos\-s\'a\-g\'at
na\-gyon pon\-to\-san l\'at\-ta.

A per\-tur\-b\'a\-ci\-\'o\-sz\'a\-m\'{\i}\-t\'as k\"o\-ze\-l\'{\i}\-t\H o\-m\'od\-sze\-r\'e\-nek ha\-son\-l\'o\-k\'ep\-pen
meg\-van\-nak a ma\-ga \'o\-ri\-\'a\-si e\-l\H o\-nye\-i. Az a\-l\'ab\-bi\-ak\-ban e\-zek\-re t\'e\-r\"unk
ki, de i\-gyek\-sz\"unk nem szem e\-l\H ol t\'e\-vesz\-te\-ni h\'at\-r\'a\-nya\-it \'es
kor\-l\'a\-ta\-it.
\subsection{A per\-tur\-b\'a\-ci\-\'o\-sz\'a\-m\'{\i}\-t\'as}
\bfr
\emph{--- Nem k\'{\i}\-v\'a\-nom pub\-li\-k\'al\-ni; csu\-p\'an fel\-jegy\-zem a t\'e\-nye\-ket,
hogy Is\-ten is tud\-jon r\'o\-luk. \\
--- Nem gon\-do\-lod, hogy Is\-ten is\-me\-ri e\-ze\-ket a t\'e\-nye\-ket? -- k\'er\-dez\-te
Hans Bet\-he. \\
--- A t\'e\-nye\-ket biz\-to\-san is\-me\-ri, de a t\'e\-nyek\-nek ezt a v\'al\-to\-za\-t\'at
le\-het, hogy nem \\
-- v\'a\-la\-szol\-ta Szi\-l\'ard Le\-\'o.}
\cite{physspeak}
\efr
A per\-tur\-b\'a\-ci\-\'o\-sz\'a\-m\'{\i}\-t\'as az el\-m\'e\-le\-ti fi\-zi\-ka e\-gyik leggyak\-rab\-ban
hasz\-n\'alt \'es leg\-hasz\-no\-sabb m\'od\-sze\-re. Klasszi\-kus me\-cha\-ni\-ka\-i
al\-kal\-ma\-z\'a\-sa\-i k\"o\-z\"ul az a\-di\-a\-ba\-ti\-kus in\-va\-ri\-\'an\-sok sz\'ar\-maz\-ta\-t\'a\-s\'at
e\-mel\-n\'enk ki \cite{goldstein}, mely sok\-kal t\"ob\-bet ad ke\-z\"unk\-be egy
egy\-sze\-r\H u\-en \'es ha\-t\'a\-so\-san al\-kal\-maz\-ha\-t\'o esz\-k\"oz\-n\'el. Az a\-di\-a\-ba\-ti\-kus
in\-va\-ri\-\'an\-sok se\-g\'{\i}\-te\-nek fe\-lis\-mer\-ni a prob\-l\'e\-ma m\'e\-lyebb me\-g\'er\-t\'e\-s\'et,
mely Di\-rac sza\-va\-i\-val a k\"o\-vet\-ke\-z\H ot je\-len\-ti:
\medskip \\
\emph{``\'Er\-tem, hogy mit je\-lent egy e\-gyen\-let, ha a me\-gol\-d\'a\-s\'at nagy
vo\-na\-lak\-ban fel tu\-dom v\'a\-zol\-ni a\-n\'el\-k\"ul is, hogy me\-gol\-da\-n\'am.''}
\cite{fey5}
\medskip \\
Egy a\-dott prob\-l\'e\-ma e\-se\-t\'en a l\'e\-nye\-ges mennyi\-s\'e\-gek
el\-k\"u\-l\"o\-n\'{\i}\-t\'e\-s\'e\-hez a di\-men\-zi\-\'o\-a\-na\-l\'{\i}\-zi\-sen k\'{\i}\-v\"ul le\-gin\-k\'abb a
per\-tur\-b\'a\-ci\-\'o\-sz\'a\-m\'{\i}\-t\'as ny\'ujt k\'e\-nyel\-mes a\-la\-pot.

A per\-tur\-b\'a\-ci\-\'o\-sz\'a\-m\'{\i}\-t\'as i\-ga\-zi al\-kal\-ma\-z\'a\-si te\-r\"u\-le\-te a
kvan\-tum\-fi\-zi\-ka. T\"or\-t\'e\-ne\-ti\-leg en\-n\'el sok\-kal t\"obb\-r\H ol is sz\'o van: a
kvan\-tu\-mel\-m\'e\-let el\-s\H o meg\-fo\-gal\-ma\-z\'a\-sa a klasszi\-kus me\-cha\-ni\-ka
a\-di\-a\-ba\-ti\-kus in\-va\-ri\-\'an\-sa\-i\-ra \'e\-p\"ult -- Max Born klasszi\-kus
tan\-k\"ony\-ve \cite{born} sze\-rint
\medskip \\
\emph{``k\'e\-zen\-fek\-v\H o a fel\-te\-v\'es, hogy csak az a\-di\-a\-ba\-ti\-ku\-san
in\-va\-ri\-\'ans mennyi\-s\'e\-gek kvan\-t\'al\-ha\-t\'ok.''}
\medskip \\
B\'ar a kvan\-tum\-me\-cha\-ni\-ka ma m\'ar m\'as a\-la\-po\-kon \'all, a
per\-tur\-b\'a\-ci\-\'o\-sz\'a\-m\'{\i}\-t\'as al\-kal\-ma\-z\'a\-sa\-i\-nak je\-len\-t\H o\-s\'e\-g\'et nem le\-het
t\'ul\-be\-cs\"ul\-ni. T\'ul az egy\-sze\-r\H u\-en al\-kal\-maz\-ha\-t\'o sz\'a\-m\'{\i}\-t\'a\-si m\'od\-szer
le\-ny\H u\-g\"o\-z\H o nu\-me\-ri\-kus e\-red\-m\'e\-nye\-in, az e\-red\-m\'e\-nye\-ket ``\'ert\-j\"uk'' is:
tud\-juk, mi\-lyen ef\-fek\-tu\-sok fe\-le\-l\H o\-sek az e\-lekt\-ron a\-no\-m\'a\-lis m\'ag\-ne\-ses
mo\-men\-tu\-m\'a\-\'ert -- \'es ez l\'e\-nye\-ge\-sen fon\-to\-sabb, mint ma\-g\'a\-nak a
nu\-me\-ri\-kus \'er\-t\'ek\-nek 10 ti\-ze\-des\-jegy he\-lyett 20-ra va\-l\'o is\-me\-re\-te.

A per\-tur\-b\'a\-ci\-\'o\-sz\'a\-m\'{\i}\-t\'as egy ``k\"ony\-ve\-l\'e\-si m\'od\-szer'', mely
sz\'a\-mos e\-set\-ben na\-gyon ha\-t\'e\-ko\-nyan m\H u\-k\"o\-dik. K\"o\-ze\-l\'{\i}\-t\H o
m\'od\-sze\-re\-ink nem sz\"uk\-s\'eg\-sze\-r\H u\-ek; Schr\"o\-din\-ger \'es He\-i\-sen\-berg a
kvan\-tum\-me\-cha\-ni\-ka k\'et ek\-vi\-va\-lens meg\-fo\-gal\-ma\-z\'a\-s\'at tel\-je\-sen el\-t\'e\-r\H o
ma\-te\-ma\-ti\-ka\-i a\-lap\-ra \'e\-p\'{\i}\-tet\-te, \'{\i}gy bi\-zo\-nyo\-san sz\'a\-mos al\-ter\-na\-t\'{\i}v
m\'od\-ja le\-het\-s\'e\-ges a fi\-zi\-ka\-i va\-l\'o\-s\'ag fel\-t\'er\-k\'e\-pe\-z\'e\-s\'e\-nek. A
per\-tur\-b\'a\-ci\-\'o\-sz\'a\-m\'{\i}\-t\'as te\-h\'at egy m\'od\-szer a sok k\"o\-z\"ul, mely
a\-zon\-ban ma\-i szem\-l\'e\-le\-t\"unk\-h\"oz na\-gyon j\'ol il\-lesz\-ke\-dik.

\section{Mi\-kor al\-kal\-maz\-ha\-t\'o a per\-tur\-b\'a\-ci\-\'o\-sz\'a\-m\'{\i}\-t\'as?
\label{pertalk}}
\fancyhead[CO]{\hst{\thesection \quad Mi\-kor al\-kal\-maz\-hat\'o a
per\-turb\'a\-ci\'osz\'am\'{\i}t\'as?}}
A per\-tur\-b\'a\-ci\-\'o\-sz\'a\-m\'{\i}\-t\'as, mint sor\-fej\-t\'es, ak\-kor al\-kal\-maz\-ha\-t\'o
ha\-t\'e\-ko\-nyan, ha az egy\-m\'ast k\"o\-ve\-t\H o ren\-dek\-b\H ol a\-d\'o\-d\'o j\'a\-ru\-l\'ek
e\-r\H o\-tel\-je\-sen cs\"ok\-ken; ek\-kor a per\-tur\-b\'a\-ci\-\'os sort a\-dott rend\-ben
le\-v\'ag\-va gya\-kor\-la\-ti c\'el\-ja\-ink\-ra meg\-fe\-le\-l\H o\-en pon\-tos e\-red\-m\'enyt
kap\-ha\-tunk, mely\-nek hi\-b\'a\-j\'at is e\-l\'eg pon\-to\-san meg tud\-juk be\-cs\"ul\-ni.
Ez azt is je\-len\-ti, hogy ha a per\-tur\-b\'a\-ci\-\'os sor egy\-m\'ast k\"o\-ve\-t\H o
tag\-ja\-i nem csen\-ge\-nek le e\-l\'eg gyor\-san, ak\-kor a
per\-tur\-b\'a\-ci\-\'o\-sz\'a\-m\'{\i}\-t\'as e\-red\-m\'e\-ny\'et nem te\-kint\-het\-j\"uk
meg\-b\'{\i}z\-ha\-t\'o\-nak.

Az e\-lekt\-ro\-gyen\-ge f\'a\-zi\-s\'at\-me\-net vizs\-g\'a\-la\-ta so\-r\'an s\'u\-lyos
inf\-ra\-v\"o\-r\"os prob\-l\'e\-m\'ak l\'ep\-nek fel a ma\-gas h\H o\-m\'er\-s\'ek\-le\-t\H u
szim\-met\-ri\-kus f\'a\-zis\-ban; az egy-hu\-rok-ren\-d\H u sz\'a\-mo\-l\'as e\-red\-m\'e\-ny\'e\-hez
a k\'et-hu\-rok-ren\-d\H u $\ordo{100\%}$-os kor\-rek\-ci\-\'ot ad \cite{fod-heb}.
\'Igy a f\'a\-zi\-s\'at\-me\-net vizs\-g\'a\-la\-t\'a\-hoz nem\-per\-tur\-ba\-t\'{\i}v esz\-k\"o\-z\"ok\-h\"oz
kell fo\-lya\-mod\-nunk. A\-zon\-ban az inf\-ra\-v\"o\-r\"os prob\-l\'e\-m\'ak csak a
bo\-zo\-ni\-kus szek\-tor\-ban l\'ep\-nek fel -- \'{\i}gy a fer\-mi\-o\-ni\-kus szek\-tor
per\-tur\-ba\-t\'{\i}v ke\-ze\-l\'e\-se le\-het\-s\'e\-ges.

Va\-la\-mennyi\-re ha\-son\-l\'o hely\-zet \'all e\-l\H o a MSSM-ben: az egy\-hu\-rok-ren\-d\H u
sz\'a\-mo\-l\'as\-hoz k\'e\-pest a k\'et\-hu\-rok-ren\-d\H u kor\-rek\-ci\-\'o nagy. K\"onnyen
meggy\H o\-z\H od\-he\-t\"unk a\-zon\-ban ar\-r\'ol, hogy ez kis sz\'a\-m\'u gr\'af
sz\'am\-l\'a\-j\'a\-ra \'{\i}r\-ha\-t\'o, me\-lyek az e\-r\H os szek\-tor ve\-ze\-t\H o rend\-j\'et
je\-len\-tik, \'{\i}gy \'essze\-r\H u a fel\-te\-v\'es, hogy a pe\-re\-tur\-b\'a\-ci\-\'o\-sz\'a\-m\'{\i}\-t\'as
ma\-ga\-sabb rend\-je\-i\-ben az egy\-m\'ast k\"o\-ve\-t\H o ren\-dek j\'a\-ru\-l\'e\-ka egy\-re
ki\-sebb lesz.

Jel\-leg\-ze\-tes nem\-per\-tur\-ba\-t\'{\i}v ef\-fek\-tu\-sok is fel\-l\'ep\-het\-nek; er\-re
a\-na\-l\'o\-gi\-a\-k\'ent az $f(x) = \exp(-1/x^2)$ f\"ugg\-v\'eny 0 k\"o\-r\"u\-li
Tay\-lor-sor\-fej\-t\'e\-s\'et te\-kint\-het\-j\"uk: a f\"ugg\-v\'eny \"osszes 0 pont\-be\-li
de\-ri\-v\'alt\-ja 0, \'{\i}gy $x$ 0-t\'ol k\"u\-l\"on\-b\"o\-z\H o \'er\-t\'e\-ke\-i\-re is $f(x)=0$
kel\-le\-ne hogy le\-gyen. Az e\-lekt\-ro\-gyen\-ge f\'a\-zi\-s\'at\-me\-net so\-r\'an ge\-ne\-r\'alt
ba\-ri\-on-a\-szim\-met\-ri\-a csak nem-per\-tur\-ba\-t\'{\i}v m\'od\-sze\-rek\-kel ke\-zel\-he\-t\H o,
u\-gya\-nis a stan\-dard mo\-dell ke\-re\-t\'en be\-l\"ul nem raj\-zol\-ha\-t\'o fel exp\-li\-cit
ba\-ri\-on\-sz\'am-s\'er\-t\H o Feyn\-man-gr\'af.

A fen\-ti\-ek a\-lap\-j\'an nyil\-v\'an\-va\-l\'o, hogy a ba\-ri\-o\-ge\-n\'e\-zis prob\-l\'e\-ma\-k\"o\-re
a stan\-dard mo\-dell ke\-re\-t\'e\-ben ki\-z\'ar\-l\'o\-ag per\-tur\-ba\-t\'{\i}v m\'od\-sze\-rek\-kel
nem ta\-nul\-m\'a\-nyoz\-ha\-t\'o ki\-e\-l\'e\-g\'{\i}\-t\H o\-en.

K\'e\-zen\-fek\-v\H o m\'od\-szer len\-ne a stan\-dard mo\-dell n\'egy\-di\-men\-zi\-\'os t\'e\-ri\-d\H o
r\'a\-cson va\-l\'o ke\-ze\-l\'e\-se -- a\-zon\-ban a Ni\-el\-sen--Ni\-no\-yi\-ma-t\'e\-tel
\'er\-tel\-m\'e\-ben a fer\-mi\-o\-nok eg\-zakt ki\-r\'a\-lis szim\-met\-ri\-\'a\-ja nem
va\-l\'o\-s\'{\i}t\-ha\-t\'o meg a r\'a\-cson \cite{niedermayer}. Az u\-t\'ob\-bi \'e\-vek\-ben a
fer\-mi\-o\-nok r\'acs\-t\'e\-rel\-m\'e\-le\-ti ke\-ze\-l\'e\-se so\-kat fej\-l\H o\-d\"ott, \'es ma m\'ar
l\'e\-te\-zik o\-lyan el\-j\'a\-r\'as, mellyel a ki\-r\'a\-lis szim\-met\-ri\-a i\-gen pon\-to\-san
meg\-va\-l\'o\-s\'{\i}t\-ha\-t\'o, a\-zon\-ban ez (\'es min\-den m\'as fer\-mi\-o\-ni\-kus
r\'acs\-ha\-t\'ast ma\-g\'a\-ban fog\-la\-l\'o el\-j\'a\-r\'as) o\-lyan nagy\-m\'e\-re\-t\H u
g\'e\-pi\-d\H o-n\"o\-ve\-ke\-d\'est von ma\-ga u\-t\'an, mely a gya\-kor\-la\-ti al\-kal\-ma\-z\'ast
e\-gye\-l\H o\-re ki\-z\'ar\-ja. Ko\-r\'ab\-bi meg\-jegy\-z\'e\-se\-ink a\-lap\-j\'an a\-zon\-ban a
fer\-mi\-o\-ni\-kus szek\-tor nem\-per\-tur\-ba\-t\'{\i}v ke\-ze\-l\'e\-s\'e\-re nincs sz\"uk\-s\'eg, \'{\i}gy
a bo\-zo\-ni\-kus (SU(2)--Higgs) szek\-tor r\'acs\-ra t\'e\-te\-le mel\-lett a
fer\-mi\-o\-no\-kat (\'es az U(1) szek\-tort) per\-tur\-ba\-t\'{\i}\-ve ke\-zel\-het\-j\"uk. Ez a
m\'od\-szer az e\-lekt\-ro\-gyen\-ge f\'a\-zi\-s\'at\-me\-net vizs\-g\'a\-la\-t\'a\-nak e\-gyik
ki\-dol\-go\-zott le\-he\-t\H o\-s\'e\-ge; a m\'a\-sik gyak\-ran hasz\-n\'alt m\'od\-szer a
per\-tur\-ba\-t\'{\i}v e\-le\-me\-ket szin\-t\'en tar\-tal\-ma\-z\'o \emph{di\-men\-zi\-\'os
re\-duk\-ci\-\'o} m\'od\-sze\-re. (U\-t\'ob\-bi\-ra r\'esz\-le\-tes \"ossze\-fog\-la\-l\'ast ad
\cite{jak, gin, app2}, \'{\i}gy er\-re r\'esz\-le\-te\-sen nem t\'e\-r\"unk ki.)

Az e\-lekt\-ro\-gyen\-ge f\'a\-zi\-s\'at\-me\-ne\-tet sz\'a\-mos per\-tur\-ba\-t\'{\i}v
meg\-k\"o\-ze\-l\'{\i}\-t\'e\-sen a\-la\-pu\-l\'o mun\-ka is t\'ar\-gyal\-ja. B\'ar tud\-juk, hogy a
per\-tur\-ba\-t\'{\i}v e\-red\-m\'e\-nyek \"on\-ma\-guk\-ban nem e\-le\-gen\-d\H o\-ek a f\'a\-zi\-s\'at\-me\-net
le\-\'{\i}\-r\'a\-s\'a\-ra, bi\-zo\-nyos tar\-to\-m\'a\-nyok\-ban j\'ol m\H u\-k\"od\-het a
per\-tur\-b\'a\-ci\-\'o\-sz\'a\-m\'{\i}\-t\'as. A per\-tur\-ba\-t\'{\i}v \'es nem\-per\-tur\-ba\-t\'{\i}v
e\-red\-m\'e\-nyek \"ossze\-ve\-t\'e\-se el\-ve\-zet\-het egy o\-lyan fi\-zi\-ka\-i me\-g\'er\-t\'es\-hez,
mely a\-lap\-j\'an nagy biz\-ton\-s\'ag\-gal tip\-pel\-het\-j\"uk meg, hogy egy
bo\-nyo\-lul\-tabb mo\-dell\-ben (pl.\ MSSM) -- mely\-ben e\-gye\-l\H o\-re nem \'all
ren\-del\-ke\-z\'e\-s\"unk\-re t\'ul sok r\'acsszi\-mu\-l\'a\-ci\-\'os e\-red\-m\'eny --, mi\-lyen
pa\-ra\-m\'e\-ter-tar\-to\-m\'any\-ban b\'{\i}z\-ha\-tunk meg a per\-tur\-ba\-t\'{\i}v j\'os\-la\-tok\-ban.
En\-nek fel\-de\-r\'{\i}\-t\'e\-se a r\'acsszi\-mu\-l\'a\-ci\-\'ok\-ban vizs\-g\'alt
pa\-ra\-m\'e\-ter-tar\-to\-m\'any meg\-v\'a\-lasz\-t\'a\-sa szem\-pont\-j\'a\-b\'ol i\-gen nagy
je\-len\-t\H o\-s\'e\-g\H u.

A k\"o\-vet\-ke\-z\H ok\-ben ar\-ra fo\-gok te\-h\'at bi\-zo\-ny\'{\i}\-t\'e\-kot ke\-res\-ni, hogy az
e\-lekt\-ro\-gyen\-ge f\'a\-zi\-s\'at\-me\-net v\'eg\-pont\-j\'a\-t\'ol t\'a\-vol a
per\-tur\-b\'a\-ci\-\'o\-sz\'a\-m\'{\i}\-t\'as j\'ol m\H u\-k\"o\-dik.

\section{A f\'a\-zi\-s\'at\-me\-net ter\-mo\-di\-na\-mi\-ka\-i jel\-lem\-z\H o\-i}
\fancyhead[CO]{\hst{\thesection \quad A f\'a\-zis\'at\-me\-net
ter\-mo\-di\-na\-mi\-ka\-i jel\-lemz\H{o}i}}
Eb\-ben a sza\-kasz\-ban az e\-lekt\-ro\-gyen\-ge f\'a\-zi\-s\'at\-me\-net per\-tur\-ba\-t\'{\i}v \'es
nem\-per\-tur\-ba\-t\'{\i}v \'u\-ton meg\-ha\-t\'a\-ro\-zott ter\-mo\-di\-na\-mi\-ka\-i jel\-lem\-z\H o\-it
ha\-son\-l\'{\i}\-tom \"ossze.

\subsubsection*{Per\-tur\-ba\-t\'{\i}v meg\-k\"o\-ze\-l\'{\i}\-t\'es}
Az SU(2)--Higgs-mo\-dell k\'et\-hu\-rok-rend\-ben is\-mert v\'e\-ges
h\H o\-m\'er\-s\'ek\-le\-t\H u ef\-fek\-t\'{\i}v po\-ten\-ci\-\'al\-j\'at vizs\-g\'al\-juk \cite{fod-heb}.
A Higgs-t\"o\-me\-get a pro\-pa\-g\'a\-tor p\'o\-lu\-sa, a\-zaz a $p^2 - M^2= \Pi(p^2)$
e\-gyen\-let me\-gol\-d\'a\-sa ad\-ja -- a\-hol $\Pi(p^2)$ a Higgs-sa\-j\'a\-te\-ner\-gi\-a.
Az ef\-fek\-t\'{\i}v po\-ten\-ci\-\'a\-lon a\-la\-pu\-l\'o meg\-k\"o\-ze\-l\'{\i}\-t\'es ke\-re\-t\'e\-ben
meg\-mu\-tat\-ha\-t\'o \cite{arn}, hogy a fen\-ti disz\-per\-zi\-\'os re\-l\'a\-ci\-\'o\-ban a
$\Pi(p^2)$ sa\-j\'a\-te\-ner\-gi\-a $\Pi(0)$-lal va\-l\'o he\-lyet\-te\-s\'{\i}\-t\'e\-se csak
$g^5 v^2 $ren\-d\H u kor\-rek\-ci\-\'ot je\-lent,\footnote{$v$ a Higgs-t\'er
v\'a\-ku\-um-v\'ar\-ha\-t\'o \'er\-t\'e\-ke 0 h\H o\-m\'er\-s\'ek\-le\-ten} te\-h\'at $g^4$ ren\-d\H u
pon\-tos\-s\'a\-got c\'el\-z\'o sz\'a\-m\'{\i}\-t\'a\-sok\-ban al\-kal\-maz\-ha\-t\'o. Ek\-kor a
fag\-r\'af-szin\-t\H u
\beq
V_{\mrm{fa}} = \frac12 m^2 \varphi^2 + \frac14 \lambda \varphi^4
\enq
po\-ten\-ci\-\'al\-hoz a\-d\'o\-d\'o kor\-rek\-ci\-\'o
\beq
\delta V={\varphi^2 \over 2} \left( \delta m^2+ {1 \over 2\beta^2}
\delta\lambda\right) + {\delta\lambda \over 4} \varphi^4,
\enq
a\-hol
\beq
\delta m^2 = \frac{9g^4v^2}{256 \pi^2}, \qquad
\delta\lambda=-\frac{9g^4}{256\pi^2}\left(\log\frac{M_W^2}{\mu^2}+
\frac23 \right).
\end{equation}
${\mu}$ a re\-nor\-m\'a\-l\'a\-si sk\'a\-la, $M_W$ pe\-dig a 0 h\H o\-m\'er\-s\'ek\-le\-ten
m\'ert W-t\"o\-meg.

A per\-tur\-ba\-t\'{\i}v e\-red\-m\'e\-nye\-ket a \ms \ csa\-to\-l\'a\-si \'al\-lan\-d\'ot a
szta\-ti\-kus po\-ten\-ci\-\'al\-b\'ol sz\'ar\-maz\-ta\-tott $g_R$ csa\-to\-l\'a\-si \'al\-lan\-d\'o\-val
\"ossze\-kap\-cso\-l\'o (\ref{cd}) e\-gyen\-let a\-lap\-j\'an kor\-ri\-g\'al\-tuk.
\subsubsection*{Nem\-per\-tur\-ba\-t\'{\i}v meg\-k\"o\-ze\-l\'{\i}\-t\'es}
A nem\-per\-tur\-ba\-t\'{\i}v e\-red\-m\'e\-nyek a\-lap\-j\'a\-ul a n\'egy\-di\-men\-zi\-\'os
SU(2)--Higgs-mo\-dell\-ben v\'eg\-re\-haj\-tott szi\-mu\-l\'a\-ci\-\'ok\-ra \cite{som94,
csik99, csik96, csik98, cikk15} a\-la\-poz\-zuk; a fer\-mi\-o\-nok \'es az U(1)
szek\-tor per\-tur\-ba\-t\'{\i}v kor\-rek\-ci\-\'o\-val ve\-he\-t\H ok fi\-gye\-lem\-be. A
Higgs-r\'e\-szecs\-ke t\"o\-me\-g\'et a kor\-re\-l\'a\-ci\-\'os f\"ugg\-v\'eny le\-csen\-g\'e\-se
ha\-t\'a\-roz\-za meg. A szi\-mu\-l\'a\-ci\-\'ok so\-r\'an $L_t=2,3,4,5$ ki\-ter\-je\-d\'e\-s\H u
v\'e\-ges h\H o\-m\'er\-s\'ek\-le\-t\H u r\'a\-cso\-kat vizs\-g\'al\-tak; a m\'ert a\-da\-tok
ki\-\'er\-t\'e\-ke\-l\'e\-s\'e\-re jackk\-ni\-fe \'es bo\-otst\-rap tech\-ni\-k\'ak al\-kal\-ma\-z\'a\-s\'a\-val
t\"or\-t\'ent \cite{efron}. A kon\-ti\-nu\-um\-be\-li ha\-t\'a\-r\'er\-t\'e\-kek
me\-g\'al\-la\-p\'{\i}\-t\'a\-sa a v\'e\-ges r\'a\-cs\'al\-lan\-d\'o mel\-lett meg\-ha\-t\'a\-ro\-zott
ter\-mo\-di\-na\-mi\-ka\-i mennyi\-s\'e\-gek\-b\H ol a bo\-zo\-ni\-kus el\-m\'e\-let\-re jel\-lem\-z\H o
$1/a^2$-s v\'e\-ges r\'a\-cs\'al\-lan\-d\'o-kor\-rek\-ci\-\'ok fi\-gye\-lem\-be\-v\'e\-te\-l\'e\-vel
t\"or\-t\'ent.

Kis Higgs-t\"o\-me\-gek e\-se\-t\'en a f\'a\-zi\-s\'at\-me\-net e\-r\H o\-sen el\-s\H o\-ren\-d\H u, a
kor\-re\-l\'a\-ci\-\'os hosszak nem t\'ul na\-gyok, \'{\i}gy kb.\ 50 GeV-ig a
szi\-mu\-l\'a\-ci\-\'ok szim\-met\-ri\-kus t\'e\-ri\-d\H o-r\'a\-cso\-kat vizs\-g\'al\-tak. A
Higgs-t\"o\-me\-get n\"o\-vel\-ve a kor\-re\-l\'a\-ci\-\'os hosszak is n\H o\-nek, a\-mi
a\-ni\-zot\-r\'op r\'a\-csok hasz\-n\'a\-la\-t\'at te\-szi sz\"uk\-s\'e\-ges\-s\'e \cite{anizo}.
\subsubsection*{A vizs\-g\'alt ter\-mo\-di\-na\-mi\-ka\-i jel\-lem\-z\H ok}
\begin{itemize}
\item \emph{kri\-ti\-kus h\H o\-m\'er\-s\'ek\-let} ($T_c$) -- a\-hol az ef\-fek\-t\'{\i}v
po\-ten\-ci\-\'al\-nak k\'et de\-ge\-ne\-r\'alt mi\-ni\-mu\-ma van
\item \emph{a rend\-pa\-ra\-m\'e\-ter ug\-r\'a\-sa} ($\varphi_+$)
\item \emph{l\'a\-tens h\H o} -- az e\-ner\-gi\-a\-s\H u\-r\H u\-s\'eg disz\-kon\-ti\-nu\-i\-t\'a\-sa
($Q$)
\item \emph{fe\-l\"u\-le\-ti fe\-sz\"ult\-s\'eg} ($\sigma$) -- a f\'a\-zis\-ha\-t\'ar k\'et
ol\-da\-l\'a\-nak sza\-ba\-de\-ner\-gi\-a-s\H u\-r\H u\-s\'eg k\"u\-l\"onb\-s\'e\-ge
\end{itemize}

Az \"ossze\-ha\-son\-l\'{\i}\-t\'as e\-red\-m\'e\-nye\-it a \ref{termo} t\'ab\-l\'a\-zat
fog\-lal\-ja \"ossze; a vizs\-g\'alt ter\-mo\-di\-na\-mi\-ka\-i mennyi\-s\'e\-gek a kri\-ti\-kus
h\H o\-m\'er\-s\'ek\-let meg\-fe\-le\-l\H o hat\-v\'a\-ny\'a\-val van\-nak di\-men\-zi\-\'ot\-la\-n\'{\i}t\-va. A
z\'a\-r\'o\-jel\-ben sze\-rep\-l\H o sz\'a\-mok szo\-k\'as sze\-rint az u\-tol\-s\'o ki\-\'{\i}rt
ti\-ze\-des\-jegy-egy\-s\'e\-gek\-ben m\'ert hi\-b\'a\-kat je\-len\-tik.

\begin{table}
\begin{center}
\begin{tabular}{|c|}
\hline
 $M_H$\\ \hline $g_R^2$  \\
 \hline
  \begin{tabular}{c|c}
  $T_c/M_H$ \hspace{-5.5pt} &
   \begin{tabular}{c}
   pert \\
   \hline
   nem\-pert
  \end{tabular}
 \end{tabular} \\
 \hline
  \begin{tabular}{c|c}
 $\varphi_+/T_c$ \hspace{-1.5pt} &
  \begin{tabular}{c}
   pert \\
   \hline
   nem\-pert
  \end{tabular}
 \end{tabular} \\
 \hline
  \begin{tabular}{c|c}
  $Q/T_c^4$ \hspace{1.9pt} &
   \begin{tabular}{c}
   pert \\
   \hline
   nem\-pert
   \end{tabular}
  \end{tabular} \\
\hline
\begin{tabular}{c|c}
$\sigma/T_c^3$ \hspace{4.0pt} &
  \begin{tabular}{c}
  pert \\
  \hline
  nem\-pert
\end{tabular}
\end{tabular}\\
\hline
\end{tabular}
\begin{tabular}{|c|c|c|c|}
\hline
16.4(7) 	 & 33.7(10)	    & 47.6(16)	       & 66.5(14) \\
\hline
0.561(6)	 & 0.585(9)	    & 0.585(7)	       & 0.582(7) \\
\hline
2.72(3) 	 & 2.28(1)	    & 2.15(2)	       & 1.99(2)  \\
\hline
2.34(5) 	 & 2.15(4)	    & 2.10(5)	       & 1.93(7)  \\
\hline
4.30(23)	 & 1.58(7)	    & 0.97(4)	       & 0.65(2)  \\
\hline
4.53(26)	 & 1.65(14)	    & 1.00(6)	       & 0 \\
\hline
0.97(7) 	 & 0.22(2)	    & 0.092(6)	       & 0.045(2) \\
\hline
1.57(37)	 & 0.24(3)	    & 0.12(2)	       & 0 \\
\hline
0.70(10)	 & 0.067(6)	    & 0.022(2)	       & 0.0096(5) \\
\hline
0.77(11)	 & 0.053(5)	    & 0.008(2)	       & 0 \\
\hline
\end{tabular}
\caption{\label{termo}
A f\'a\-zi\-s\'at\-me\-net per\-tur\-ba\-t\'{\i}v \'es nem\-per\-tur\-ba\-t\'{\i}v \'u\-ton
meg\-ha\-t\'a\-ro\-zott ter\-mo\-di\-na\-mi\-ka\-i jel\-lem\-z\H o\-i\-nek \"ossze\-ha\-son\-l\'{\i}\-t\'a\-sa.}
\end{center}
\end{table}

A per\-tur\-ba\-t\'{\i}v e\-red\-m\'e\-nyek hi\-b\'a\-i a nem\-per\-tur\-ba\-t\'{\i}v e\-red\-m\'e\-nyek\-kel
va\-l\'o \"ossze\-ve\-t\'es\-b\H ol a\-d\'od\-nak. A r\'acsszi\-mu\-l\'a\-ci\-\'ok\-ban a csa\-to\-l\'a\-si
\'al\-lan\-d\'o \'es a Higgs-t\"o\-meg csak bi\-zo\-nyos pon\-tos\-s\'ag\-gal m\'er\-he\-t\H o;
\'{\i}gy e\-zek\-nek a mennyi\-s\'e\-gek\-nek a per\-tur\-ba\-t\'{\i}v meg\-fe\-le\-l\H ok\-kel va\-l\'o
\"ossze\-e\-gyez\-te\-t\'e\-se e\-red\-m\'e\-nye\-zi azt, hogy a per\-tur\-ba\-t\'{\i}v j\'os\-lat is
in\-k\'abb egy in\-ter\-val\-lum, mint egy konk\-r\'et \'er\-t\'ek.

\section{A per\-tur\-ba\-t\'{\i}v \'es nem\-per\-tur\-ba\-t\'{\i}v e\-red\-m\'e\-nyek \"ossze\-ve\-t\'e\-se}
\fancyhead[CO]{\hst{\thesection \quad A per\-tur\-ba\-t\'{\i}v \'es
nem\-per\-tur\-ba\-t\'{\i}v e\-red\-m\'e\-nyek \"ossze\-ve\-t\'e\-se}}
Ha e\-gyet\-len pa\-ra\-m\'e\-ter\-rel k\'{\i}\-v\'an\-juk jel\-le\-mez\-ni, hogy mi\-lyen
m\'er\-t\'ek\-ben e\-gyez\-tet\-he\-t\H ok \"ossze a per\-tur\-ba\-t\'{\i}v \'es nem\-per\-tur\-ba\-t\'{\i}v
e\-red\-m\'e\-nyek, c\'el\-sze\-r\H u az a\-l\'ab\-bi de\-fi\-n\'{\i}\-ci\-\'ot al\-kal\-maz\-nunk:
\beq
\mathrm{pull} = \frac{
{\mathrm{perturbat\acute{\i}v\ \acute{a}tlag}} -
{\mathrm{nemperturbat\acute{\i}v\ \acute{a}tlag}}}
{{\mathrm{perturbat\acute{\i}v\ e\-redm\acute{e}ny\ hib\acute{a}ja}} +
{\mathrm{nemperturbat\acute{\i}v\ e\-redm\acute{e}ny\ hib\acute{a}ja}}}
\enq
A \ref{termo} t\'ab\-l\'a\-zat\-ban sze\-rep\-l\H o mennyi\-s\'e\-gek \emph{pull}
pa\-ra\-m\'e\-te\-re\-i\-re a k\"o\-vet\-ke\-z\H o je\-l\"o\-l\'est ve\-zet\-j\"uk be:
\medskip

$P_T = T_c/M_H$ pull\-ja; \quad $P_\phi = \varphi_+/T_c$ pull\-ja; \quad
$P_Q = Q / T_c^4$ pull\-ja; \quad $P_\sigma = \sigma / T_c^3$ pull\-ja.
\medskip \\
A k\"u\-l\"on\-b\"o\-z\H o Higgs-t\"o\-me\-gek\-re a\-d\'o\-d\'o \emph{pull}o\-kat a \ref{pull}
t\'ab\-l\'a\-zat \'es a \ref{pullabra} \'ab\-ra fog\-lal\-ja \"ossze.
\begin{table}[htb]
\begin{center}
\begin{tabular}{|c|c|c|c|c|}
\hline
$m_H$ (GeV) & 16.4(7) & 33.7(10) & 47.6(16) & 66.5(14) \\
\hline
$P_T$	    & 4.75    & 2.60	 & 0.71     & 0.67 \\
\hline
$P_\varphi$ & 0.47    & -0.33	 & -0.3     & 32.5 \\
\hline
$P_Q$	    & -1.36   & -0.4	& -1.08    & 22.5 \\
\hline
$P_\sigma$  & -0.33   & 1.27	 & 3.5	    & 19.2 \\
\hline
\end{tabular}
\caption{\label{pull}
A k\"u\-l\"on\-b\"o\-z\H o Higgs-t\"o\-me\-gek\-hez tar\-to\-z\'o \emph{pull} \'er\-t\'e\-kek.}
\end{center}
\end{table}
\bef[ht]
\bc
\hspace{-1cm}
\vspace{0.5cm}
\epsfig{file=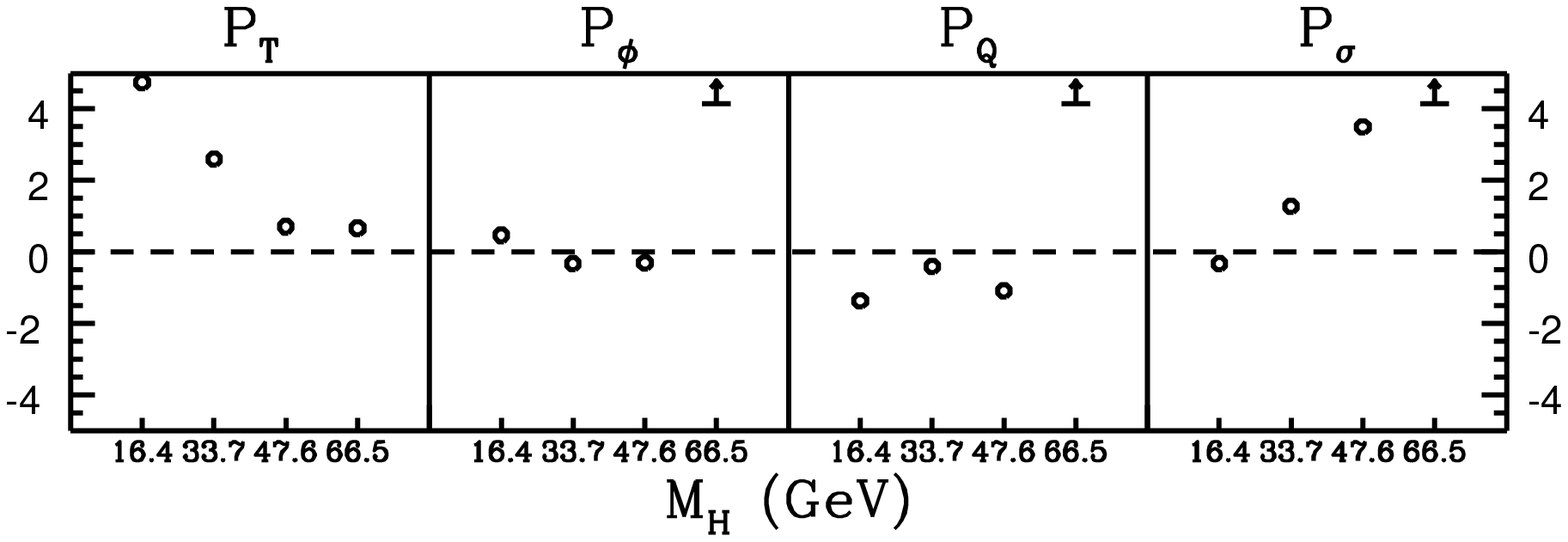,width=16cm}
\caption{\label{pullabra}
A n\'egy \emph{pull} pa\-ra\-m\'e\-ter a Higgs-t\"o\-meg f\"ugg\-v\'e\-ny\'e\-ben.
A nyi\-lak a $[-5,5]$ in\-ter\-val\-lu\-mon k\'{\i}\-v\"ul e\-s\H o \'er\-t\'e\-ke\-ket jel\-zik.}
\ec
\enf

Nagy Higgs-t\"o\-me\-gek e\-se\-t\'en a pull-\'er\-t\'e\-kek e\-r\H o\-tel\-je\-sen meg\-n\H o\-nek.
Itt a per\-tur\-ba\-t\'{\i}v \'es nem\-per\-tur\-ba\-t\'{\i}v e\-red\-m\'e\-nyek egy\-m\'as\-nak
el\-lent\-mon\-d\'o\-ak: az SU(2)--Higgs-mo\-dell 4-di\-men\-zi\-\'os Mon\-te Car\-lo
szi\-mu\-l\'a\-ci\-\'o\-i \'es a di\-men\-zi\-\'os re\-duk\-ci\-\'o\-val ka\-pott h\'a\-rom\-di\-men\-zi\-\'os
el\-m\'e\-let szi\-mu\-l\'a\-ci\-\'o\-i e\-gya\-r\'ant 65 GeV k\"o\-r\"ul j\'o\-sol\-j\'ak a
f\'a\-zi\-s\'at\-me\-net v\'eg\-pont\-j\'at, mely a fer\-mi\-no\-nok \'es az U(1) fak\-tor
per\-tur\-ba\-t\'{\i}v fi\-gye\-lem\-be\-v\'e\-te\-l\'e\-vel a tel\-jes stan\-dard mo\-dell\-re 72 GeV
k\"o\-r\"u\-li \'er\-t\'e\-ket ad. Ez\-zel szem\-ben a per\-tur\-b\'a\-ci\-\'o\-sz\'a\-m\'{\i}\-t\'a\-son
a\-la\-pu\-l\'o meg\-k\"o\-ze\-l\'{\i}\-t\'es sze\-rint tet\-sz\H o\-le\-ge\-sen nagy Higgs-t\"o\-meg
e\-se\-t\'en is el\-s\H o\-ren\-d\H u a f\'a\-zi\-s\'at\-me\-net.

\emph{A pri\-o\-ri} nem tud\-tunk e\-l\'eg e\-r\H os \'er\-ve\-ket fel\-hoz\-ni a\-mel\-lett,
hogy a per\-tur\-b\'a\-ci\-\'o\-sz\'a\-m\'{\i}\-t\'as al\-kal\-mas len\-ne az e\-lekt\-ro\-gyen\-ge
f\'a\-zi\-s\'at\-me\-net le\-\'{\i}\-r\'a\-s\'a\-ra. Az egy\-m\'ast k\"o\-ve\-t\H o ren\-dek
$\ordo{100\%}$-os kor\-rek\-ci\-\'ot je\-lent\-het\-nek -- \'{\i}gy sem\-mi meg\-le\-p\H o
sincs ab\-ban, hogy nagy Higgs-t\"o\-me\-gek e\-se\-t\'en a per\-tur\-ba\-t\'{\i}v
meg\-k\"o\-ze\-l\'{\i}\-t\'es nem m\H u\-k\"o\-dik \cite{buch-fod-heb}. C\'e\-lom in\-k\'abb az
volt, hogy o\-lyan pa\-ra\-m\'e\-ter-tar\-to\-m\'anyt ke\-res\-sek, a\-hol a
per\-tur\-b\'a\-ci\-\'o\-sz\'a\-m\'{\i}\-t\'as is m\H u\-k\"o\-dik: a ka\-pott e\-red\-m\'e\-nyek sze\-rint
az 50 GeV a\-lat\-ti tar\-to\-m\'any i\-lyen.

A $P_T$ mennyi\-s\'eg\-re a fen\-ti \'al\-l\'{\i}\-t\'as nem tel\-je\-s\"ul; itt a
per\-tur\-ba\-t\'{\i}v e\-red\-m\'e\-nyek a Higgs-t\"o\-meg n\"o\-ve\-ked\-t\'e\-vel egy\-re
pon\-to\-sab\-bak lesz\-nek. En\-nek o\-ka az, hogy az $M_H$ mennyi\-s\'eg\-ben
fel\-l\'e\-p\H o h\H o\-m\'er\-s\'ek\-let-in\-teg\-r\'a\-lok\-ra a $g^4, \lambda^2$ ren\-d\H u
per\-tur\-b\'a\-ci\-\'o\-sz\'a\-m\'{\i}\-t\'as az eb\-ben a Higgs-t\"o\-meg tar\-to\-m\'any\-ban j\'ol
m\H u\-k\"o\-d\H o ma\-gas h\H o\-m\'er\-s\'ek\-le\-t\H u sor\-fej\-t\'es\-sel j\'ol
\"ossze\-e\-gyez\-tet\-he\-t\H o e\-red\-m\'enyt ad \cite{fod-heb}. A per\-tur\-ba\-t\'{\i}v \'es
nem\-per\-tur\-ba\-t\'{\i}v m\'od\-sze\-rek\-kel e\-gya\-r\'ant j\'ol ke\-zel\-he\-t\H o $T_c/M_H$-t
mennyi\-s\'eg Higgs-t\"o\-meg-f\"ug\-g\'e\-se a k\"o\-vet\-ke\-z\H o m\'a\-sod\-fo\-k\'u
f\"ugg\-v\'ennyel \'{\i}r\-ha\-t\'o le:
\beq
\frac{T_c}{M_H}=2.494-0.842 R_{HW} + 0.223 R_{HW}^2.
\enq

V\'e\-ge\-ze\-t\"ul a csa\-to\-l\'a\-si \'al\-lan\-d\'ok kap\-cso\-la\-t\'a\-nak m\'eg egy\-faj\-ta
al\-kal\-ma\-z\'a\-s\'a\-ra t\'er\-n\'ek ki. Az e\-lekt\-ro\-gyen\-ge f\'a\-zi\-s\'at\-me\-ne\-ti
v\'eg\-pont \'er\-t\'e\-k\'e\-nek meg\-ha\-t\'a\-ro\-z\'a\-sa\-kor az SU(2)--Higgs-mo\-dell
vizs\-g\'a\-la\-t\'at a fer\-mi\-o\-no\-kat \'es az U(1) szek\-tort fi\-gye\-lem\-be ve\-v\H o
per\-tur\-ba\-t\'{\i}v l\'e\-p\'es e\-g\'e\-sz\'{\i}\-tet\-te ki. A per\-tur\-ba\-t\'{\i}v \'es
nem\-per\-tur\-ba\-t\'{\i}v m\'od\-sze\-rek ke\-ve\-re\-d\'e\-se k\'et\-f\'e\-le csa\-to\-l\'a\-si \'al\-lan\-d\'o
de\-fi\-n\'{\i}\-ci\-\'o hasz\-n\'a\-la\-t\'at k\"o\-ve\-tel\-te meg. A v\'eg\-pont\-ra ka\-pott
$72.4 \pm 1.7$ GeV \'er\-t\'ek \cite{csik99} hi\-b\'a\-ja a csa\-to\-l\'a\-si
\'al\-lan\-d\'ok k\"o\-z\"ot\-ti kap\-cso\-lat\-tal cs\"ok\-kent\-he\-t\H o; \'uj \'er\-t\'ek\-k\'ent
$72.1 \pm 1.4$ GeV a\-d\'o\-dik \cite{cikk1}. Ez a h\'a\-rom\-di\-men\-zi\-\'os
e\-red\-m\'e\-nyek\-kel \cite{kaj2, buchm} \"ossz\-hang\-ban \'all, \'es min\-tegy $20
\sigma$ bi\-zo\-nyos\-s\'ag\-gal ki\-z\'ar\-ja a stan\-dard mo\-dell\-be\-li e\-lekt\-ro\-gyen\-ge
f\'a\-zi\-s\'at\-me\-ne\-tet.

A n\'egy\-di\-men\-zi\-\'os \'es h\'a\-rom\-di\-men\-zi\-\'os e\-red\-m\'e\-nyek to\-v\'ab\-bi
\"ossze\-ve\-t\'e\-se le\-het\-s\'e\-ges a f\'a\-zis\-di\-ag\-ra\-mok \"ossze\-ha\-son\-l\'{\i}\-t\'a\-sa
r\'e\-v\'en. A csa\-to\-l\'a\-si \'al\-lan\-d\'ok k\"oz\-ti kap\-cso\-lat\-b\'ol a per\-tur\-ba\-t\'{\i}v
\'es nem\-per\-tur\-ba\-t\'{\i}v l\'e\-p\'e\-sek e\-gy\"ut\-tes je\-len\-l\'e\-t\'e\-b\H ol fa\-ka\-d\'o
hi\-b\'at ki\-k\"u\-sz\"o\-b\"ol\-ve a\-d\'o\-dik a \ref{laineabr} di\-ag\-ram \cite{la}.
\bef[ht]
\bc
\epsfig{file=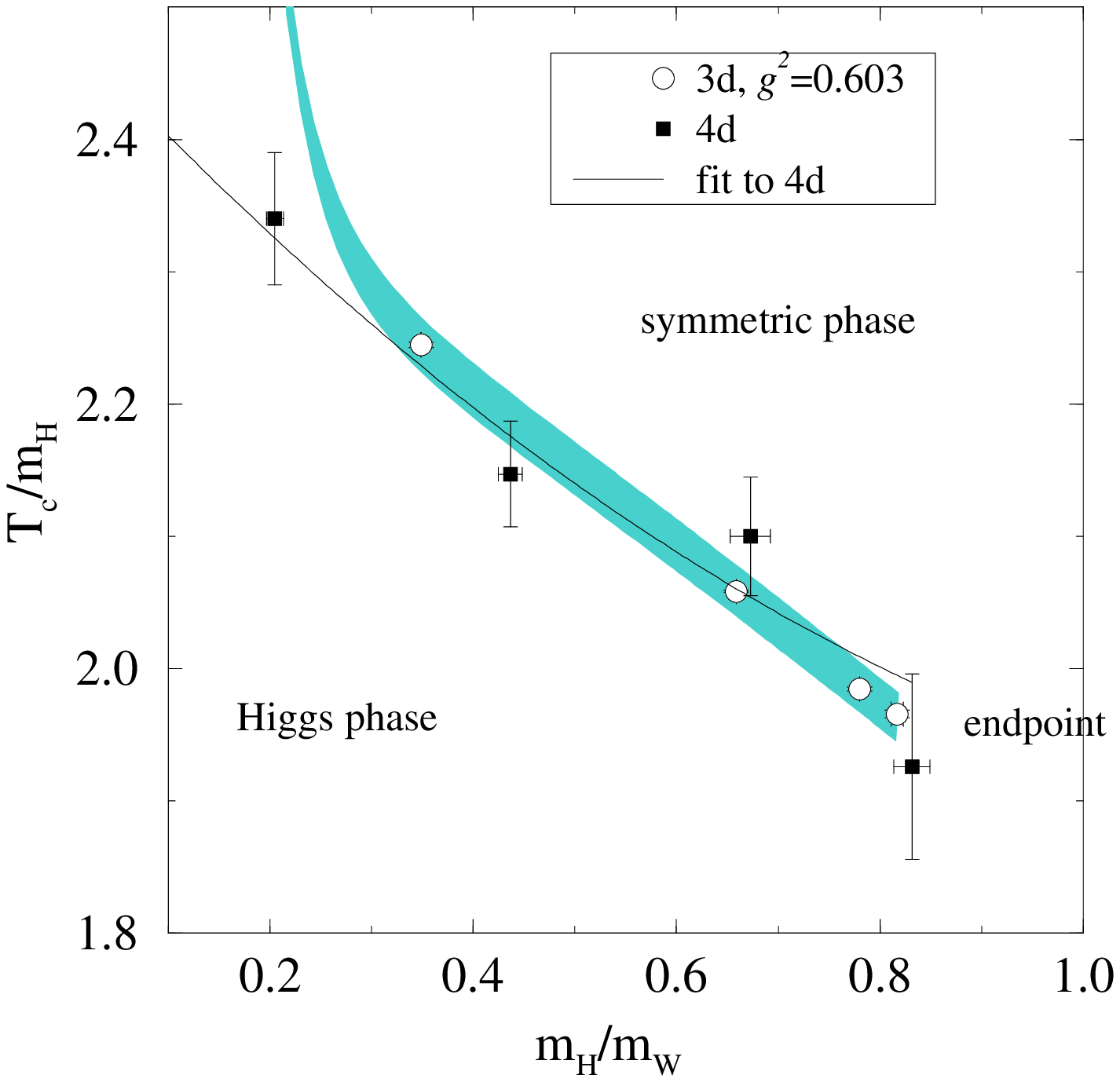,width=10cm} \label{laineabr}
\caption{A h\'a\-rom \'es n\'egy\-di\-men\-zi\-\'os szi\-mu\-l\'a\-ci\-\'ok f\'a\-zis\-di\-ag\-ram\-ja\-i.
A 4D e\-red\-m\'e\-nye\-ket n\'egy\-ze\-tek\-kel, a 3D e\-red\-m\'e\-nye\-ket sa\-t\'{\i}\-ro\-zott
vo\-nal\-lal je\-l\"ol\-t\"uk.}
\ec
\enf
A k\'et meg\-k\"o\-ze\-l\'{\i}\-t\'es te\-h\'at egy\-m\'as\-sal j\'o e\-gye\-z\'est mu\-tat; a
h\'a\-rom\-di\-men\-zi\-\'os e\-red\-m\'e\-nyek kis Higgs-t\"o\-me\-gek e\-se\-t\'en pon\-tat\-lan\-n\'a
v\'al\-nak. Ez nem is meg\-le\-p\H o, hi\-szen a ma\-gas h\H o\-m\'er\-s\'ek\-le\-t\H u
sor\-fej\-t\'es eb\-ben a tar\-to\-m\'any\-ban ke\-v\'es\-b\'e m\H u\-k\"o\-dik; a kis
Higgs-t\"o\-me\-gek e\-se\-t\'en je\-len\-t\H os, n\'egy\-di\-men\-zi\-\'os m\'od\-sze\-rek\-kel
t\'ar\-gyal\-ha\-t\'o Co\-le\-man--We\-in\-berg-tar\-to\-m\'any\-r\'ol pe\-dig a
h\'a\-rom\-di\-men\-zi\-\'os el\-j\'a\-r\'as nem k\'e\-pes sz\'a\-mot ad\-ni. A\-zon\-ban \emph{a
pri\-o\-ri} nem tud\-hat\-tuk, hol van az a tar\-to\-m\'any, a\-hol a
h\'a\-rom\-di\-men\-zi\-\'os m\'od\-sze\-rek meg\-b\'{\i}z\-ha\-t\'o\-v\'a v\'al\-nak -- ad ab\-sur\-dum
le\-het\-tek vol\-na a n\'egy\-di\-men\-zi\-\'os m\'od\-sze\-rek\-kel jel\-zett f\'a\-zi\-s\'at\-me\-ne\-ti
v\'eg\-pon\-ton t\'ul is.
\bigskip

Le\-sz\'a\-mol\-tunk te\-h\'at a per\-tur\-b\'a\-ci\-\'o\-sz\'a\-m\'{\i}\-t\'as\-sal -- \'es ve\-le e\-gy\"utt
o\-da\-lett az el\-s\H o\-ren\-d\H u e\-lekt\-ro\-gyen\-ge f\'a\-zi\-s\'at\-me\-net
per\-tur\-b\'a\-ci\-\'o\-sz\'a\-m\'{\i}\-t\'a\-son a\-la\-pu\-l\'o szem\-l\'e\-le\-tes k\'e\-pe is. Nem
si\-ke\-r\"ult az u\-ni\-ver\-zum ba\-ri\-on-a\-szim\-met\-ri\-\'a\-j\'a\-nak ma\-gya\-r\'a\-za\-t\'a\-hoz
sz\"uk\-s\'e\-ges ne\-me\-gyen\-s\'u\-lyi fo\-lya\-ma\-tok je\-len\-l\'e\-t\'et a r\'e\-szecs\-ke\-fi\-zi\-ka\-i
stan\-dard mo\-dell\-j\'en be\-l\"ul ki\-mu\-tat\-ni. Bo\-nyo\-lul\-tabb mo\-dell\-re van
sz\"uk\-s\'e\-g\"unk -- \'{\i}gy a k\'{\i}\-s\'er\-le\-ti\-leg min\-ded\-dig a\-l\'a nem t\'a\-masz\-tott
el\-m\'e\-le\-ti konst\-ruk\-ci\-\'ok k\"o\-z\"ul a legp\-rag\-ma\-ti\-ku\-sabb\-nak te\-kin\-tett
mi\-ni\-m\'a\-lis szu\-per\-szim\-met\-ri\-kus stan\-dard mo\-dell ke\-re\-t\'e\-ben foly\-tat\-juk a
ba\-ri\-o\-ge\-n\'e\-zis u\-t\'a\-ni haj\-sz\'at.

\chapter{A szu\-per\-szim\-met\-ri\-kus stan\-dard mo\-dell}
\fancyhead[CE]{\hst{\thechapter{}.\ fe\-je\-zet \quad A szu\-per\-szim\-met\-ri\-kus
stan\-dard mo\-dell}}
\section{A h\'o\-pe\-hely}
\bfr
\emph{Mi\-t\H ol len\-ne a szim\-met\-ri\-\'a\-nak b\'ar\-mi\-f\'e\-le je\-len\-t\H o\-s\'e\-ge?} \\
(Ma\-o Ce-tung) \cite{physspeak}
\efr

A\-mi\-kor egy h\'o\-peh\-lyet
fi\-zi\-ka\-i szem\-sz\"og\-b\H ol k\'{\i}\-v\'a\-nunk le\-\'{\i}r\-ni, a \emph{szimmetria} \'es a
\emph{frakt\'al} fo\-gal\-ma el\-ke\-r\"ul\-he\-tet\-le\-n\"ul fel\-buk\-kan.

\bef[hbt]
\bc
\epsfig{file=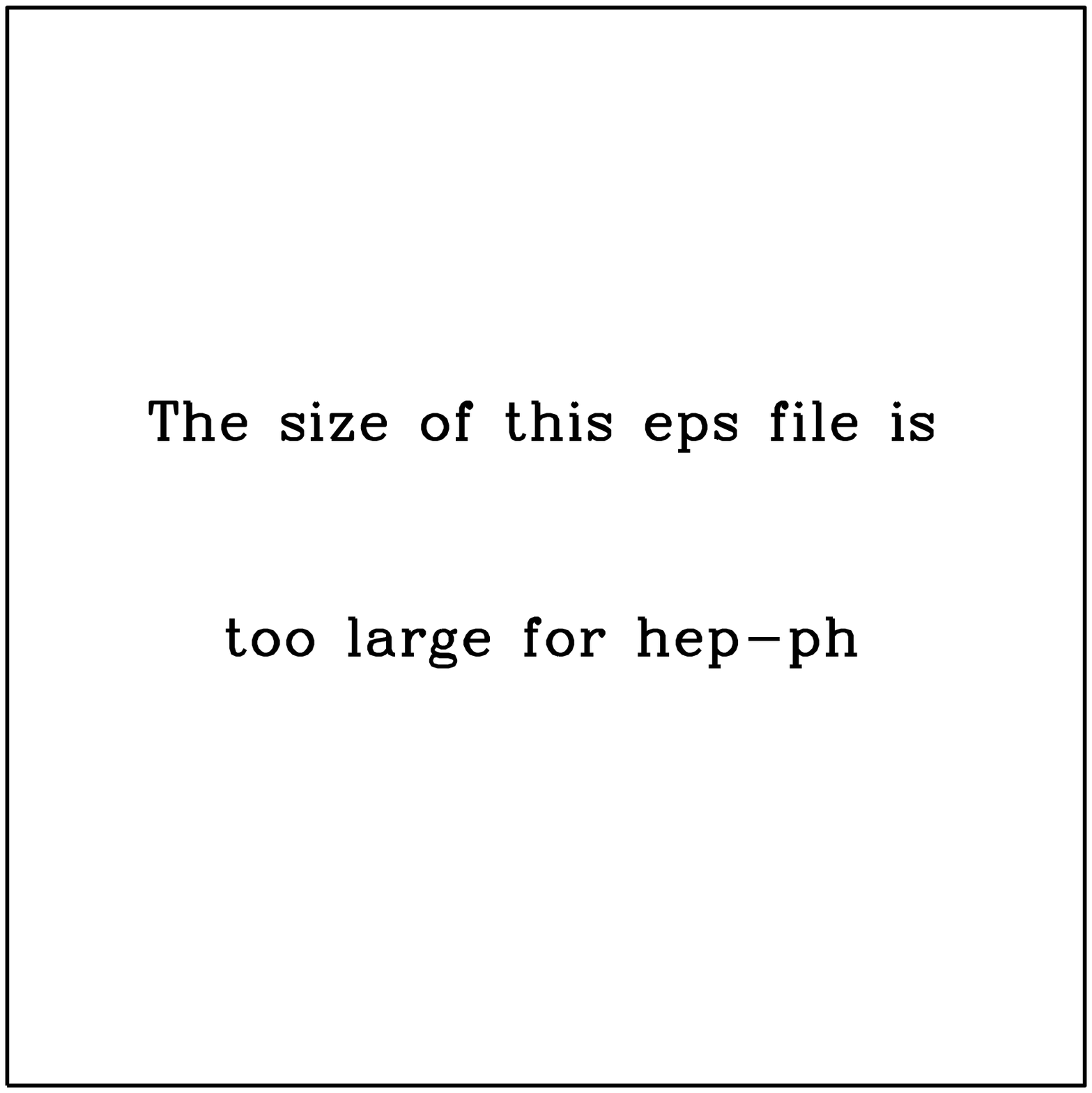,width=10cm}
\caption{Frak\-t\'al\-sze\-r\H u, hat\-fo\-g\'a\-s\'u szim\-met\-ri\-\'at mu\-ta\-t\'o h\'o\-pe\-hely
\cite{bentley}}
\ec
\enf

A fen\-ti \'ab\-r\'an j\'ol ki\-ve\-he\-t\H o hat\-fo\-g\'a\-s\'u szim\-met\-ri\-a a\-la\-pos
ta\-nul\-m\'a\-nyo\-z\'a\-sa a szi\-l\'ard\-test\-fi\-zi\-ka fon\-tos fe\-je\-ze\-te;
\"ot\-sz\"og\-le\-t\H u a\-na\-l\'og\-j\'a\-val a s\'{\i}k kv\'a\-zi\-pe\-ri\-o\-di\-kus le\-fe\-d\'e\-se is
meg\-va\-l\'o\-s\'{\i}t\-ha\-t\'o \cite{penrose}, mely a kv\'a\-zik\-ris\-t\'a\-lyok r\'e\-v\'en
\'u\-jabb, nem\-v\'art fiz\-ka\-i m\'ely\-s\'e\-ge\-ket nyit meg. A szim\-met\-ri\-a
r\'e\-szecs\-ke\-fi\-zi\-ka\-i al\-kal\-ma\-z\'a\-sa\-i\-nak hasz\-nos\-s\'a\-ga a\-lig\-ha
t\'ul\-be\-cs\"ul\-he\-t\H o -- er\-re a k\"o\-vet\-ke\-z\H o sza\-kasz\-ban t\"obb p\'el\-d\'at is
fo\-gunk l\'at\-ni.

A frak\-t\'a\-lok ha\-son\-l\'o\-k\'epp na\-gyon messzi\-re ve\-zet\-nek. K\"o\-z\"os
ma\-te\-ma\-ti\-ka\-i a\-la\-pot ny\'uj\-ta\-nak o\-lyan t\'a\-vo\-li te\-r\"u\-le\-tek sz\'a\-m\'a\-ra, mint
a f\"old\-rajz, a n\"o\-v\'eny\-vi\-l\'ag, vagy bi\-zo\-nyos sz\'a\-m\'{\i}\-t\'as\-tech\-ni\-ka\-i
prob\-l\'e\-m\'ak; a sz\'a\-m\'{\i}\-t\'o\-g\'e\-pes k\'ep\-z\H o\-m\H u\-v\'e\-szet r\'e\-v\'en pe\-dig to\-v\'ab\-bi
\'er\-de\-kes k\'er\-d\'e\-se\-ket vet\-nek fel.

A h\'o\-pe\-hely a\-zon\-ban se nem frak\-t\'al, se nem szim\-met\-ri\-kus: a\-to\-mi
sk\'a\-l\'an vizs\-g\'al\-va az \"on\-ha\-son\-l\'o szer\-ke\-zet nem tart\-ha\-t\'o fenn, \'es
kel\-l\H o a\-la\-pos\-s\'ag\-gal meg\-vizs\-g\'al\-va a szim\-met\-ri\-a sem bi\-zo\-nyul
t\"o\-k\'e\-le\-tes\-nek. A szim\-met\-ri\-a, a frak\-t\'al, \'es a mak\-rosz\-ko\-pi\-kus tes\-tek
vizs\-g\'a\-la\-t\'a\-b\'ol le\-sz\H urt to\-v\'ab\-bi se\-g\'ed\-fo\-gal\-ma\-ink, me\-lyek
t\"o\-k\'e\-let\-len t\H o\-r\H ol fa\-kad\-nak, \'o\-ri\-\'a\-si se\-g\'{\i}t\-s\'e\-get ny\'uj\-ta\-nak
fi\-zi\-ka\-i fo\-gal\-ma\-ink ki\-a\-la\-k\'{\i}\-t\'a\-s\'a\-ban. Rend\-sze\-re\-z\'est tesz\-nek
le\-he\-t\H o\-v\'e, mely e\-se\-ten\-k\'ent nem t\"o\-k\'e\-le\-tes, de k\"onnyen
\'at\-l\'at\-ha\-t\'o \'es me\-g\'ert\-he\-t\H o. Ez az\-zal is j\'ar, hogy a va\-l\'o\-s\'ag
le\-\'{\i}\-r\'a\-sa he\-lyett \'al\-ta\-l\'a\-ban an\-nak i\-de\-a\-li\-z\'alt k\'e\-p\'e\-vel kell
fog\-lal\-koz\-nunk. Sze\-ren\-cs\'es e\-set\-ben az el\-t\'e\-r\'es a mo\-dell
fi\-no\-m\'{\i}\-t\'a\-s\'a\-val cs\"ok\-kent\-he\-t\H o, \'{\i}gy a gya\-kor\-lat\-ban j\'ol
al\-kal\-maz\-ha\-t\'o k\"o\-ze\-l\'{\i}\-t\H o\-m\'od\-szert ka\-punk.

A mik\-ro\-vi\-l\'ag\-ban, szu\-ba\-to\-mi szin\-ten ma\-i tu\-d\'a\-sunk sze\-rint
\emph{l\'e\-te\-zik }t\"o\-k\'e\-le\-te\-sen meg\-va\-l\'o\-su\-l\'o szim\-met\-ri\-a (CPT, il\-let\-ve
pl.\ az e\-r\H os k\"ol\-cs\"on\-ha\-t\'as e\-se\-t\'e\-ben C, P, T k\"u\-l\"on--k\"u\-l\"on is).
Bi\-zo\-nyos szim\-met\-ri\-\'ak meg\-le\-p\H o m\'o\-don s\'e\-r\"ul\-nek -- a pa\-ri\-t\'as\-s\'er\-t\'es
il\-let\-ve a CP-s\'er\-t\'es gon\-do\-la\-ta \'epp e\-mi\-att na\-gyon ke\-ve\-sek\-ben \"ot\-l\"ott
fel 1956 e\-l\H ott -- m\'a\-sok ta\-l\'an \'ep\-pen meg\-le\-p\H o m\'o\-don va\-l\'o\-sul\-nak
meg. Egy i\-lyen le\-he\-t\H o\-s\'e\-get, a szu\-per\-szim\-met\-ri\-\'at vizs\-g\'a\-lunk meg az
a\-l\'ab\-bi\-ak\-ban.

\section{Szuperszimmetria}
\fancyhead[CO]{\hst{\thesection \quad Szu\-per\-szim\-met\-ri\-a}}
A r\'e\-szecs\-ke\-fi\-zi\-ka stan\-dard mo\-dell\-je a hu\-sza\-dik sz\'a\-za\-di el\-m\'e\-le\-ti
fi\-zi\-ka e\-gyik leg\-ki\-e\-mel\-ke\-d\H obb v\'{\i}v\-m\'a\-nya. E\-gyes j\'os\-la\-ta\-i -- min\-de\-nek
e\-l\H ott az e\-lekt\-ron a\-no\-m\'a\-lis m\'ag\-ne\-ses mo\-men\-tu\-m\'a\-ra vo\-nat\-ko\-z\'o
sz\'a\-m\'{\i}\-t\'a\-sok -- a tu\-do\-m\'any\-t\"or\-t\'e\-net\-ben e\-gye\-d\"u\-l\'al\-l\'o\-an pon\-to\-sak;
e\-zen fe\-l\"ul a mo\-dell a n\'egy a\-lap\-ve\-t\H o k\"ol\-cs\"on\-ha\-t\'as k\"o\-z\"ul h\'ar\-mat
esz\-t\'e\-ti\-ku\-san egy\-s\'e\-ges ke\-ret\-be fog\-lal, \'es ez\-zel nagy l\'e\-p\'est tesz a
mo\-dern el\-m\'e\-le\-ti fi\-zi\-ku\-sok ``Szent Gr\'al\-j\'a\-nak'' te\-kin\-tett
``Min\-den\-s\'eg El\-m\'e\-le\-t\'e\-nek'' me\-gal\-ko\-t\'a\-sa fe\-l\'e.

A stan\-dard mo\-dell e\-gyik a\-lap\-ve\-t\H o vo\-n\'a\-sa szim\-met\-ri\-kus jel\-le\-ge: m\'{\i}g a
Nyol\-cas \'Ut e\-se\-t\'e\-ben az el\-m\'e\-let \'al\-tal meg\-j\'o\-solt $\Omega^{--}$
ba\-ri\-on fel\-fe\-de\-z\'e\-se \'o\-ri\-\'a\-si szen\-z\'a\-ci\-\'ot ka\-vart, a top kvark
l\'e\-te\-z\'e\-s\'et m\'ar fel\-fe\-de\-z\'e\-se e\-l\H ott is tel\-je\-sen bi\-zo\-nyos\-ra le\-he\-tett
ven\-ni.

A\-zon\-ban b\'ar\-mi\-lyen ki\-v\'a\-l\'o\-an m\H u\-k\"o\-dik is a stan\-dard mo\-dell, nem ez a
v\'eg\-le\-ges el\-m\'e\-let. En\-nek e\-gyik leg\-k\'e\-zen\-fek\-v\H obb bi\-zo\-ny\'{\i}\-t\'e\-ka az
u\-ni\-ver\-zum ba\-ri\-on-a\-szim\-met\-ri\-\'a\-ja; e\-zen k\'{\i}\-v\"ul esz\-t\'e\-ti\-ka\-i
prob\-l\'e\-m\'ak is fel\-me\-r\"ul\-nek: a stan\-dard mo\-dell\-ben 19 pa\-ra\-m\'e\-ter
sze\-re\-pel, a\-mi t\'ul nagy sz\'am egy a\-lap\-ve\-t\H o el\-m\'e\-let e\-se\-t\'en; a
h\'a\-rom\-faj\-ta k\"ol\-cs\"on\-ha\-t\'as nincs kel\-l\H o\-k\'epp e\-gye\-s\'{\i}t\-ve, az er\-re
hi\-va\-tott nagy e\-gye\-s\'{\i}\-tett el\-m\'e\-le\-te\-ket (GU\-Tok) pe\-dig a
hi\-e\-rar\-chi\-a-prob\-l\'e\-ma te\-szi ne\-he\-zen el\-fo\-gad\-ha\-t\'o\-v\'a, stb.

Min\-de\-zen prob\-l\'e\-m\'ak el\-le\-n\'e\-re a stan\-dard mo\-dell \'er\-t\'e\-k\'et ne\-h\'ez
t\'ul\-be\-cs\"ul\-ni. Ha p\'ar\-hu\-za\-mot k\'{\i}\-v\'a\-nunk von\-ni, a New\-ton-f\'e\-le
me\-cha\-ni\-k\'at, majd az ezt to\-v\'abb\-fej\-lesz\-t\H o \'al\-ta\-l\'a\-nos
re\-la\-ti\-vi\-t\'a\-sel\-m\'e\-le\-tet e\-mel\-n\'enk ki: ha\-son\-l\'o m\'o\-don (b\'ar nem
fel\-t\'et\-le\-n\"ul ha\-son\-l\'o m\'er\-t\'ek\-ben) v\'a\-lik bo\-nyo\-lul\-tab\-b\'a min\-den\-faj\-ta
sz\'a\-m\'{\i}\-t\'a\-si prob\-l\'e\-ma, ha a stan\-dard mo\-dell\-r\H ol va\-la\-me\-lyik \'uj,
\'al\-ta\-l\'a\-no\-sabb el\-m\'e\-let-je\-l\"olt\-re t\'e\-r\"unk \'at. A p\'ar\-hu\-zam en\-n\'el
m\'e\-lyebb: a\-hogy E\-ins\-te\-in is egy a\-lap\-ja\-i\-ban egy\-sze\-r\H ubb, esz\-t\'e\-ti\-kus
el\-m\'e\-let\-tel l\'e\-pett t\'ul a new\-to\-ni fi\-zi\-k\'an, \'ugy a stan\-dard mo\-dell
leg\-va\-l\'o\-sz\'{\i}\-n\H ubb u\-t\'od\-j\'a\-nak te\-kin\-tett szu\-per\-szim\-met\-ri\-kus mo\-del\-lek
is ezt te\-szik.

Esz\-t\'e\-ti\-kus, te\-h\'at va\-la\-mi\-lyen (ne\-he\-zen k\"o\-r\"ul\-ha\-t\'a\-rol\-ha\-t\'o)
\'er\-te\-lem\-ben egy\-sze\-r\H u el\-m\'e\-le\-tet sze\-ret\-n\'enk. Ez a t\"o\-rek\-v\'es,
mond\-hat\-ni, e\-gyi\-d\H os a tu\-do\-m\'annyal: a\-mi\-kor A\-risz\-to\-te\-l\'esz
fi\-lo\-z\'o\-fi\-\'a\-j\'a\-ban meg\-ve\-tet\-te a mo\-dern ter\-m\'e\-szet\-tu\-do\-m\'a\-nyok a\-lap\-ja\-it,
az egy\-sze\-r\H u\-s\'eg \'es az esz\-t\'e\-ti\-ka k\"o\-ve\-tel\-m\'e\-nye\-it r\'ot\-ta ki a
csil\-la\-g\'a\-szat\-ra is: a boly\-g\'ok t\"o\-k\'e\-le\-tes k\"or\-p\'a\-ly\'ak men\-t\'en
v\'eg\-zik \'al\-lan\-d\'o, v\'al\-to\-zat\-lan moz\-g\'a\-su\-kat. Ez az esz\-t\'e\-ti\-ka\-i
k\"o\-ve\-tel\-m\'eny o\-lyan b\'ek\-ly\'o\-nak bi\-zo\-nyult, mely\-nek le\-ved\-l\'e\-s\'e\-hez majd
k\'et \'e\-vez\-red kel\-lett. Kep\-ler sza\-ba\-d\'{\i}\-tot\-ta meg a boly\-g\'o\-kat a
k\"or\-moz\-g\'as l\'an\-c\'a\-t\'ol \'es ren\-del\-te \H o\-ket -- sa\-j\'at n\'e\-ze\-te sze\-rint
vissza\-ta\-sz\'{\i}\-t\'o -- el\-lip\-szis\-p\'a\-ly\'ak\-ra, mi\-k\"oz\-ben egy m\'eg
hold\-k\'o\-ro\-sabb li\-d\'erc\-f\'enyt ker\-ge\-tett: egy o\-lyan esz\-t\'e\-ti\-kus
nap\-rend\-szer-mo\-dellt, mely\-ben a mi\-\'er\-tek is v\'a\-laszt kap\-nak, mely\-ben a
fi\-zi\-ka\-i vi\-l\'a\-got az \"ot pla\-t\'o\-i test for\-m\'a\-ja ha\-t\'a\-roz\-za meg.
E mo\-dell sze\-rint ha a Nap k\"o\-r\"ul k\"or\-p\'a\-ly\'an ke\-rin\-g\H o boly\-g\'ok
su\-ga\-r\'a\-val g\"om\-b\"o\-ket raj\-zo\-lunk, a(z ak\-kor is\-mert) hat boly\-g\'o
g\"omb\-je k\"o\-z\'e be\-il\-leszt\-he\-t\H o az \"ot sza\-b\'a\-lyos test oly m\'o\-don, hogy
a\-zok az e\-gyik g\"omb k\"o\-r\'e \'es a r\'a\-k\"o\-vet\-ke\-z\H o g\"omb bel\-se\-j\'e\-be
le\-gye\-nek \'{\i}r\-va. (A\-mi\-b\H ol j\'ol l\'at\-szik, hogy a zse\-ni\-\'a\-lis
fel\-fe\-de\-z\'e\-sek\-hez j\'o a\-dag sze\-ren\-cse \'es meg\-fe\-le\-l\H o i\-d\H o\-z\'{\i}\-t\'es is
kell: ha Kep\-ler i\-de\-j\'en az U\-r\'a\-nusz m\'ar is\-mert lett vol\-na, Kep\-ler
fe\-j\'e\-be a\-lig\-ha f\'esz\-kel\-te vol\-na be ma\-g\'at a fen\-ti k\'ep.)
\\
Kep\-ler el\-lip\-szis\-p\'a\-ly\'a\-it ma m\'ar esz\-t\'e\-ti\-kus\-nak tart\-juk: a
New\-ton-f\'e\-le le\-\'{\i}\-r\'as sze\-rint u\-gya\-nis csak k\'et\-f\'e\-le cent\-r\'a\-lis
e\-r\H o\-t\'er e\-se\-t\'en ka\-punk min\-den e\-set\-ben z\'art p\'a\-ly\'a\-kat, e\-zek k\"o\-z\"ul
az e\-gyik a Kep\-ler-prob\-l\'e\-ma.

\'All\-jon itt m\'eg egy XX.\ sz\'a\-za\-di p\'el\-da az esz\-t\'e\-ti\-ka\-i szem\-pon\-tok
il\-luszt\-r\'a\-l\'a\-s\'a\-ra: Di\-rac p\'el\-d\'a\-ja. Sa\-j\'at be\-val\-l\'a\-sa sze\-rint
Di\-rac-ot esz\-t\'e\-ti\-ka\-i szem\-pon\-tok ve\-zet\-t\'ek h\'{\i}\-res e\-gyen\-le\-t\'e\-nek
fe\-l\'{\i}\-r\'a\-s\'a\-hoz \cite{weinberg}; a (fe\-les) spin le\-\'{\i}\-r\'a\-s\'a\-ra konst\-ru\-\'alt
e\-gyen\-let a\-zon\-ban meg\-h\"ok\-ken\-t\H o j\'os\-lat\-tal \'allt e\-l\H o: ne\-ga\-t\'{\i}v
e\-ner\-gi\-\'as \'al\-la\-po\-tok\-kal. Az e\-zek\-nek meg\-fe\-le\-l\H o ``an\-ti\-r\'e\-szecs\-k\'ek''
meg\-ta\-l\'a\-l\'a\-sa tet\-te a Di\-rac-e\-gyen\-le\-tet a r\'e\-szecs\-ke\-fi\-zi\-ka e\-gyik
leg\-fon\-to\-sabb e\-gyen\-le\-t\'e\-v\'e.

Di\-rac egy m\'a\-sik, fe\-let\-t\'ebb e\-le\-g\'ans j\'os\-la\-ta a m\'ag\-ne\-ses
mo\-no\-p\'o\-lu\-sok l\'e\-te\-z\'e\-se -- mellyel a stan\-dard mo\-dell e\-gyik nagy
rej\-t\'e\-lye, a t\"ol\-t\'es kvan\-t\'alt\-s\'a\-ga ma\-gya\-r\'az\-ha\-t\'o. Er\-r\H ol a
j\'os\-lat\-r\'ol ma\-ga Di\-rac \'{\i}gy v\'e\-le\-ke\-dett: \medskip \\
\emph{``El\-m\'e\-le\-ti szem\-pont\-b\'ol \'ugy gon\-dol\-hat\-juk, hogy a [m\'ag\-ne\-ses]
mo\-no\-p\'o\-lu\-sok\-nak a ma\-te\-ma\-ti\-ka\-i gon\-do\-lat\-me\-net sz\'ep\-s\'e\-ge mi\-att
l\'e\-tez\-ni\-\"uk kell. A sz\'a\-mos pr\'o\-b\'al\-ko\-z\'as el\-le\-n\'e\-re a\-zon\-ban
min\-de\-zi\-de\-ig nem si\-ke\-r\"ult a nyo\-muk\-ra buk\-kan\-nunk. Azt a
k\"o\-vet\-kez\-te\-t\'est kell te\-h\'at le\-von\-nunk, hogy a ma\-te\-ma\-ti\-ka sz\'ep\-s\'e\-ge
\"on\-ma\-g\'a\-ban nem e\-le\-gen\-d\H o ok ar\-ra, hogy a ter\-m\'e\-szet a sz\'o\-ban for\-g\'o
el\-m\'e\-le\-tet meg is va\-l\'o\-s\'{\i}t\-sa.''} \cite{physspeak}
\bigskip

Az esz\-t\'e\-ti\-ka\-i szem\-pon\-tok te\-h\'at le\-het\-nek j\'o i\-r\'any\-jel\-z\H ok -- de
en\-n\'el t\"ob\-bet nem \'al\-l\'{\i}t\-ha\-tunk. \'Igy az a\-l\'ab\-bi\-ak\-ban t\'ar\-gya\-lan\-d\'o
szu\-per\-szim\-met\-ri\-kus stan\-dard mo\-dellt nem t\"obb, mint egy le\-het\-s\'e\-ges
el\-m\'e\-le\-ti konst\-ruk\-ci\-\'o, mely ta\-l\'an t\'ul\-s\'a\-go\-san is nagy hang\-s\'ulyt
fek\-tet a szim\-met\-ri\-\'ak je\-len\-t\H o\-s\'e\-g\'e\-re \'es ke\-v\'es\-s\'e tesz e\-le\-get a
k\"o\-vet\-ke\-z\H o, prag\-ma\-ti\-kus k\"o\-ve\-tel\-m\'eny\-nek: \medskip \\
\emph{A fi\-zi\-ka f\H o c\'el\-ja, hogy mi\-n\'el t\"obb je\-len\-s\'e\-get \'{\i}r\-jon le
mi\-n\'el ke\-ve\-sebb v\'al\-to\-z\'o se\-g\'{\i}t\-s\'e\-g\'e\-vel.} \cite{cerncour}
\medskip

Ah\-hoz te\-h\'at, hogy a k\"o\-vet\-ke\-z\H ok\-ben t\'ar\-gya\-l\'as\-ra ke\-r\"u\-l\H o el\-m\'e\-le\-tet
ko\-mo\-lyan ve\-hes\-s\"uk, k\'{\i}\-s\'er\-le\-ti in\-di\-k\'a\-ci\-\'ok kel\-le\-nek. Sz\'a\-mos \'uj
r\'e\-szecs\-k\'et kell ta\-l\'al\-nunk a k\"o\-zel\-j\"o\-v\H o\-ben, ha a ba\-ri\-o\-ge\-n\'e\-zist
az MSSM ke\-re\-t\'e\-ben k\'{\i}\-v\'an\-juk meg\-ma\-gya\-r\'az\-ni. Az e\-s\'e\-lyek e\-gy\'al\-ta\-l\'an
nem biz\-ta\-t\'o\-ak. Sz\'a\-mos \'uj pa\-ra\-m\'e\-te\-rek ve\-ze\-t\"unk be ab\-b\'ol a
c\'el\-b\'ol, hogy az u\-ni\-ver\-zum\-ban meg\-fi\-gyelt ba\-ri\-on--fo\-ton h\'a\-nya\-dost
meg\-ma\-gya\-r\'az\-zuk. Azt is l\'at\-ni fog\-juk, hogy e\-zen
pa\-ra\-m\'e\-ter\-t\'er\-ben i\-gen ki\-csit az a tar\-to\-m\'any, a\-hol ez le\-het\-s\'e\-ges.

Sok\-kal na\-gyobb az e\-s\'e\-lye an\-nak, hogy m\'eg ha ta\-l\'a\-lunk is
szu\-per\-szim\-met\-ri\-kus r\'e\-szecs\-k\'e\-ket, a ba\-ri\-on-a\-szim\-met\-ri\-a prob\-l\'e\-ma
a\-lap\-j\'an ar\-ra a k\"o\-vet\-kez\-te\-t\'es\-re ju\-tunk: a mi\-ni\-m\'a\-lis
szu\-per\-szim\-met\-ri\-kus stan\-dard mo\-dell nem e\-l\'eg. Egy i\-lyen e\-red\-m\'enyt
vi\-szont nem\-csak e\-l\'er\-ni k\"onnyebb, ha\-nem el\-hin\-ni is.

Min\-de\-zek el\-le\-n\'e\-re a szu\-per\-szim\-met\-ri\-kus mo\-dell te\-kint\-he\-t\H o a stan\-dard
mo\-dell ma is\-mert legp\-rag\-ma\-ti\-ku\-sabb ki\-b\H o\-v\'{\i}\-t\'e\-s\'e\-nek; az pe\-dig, hogy a
szu\-per\-szim\-met\-ri\-kus mo\-del\-lek\-re jel\-lem\-z\H o e\-ner\-gi\-a\-tar\-to\-m\'a\-nyok a k\"o\-ze\-li
j\"o\-v\H o gyor\-s\'{\i}\-t\'o\-i\-ban e\-l\'er\-h\'e\-t\H o\-ek lesz\-nek, fe\-let\-t\'ebb m\'eg von\-z\'obb
f\'enyt vet az MSSM-re.

\section{A f\'a\-zi\-s\'at\-me\-net per\-tur\-ba\-t\'{\i}v vizs\-g\'a\-la\-ta}
\fancyhead[CO]{\hst{\thesection \quad A f\'a\-zis\'at\-me\-net
per\-tur\-bat\'{\i}v vizsg\'a\-la\-ta}}
Mi\-e\-l\H ott be\-le\-fog\-n\'ank az MSSM-be\-li e\-lekt\-ro\-gyen\-ge f\'a\-zi\-s\'at\-me\-net
vizs\-g\'a\-la\-t\'a\-ba, fel\-ve\-t\"unk n\'e\-h\'any, e mo\-dell vizs\-g\'a\-la\-ta mel\-lett
sz\'o\-l\'o \'er\-vet. A \ref{veff} \'ab\-r\'an l\'at\-ha\-t\'o po\-ten\-ci\-\'alt a
\beq
V_{\mrm{eff}} = A \Phi^2 - B \Phi^3 + C \Phi^4
\enq
k\"o\-ze\-l\'{\i}\-t\H o a\-lak\-ba \'{\i}r\-va a $\Phi^2$-s tag $T^2-T_C^2$-tel a\-r\'a\-nyos,
te\-h\'at ez fe\-le\-l\H os a ma\-gas h\H o\-m\'er\-s\'ek\-le\-ten hely\-re\-\'al\-l\'o szim\-met\-ri\-\'a\-\'ert,
a\-zon\-ban a f\'a\-zi\-s\'at\-me\-ne\-ti pon\-tot a\-lap\-ve\-t\H o\-en a k\'et mi\-ni\-mum k\"oz\-ti
p\'up ha\-t\'a\-roz\-za meg, mely a k\"o\-b\"os tag\-gal \'all kap\-cso\-lat\-ban
\cite{clin00}. Ah\-hoz, hogy a f\'a\-zi\-s\'at\-me\-net v\'eg\-pont\-ja ki\-jebb
tol\-ha\-t\'o le\-gyen, $B$ \'er\-t\'e\-k\'et n\"o\-vel\-ni kell. $B$ a Higgs-t\"o\-meg\-gel
\'all kap\-cso\-lat\-ban; a Higgs-t\"o\-meg n\"o\-ve\-l\'e\-se so\-r\'an az $\sqrt{m^2 +
c T^2}^3$ plaz\-ma-t\"o\-meg\-gel a\-r\'a\-nyos $B$ a\-zon\-ban cs\"ok\-ken, \'{\i}gy a
a f\'a\-zi\-s\'at\-me\-net gyen\-g\"u\-l\'e\-s\'e\-vel j\'ar e\-gy\"utt. E\-z\'ert te\-h\'at
re\-a\-lisz\-ti\-kus Higgs-t\"o\-me\-gek\-n\'el nincs i\-ga\-zi f\'a\-zi\-s\'at\-me\-net, csu\-p\'an
egy si\-ma ``cross-o\-ver''. A szu\-per\-szim\-met\-ri\-kus mo\-dell sz\'a\-mos
is\-me\-ret\-len pa\-ra\-m\'e\-te\-re a\-zon\-ban han\-gol\-ha\-t\'o \'ugy, hogy a k\'{\i}\-s\'er\-le\-ti\-leg
min\-ded\-dig ki nem z\'art Higgs-t\"o\-meg tar\-to\-m\'any\-ban is le\-het\-s\'e\-ges
le\-gyen az el\-s\H o\-ren\-d\H u f\'a\-zi\-s\'at\-me\-net. Az e\-lekt\-ro\-gyen\-ge f\'a\-zi\-s\'at\-me\-net
MSSM-en be\-l\"u\-li vizs\-g\'a\-la\-t\'a\-nak f\H o k\'er\-d\'e\-se te\-h\'at az,
hogy a pa\-ra\-m\'e\-ter\-t\'er me\-lyik (\'es mek\-ko\-ra m\'e\-re\-t\H u) r\'e\-sz\'e\-ben
va\-l\'o\-s\'{\i}t\-ha\-t\'o ez meg.

Szem\-ben a stan\-dard mo\-del\-lel, a\-hol a ba\-ri\-o\-ge\-n\'e\-zis\-hez sz\"uk\-s\'e\-ges
CP-s\'er\-t\'es t\'ul ki\-csi, az MSSM-ben ez a Sza\-ha\-rov-fel\-t\'e\-tel is
tel\-je\-s\'{\i}t\-he\-t\H o: $(2-3) \times 10^{-3}$ nagy\-s\'a\-g\'u CP-s\'er\-t\H o f\'a\-zis
is e\-le\-gen\-d\H o \cite{clin00}, mely nem ter\-m\'e\-sze\-tel\-le\-ne\-sen nagy.
\medskip

Az MSSM-be\-li e\-lekt\-ro\-gyen\-ge f\'a\-zi\-s\'at\-me\-net vizs\-g\'a\-la\-t\'a\-r\'ol sz\'o\-l\'o
el\-s\H o dol\-go\-za\-tok per\-tur\-ba\-t\'{\i}v meg\-k\"o\-ze\-l\'{\i}\-t\'e\-s\H u\-ek. E\-zek e\-red\-m\'e\-nye
sze\-rint az MSSM-ben sok\-kal e\-r\H o\-sebb f\'a\-zi\-s\'at\-me\-net le\-he\-s\'e\-ges, mint a
stan\-dard mo\-dell\-ben \cite{car97, bod, los, gu, esp, brig, esp2, carl,
clin}, k\"u\-l\"o\-n\"o\-sen ak\-kor, ha a top kvark t\"o\-me\-ge meg\-ha\-lad\-ja
jobb\-ke\-zes szu\-per\-szim\-met\-ri\-kus p\'ar\-j\'a\-nak t\"o\-me\-g\'et \cite{car96, car98}.
A ba\-ri\-o\-ge\-n\'e\-zis\-hez sz\"uk\-s\'e\-ges je\-len\-t\H os CP-s\'er\-t\'es
meg\-va\-l\'o\-s\'{\i}\-t\'a\-s\'a\-ra is na\-gyobb t\'er k\'{\i}\-n\'al\-ko\-zik a szu\-per\-szim\-met\-ri\-kus
mo\-dell\-ben \cite{fun, la99}. Az a\-l\'ab\-bi\-ak\-ban e\-l\H o\-sz\"or a per\-tur\-ba\-t\'{\i}v
meg\-k\"o\-ze\-l\'{\i}\-t\'est is\-mer\-te\-tem, mely\-nek e\-red\-m\'e\-nye\-i fel\-hasz\-n\'al\-ha\-t\'o\-ak
a n\'egy\-di\-men\-zi\-\'os r\'acsszi\-mu\-l\'a\-ci\-\'ok pa\-ra\-m\'e\-te\-re\-i\-nek
meg\-v\'a\-lasz\-t\'a\-sa\-kor \'es az ott ka\-pott e\-red\-m\'e\-nyek \'er\-t\'e\-ke\-l\'e\-se\-kor.
\subsection{A $\Phi$ i\-r\'a\-ny\'u po\-ten\-ci\-\'al}
A per\-tur\-ba\-t\'{\i}v meg\-k\"o\-ze\-l\'{\i}\-t\'es a\-lap\-j\'at k\'e\-pe\-z\H o ef\-fek\-t\'{\i}v
po\-ten\-ci\-\'al\-b\'ol in\-du\-lunk ki, egy\-hu\-rok szin\-ten \cite{car97};
az a\-l\'abb hasz\-n\'alt je\-l\"o\-l\'e\-sek is e cikk kon\-ven\-ci\-\'o\-it k\"o\-ve\-tik. A
po\-ten\-ci\-\'al a\-lak\-j\'at i\-gyek\-sz\"unk a le\-he\-t\H o le\-gegy\-sze\-r\H ubb\-re
v\'a\-lasz\-ta\-ni; a stan\-dard mo\-dell\-n\'el l\'a\-tot\-tak\-hoz ha\-son\-l\'o\-an a
fer\-mi\-o\-ni\-kus szek\-tor\-t\'ol itt is el\-te\-kin\-t\"unk. M\'er\-t\'ek\-r\"og\-z\'{\i}\-t\'es\-kor a
't~Ho\-oft--Lan\-da\-u-m\'er\-t\'e\-ket v\'a\-laszt\-juk; az
egy\-hu\-rok-szin\-ten meg\-je\-le\-n\H o di\-ver\-gen\-ci\-\'a\-kat az \ms\ m\'od\-szer\-rel
k\"u\-sz\"o\-b\"ol\-j\"uk ki. Az \ms\ s\'e\-m\'a\-ban fel\-buk\-ka\-n\'o $\mu$
t\"o\-meg\-pa\-ra\-m\'e\-tert c\'el\-sze\-r\H u lesz \'al\-ta\-l\'a\-ban az el\-m\'e\-let m\'a\-sik
e\-ner\-gi\-a-di\-men\-zi\-\'o\-j\'u pa\-ra\-m\'e\-te\-r\'e\-vel, a $T$ h\H o\-m\'er\-s\'ek\-let\-tel
a\-zo\-no\-s\'{\i}\-ta\-ni. Az SU(2)--Higgs-mo\-dell ki\-ter\-jesz\-t\'e\-s\'e\-nek
vizs\-g\'a\-la\-ta\-kor a $W$ \'es $Z$ t\"o\-me\-gek to\-v\'abb\-ra is me\-ge\-gyez\-nek, m\'as
sz\'o\-val a po\-ten\-ci\-\'al\-ban \'al\-ta\-l\'a\-nos e\-set\-ben je\-len l\'e\-v\H o $g'$
csa\-to\-l\'ast 0-nak v\'a\-laszt\-juk.

A po\-ten\-ci\-\'al leg\-je\-len\-t\H o\-sebb j\'a\-ru\-l\'e\-ka\-it a stan\-dard mo\-dell\-be\-li
r\'e\-szecs\-k\'e\-ken ($W$, $Z$, $h$, $H$ (Higgs-dub\-lett),
$\chi$ (Golds\-to\-ne-bo\-zon)) k\'{\i}\-v\"ul a top-kvark szu\-per\-szim\-met\-ri\-kus
p\'ar\-ja, $\tilde{t}_L, \tilde{t}_R$ ad\-ja. B\'ar a $t$ top kvark
j\'a\-ru\-l\'e\-ka is sz\'a\-mot\-te\-v\H o, az SU(2)--Higgs-mo\-dell\-hez ha\-son\-l\'o\-an el\-s\H o
meg\-k\"o\-ze\-l\'{\i}\-t\'es\-ben ezt nem vessz\"uk fi\-gye\-lem\-be. B\'ar a n\'egy\-di\-men\-zi\-\'os
r\'acsszi\-mu\-l\'a\-ci\-\'ok\-ban az MSSM mind\-k\'et Higgs-dub\-lett\-j\'et fi\-gye\-lem\-be
vessz\"uk, itt a k\"onnyebb\-s\'e\-g\'ert csak az e\-gyi\-ket. (A m\'a\-so\-dik
Higgs-dub\-lett el\-ha\-gy\'a\-sa a di\-men\-zi\-\'os re\-duk\-ci\-\'on a\-la\-pu\-l\'o
vizs\-g\'a\-la\-tok\-ban is el\-ter\-jedt \cite{la98}).

A po\-ten\-ci\-\'al\-ban sze\-rep\-l\H o pa\-ra\-m\'e\-te\-rek k\"o\-z\"ul az a\-l\'ab\-bi\-a\-kat kell
k\'ez\-zel be\-ten\-ni: $h_t$, $m_Q$, $m_U$, $A_t$, $\beta$ -- te\-h\'at e\-zen
pa\-ra\-m\'e\-te\-rek te\-r\'e\-ben ke\-res\-s\"uk azt a tar\-to\-m\'anyt, a\-hol a $T$
h\H o\-m\'er\-s\'ek\-let\-tel jel\-lem\-zett f\'a\-zi\-s\'at\-me\-net fi\-zi\-ka\-i\-lag \'er\-de\-kes --
p\'el\-d\'a\-ul e\-le\-get tesz a (\ref{vtpert}) fel\-t\'e\-tel\-nek.

A stan\-dard mo\-dell\-be\-li r\'e\-szecs\-k\'ek t\"o\-me\-ge\-i:
\beqar
m_W^2 & = & \frac14 g^2 \phi^2 \\
m_Z^2 & = & \frac14 g^2 \phi^2 \\
m_t^2 & = & \frac12 \sin \beta h_t^2 \phi^2 \\
m_h^2 & = & \frac12 \left[m_A^2 + m_Y^2 - \sqrt{m_A^4 + m_Y^4 - 2
      m_A^2 m_Y^2 \cos 4 \beta} \right] \\
m_H^2 & = & \frac12 \left[m_A^2 + m_Y^2 + \sqrt{m_A^4 + m_Y^4 - 2
      m_A^2 m_Y^2 \cos 4 \beta} \right]
\enqar
a\-hol $v$ a z\'e\-rus-h\H o\-m\'er\-s\'ek\-le\-t\H u v\'a\-ku\-um-v\'ar\-ha\-t\'o \'er\-t\'ek, kb.\ 246
GeV, to\-v\'ab\-b\'a
\beq
m_Y^2 = \frac18 g^2 \left(3\phi^2-v^2\right),
\enq
\'{\i}gy az $m_A \rightarrow \infty$ ha\-t\'a\-re\-set\-ben a $H$ Higgs
le\-csa\-to\-l\'o\-dik, mi\-vel t\"o\-me\-ge v\'e\-ge\-tel\-hez tart, a $h$ Higgs t\"o\-me\-ge
pe\-dig
\beq
m_h = \frac18 g^2 \cos^2 2 \beta \left(3 \phi^2
- v^2 \right)
\enq
lesz. A szi\-mu\-l\'a\-ci\-\'ok\-ban ti\-pi\-kus $m_A$ \'er\-t\'e\-kek mel\-lett (300 GeV,
150 GeV) \'ugy ta\-l\'al\-tam, hogy a na\-gyobb t\"o\-me\-g\H u Higgs i\-gen kis
m\'er\-t\'ek\-ben m\'o\-do\-s\'{\i}t\-ja a po\-ten\-ci\-\'alt: m\'eg a 150 GeV-es e\-set\-ben is
csak ez\-re\-l\'ek\-nyi kor\-rek\-ci\-\'ot o\-koz -- \'{\i}gy az $m_A \rightarrow \infty$
ha\-t\'a\-re\-set i\-gen nagy tar\-to\-m\'any\-ban j\'o k\"o\-ze\-l\'{\i}\-t\'es -- gyak\-ran ezt
fo\-gom hasz\-n\'al\-ni.

A Golds\-to\-ne-bo\-zon t\"o\-me\-g\'e\-re
\beq
m_\chi^2 = \frac18 g^2 \cos^2 2\beta \left(\phi^2 - v^2 \right)
\enq
a\-d\'o\-dik.

A transz\-ver\-z\'a\-lis sza\-bad\-s\'a\-gik fo\-ko\-kat $T$, a lon\-gi\-tu\-di\-n\'a\-li\-so\-kat $L$
in\-dex\-szel je\-l\"ol\-ve
\beq
n_{W_L} = 2, \quad
n_{W_T} = 4, \quad
n_{Z_L} = 1, \quad
n_{Z_T} = 2, \quad
n_h = 1, \quad
n_\chi = 3, \quad
\enq

A top-kvar\-kot is fi\-gye\-lem\-be ve\-v\H o e\-set\-ben $n_t = -12$ len\-ne.
A bal- \'es jobb\-ke\-zes stop-te\-rek 6--6 sza\-bad\-s\'a\-gi fo\-kot hor\-doz\-nak;
\beq
n_{\tilde{t}_L} = n_{\tilde{t}_R} = 6,
\enq
t\"o\-meg\-n\'egy\-ze\-t\"uk pe\-dig a
\beq
{\cal{M}}^2_{\tilde{t}} = \left(
\begin{array}{cc}
 m_Q^2 + m_t^2 + \frac18 g^2 \cos 2\beta \phi^2 &
 m_t \tilde{A}_t
 \\
 m_t \tilde{A}_t &
 m_U^2 + m_t^2
\end{array}
\right)
\enq
t\"o\-meg\-m\'at\-rix\-b\'ol sz\'ar\-maz\-tat\-ha\-t\'o, mely\-ben az $\tilde{A}_t$
pa\-ra\-m\'e\-tert
\beq
\tilde{A}_t = A_t - \frac{\mu}{\tan \beta}
\enq
ad\-ja meg.

V\'e\-ges h\H o\-m\'er\-s\'ek\-le\-ten a lon\-gi\-tu\-di\-n\'a\-lis t\"o\-me\-gek ter\-mi\-kus
kor\-rek\-ci\-\'ot kap\-nak, a v\'e\-ges h\H o\-m\'er\-s\'ek\-le\-t\H u t\"o\-me\-ge\-ket
fe\-l\"ul\-vo\-n\'as\-sal je\-l\"ol\-j\"uk.
A stan\-dard mo\-dell\-b\H ol is\-mert r\'e\-szecs\-k\'ek e\-se\-t\'en
\beqar
\bar m^2_{W_L}	&=& m_W^2 + \Pi_W, \\
\bar m^2_{Z_L}	&=& m_W^2 + \Pi_W, \\
\bar m^2_{h}	&=& m_h^2 + \Pi_h, \\
\bar m^2_{\chi} &=& m_\chi^2 + \Pi_\chi,
\enqar
a\-hol a v\'e\-ges h\H o\-m\'er\-s\'ek\-le\-t\H u sa\-j\'a\-te\-ner\-gi\-\'ak
\beqar
\Pi_W	 &=& \frac73 g^2 T^2, \\
\Pi_h	 &=& \frac{1}{16} g^2 \cos^2 2\beta T^2 + \frac{5}{16} g^2
	     T^2 + \frac{1}{12} h_t^2 \sin^2 \beta T^2, \\
\Pi_\chi &=& \Pi_h,
\enqar
m\'{\i}g a sto\-pok\-ra a
\beqar
\Pi_{\tilde{t}_L} &=& \frac{1}{3} g_s^2 T^2 + \frac{5}{6} g^2 T^2 +
		      \frac{1}{12} h_t^2 \left(2 + \sin^2 \beta
		      \right) T^2, \\
\Pi_{\tilde{t}_R} &=& \frac{4}{9} g_s^2 T^2 + \frac{1}{6} h_t^2
		      \left[ 1 + \sin^2 \beta \left(1 -
		      frac{\tilde{A}_t^2} {m_Q^2} \right) \right] T^2
\enqar
sa\-j\'a\-te\-ner\-gi\-\'a\-kat a
\beq
{\cal{M}}^2_{\tilde{t}} = \left(
\begin{array}{cc}
 m_Q^2 + m_t^2 + \frac18 g^2 \cos 2\beta \phi^2 + \Pi_{\tilde{t}_L} &
 m_t \tilde{A}_t
 \\
 m_t \tilde{A}_t &
 m_U^2 + m_t^2 + \Pi_{\tilde{t}_R}
\end{array}
\right)
\enq
t\"o\-meg\-m\'at\-rix\-ba t\'e\-ve an\-nak sa\-j\'a\-t\'er\-t\'e\-ke\-k\'ent kap\-juk meg a
kor\-ri\-g\'alt t\"o\-meg\-n\'egy\-ze\-te\-ket.

A fen\-ti\-ek se\-g\'{\i}t\-s\'e\-g\'e\-vel m\'ar fe\-l\'{\i}r\-ha\-t\'o az ef\-fek\-t\'{\i}v po\-ten\-ci\-\'al
\beq
V(\phi, T) = V_0 + V_1 + V_2 + \ldots
\enq
a\-lak\-ban, a\-hol $V_n$ az $n$-hu\-rok\-ren\-d\H u po\-ten\-ci\-\'al\-j\'a\-ru\-l\'ek.
A fag\-r\'af\-szin\-t\H u $V_0$ tag ki\-fe\-je\-z\'e\-se
\beq
V_0 = - \frac12 m^2(\mu) \phi^2 + \frac{1}{32} g^2 \cos^2 2\beta
\phi^4,
\enq
a\-hol
\beq
m^2 (\mu) = \frac12 m_Z^2(v) \cos^2 2\beta + \sum_i \frac{n_i}{16
\pi^2} m_i^2(v) \frac{dm_i^2 (v)}{dv^2} \left[ \log \frac{m_i^2
(v)}{\mu^2} + \frac12 - C_i \right].
\enq
Az $i$ in\-dex a $W$, $Z$, $h$, $\chi$, $\tilde t$, $\tilde T$
r\'e\-szecs\-k\'e\-ken fut v\'e\-gig. A ki\-fe\-je\-z\'es\-ben sze\-rep\-l\H o $C_i$ kons\-tans
\'er\-t\'e\-ke m\'er\-t\'ek\-bo\-zo\-nok\-ra $3/2$, a t\"ob\-bi (ska\-l\'ar)r\'e\-szecs\-k\'e\-re
$5/6$.

Az egy\-hu\-rok\-ren\-d\H u j\'a\-ru\-l\'ek a m\'er\-t\'ek\-bo\-zo\-nok lon\-gi\-tu\-di\-n\'a\-lis
kom\-po\-nen\-s\'e\-nek, il\-let\-ve a $h$ Higgs, $\chi$ Golds\-to\-ne-bo\-zon \'es a
$\tilde t$ k\"onny\H u stop $n=0$ m\'o\-du\-sa\-i\-nak da\-isy-fe\-l\"osszeg\-z\'e\-s\'e\-b\H ol
kap\-ha\-t\'o meg \cite{car97}:
\beq
V_1 = \sum_i \frac{n_i}{64 \pi^2} M_i^4 \left( \log \frac{M_i^2}
{\mu^2} - C_i \right) + \sum_i \frac{n_i}{2 \pi^2} J^{(i)} T^4
\label{V_1}
\enq
Az $i$ in\-dex az e\-l\H o\-z\H o r\'e\-szecs\-k\'e\-ken fut v\'e\-gig; a $W$ \'es $Z$
e\-se\-t\'e\-ben a\-zon\-ban k\"u\-l\"on kell ke\-zel\-ni a lon\-gi\-tu\-di\-n\'a\-lis m\'o\-du\-so\-kat
($W_L$, $Z_L$) \'es a transz\-ver\-z\'a\-li\-sa\-kat ($W_T$, $Z_T$). A
ki\-fe\-je\-z\'es\-ben sze\-rep\-l\H o $M_i$ t\"o\-me\-gek a ko\-r\'ab\-ban de\-fi\-ni\-\'alt $m_i$
\'es $\bar m_i$ t\"o\-me\-gek\-kel a\-zo\-no\-sak, konk\-r\'e\-tan
\beq
M_i = \left\{
\begin{array}{l}
 m_i, \quad i = W_L, Z_L, W_T, Z_T, \tilde T, \\
 \bar m_i, \quad i = h, \chi, \tilde t.
\end{array}
\right.
\enq
A $J^{(i)}$ ter\-mi\-kus j\'a\-ru\-l\'ek
\beq
J^{(i)} = \left\{
\begin{array}{l}
 J_B(m_i^2) - \frac{\pi}{6} \left( \bar m_i^2 - m_i^2 \right),
  \quad i = W_L, Z_L, \\
 J_B(m_i^2), \quad i = W_T, Z_T, \tilde T, \\
 J_B(\bar m_i^2), \quad i = h, \chi, \tilde t,
\end{array}
\right.
\enq
a\-hol a bo\-zo\-ni\-kus ter\-mi\-kus in\-teg\-r\'al
\beq
J_B(y^2) = \int_0^{\infty} dx \ x^2 \log \left( 1 - e^{- \sqrt{x^2 +
y^2}} \right).
\enq
Mi\-vel a \emph{Maple} prog\-ram\-mal k\'{\i}\-v\'a\-nom a f\'a\-zi\-s\'at\-me\-ne\-tet
vizs\-g\'al\-ni, c\'el\-sze\-r\H u az e\-l\H o\-z\H o in\-teg\-r\'al\-ki\-fe\-je\-z\'es he\-lyett
k\"onnyeb\-ben ke\-zel\-he\-t\H o a\-la\-kot ke\-res\-ni. Az i\-ro\-da\-lom\-b\'ol \cite{boyd}
is\-mert \"ossze\-f\"ug\-g\'es
\beq
J_B(y^2) \approx \frac{- \pi^4}{45} + \frac{\pi^2}{12} y^2 -
\frac{\pi}{6} y^3 - \frac{y^4}{32} \left(\log y^2 - 5.4076 \right) +
0.00031 y^6 \label{jb1}
\enq
az $y<1$ tar\-to\-m\'a\-nyon a\-j\'an\-lott leg\-jobb k\"o\-ze\-l\'{\i}\-t\'es, $y>1$ e\-se\-t\'en az
en\-n\'el l\'e\-nye\-ge\-sen ne\-h\'ez\-ke\-sebb Bes\-sel-f\"ugg\-v\'e\-nyek\-kel fe\-l\'{\i}rt
\beq
J_B(y^2) = - \sum_{n=1}^\infty \frac{y^2 K_2(ny)}{n^2} \label{jb2}
\enq
\"ossze\-f\"ug\-g\'es sze\-re\-pel, a\-zon\-ban a (\ref{jb1}) ki\-fe\-je\-z\'es m\'eg az
$y>1$ tar\-to\-m\'any egy r\'e\-sz\'en is j\'ol k\"o\-ze\-l\'{\i}\-ti az eg\-zakt \'er\-t\'e\-ket:

$y=2.0$ e\-se\-t\'en az eg\-zakt --1.03324 \'er\-t\'ek he\-lyett --1.03307,

$y=2.5$ e\-se\-t\'en az eg\-zakt --7.6765 \'er\-t\'ek he\-lyett --7.6757,

$y=3.0$ e\-se\-t\'en az eg\-zakt --0.5574 \'er\-t\'ek he\-lyett --0.5473 a\-d\'o\-dik.
\\
\'Igy a ne\-h\'ez\-kes (\ref{jb2}) for\-mu\-la hasz\-n\'a\-la\-ta az \'al\-ta\-lunk
vizs\-g\'alt tar\-to\-m\'a\-nyon ki\-ke\-r\"ul\-he\-t\H o.

A\-mennyi\-ben az $y$ ar\-gu\-men\-tum k\'ep\-ze\-tes, \'ugy az $y/i<1$ tar\-to\-m\'a\-nyon
a\-j\'an\-lott
\beq
J_B(y^2) \approx \frac{- \pi^4}{45} + \frac{\pi^2}{12} y^2 + y^3
\left[ \frac49 - \frac13 \log(2y) \right] - \frac{y^4}{32} \left(\log
y^2 - 5.4076 \right) + \frac{y^5}{180} - 0.00029 y^6 \label{jb3}
\enq
k\'ep\-let hasz\-n\'al\-ha\-t\'o -- mely $y/i$ 1-n\'el na\-gyobb \'er\-t\'e\-ke\-i\-re is j\'o
k\"o\-ze\-l\'{\i}\-t\'es az \'al\-ta\-lunk vizs\-g\'alt tar\-to\-m\'any\-ban. Mint\-hogy a\-zon\-ban a
n\'egy\-di\-men\-zi\-\'os szi\-mu\-l\'a\-ci\-\'ok\-ban hasz\-n\'a\-la\-tos $m_Q$, $m_U$ \'er\-t\'e\-kek
mel\-lett a stop\-t\"o\-me\-gek nem v\'al\-nak k\'ep\-ze\-tes\-s\'e, e\-z\'ert a
sz\'a\-mo\-l\'a\-sok so\-r\'an ezt a k\'ep\-le\-tet nem kel\-lett al\-kal\-maz\-nom.
\subsection{Az U i\-r\'a\-ny\'u po\-ten\-ci\-\'al}
Az MSSM k\'et\-hu\-rok\-ren\-d\H u pon\-ten\-ci\-\'al\-j\'a\-ak vizs\-g\'a\-la\-ta\-kor a
stop-szek\-tor\-ban v\'a\-rat\-lan f\'a\-zi\-s\'at\-me\-ne\-tet \'all e\-l\H o \cite{bod}:
az $\langle U^\dag U \rangle$ o\-pe\-r\'a\-tor z\'e\-rus\-t\'ol k\"u\-l\"on\-b\"o\-z\H o
v\'a\-ku\-um v\'ar\-ha\-t\'o \'er\-t\'e\-ket kap\-hat. Ez ak\-kor k\"o\-vet\-ke\-zik be, a\-mi\-kor
$m_U^2$ \'er\-t\'e\-k\'et kel\-l\H o\-en ne\-ga\-t\'{\i}v -- te\-h\'at a ba\-ri\-o\-ge\-n\'e\-zis \'al\-tal is
pre\-fe\-r\'alt tar\-to\-m\'any\-ban. Az \'ugy\-ne\-ve\-zett sz\'{\i}n\-s\'er\-t\H o f\'a\-zi\-s\'at\-me\-net
fel\-ve\-tet\-te egy k\'et\-l\'ep\-cs\H os f\'a\-zi\-s\'at\-me\-net le\-he\-t\H o\-s\'e\-g\'et is
\cite{bod, kus96}, mely\-ben az u\-ni\-ver\-zum h\H u\-l\'e\-se so\-r\'an e\-l\H o\-sz\"or
sz\'{\i}n\-s\'er\-t\H o f\'a\-zis\-ba, majd on\-nan sz\'{\i}n\-szim\-met\-ri\-kus, de nem-0
Higgs-v\'ar\-ha\-t\'o \'er\-t\'e\-k\H u szim\-met\-ri\-s\'er\-t\H o f\'a\-zis\-ba jut a rend\-szer.
A\-la\-po\-sabb vizs\-g\'a\-la\-tok ki\-mu\-tat\-t\'ak, hogy en\-nek a le\-he\-t\H o\-s\'eg\-nek nincs
koz\-mo\-l\'o\-gi\-a\-i je\-len\-t\H o\-s\'e\-ge, hi\-szen a ba\-ri\-on\-sz\'am\-s\'er\-t\'es\-hez
sz\"uk\-s\'e\-ges bu\-bo\-r\'ek\-k\'ep\-z\H o\-d\'es se\-bes\-s\'e\-ge nagy\-s\'ag\-ren\-di\-leg t\'ul ki\-csi
len\-ne \cite{clin-gdm99}. A le\-he\-t\H o\-s\'e\-get a\-zon\-ban c\'el\-sze\-r\H u mind az
ef\-fek\-t\'{\i}v po\-ten\-ci\-\'al ny\'uj\-tot\-ta ke\-re\-ten be\-l\"ul, mind a
r\'acsszi\-mu\-l\'a\-ci\-\'ok se\-g\'{\i}t\-s\'e\-g\'e\-vel meg\-vizs\-g\'al\-ni, egy\-r\'eszt hogy az
\'{\i}gy ka\-pott f\'a\-zis\-di\-ag\-ra\-mo\-kat \"ossze\-ha\-son\-l\'{\i}t\-has\-suk, m\'as\-r\'eszt hogy
a f\'a\-zi\-s\'at\-me\-ne\-tek ki\-t\"un\-te\-tett pont\-ja\-i (pl.\ h\'ar\-mas\-pont) r\'e\-v\'en
job\-ban \"ossze\-vet\-hes\-s\"uk a k\'et meg\-k\"o\-ze\-l\'{\i}\-t\'est.

A $\Phi$ i\-r\'a\-ny\'u po\-ten\-ci\-\'al\-hoz na\-gyon ha\-son\-l\'o az $U$ i\-r\'a\-ny\'u, \'{\i}gy
ezt r\'esz\-le\-te\-sen nem \'{\i}r\-juk ki -- a po\-ten\-ci\-\'al r\'esz\-le\-tes a\-lak\-ja
\cite{car97}-ben meg\-ta\-l\'al\-ha\-t\'o. Az $U$ i\-r\'a\-ny\'u ef\-fek\-t\'{\i}v
po\-ten\-ci\-\'al\-hoz a glu\-o\-nok, a kvar\-kok szu\-per\-szim\-met\-ri\-kus p\'ar\-ja\-i, a
stan\-dard mo\-dell\-be\-li Higgs-dub\-lett \'es a bal\-ke\-zes (har\-ma\-dik
ge\-ne\-r\'a\-ci\-\'os) skvar\-kok ke\-ve\-re\-d\'e\-s\'e\-b\H ol a\-d\'o\-d\'o (4--4) ne\-h\'ez \'es
k\"onny\H u ska\-l\'a\-rok ad\-nak j\'a\-ru\-l\'e\-kot. A fel\-so\-rolt r\'e\-szecs\-k\'ek z\'e\-rus
h\H o\-m\'er\-s\'ek\-le\-t\H u t\"o\-me\-ge\-i a $\Phi$ i\-r\'a\-ny\'u po\-ten\-ci\-\'al e\-se\-t\'e\-ben
l\'a\-tot\-tak\-hoz ha\-son\-l\'o v\'e\-ges h\H o\-m\'er\-s\'ek\-le\-t\H u kor\-rek\-ci\-\'ot kap\-nak. A
po\-ten\-ci\-\'al
\beq
V(U, T) = V_0 + V_1 + V_2 + \ldots
\enq
a\-lak\-ba \'{\i}r\-ha\-t\'o, \'es csak a fag\-r\'af- \'es az egy\-hu\-rok szin\-t\H u ta\-got
tart\-juk meg. A fag\-r\'af szin\-t\H u tag
\beq
V_0 = m_U^2(\mu) U^2 + \frac16 g_s^2 U^4,
\enq
mely\-ben
\beq
m^2_U(\mu) = m_U^2 - \sum_i \frac{n_i}{32 \pi^2} m_i^2(u)
\frac{dm_i^2 (u)}{du^2} \left[ \log \frac{m_i^2(u)}{\mu^2} + \frac12
- C_i \right].
\enq
Az egy\-hu\-rok\-ren\-d\H u kor\-rek\-ci\-\'o k\'ep\-le\-te (\ref{V_1})-gyel tel\-je\-sen a\-zo\-nos,
\beq
V_1 = \sum_i \frac{n_i}{64 \pi^2} M_i^4 \left( \log \frac{M_i^2}
{\mu^2} - C_i \right) + \sum_i \frac{n_i}{2 \pi^2} J^{(i)} T^4,
\enq
az \"osszeg\-z\H o\-in\-dex a\-zon\-ban az itt re\-le\-v\'ans r\'e\-szecs\-k\'e\-ken fut v\'e\-gig.

J\'ol l\'at\-ha\-t\'o, hogy a k\'et k\"u\-l\"on\-b\"o\-z\H o i\-r\'a\-ny\'u po\-ten\-ci\-\'al
va\-la\-mennyi\-re f\"ug\-get\-len egy\-m\'as\-t\'ol.
\section{Per\-tur\-ba\-t\'{\i}v e\-red\-m\'e\-nyek}
\fancyhead[CO]{\hst{\thesection \quad Per\-tur\-bat\'{\i}v e\-redm\'e\-nyek}}
A f\'a\-zi\-s\'at\-me\-ne\-tet a \emph{Maple} prog\-ram se\-g\'{\i}t\-s\'e\-g\'e\-vel vizs\-g\'al\-tam.
El\-s\H od\-le\-ges c\'e\-lom egy f\'a\-zis\-di\-ag\-ram fel\-v\'e\-te\-le volt, mely
r\"og\-z\'{\i}\-tett pa\-ra\-m\'e\-te\-rek e\-se\-t\'en az $m_U-T$ di\-ag\-ra\-mon ha\-t\'a\-roz\-za meg a
k\"u\-l\"on\-b\"o\-z\H o f\'a\-zi\-sok (szim\-met\-ri\-kus, Higgs-, sz\'{\i}n\-s\'er\-t\H o-)
hely\-ze\-t\'et. En\-nek se\-g\'{\i}t\-s\'e\-g\'e\-vel meg\-ha\-t\'a\-roz\-ha\-t\'o a h\'ar\-mas\-pont, mely
az a\-dott pa\-ra\-m\'e\-te\-rek\-re jel\-lem\-z\H o fi\-zi\-ka\-i pont, \'{\i}gy a nem\-per\-tur\-ba\-t\'{\i}v
e\-red\-m\'e\-nyek ki\-\'er\-t\'e\-ke\-l\'e\-s\'e\-hez hasz\-nos se\-g\'e\-desz\-k\"oz.
\subsection{Param\'eterv\'alaszt\'as}
K\'et\-f\'e\-le pa\-ra\-m\'e\-ter-hal\-mazt hasz\-n\'al\-tam: az e\-gyik me\-ge\-gye\-zik a
r\'acsszi\-mu\-l\'a\-ci\-\'ok\-ban ha\-sz\'alt\-tal, a m\'a\-sik az a\-l\'ab\-bi\-ak\-ban
ki\-fej\-t\'es\-re ke\-r\"u\-l\H o al\-ter\-na\-t\'{\i}v per\-tur\-ba\-t\'{\i}v meg\-k\"o\-ze\-l\'{\i}\-t\'e\-s\'e\-vel
\cite{jak2}. Az e\-ner\-gi\-a\-di\-men\-zi\-\'o\-j\'u  mennyi\-s\'e\-ge\-ket TeV-ben m\'er\-ve a
k\'et pa\-ra\-m\'e\-ter\-hal\-maz:

\begin{center}
\begin{tabular}{||l|c|c|c|c|c|c|c||}
\hline
& $v$	& $g$	  & $g_s$ & $\beta$ & $h_t$ & $\mu$ & $A_t$ \\
\hline
r\'acsszi\-mul\'a\-ci\'o &
0.246	& 0.64807 & 0.793 & 1.40565 & 1     & T     & 0     \\
pert.\ megk\"o\-zel\'{\i}t\'es &
0.246	& 0.66	  & 0.793 & 1.19235 & 0.95  & 0.08  & 0     \\
\hline
\end{tabular}
\end{center}

Az $A_t$ pa\-ra\-m\'e\-ter eg\-zakt z\'e\-rus \'er\-t\'e\-k\'et a \emph{Maple} nem tud\-ta
ke\-zel\-ni, \'{\i}gy e\-he\-lyett $10^{-30}$-nal sz\'a\-mol\-tam.
Az $m_A$ t\"o\-meg\-pa\-ra\-m\'e\-tert $\infty$-r\H ol 0.300, il\-let\-ve 0.150 TeV-re
v\'al\-toz\-tat\-va a po\-ten\-ci\-\'al \'er\-t\'e\-ke ez\-re\-l\'ek-szin\-t\H u kor\-rek\-ci\-\'ot ka\-pott
-- \'{\i}gy eb\-ben az \'er\-t\'ek\-tar\-to\-m\'any\-ban el\-te\-kint\-he\-t\"unk az
$m_A$-f\"ug\-g\'es\-t\H ol.
Az $m_Q$ pa\-ra\-m\'e\-ter r\'acsszi\-mu\-l\'a\-ci\-\'os \'er\-t\'e\-ke 0.300
TeV. A m\'a\-sik per\-tur\-ba\-t\'{\i}v meg\-k\"o\-ze\-l\'{\i}\-t\'es\-ben hasz\-n\'alt 0.07 TeV
\'er\-t\'ek\-n\'el v\'eg\-zett vizs\-g\'a\-la\-tok\-n\'al a po\-ten\-ci\-\'al\-ban szin\-gu\-l\'a\-ris
pon\-tok buk\-kan\-tak fel. Er\-re a vi\-sel\-ke\-d\'es\-re k\"u\-l\"on ki fo\-gok t\'er\-ni.

A szim\-met\-ri\-kus \'es Higgs-f\'a\-zis k\"oz\-ti \'at\-me\-net kri\-ti\-kus
h\H o\-m\'er\-s\'ek\-le\-t\'e\-nek meg\-ha\-t\'a\-ro\-z\'a\-s\'a\-hoz $m_U$ \'er\-t\'e\-k\'et
r\"og\-z\'{\i}\-tet\-tem, a $T$ h\H o\-m\'er\-s\'ek\-le\-tet pe\-dig ad\-dig v\'al\-toz\-tat\-tam, m\'{\i}g
a $\Phi$ t\'er f\"ugg\-v\'e\-ny\'e\-ben \'ab\-r\'a\-zolt $\Phi$ i\-r\'a\-ny\'u po\-ten\-ci\-\'al
k\'et de\-ge\-ne\-r\'alt mi\-ni\-mum\-mal nem ren\-del\-ke\-zett. A kri\-ti\-kus
h\H o\-m\'er\-s\'ek\-le\-tet 5 ti\-ze\-des\-jegy pon\-tos\-s\'ag\-gal ha\-t\'a\-roz\-tam meg.
\bef[ht]
\bc
\epsfig{file=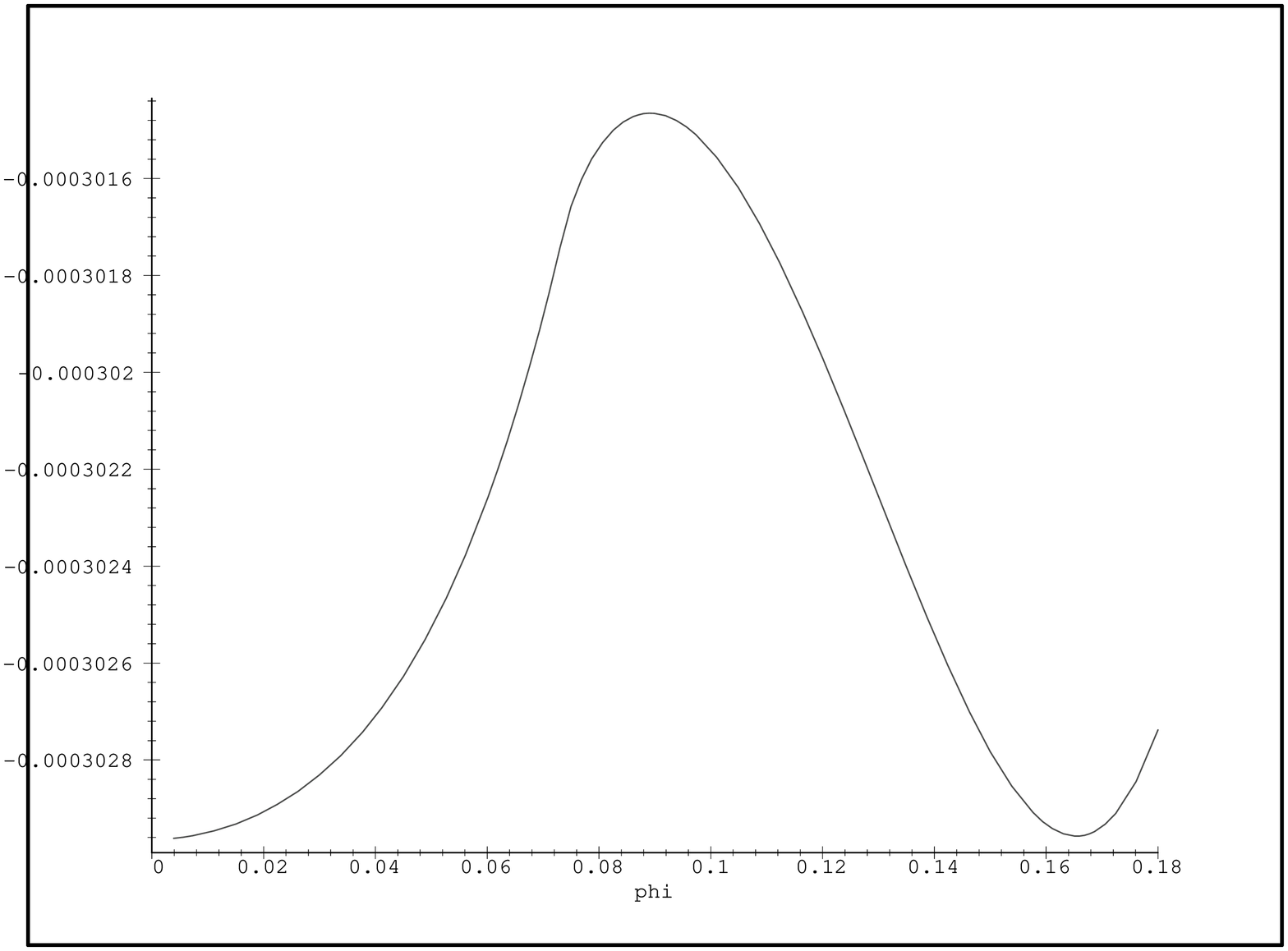,angle=270,width=8.3cm}\,
\epsfig{file=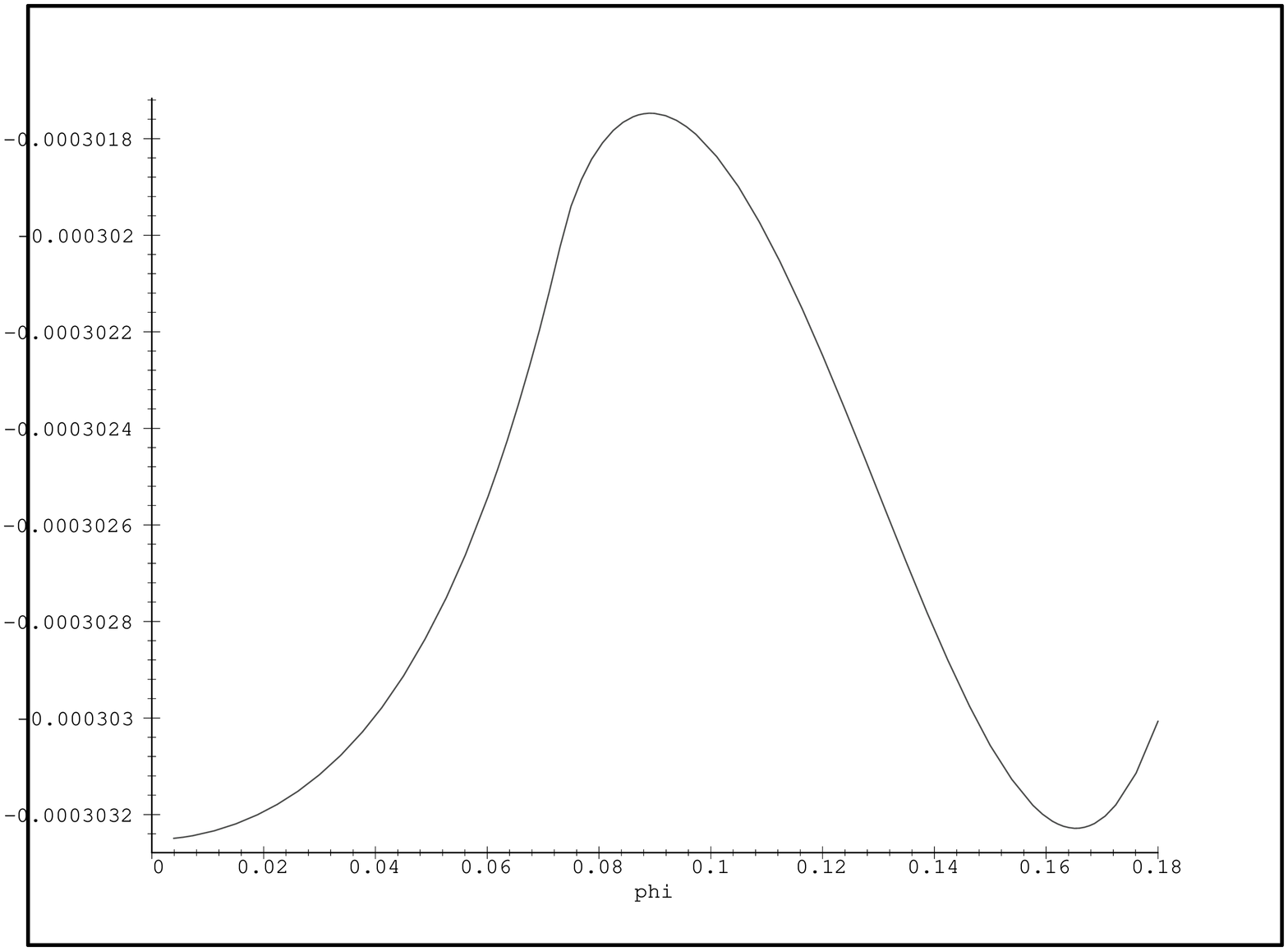,angle=270,width=8.3cm}
\caption{A $\Phi$ i\-r\'a\-ny\'u po\-ten\-ci\-\'al a f\'a\-zi\-s\'at\-me\-ne\-ti pont\-ban. A bal
ol\-da\-li \'ab\-r\'an $T=0.10990$ TeV, a jobb ol\-da\-lin $T=0.10991$ TeV}
\ec
\enf

TeV egy\-s\'e\-gek\-ben \'ab\-r\'a\-zol\-tam a f\"ugg\-v\'e\-nye\-ket; a fen\-ti \'ab\-r\'an $m_U
= 0.1(\mrm{TeV}) \cdot \sqrt{-1}$. Az \'ab\-r\'a\-r\'ol le\-ol\-vas\-ha\-t\'o, hogy a
Higgs-t\'er v\'a\-ku\-um v\'ar\-ha\-t\'o \'er\-t\'e\-ke (0.162 TeV) meg\-ha\-lad\-ja a
kri\-ti\-kus h\H o\-m\'er\-s\'ek\-le\-tet (0.1099 TeV), te\-h\'at a ba\-ri\-o\-ge\-n\'e\-zis
sz\"uk\-s\'e\-ges (\ref{vtpert}) fel\-t\'e\-te\-le eb\-ben a pont\-ban a
le\-egy\-sze\-r\H u\-s\'{\i}\-tett mo\-dell per\-tur\-ba\-t\'{\i}v vizs\-g\'a\-la\-t\'a\-nak ke\-re\-t\'e\-ben
tel\-je\-s\"ul.

Nem k\'{\i}\-v\'a\-nom a\-zon\-ban e\-gye\-l\H o\-re fel\-t\'er\-k\'e\-pez\-ni azt a tar\-to\-m\'anyt,
a\-hol ez az \"ossze\-f\"ug\-g\'es tel\-je\-s\"ul. A fen\-ti\-ek\-b\H ol l\'at\-ha\-t\'o\-an ez nem
len\-ne t\'ul nagy fe\-la\-dat: az e\-gyes f\'a\-zi\-s\'at\-me\-ne\-ti pon\-tok\-ban
k\"oz\-vet\-le\-n\"ul le\-ol\-vas\-ha\-t\'o, tel\-je\-s\"ul-e a fel\-t\'e\-tel. \'Igy a
Higgs-t\"o\-me\-get is v\'al\-toz\-tat\-va az $m_h-m_U$ s\'{\i}\-kon k\"onnyen
ki\-je\-l\"ol\-he\-t\H o a ba\-ri\-o\-ge\-n\'e\-zis szem\-pont\-j\'a\-b\'ol re\-le\-v\'ans tar\-to\-m\'any.
Et\-t\H ol a\-zon\-ban e\-gye\-l\H o\-re el\-te\-kin\-tek: a per\-tur\-b\'a\-ci\-\'o\-sz\'a\-m\'{\i}\-t\'as
i\-lyen le\-egy\-sze\-r\H u\-s\'{\i}\-tett mo\-dell\-re al\-kal\-maz\-va a\-lig\-ha szol\-g\'al\-tat\-na
kel\-l\H o pon\-tos\-s\'a\-g\'u a\-da\-to\-kat.

\'Er\-de\-mes vi\-szont fel\-fi\-gyel\-ni pl.\ az $m_U/i$ n\"o\-ve\-ke\-d\'e\-se \'es a
f\'a\-zi\-s\'at\-me\-net e\-r\H o\-s\"o\-d\'e\-se (te\-h\'at a Higgs-t\'er ug\-r\'a\-s\'a\-nak
n\"o\-ve\-ke\-d\'e\-se) k\"oz\-ti kap\-cso\-lat\-ra: m\'ar ez az egy\-sze\-r\H u mo\-dell is
mu\-tat\-ja, hogy a nagy ne\-ga\-t\'{\i}v $m_U^2$ tar\-to\-m\'any lesz ked\-ve\-z\H obb a
ba\-ri\-o\-ge\-n\'e\-zis\-hez. \smallskip

Az $U$ i\-r\'a\-ny\'u po\-ten\-ci\-\'alt ha\-son\-l\'o m\'o\-don vizs\-g\'al\-tam, eb\-ben a\-zon\-ban
a $\Phi$ t\'er imp\-li\-ci\-ten sze\-re\-pelt. Mi\-vel a sz\'{\i}n\-s\'er\-t\H o f\'a\-zis\-ban a
Higgs-t\'er v\'ar\-ha\-t\'o \'er\-t\'e\-ke 0 \cite{clin-gdm99}, e\-z\'ert itt $\Phi=0$
sze\-re\-pel a po\-ten\-ci\-\'al\-ban. Meg\-fe\-le\-l\H o $m_U$ \'er\-t\'e\-kek mel\-lett az $U$
i\-r\'a\-ny\'u po\-ten\-ci\-\'al mi\-ni\-mu\-ma kis h\H o\-m\'er\-s\'ek\-le\-ten a 0-ban van
(szim\-met\-ri\-kus f\'a\-zis); a h\H o\-m\'er\-s\'ek\-let n\"o\-ve\-l\'e\-s\'e\-vel a sz\'{\i}n\-s\'er\-t\H o
f\'a\-zis\-ba ju\-tunk, majd m\'eg to\-v\'abb n\"o\-vel\-ve a h\H o\-m\'er\-s\'ek\-le\-tet \'uj\-ra
a szim\-met\-ri\-kus f\'a\-zis\-ba ke\-r\"u\-l\"unk. \'Igy te\-h\'at k\'et --
\"ossze\-csat\-la\-ko\-z\'o -- g\"or\-b\'et kell meg\-ha\-t\'a\-roz\-nunk. Sz\'{\i}n\-s\'er\-t\H o
f\'a\-zi\-s\'at\-me\-net csak $m_U^2$ kel\-l\H o\-en nagy ab\-szo\-l\'ut \'er\-t\'e\-k\H u
ne\-ga\-t\'{\i}v \'er\-t\'e\-ke\-i\-re a\-la\-kul\-hat ki; a vizs\-g\'a\-la\-tok so\-r\'an $m_U^2 \approx
-(0.1 \mathrm{TeV})^2$ k\"o\-r\"ul je\-lent meg a sz\'{\i}n\-s\'er\-t\H o \'at\-me\-net. A
sz\'{\i}n\-s\'er\-t\H o \'at\-me\-ne\-tet mu\-ta\-t\'o leg\-na\-gyobb (a\-zaz leg\-ki\-sebb ab\-szo\-l\'ut
\'er\-t\'e\-k\H u) $m_U^2$ \'er\-t\'e\-k\'et meg\-ha\-t\'a\-roz\-tam.
\bef[ht]
\bc
\epsfig{file=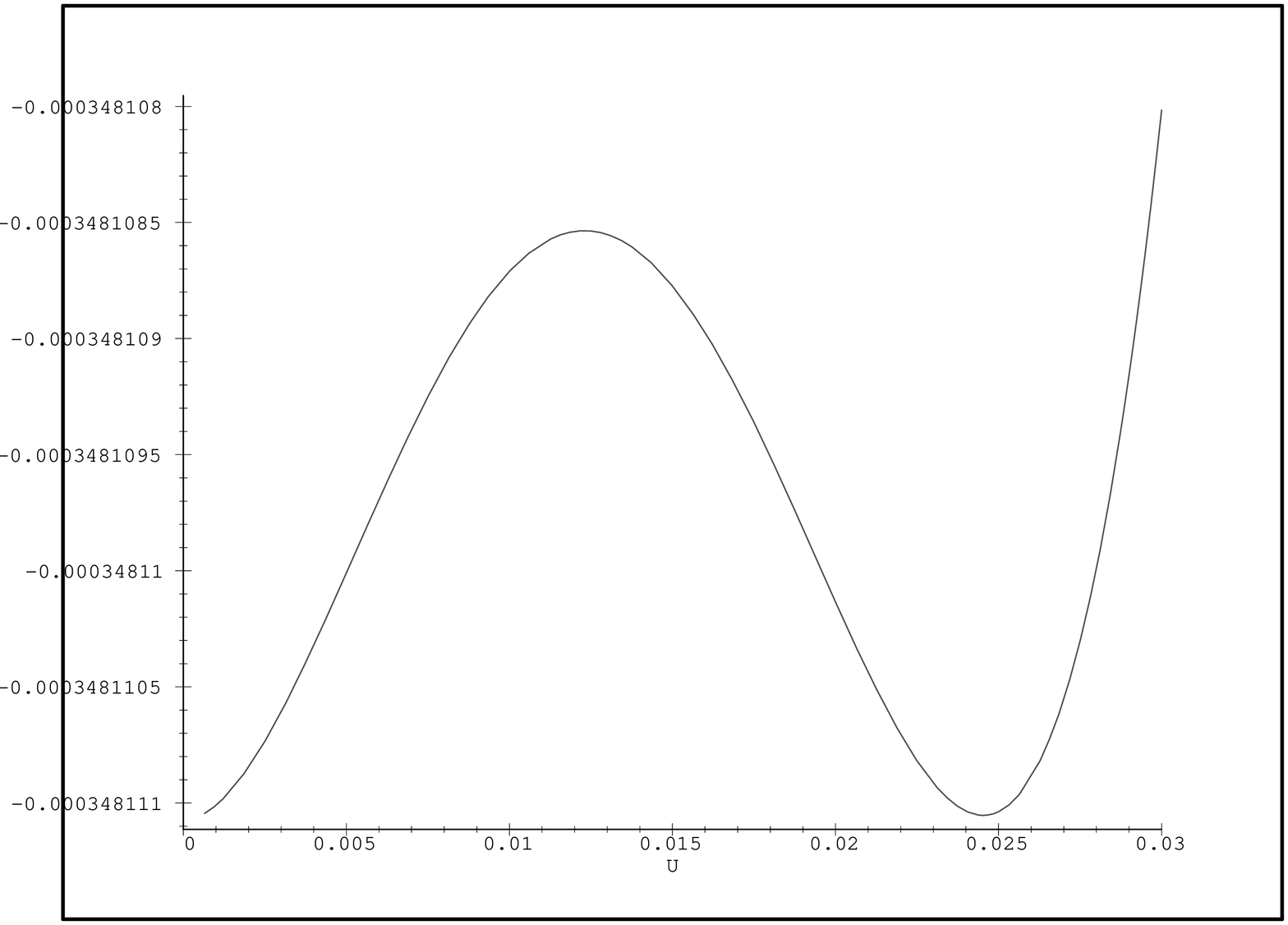,angle=270,width=8.3cm}\,
\epsfig{file=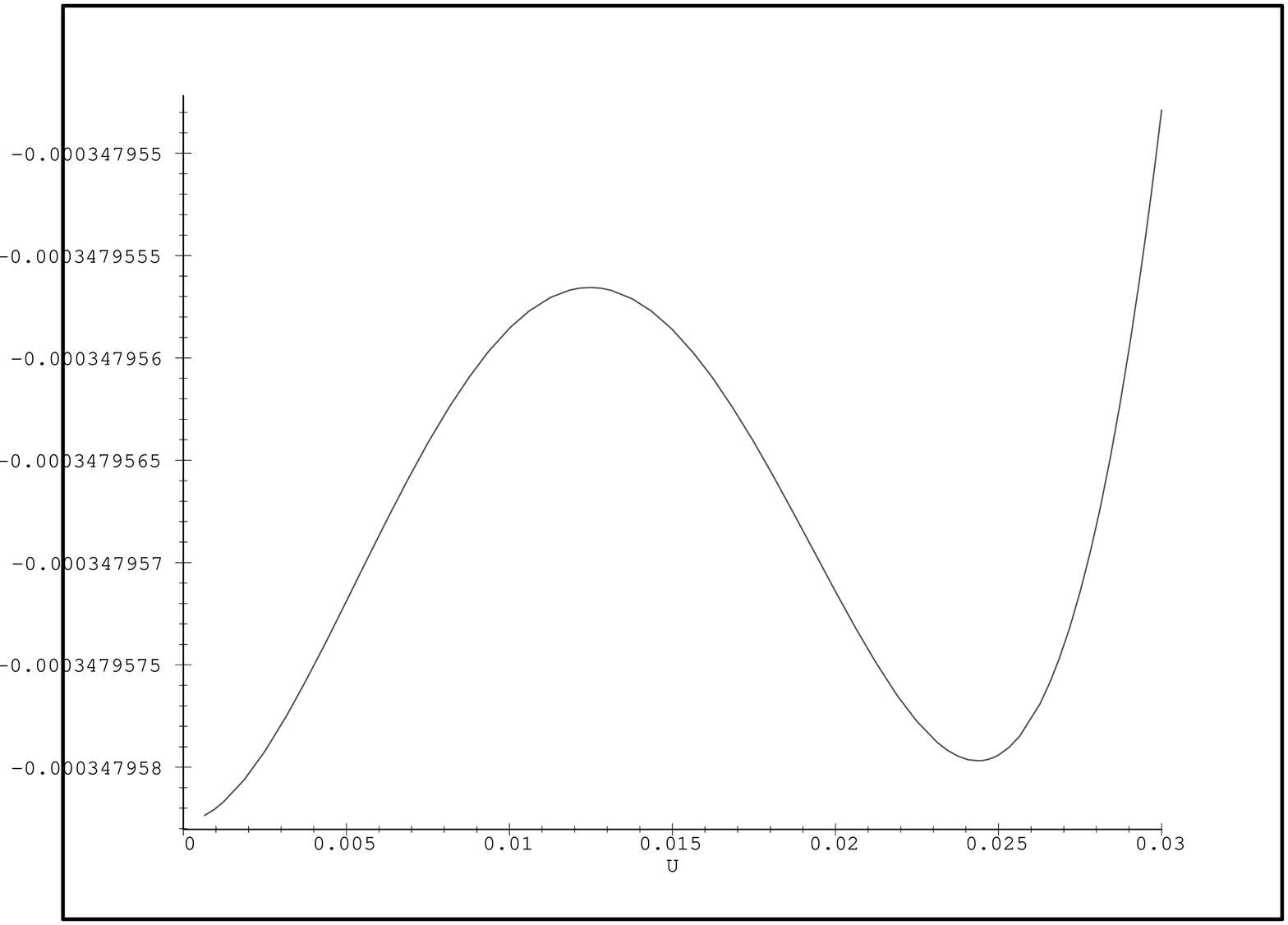,angle=270,width=8.3cm}
\caption{Az $U$ i\-r\'a\-ny\'u po\-ten\-ci\-\'al a sz\'{\i}n\-s\'er\-t\H o f\'a\-zi\-s\'at\-me\-ne\-ti
pont\-ban. A bal ol\-da\-li \'ab\-r\'an $T=0.10472$ TeV, a jobb ol\-da\-lin $T=0.10473$
TeV.}
\ec
\enf
\subsection{Eredm\'enyek}
E\-red\-m\'e\-nye\-in\-ket a \ref{fazpert} \'es a \ref{pararacs} t\'ab\-l\'a\-zat
fog\-lal\-ja \"ossze.
A sz\'a\-mo\-l\'as\-hoz hasz\-n\'alt pa\-ra\-m\'e\-te\-rek a r\'acsszi\-mu\-l\'a\-ci\-\'ok\-ban
hasz\-n\'al\-tak\-kal e\-gyez\-nek meg. A szim\-met\-ri\-kus \'es a Higgs-f\'a\-zis k\"oz\-ti
\'at\-me\-net kri\-ti\-kus h\H o\-m\'er\-s\'ek\-le\-t\'et $T_1$, a szim\-met\-ri\-kus \'es a
sz\'{\i}n\-s\'er\-t\H o f\'a\-zi\-sok k\"oz\-ti \'at\-me\-net kri\-ti\-kus h\H o\-m\'er\-s\'ek\-le\-te\-it
$T_2$ je\-l\"o\-li.
\begin{table}[htb]
\begin{center}
\begin{tabular}{||r|c|c||}
\hline
$m_U$  & $T_1$ & $\langle \phi \rangle$ \\
\hline
0.10   & 0.12797 & 0.040 \\
0.08   & 0.12710 & 0.044 \\
0.06   & 0.12456 & 0.048 \\
0.04   & 0.12239 & 0.054 \\
0.02   & 0.12089 & 0.059 \\
0.00   & 0.12036 & 0.061 \\
0.02i  & 0.11979 & 0.063 \\
0.04i  & 0.11796 & 0.071 \\
0.06i  & 0.11424 & 0.094 \\
0.08i  & 0.10615 & 0.142 \\
0.10i  & 0.09950 & \\
0.105i & 0.10037 & \\
0.107i & 0.10074 & \\
0.120i & 0.10338 & \\
\hline
\end{tabular} \qquad
\begin{tabular}{||r|c|c||}
\hline
$m_U$	& $T_2^>$ & $T_2^<$ \\
\hline
0.0901i & 0.04383 & 0.03958 \\
0.091i	& 0.05221 & 0.03352 \\
0.095i	& 0.06934 & 0.02601 \\
0.10i	& 0.08419 & 0.02196 \\
0.105i	& 0.09649 & --	    \\
0.107i	& 0.10098 & --	    \\
0.110i	& 0.10737 & --	    \\
\hline
\end{tabular}
\caption{A `per\-tur\-ba\-t\'{\i}v' pa\-ra\-m\'e\-te\-rek mel\-let\-ti f\'a\-zi\-s\'at\-me\-ne\-ti
pon\-tok. \label{fazpert}}
\end{center}
\end{table}
\begin{table}[htb]
\begin{center}
\begin{tabular}{||r|c|c||}
\hline
$m_U$  & $T_1$ & $\langle \phi \rangle$ \\
\hline
0.20   & 0.14895 & 0.030  \\
0.15   & 0.14324 & 0.033  \\
0.10   & 0.13707 & 0.038  \\
0.05   & 0.13149 & 0.049  \\
0.03   & 0.12995 & 0.055  \\
0.00   & 0.12900 & 0.059  \\
0.02i  & 0.12855 & 0.061  \\
0.04i  & 0.12713 & 0.069  \\
0.05i  & 0.12598 & 0.076  \\
0.06i  & 0.12442 & 0.086  \\
0.08i  & 0.11961 & 0.118  \\
0.10i  & 0.10991 & 0.163  \\
\hline
\end{tabular} \qquad
\begin{tabular}{||r|c|c||}
\hline
$m_U$	& $T_2^>$ & $T_2^<$ \\
\hline
0.091i	& 0.14702 & 0.13875 \\
0.092i	& 0.15790 & 0.12969 \\
0.095i	& 0.17366 & 0.12122 \\
0.10i	& 0.19137 & 0.10472 \\
0.12i	& 0.24167 & 0.07939 \\
0.15i	& 0.30270 &	    \\
\hline
\end{tabular}
\caption{A r\'acsszi\-mu\-l\'a\-ci\-\'ok\-ban hasz\-n\'alt pa\-ra\-m\'e\-te\-rek mel\-let\-ti
f\'a\-zi\-s\'at\-me\-ne\-ti pon\-tok} \label{pararacs}
\end{center}
\end{table}

%
\subsection{N\'e\-h\'any fur\-csa\-s\'ag}
A pa\-ra\-m\'e\-te\-rek bi\-zo\-nyos meg\-v\'a\-lasz\-t\'a\-sa e\-se\-t\'en a po\-ten\-ci\-\'al\-g\"or\-b\'en
t\"o\-r\'e\-sek buk\-kan\-hat\-nak fel, mint a \ref{furcsa} \'ab\-r\'an.
\bef[ht]
\bc
\epsfig{file=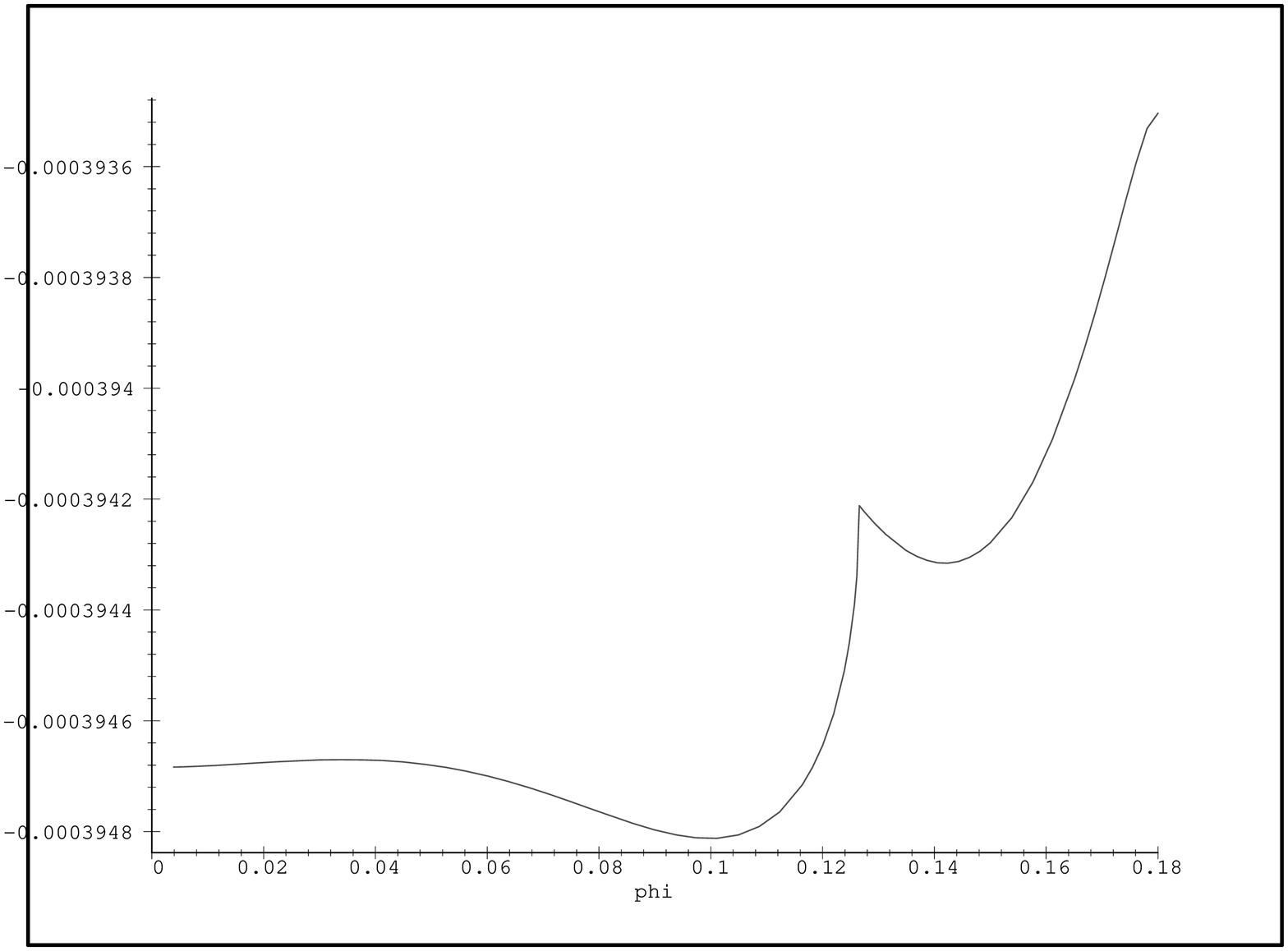,angle=270,width=8.3cm}\,
\epsfig{file=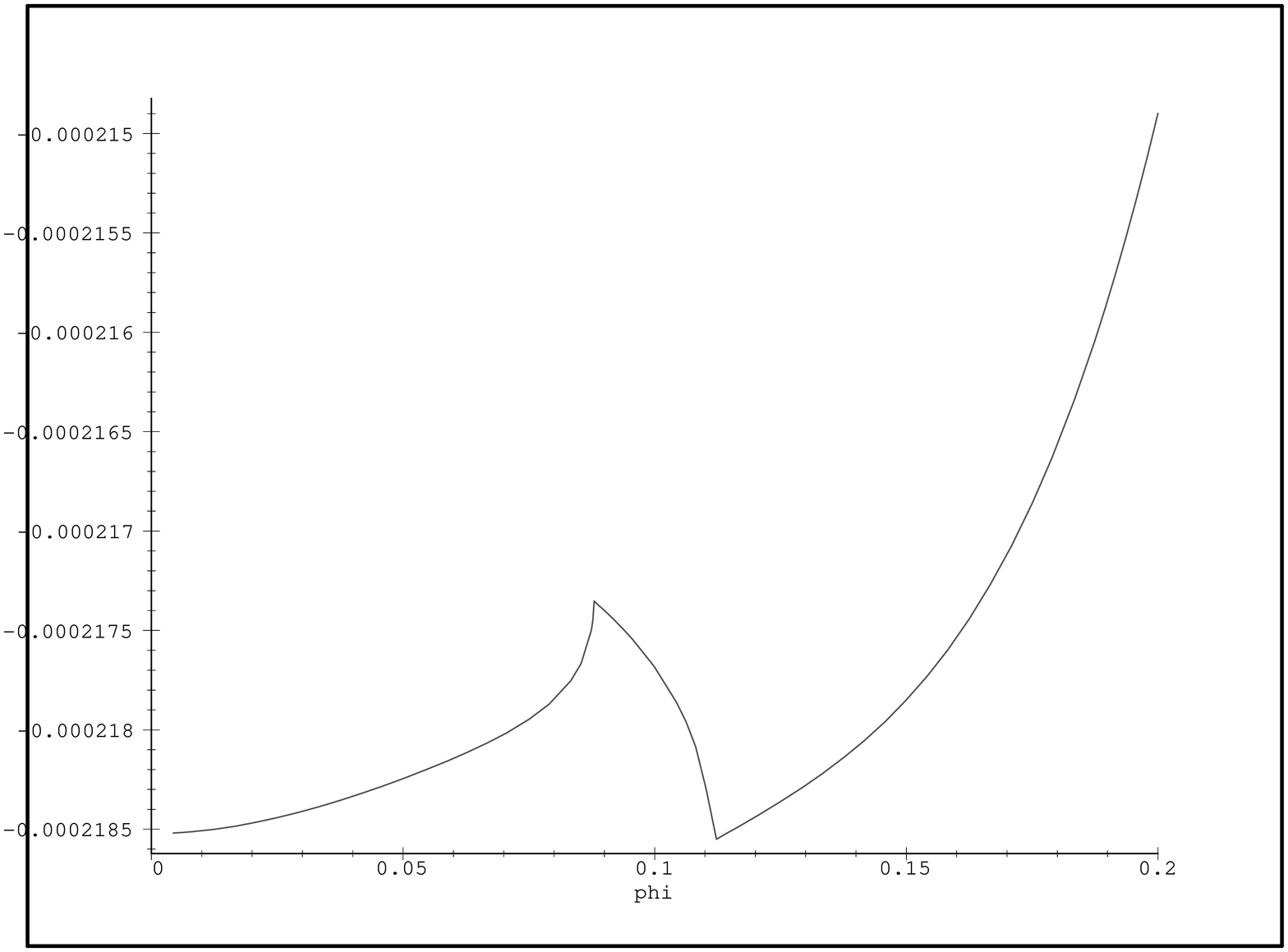,angle=270,width=8.3cm}
\caption{Bi\-zo\-nyos pa\-ra\-m\'e\-te\-rek\-n\'el a $\Phi$ i\-r\'a\-ny\'u po\-ten\-ci\-\'al\-ban
szin\-gu\-l\'a\-ris pon\-tok buk\-kan\-nak fel \label{furcsa}}
\ec
\enf
Az i\-lyen t\"o\-r\'es\-pon\-tok l\'e\-te az i\-ro\-da\-lom\-b\'ol is\-mert: jel\-lem\-z\H o\-en ak\-kor
szo\-kott fel\-l\'ep\-ni, ha a sz\'{\i}n\-s\'er\-t\H o f\'a\-zi\-s\'at\-me\-net kri\-ti\-kus
h\H o\-m\'er\-s\'ek\-le\-te a\-la\-cso\-nyabb, mint a Higgs-f\'a\-zi\-s\'at\-me\-ne\-t\'e. A fen\-ti
\'ab\-r\'at az $m_Q=70$ GeV mel\-lett kap\-tuk; a\-mennyi\-ben $m_Q$ \'er\-t\'e\-k\'et
n\"o\-vel\-j\"uk, a szin\-gu\-la\-ri\-t\'as m\'eg a r\'acsszi\-mu\-l\'a\-ci\-\'ok\-ban hasz\-n\'alt \'er\-t\'ek
e\-l\'e\-r\'e\-se e\-l\H ott ki\-si\-mul.

T\"obb\-l\'ep\-cs\H os f\'a\-zi\-s\'at\-me\-net is ki\-a\-la\-kul\-hat, pl.\ a
r\'acsszi\-mu\-l\'a\-ci\-\'ok\-ban hasz\-n\'alt pa\-ra\-m\'e\-te\-rek \'es $T_1=0.09500$ TeV
v\'a\-lasz\-t\'a\-sa mel\-lett, a\-mint azt a \ref{tobblep} \'ab\-ra mu\-tat\-ja.
\bef[ht]
\bc
\epsfig{file=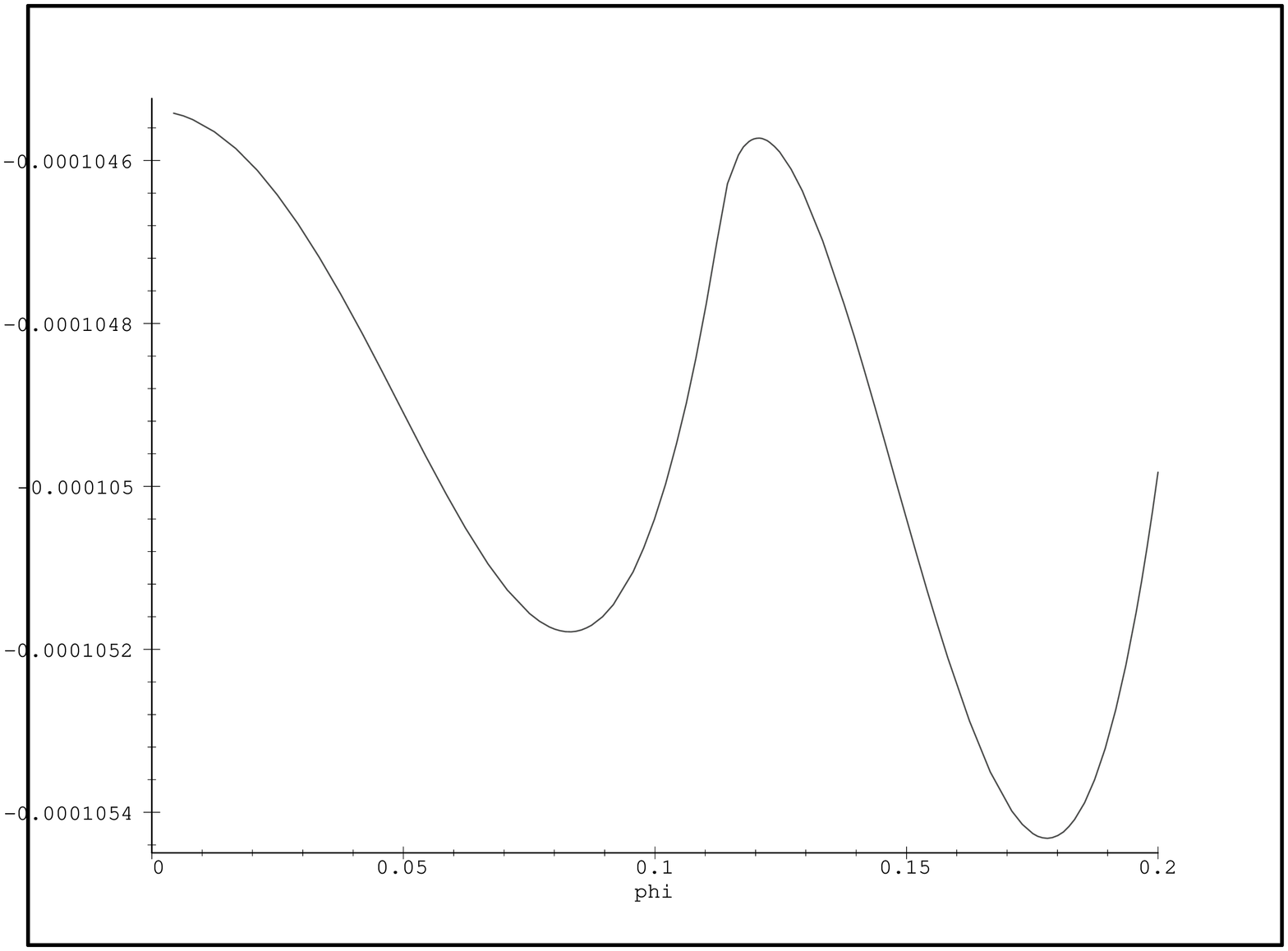,angle=270,width=8.3cm}\,
\epsfig{file=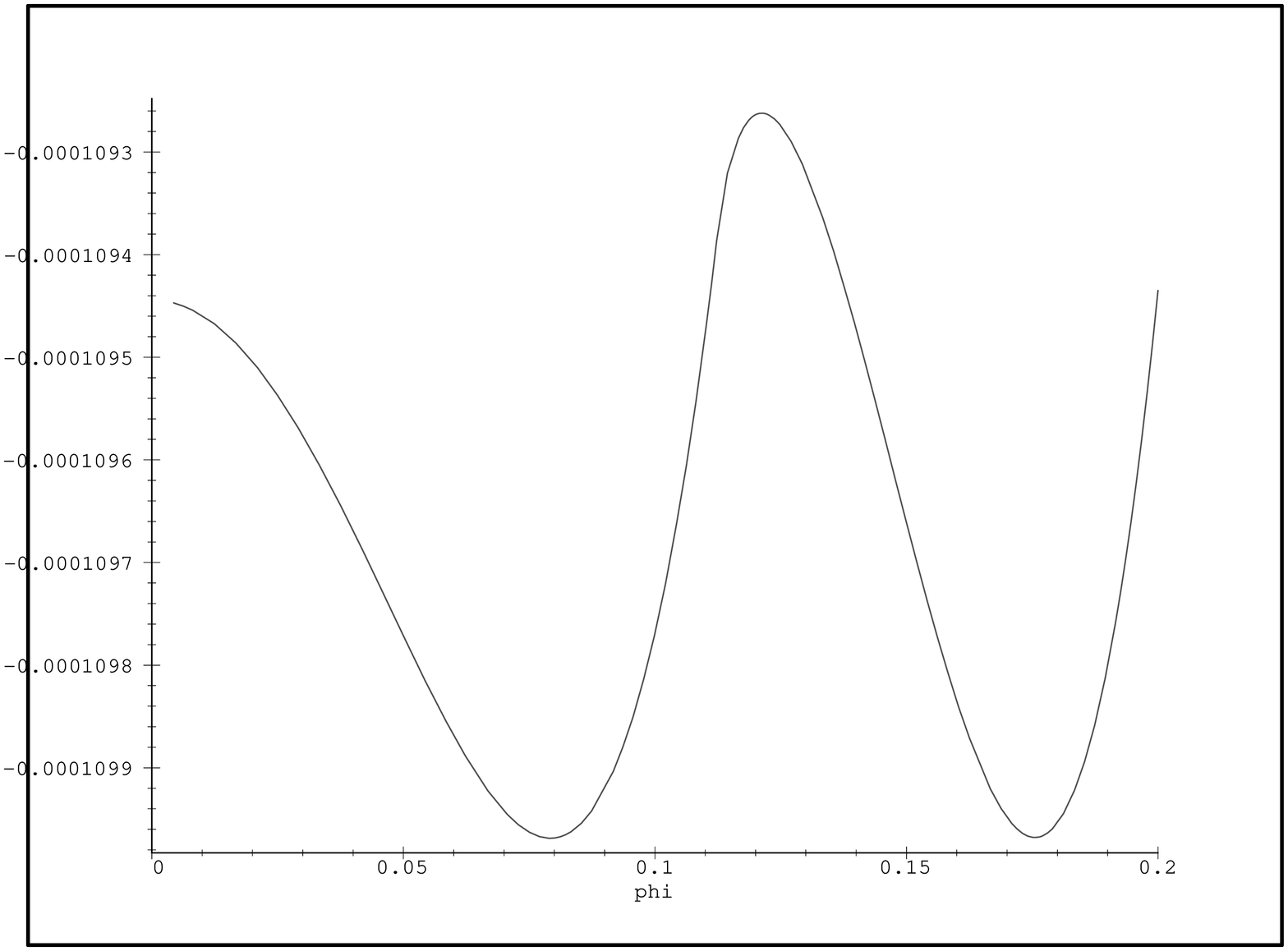,angle=270,width=8.3cm}
\caption{Bi\-zo\-nyos pa\-ra\-m\'e\-te\-rek mel\-lett a\-k\'ar t\"obb\-l\'ep\-cs\H os
f\'a\-zi\-s\'at\-me\-net is le\-j\'at\-sz\'od\-hat \label{tobblep}}
\ec
\enf
A\-zon\-ban mind\-k\'et e\-set o\-lyan pa\-ra\-m\'e\-te\-rek mel\-lett va\-l\'o\-sul meg,
a\-me\-lyek a sz\'{\i}n\-s\'er\-t\H o f\'a\-zi\-s\'at\-me\-ne\-tet is le\-he\-t\H o\-v\'e te\-szik, \'{\i}gy
a fi\-zi\-ka\-i tar\-to\-m\'a\-nyon k\'{\i}\-v\"ul va\-gyunk -- a fen\-ti \'ab\-r\'ak te\-h\'at
\'er\-de\-kes jel\-le\-g\"u\-k\"on k\'{\i}\-v\"ul kis je\-len\-t\H o\-s\'eg\-gel b\'{\i}r\-nak.
\section{A per\-tur\-ba\-t\'{\i}v e\-red\-m\'e\-nyek meg\-b\'{\i}z\-ha\-t\'o\-s\'a\-ga}
\fancyhead[CO]{\hst{\thesection \quad A per\-tur\-bat\'{\i}v e\-redm\'e\-nyek
megb\'{\i}zhat\'os\'a\-ga}}
A fen\-ti\-ek csak az el\-s\H o l\'e\-p\'est je\-len\-tik a mi\-ni\-m\'a\-lis
szu\-per\-szim\-met\-ri\-kus stan\-dard mo\-dell\-be\-li e\-lekt\-ro\-gyen\-ge f\'a\-zi\-s\'at\-me\-net
per\-tur\-ba\-t\'{\i}v vizs\-g\'a\-la\-t\'a\-ban. Ez a meg\-k\"o\-ze\-l\'{\i}\-t\'es ki\-ter\-jedt
i\-ro\-da\-lom\-mal b\'{\i}r, vizs\-g\'a\-la\-ta ma is \'e\-l\'enk ku\-ta\-t\'as t\'ar\-gya. \'Igy
c\'el\-sze\-r\H u leg\-fon\-to\-sabb j\'os\-la\-ta\-it r\"o\-vi\-den \"ossze\-fog\-lal\-ni, mint\-hogy
e\-zek e\-r\H o\-sen mo\-ti\-v\'al\-j\'ak a nem\-per\-tur\-ba\-t\'{\i}v m\'od\-sze\-rek ki\-dol\-go\-z\'a\-s\'at.

A per\-tur\-ba\-t\'{\i}v e\-red\-m\'e\-nyek e\-gy\'er\-tel\-m\H u\-en ar\-ra u\-tal\-nak, hogy az
MSSM-be\-li f\'a\-zi\-s\'at\-me\-net l\'e\-nye\-ge\-sen e\-r\H o\-tel\-je\-sebb, mint a stan\-dard
mo\-dell\-be\-li. A k\"onnyeb\-bik Higgs t\"o\-me\-ge a\-k\'ar 105 GeV is le\-het, a
jobb\-ke\-zes stop t\"o\-me\-ge pe\-dig majd\-nem a top-t\"o\-me\-gig k\'usz\-hat fel -- a
f\'a\-zi\-s\'at\-me\-net m\'eg ek\-kor is e\-l\'eg e\-r\H o\-sen el\-s\H o\-ren\-d\H u. Ez a
pa\-ra\-m\'e\-ter\-tar\-to\-m\'any az LHC il\-let\-ve a Te\-vat\-ron k\'{\i}\-s\'er\-le\-te\-i\-ben a
k\"o\-ze\-li j\"o\-v\H o\-ben fel lesz de\-r\'{\i}t\-ve \cite{car97}.

A tisz\-t\'an per\-tur\-ba\-t\'{\i}v le\-\'{\i}\-r\'as a\-zon\-ban -- a stan\-dard mo\-dell\-hez
ha\-son\-l\'o\-an -- nem tel\-je\-sen ki\-e\-l\'e\-g\'{\i}\-t\H o. A fen\-ti, fe\-let\-t\'ebb
egy\-sze\-r\H u\-s\'{\i}\-tett le\-\'{\i}\-r\'as\-b\'ol ki\-hagy\-tuk a fer\-mi\-o\-no\-kat, me\-lye\-ket egy
e\-set\-le\-ges to\-v\'ab\-bi per\-tur\-ba\-t\'{\i}v l\'e\-p\'es\-sel fi\-gye\-lem\-be ve\-he\-t\"unk -- a
SM-be\-li n\'egy\-di\-men\-zi\-\'os szi\-mu\-l\'a\-ci\-\'ok min\-t\'a\-j\'a\-ra. Ez a\-zon\-ban egy \'uj
prob\-l\'e\-m\'at vet fel.

A szu\-per\-szim\-met\-ri\-kus mo\-del\-lek a\-lap\-ve\-t\H o vo\-n\'a\-sa a fer\-mi\-o\-nok \'es a
bo\-zo\-nok k\"o\-z\"ott fen\-n\'al\-l\'o szim\-met\-ri\-a. En\-nek meg\-bon\-t\'a\-sa el\-lent\-mond a
mo\-dell l\'et\-re\-ho\-z\'a\-s\'at szor\-gal\-ma\-z\'o esz\-t\'e\-ti\-ka\-i a\-la\-pok\-nak -- a\-zon\-ban
ez a l\'e\-p\'es prag\-ma\-ti\-kus szem\-pon\-tok\-kal k\"onnyen in\-do\-kol\-ha\-t\'o.
P\'el\-d\'a\-ul a \ref{nonper}.\ fe\-je\-zet\-ben is\-mer\-te\-t\'es\-re ke\-r\"u\-l\H o
r\'acsszi\-mu\-l\'a\-ci\-\'ok e\-se\-t\'en a stan\-dard mo\-dell\-be\-li vizs\-g\'a\-la\-tok
a\-lap\-j\'an azt v\'ar\-juk, hogy a bo\-zo\-ni\-kus szek\-tor\-ban buk\-kan\-nak csak fel
o\-lyan ne\-h\'e\-zs\'e\-gek, me\-lyek a r\'acsszi\-mu\-l\'a\-ci\-\'o hasz\-n\'a\-la\-t\'at
sz\"uk\-s\'e\-ges\-s\'e te\-szik. (A szu\-per\-szim\-met\-ri\-a r\'acs\-ra t\'e\-te\-le et\-t\H ol
f\"ug\-get\-le\-n\"ul is vet fel prob\-l\'e\-m\'a\-kat: szu\-per\-szim\-met\-ri\-kus
kon\-ti\-nu\-um-t\'e\-rel\-m\'e\-let\-be\-li Lag\-ran\-ge-f\"ugg\-v\'eny r\'acs\-ra\-t\'e\-te\-le\-kor a
bo\-zo\-ni\-kus v\'al\-to\-z\'ok\-ra e\-l\H o\-\'{\i}rt pe\-ri\-o\-di\-kus \'es a fer\-mi\-o\-ni\-kus
v\'al\-to\-z\'ok\-ra e\-l\H o\-\'{\i}rt an\-ti\-pe\-ri\-o\-di\-kus ha\-t\'ar\-fel\-t\'e\-te\-lek nem
szu\-per\-szim\-met\-ri\-kus r\'acs\-ha\-t\'as\-ra fog\-nak ve\-zet\-ni; a szu\-per\-szim\-met\-ri\-a
csak a v\'eg\-te\-len t\'er\-fo\-ga\-t\'u ha\-t\'a\-re\-set\-ben \'all\-hat\-na hely\-re, \'{\i}gy a
kis i\-d\H o\-i\-r\'a\-ny\'u r\'acs\-ki\-ter\-je\-d\'es\-sel jel\-lem\-zett v\'e\-ges
h\H o\-m\'er\-s\'ek\-le\-t\H u e\-set\-ben a szu\-per\-szim\-met\-ri\-a min\-dig s\'e\-r\"ul.)

Ne\-he\-zebb prob\-l\'e\-m\'at je\-lent az, hogy a szu\-per\-szim\-met\-ri\-a, mint a
szim\-met\-ri\-a\-k\"o\-ve\-tel\-m\'e\-nyek \'al\-ta\-l\'a\-ban, meg\-szo\-r\'{\i}\-t\'ast r\'o ki bi\-zo\-nyos
mennyi\-s\'e\-gek (jel\-lem\-z\H o\-en t\"o\-me\-gek) re\-nor\-m\'a\-l\'a\-s\'a\-ra -- a\-hogy a
szo\-k\'a\-sos Slav\-nov--Tay\-lor a\-zo\-nos\-s\'a\-gok biz\-to\-s\'{\i}\-tot\-t\'ak, hogy a
fo\-ton\-t\"o\-meg re\-nor\-m\'a\-l\'as u\-t\'an is 0 ma\-rad. A t\"o\-meg\-re\-nor\-m\'a\-l\'as
prob\-l\'e\-m\'a\-ja, ha\-son\-l\'o k\"on\-t\"os\-ben, a r\'acsszi\-mu\-l\'a\-ci\-\'ok kap\-cs\'an
is\-m\'et e\-l\H o\-ke\-r\"ul, \'es ott is ko\-moly fej\-f\'a\-j\'ast o\-koz.

Ez a fen\-ti\-ek\-ben is j\'ol l\'at\-ha\-t\'o m\'o\-don je\-lent\-ke\-zik: a be\-me\-n\H o
pa\-ra\-m\'e\-te\-rek \'al\-lan\-d\'o \'er\-t\'e\-ken tar\-t\'a\-sa mel\-lett a k\"u\-l\"on\-b\"o\-z\H o $m_U$
\'er\-t\'e\-kek\-hez tar\-to\-z\'o f\'a\-zi\-s\'at\-me\-ne\-ti pon\-tok\-ban a Higgs-W t\"o\-me\-ga\-r\'any
m\'as \'es m\'as, \'{\i}gy egy e\-set\-le\-ges f\'a\-zis\-di\-ag\-ram fel\-v\'e\-te\-l\'e\-hez a fen\-ti
t\'ab\-l\'a\-zat e\-red\-m\'e\-nye\-i nem meg\-fe\-le\-l\H o\-ek. Az a\-l\'ab\-bi\-ak\-ban r\"o\-vid
in\-di\-k\'a\-ci\-\'ot a\-dunk ar\-ra, ho\-gyan ke\-zel\-he\-t\H o ez a prob\-l\'e\-ma.

El\-s\H o l\'e\-p\'es\-k\'ent az $m_W$ z\'e\-rus h\H o\-m\'er\-s\'ek\-le\-ten m\'ert t\"o\-meg
r\"og\-z\'{\i}\-t\'e\-se sz\"uk\-s\'e\-ges. Mint\-hogy ez $g^2 \phi^2$-tel a\-r\'a\-nyos, \'es
a $g$ csa\-to\-l\'a\-si \'al\-lan\-d\'ot fi\-zi\-ka\-i \'er\-t\'e\-k\'en k\'{\i}\-v\'an\-juk r\"og\-z\'{\i}\-te\-ni,
a $\phi^2$ mennyi\-s\'e\-get is  \'al\-lan\-d\'o\-nak kell tar\-ta\-ni. Ez
biz\-to\-s\'{\i}t\-ha\-t\'o \'ugy, hogy az el\-m\'e\-let\-ben sze\-rep\-l\H o pa\-ra\-m\'e\-te\-rek
k\"o\-z\"ul az e\-gyi\-ket meg\-fe\-le\-l\H o\-k\'epp v\'al\-toz\-tat\-juk --  c\'el\-sze\-r\H u
v\'a\-lasz\-t\'as a $\mu$ re\-nor\-m\'a\-l\'a\-si sk\'a\-la. K\"o\-vet\-ke\-z\H o l\'e\-p\'es\-k\'ent
egy m\'a\-sik pa\-ra\-m\'e\-ter al\-kal\-mas han\-go\-l\'a\-s\'a\-val r\"og\-z\'{\i}t\-het\-j\"uk  a
k\"onny\H u Higgs-bo\-zon $m_h$ t\"o\-me\-g\'et, majd ne\-h\'ez p\'ar\-j\'a\-\'et,
$m_H$-\'et egy har\-ma\-dik pa\-ra\-m\'e\-ter han\-go\-l\'a\-sa se\-g\'{\i}t\-s\'e\-g\'e\-vel, \'es
\'{\i}gy to\-v\'abb. Az el\-j\'a\-r\'as e\-red\-m\'e\-nye\-i\-re r\'esz\-le\-te\-seb\-ben a \ref{kre}
sza\-kasz\-ban t\'e\-rek ki.

A\-zon\-ban a fen\-ti a\-da\-tok\-b\'ol e\-n\'el\-k\"ul is k\"onnyen le\-ol\-vas\-ha\-t\'o n\'e\-h\'any
l\'e\-nye\-ges ten\-den\-ci\-a. A $\langle \phi \rangle$ v\'ar\-ha\-t\'o \'er\-t\'ek
e\-r\H o\-tel\-je\-sen n\H o, ha $m_U^2$ nagy ne\-ga\-t\'{\i}v \'er\-t\'e\-ke\-ket vesz
fel (te\-h\'at ha a stop\-t\"o\-meg sz\'a\-mot\-te\-v\H o\-en ki\-sebb, mint a top-t\"o\-meg),
eb\-ben az e\-set\-ben n\'e\-h\'any pont\-ban l\'at\-tuk, hogy fe\-n\'all a  koz\-mo\-l\'o\-gi\-a\-i
je\-len\-t\H o\-s\'e\-g\H u $\langle \phi \rangle / T_c > 1$ e\-gyen\-l\H ot\-len\-s\'eg. $m_U^2$
meg\-fe\-le\-l\H o\-en nagy ne\-ga\-t\'{\i}v \'er\-t\'e\-ke\-i\-n\'el sz\'{\i}n\-s\'er\-t\H o f\'a\-zi\-s\'at\-me\-net
j\'at\-sz\'od\-hat le; a fen\-ti be\-me\-n\H o pa\-ra\-m\'e\-te\-rek e\-se\-t\'e\-ben a sz\'{\i}n\-s\'er\-t\H o
f\'a\-zis $m_U \approx 0.09 i$ k\"or\-ny\'e\-k\'en je\-le\-nik meg.

A fi\-zi\-ka\-i pa\-ra\-m\'e\-te\-rek r\"og\-z\'{\i}\-tett \'er\-t\'e\-ken tar\-t\'a\-s\'an k\'{\i}\-v\"ul fel\-me\-r\"ul
m\'eg az a ne\-h\'e\-zs\'eg is, hogy az al\-kal\-ma\-zott per\-tur\-b\'a\-ci\-\'o\-sz\'a\-m\'{\i}\-t\'as nem
megy t\'ul az egy-hu\-rok ren\-den, \'es is\-mert, hogy a bi\-zo\-nyos k\'et\-hu\-rok
ren\-d\H u gr\'a\-fok j\'a\-ru\-l\'e\-ka je\-len\-t\H os \cite{car97}. A m\'a\-so\-dik hu\-rok-rend a
f\'a\-zi\-s\'at\-me\-net e\-r\H o\-s\"o\-d\'e\-s\'et jel\-zi. Ko\-moly \'er\-vek sz\'ol\-nak a\-mel\-lett, hogy
a per\-tur\-b\'a\-ci\-\'os sor to\-v\'ab\-bi rend\-je\-i sz\'e\-pen cs\"ok\-ken\-nek, mi\-\'al\-tal a
k\'et\-hu\-rok-ren\-d\H u e\-red\-m\'eny meg\-b\'{\i}z\-ha\-t\'o. Nyil\-v\'an\-va\-l\'o a\-zon\-ban a
nem\-per\-tur\-ba\-t\'{\i}v meg\-k\"o\-ze\-l\'{\i}\-t\'es sz\"uk\-s\'e\-ges\-s\'e\-ge; a fen\-ti egy\-sze\-r\H u
per\-tur\-ba\-t\'{\i}v vizs\-g\'a\-lat e hasz\-nos i\-r\'any\-jel\-z\H ok\-nek bi\-zo\-nyul majd a
vizs\-g\'a\-lan\-d\'o pa\-ra\-m\'e\-ter\-tar\-to\-m\'any ki\-v\'a\-lasz\-t\'a\-s\'a\-ban,

\section{Di\-men\-zi\-\'os re\-duk\-ci\-\'o\-val ka\-pott e\-red\-m\'e\-nyek}
\fancyhead[CO]{\hst{\thesection \quad Di\-men\-zi\'os re\-duk\-ci\'o\-val ka\-pott
e\-redm\'e\-nyek}}
A \ref{pnp} sza\-kasz\-ban em\-l\'{\i}\-tett nem\-per\-tur\-ba\-t\'{\i}v m\'od\-sze\-rek k\"o\-z\"ul
el\-s\H o\-k\'ent a di\-men\-zi\-\'os re\-duk\-ci\-\'on \cite{jak, gin, app2}
a\-la\-pu\-l\'o meg\-k\"o\-ze\-l\'{\i}\-t\'est dol\-goz\-t\'ak ki \cite{la98, far}, mi\-vel a
n\'egy\-di\-men\-zi\-\'os szi\-mu\-l\'a\-ci\-\'ok l\'e\-nye\-ge\-sen na\-gyobb g\'e\-pi\-d\H ot
i\-g\'e\-nyel\-nek -- b\'ar \'al\-ta\-l\'a\-ban nem o\-lyan nagy m\'er\-t\'ek\-ben, mint a
stan\-dard mo\-dell ke\-re\-t\'e\-ben. En\-nek o\-ka az, hogy a f\'a\-zi\-s\'at\-me\-net
l\'e\-nye\-ge\-sen e\-r\H o\-sebb, \'{\i}gy a n\'egy\-di\-men\-zi\-\'os szi\-mu\-l\'a\-ci\-\'o\-ban
fel\-buk\-ka\-n\'o jel\-lem\-z\H o hossz\'u\-s\'a\-gok nagy\-s\'ag\-ren\-di\-leg nem na\-gyob\-bak,
mint a kri\-ti\-kus h\H o\-m\'er\-s\'ek\-let re\-cip\-ro\-ka, $T_c^{-1}$. \'Igy a
n\'egy\-di\-men\-zi\-\'os szi\-mu\-l\'a\-ci\-\'ok, ha ne\-he\-zen is, de meg\-va\-l\'o\-s\'{\i}t\-ha\-t\'ok --
b\'ar eh\-hez \'o\-ri\-\'a\-si sz\'a\-m\'{\i}\-t\'o\-g\'e\-pes ka\-pa\-ci\-t\'as sz\"uk\-s\'e\-ges. E\-zen
ne\-h\'e\-zs\'e\-gek je\-len\-tik a k\"o\-vet\-ke\-z\H o fe\-je\-ze\-tek k\"oz\-pon\-ti
k\'er\-d\'es\-k\"o\-r\'et. (Az ``egy\-sze\-r\H ubb'', di\-men\-zi\-\'os re\-duk\-ci\-\'on a\-la\-pu\-l\'o
sz\'a\-m\'{\i}\-t\'a\-sok e\-red\-m\'e\-nye\-it k\"oz\-l\H o \cite{la98} dol\-go\-zat\-hoz sz\"uk\-s\'e\-ges
szi\-mu\-l\'a\-ci\-\'ok g\'e\-pi\-de\-je 7.5 n\'o\-dus-\'ev volt egy Cray T3E t\'{\i}\-pu\-s\'u
szu\-per\-sz\'a\-m\'{\i}\-t\'o\-g\'e\-pen).

A tel\-jes pa\-ra\-m\'e\-ter\-t\'er fel\-t\'er\-k\'e\-pe\-z\'e\-se ter\-m\'e\-sze\-te\-sen le\-he\-tet\-len
fe\-la\-dat, \'{\i}gy k\'e\-zen\-fek\-v\H o \"ot\-let a per\-tur\-ba\-t\'{\i}v j\'os\-la\-tok \'al\-tal
fa\-vo\-ri\-z\'alt pa\-ra\-m\'e\-ter\-tar\-to\-m\'any vizs\-g\'a\-la\-ta. Az \'{\i}gy ka\-pott
h\'a\-rom\-di\-men\-zi\-\'os e\-red\-m\'e\-nyek azt jel\-zik, hogy az e\-l\H o\-z\H o\-ek\-ben
me\-ga\-dott per\-tur\-ba\-t\'{\i}v e\-red\-m\'e\-nyek j\'ok: a nem\-per\-tur\-ba\-t\'{\i}v vizs\-g\'a\-lat
az ott ta\-l\'at fel\-s\H o t\"o\-meg\-kor\-l\'a\-to\-kat m\'eg messzebb tol\-ja, \'{\i}gy a
per\-tur\-ba\-t\'{\i}v j\'os\-la\-to\-kat \emph{konzervat\'{\i}v}nak le\-het ne\-vez\-ni.
M\'as\-k\'ent fo\-gal\-maz\-va: a f\'a\-zi\-s\'at\-me\-net e\-r\H os\-s\'e\-ge -- mely\-nek
jel\-lem\-z\H o\-je le\-het a l\'a\-tens h\H o -- nem\-per\-tur\-ba\-t\'{\i}\-ve l\'e\-nye\-ge\-sen
na\-gyobb\-nak a\-d\'o\-dik, mint a\-hogy a k\'et\-hu\-rok-ren\-d\H u
per\-tur\-b\'a\-ci\-\'o\-sz\'a\-m\'{\i}\-t\'as j\'o\-sol\-ta \cite{la98}.

A sz\'{\i}n\-s\'er\-t\H o f\'a\-zi\-s\'at\-me\-net a pa\-ra\-m\'e\-te\-rek meg\-fe\-le\-l\H o
meg\-v\'a\-lasz\-t\'a\-s\'a\-n\'al je\-len van; en\-nek sz\"uk\-s\'e\-ges fel\-t\'e\-te\-le az, hogy
a stop-t\"o\-meg e\-le\-gen\-d\H o\-en ki\-csi le\-gyen. Eb\-ben az e\-set\-ben
k\'et\-l\'ep\-cs\H os f\'a\-zi\-s\'at\-me\-net is
meg\-va\-l\'o\-sul\-hat.

Az MSSM pa\-ra\-m\'e\-ter\-te\-r\'e\-ben te\-h\'at van o\-lyan tar\-to\-m\'any mely\-ben
meg\-va\-l\'o\-sul\-hat a ba\-ri\-o\-ge\-n\'e\-zis -- en\-nek je\-len\-t\H o\-s\'e\-g\'et a\-lig\-ha kell
hang\-s\'u\-lyoz\-ni. K\'e\-zen\-fek\-v\H o te\-h\'at, hogy u\-gya\-nezt a le\-he\-t\H o\-s\'e\-get egy
m\'a\-sik n\'e\-z\H o\-pont\-b\'ol, a n\'egy\-di\-men\-zi\-\'os szi\-mu\-l\'a\-ci\-\'ok
szem\-pont\-j\'a\-b\'ol is a\-la\-po\-san meg\-vizs\-g\'al\-juk.

\chapter{A PMS szu\-per\-sz\'a\-m\'{\i}\-t\'o\-g\'ep}
\fancyhead[CE]{\hst{\thechapter{}.\ fe\-je\-zet \quad A PMS
szu\-persz\'am\'{\i}t\'og\'ep}}
\bfr
\emph{``Egy\-szer I.I.~Ra\-bi meg is je\-gyez\-te: \\
az e\-u\-r\'o\-pa\-i k\'{\i}\-s\'er\-le\-ti fi\-zi\-ku\-sok nem tud\-nak
egy hosszabb sz\'a\-mosz\-lo\-pot \"ossze\-ad\-ni, \\
az el\-m\'e\-le\-ti\-ek pe\-dig nem tud\-j\'ak meg\-k\"ot\-ni
a sa\-j\'at ci\-p\H o\-f\H u\-z\H o\-j\"u\-ket.''} \\
(Le\-on Le\-der\-man, \emph{Az is\-te\-ni a-tom})
\efr

Az MSSM-be\-li e\-lekt\-ro\-gyen\-ge f\'a\-zi\-s\'at\-me\-net e\-l\H o\-z\H o fe\-je\-zet\-ben t\'ar\-gyalt
per\-tur\-ba\-t\'{\i}v vizs\-g\'a\-la\-ta nem \'uj\-ke\-le\-t\H u. A j\'os\-la\-tok fenn\-tar\-t\'as
n\'el\-k\"u\-li el\-fo\-ga\-d\'a\-s\'a\-hoz a\-zon\-ban az sz\"uk\-s\'e\-ges, hogy nem\-per\-tur\-ba\-t\'{\i}v
m\'od\-sze\-rek\-kel is meg\-vizs\-g\'al\-juk a f\'a\-zi\-s\'at\-me\-ne\-tet, \'es az
e\-l\H o\-z\H o\-ek\-kel \"ossz\-hang\-ban \'al\-l\'o e\-red\-m\'e\-nyek\-re jus\-sunk.

Nem\-per\-tur\-ba\-t\'{\i}v m\'od\-szer a\-latt ter\-m\'e\-sze\-te\-sen to\-v\'abb\-ra sem tisz\-t\'an
nem\-per\-tur\-ba\-t\'{\i}v m\'od\-szert \'er\-t\"unk -- a\-mi\-lyen a tel\-jes MSSM
n\'egy\-di\-men\-zi\-\'os r\'acsszi\-mu\-l\'a\-ci\-\'o\-ja len\-ne --, ha\-nem a stan\-dard mo\-dell
e\-se\-t\'e\-ben l\'a\-tott k\'et m\'od\-szert, a di\-men\-zi\-\'os re\-duk\-ci\-\'on a\-la\-pu\-l\'o
ef\-fek\-t\'{\i}v po\-ten\-ci\-\'al\-ra a\-la\-po\-zott m\'od\-szert, il\-let\-ve az MSSM bo\-zo\-ni\-kus
szek\-to\-r\'a\-nak n\'egy\-di\-men\-zi\-\'os r\'acsszi\-mu\-l\'a\-ci\-\'o\-j\'at. Ez a fe\-je\-zet
az u\-t\'ob\-bi m\'od\-szer meg\-va\-l\'o\-s\'{\i}\-t\'a\-s\'a\-nak ne\-h\'e\-zs\'e\-ge\-it pr\'o\-b\'al\-ja
be\-mu\-tat\-ni. A n\'egy\-di\-men\-zi\-\'os szi\-mu\-l\'a\-ci\-\'ok g\'e\-pi\-d\H o-i\-g\'e\-nye m\'ar
a stan\-dard mo\-dell ke\-re\-te\-in be\-l\"ul is l\'e\-nye\-ge\-sen na\-gyobb volt, mint a
h\'a\-rom\-di\-men\-zi\-\'os re\-du\-k\'alt mo\-del\-l\'e \cite{far}, a\-mi azt is
je\-len\-tet\-te, hogy a n\'egy\-di\-men\-zi\-\'os szi\-mu\-l\'a\-ci\-\'ok kel\-l\H o pon\-tos\-s\'a\-g\'u
el\-v\'eg\-z\'e\-s\'e\-hez '90-es \'e\-vek k\"o\-ze\-p\'e\-ig v\'ar\-ni kel\-lett: az ak\-ko\-ri
leg\-na\-gyobb ka\-pa\-ci\-t\'a\-s\'u sz\'a\-m\'{\i}\-t\'o\-g\'e\-pek m\'ar l\'e\-nye\-g\'e\-ben meg tud\-tak
bir\-k\'oz\-ni a fe\-la\-dat\-tal.

A stan\-dard mo\-dell szu\-per\-szim\-met\-ri\-kus ki\-ter\-jesz\-t\'e\-s\'e\-nek vizs\-g\'a\-la\-ta
l\'e\-nye\-ge\-sen \"ossze\-tet\-tebb. Ez j\'ol l\'at\-szik m\'ar a ha\-t\'as
szi\-mu\-l\'a\-ci\-\'ok\-ban al\-kal\-ma\-zott a\-lak\-j\'a\-b\'ol is, mely\-net a (\ref{racslag1})
-- (\ref{racslag2}) k\'ep\-le\-tek k\"o\-z\"ott \'{\i}r\-tunk ki tel\-jes\-s\'e\-g\'e\-ben.
A leg\-f\H obb ne\-he\-z\'{\i}\-t\'est az je\-len\-ti a stan\-dard mo\-dell e\-se\-t\'e\-hez
k\'e\-pest, hogy m\'{\i}g ott az el\-m\'e\-let e\-gyet\-len is\-me\-ret\-len pa\-ra\-m\'e\-te\-re
a Higgs-bo\-zon t\"o\-me\-ge, ad\-dig itt sz\'a\-mos k\'{\i}\-s\'er\-le\-ti\-leg min\-ded\-dig meg
nem ha\-t\'a\-ro\-zott pa\-ra\-m\'e\-ter van je\-len. \'Igy a vizs\-g\'a\-lan\-d\'o
pa\-ra\-m\'e\-ter\-t\'er sok di\-men\-zi\-\'os, mely\-nek tel\-jes fel\-t\'er\-k\'e\-pe\-z\'e\-se a
k\"o\-zel\-j\"o\-v\H o\-ben re\-m\'eny\-te\-len fe\-la\-dat.

Sze\-ren\-cs\'e\-re ren\-del\-ke\-z\'e\-s\"unk\-re \'all\-nak a per\-tur\-ba\-t\'{\i}v e\-red\-m\'e\-nyek,
me\-lyek e\-r\H o\-tel\-je\-sen le\-sz\H u\-k\'{\i}\-tik a fi\-zi\-ka\-i szem\-pont\-b\'ol \'er\-de\-kes
pa\-ra\-m\'e\-ter\-tar\-to\-m\'a\-nyo\-kat. En\-nek el\-le\-n\'e\-re nyil\-v\'an\-va\-l\'o, hogy a
n\'egy\-di\-men\-zi\-\'os szi\-mu\-l\'a\-ci\-\'ok\-hoz sz\"uk\-s\'e\-ges sz\'a\-m\'{\i}\-t\'o\-g\'e\-pes
ka\-pa\-ci\-t\'as ret\-ten\-t\H o nagy -- a je\-len k\"o\-r\"ul\-m\'e\-nyek k\"o\-z\"ott
k\'{\i}\-n\'al\-ko\-z\'o e\-gyet\-len me\-gol\-d\'as te\-h\'at egy ol\-cs\'o
szu\-per\-sz\'a\-m\'{\i}\-t\'o\-g\'ep \'e\-p\'{\i}\-t\'e\-se. Ez a ne\-h\'e\-zs\'eg egy\-ben egy m\'a\-sik
fej\-t\"o\-r\H o me\-gol\-d\'a\-s\'a\-ul szol\-g\'al: ho\-gyan h\'{\i}v\-juk a me\-g\'e\-p\"u\-l\H o
monst\-ru\-mot? A fen\-ti\-ek a\-lap\-j\'an a v\'a\-lasz\-t\'as a PMS n\'ev\-re e\-sett, mely
a \emph{Po\-or Man's Su\-per\-com\-pu\-ter} sza\-va\-kat r\"o\-vi\-d\'{\i}\-ti -- \'ep\-p\'ugy,
mint a sz\'a\-m\'{\i}\-t\'o\-g\'ep\-re u\-gya\-nennyi\-re jel\-lem\-z\H o, az e\-l\H o\-z\H o\-n\'el
ke\-v\'es\-b\'e k\'ep\-le\-te\-sen \'er\-ten\-d\H o \emph{Pa\-ral\-lel Mul\-tip\-ro\-ces\-sor
Su\-per\-com\-pu\-ter}-t, stb.

\section{A PMS fe\-l\'e\-p\'{\i}\-t\'e\-se}
A szu\-per\-sz\'a\-m\'{\i}\-t\'o\-g\'e\-pek \'ar/tel\-je\-s\'{\i}t\-m\'eny a\-r\'a\-nya l\'e\-nye\-ge\-sen
na\-gyobb, mint a sze\-m\'e\-lyi sz\'a\-m\'{\i}\-t\'o\-g\'e\-pe\-k\'e. \'Igy sok PC meg\-fe\-le\-l\H o
\"ossze\-kap\-cso\-l\'a\-s\'a\-val na\-gyon is ver\-seny\-k\'e\-pes,
szu\-per\-sz\'a\-m\'{\i}\-t\'o\-g\'ep \'e\-p\'{\i}t\-he\-t\H o. A sze\-m\'e\-lyi sz\'a\-m\'{\i}\-t\'o\-g\'e\-pek\-b\H ol
fe\-l\'e\-p\"u\-l\H o szu\-per\-sz\'a\-m\'{\i}\-t\'o\-g\'ep to\-v\'ab\-bi e\-l\H o\-nye, hogy az \'u\-jabb,
na\-gyobb tel\-je\-s\'{\i}t\-m\'e\-ny\H u al\-kat\-r\'e\-szek\-re va\-l\'o \'at\-t\'e\-r\'es k\"onnyen
meg\-va\-l\'o\-s\'{\i}t\-ha\-t\'o, \'{\i}gy m\'as szu\-per\-sz\'a\-m\'{\i}\-t\'o\-g\'e\-pek\-kel szem\-ben a
ver\-seny\-k\'e\-pes tel\-je\-s\'{\i}t\-m\'eny -- meg\-fe\-le\-l\H o a\-nya\-gi r\'a\-for\-d\'{\i}\-t\'as\-sal --
k\"onnyen fenn\-tart\-ha\-t\'o.

A g\'ep fe\-l\'e\-p\'{\i}\-t\'e\-s\'et a vizs\-g\'alt fi\-zi\-ka\-i prob\-l\'e\-ma, az MSSM-be\-li
e\-lekt\-ro\-gyen\-ge f\'a\-zi\-s\'at\-me\-net szi\-mu\-l\'a\-ci\-\'o\-ja ha\-t\'a\-roz\-za meg. Az e\-gyik
le\-he\-t\H o\-s\'eg a k\"o\-vet\-ke\-z\H o: a k\"oz\-pon\-ti g\'ep \'al\-tal i\-r\'a\-ny\'{\i}\-tott e\-gyes
PC-k u\-gya\-nazt a prog\-ra\-mot fut\-tat\-j\'ak k\"u\-l\"on\-b\"o\-z\H o pa\-ra\-m\'e\-te\-rek
mel\-lett (\emph{sing\-le pro\-cess mul\-tip\-le da\-ta}, SPMD \"u\-zem\-m\'od),
\'{\i}gy nem t\'ul nagy m\'e\-re\-t\H u r\'a\-csok e\-se\-t\'en k\"onnyen jut\-ha\-tunk
kel\-l\H o\-en nagy sta\-tisz\-ti\-k\'a\-hoz. A 0 r\'a\-cs\'al\-lan\-d\'o\-j\'u e\-set\-re
(kon\-ti\-nu\-um-li\-mesz\-re) va\-l\'o ext\-ra\-po\-l\'a\-l\'as kis m\'e\-re\-t\H u r\'a\-csok\-ra
m\'ert a\-da\-to\-kat is i\-g\'e\-nyel; e\-zek fel\-v\'e\-te\-l\'e\-hez ez a meg\-k\"o\-ze\-l\'{\i}\-t\'e\-si
m\'od a leg\-c\'el\-sze\-r\H ubb.

A sz\'a\-m\'{\i}\-t\'o\-g\'ep nagy ka\-pa\-ci\-t\'a\-sa na\-gyobb r\'a\-csok\-n\'al je\-lent i\-ga\-z\'an
nagy e\-l\H onyt; m\'{\i}g egy kis sz\'a\-m\'{\i}\-t\'o\-g\'ep\-n\'el nem ke\-r\"ul\-he\-t\H o el a
me\-rev-le\-mez\-re t\"or\-t\'e\-n\H o ki\-\'{\i}\-r\'as \'es az ar\-r\'ol t\"or\-t\'e\-n\H o
be\-ol\-va\-s\'as, ad\-dig a szu\-per\-sz\'a\-m\'{\i}\-t\'o\-g\'ep e\-se\-t\'e\-ben a nagy me\-m\'o\-ri\-a itt
ki\-hasz\-n\'al\-ha\-t\'o. Eh\-hez nyil\-v\'an a szu\-per\-sz\'a\-m\'{\i}\-t\'o\-g\'ep\-be \'e\-p\'{\i}\-tett PC
e\-le\-mek\-nek egy\-s\'eg\-k\'ent kell m\H u\-k\"od\-ni\-\"uk, te\-h\'at a k\"u\-l\"on\-b\"o\-z\H o
n\'o\-du\-sok k\"o\-z\"ot\-ti kom\-mu\-ni\-k\'a\-ci\-\'o\-ra van sz\"uk\-s\'eg. A kel\-l\H o\-en
ha\-t\'e\-kony kom\-mu\-ni\-k\'a\-ci\-\'o meg\-va\-l\'o\-s\'{\i}\-t\'a\-sa volt a PMS prog\-ram
leg\-k\'e\-nye\-sebb f\'a\-zi\-sa. Mi\-e\-l\H ott er\-re ki\-t\'er\-n\'enk, n\'ez\-z\"uk meg nagy
vo\-na\-lak\-ban, hogy is kell el\-ren\-dez\-ni a szu\-per\-sz\'a\-m\'{\i}\-t\'o\-g\'e\-pet
\"ossze\-te\-v\H o PC-ket.

V\'e\-ges h\H o\-m\'er\-s\'ek\-le\-t\H u szi\-mu\-l\'a\-ci\-\'ok e\-se\-t\'en a h\H o\-m\'er\-s\'ek\-let a r\'acs
i\-d\H o\-i\-r\'a\-ny\'u ki\-ter\-je\-d\'e\-s\'e\-nek re\-cip\-ro\-ka ha\-t\'a\-roz\-za meg. Ez a stan\-dard
mo\-dell\-be\-li szi\-mu\-l\'a\-ci\-\'ok e\-se\-t\'en 2--5-ig ter\-jedt. A t\'e\-ri\-r\'a\-ny\'u
ki\-ter\-je\-d\'es en\-n\'el \'al\-ta\-l\'a\-ban (l\'e\-nye\-ge\-sen) na\-gyobb: a tel\-jes PMS
g\'e\-pen e\-gyet\-len r\'a\-csot fe\-loszt\-va a $48^3 \times 4$-es r\'acs ti\-pi\-kus\-nak
mond\-ha\-t\'o.

A PC-k k\"o\-z\"ott te\-h\'at a h\'a\-rom t\'e\-ri\-r\'any\-nak meg\-fe\-le\-l\H o
kom\-mu\-ni\-k\'a\-ci\-\'ot c\'el\-sze\-r\H u biz\-to\-s\'{\i}\-ta\-ni. Ez k\'et g\'ep k\"o\-z\"ott \'ugy
t\"or\-t\'e\-nik meg, hogy az e\-gyes g\'e\-pek egy r\'acs fe\-l\'et vizs\-g\'al\-j\'ak,
\'es a vizs\-g\'alt f\'el-r\'acs pe\-re\-m\'en le\-v\H o a\-da\-to\-kat, me\-lyek
el\-s\H o-szom\-sz\'ed-k\"ol\-cs\"on\-ha\-t\'as r\'e\-v\'en a m\'a\-sik g\'e\-pen vizs\-g\'alt
f\'el-r\'acs ha\-t\'a\-ro\-l\'o s\'{\i}k\-j\'an le\-v\H o a\-da\-tok\-kal kap\-cso\-l\'od\-nak \"ossze,
to\-v\'ab\-b\'{\i}\-ta\-ni kell a m\'a\-sik g\'ep\-re. Egy\-sze\-r\H u meg\-gon\-do\-l\'a\-sok a\-lap\-j\'an
nyil\-v\'an\-va\-l\'o, hogy l\'e\-nye\-g\'e\-ben k\"o\-b\"os r\'a\-csok e\-se\-t\'e\-ben c\'el\-sze\-r\H u
a h\'a\-rom t\'e\-ri\-r\'anyt e\-gyen\-l\H o m\'er\-t\'ek\-ben fe\-losz\-ta\-ni; a k\"u\-l\"on\-b\"o\-z\H o
g\'e\-pek k\"o\-z\"ot\-ti a\-da\-t\'at\-vi\-te\-li i\-g\'eny \'{\i}gy cs\"ok\-kent\-he\-t\H o
mi\-ni\-m\'a\-lis\-ra.

A szi\-mu\-l\'a\-ci\-\'ok\-ban vizs\-g\'alt r\'a\-csok t\'er\-sze\-r\H u m\'e\-re\-t\'et nagy
r\'acs\-ki\-ter\-je\-d\'e\-sek e\-se\-t\'en in\-k\'abb p\'a\-ros\-nak szo\-k\'as v\'a\-lasz\-ta\-ni, \'{\i}gy
az egy i\-r\'any\-ba egy\-m\'as mel\-l\'e he\-lye\-zen\-d\H o g\'e\-pek sz\'a\-m\'at is
c\'el\-sze\-r\H ubb p\'a\-ros\-nak ven\-ni. I\-r\'a\-nyon\-k\'ent 4 g\'e\-pet el\-he\-lyez\-ve
te\-h\'at 64 g\'ep sz\"uk\-s\'e\-ges. A ren\-del\-ke\-z\'es\-re \'al\-l\'o \"osszeg\-b\H ol 32 PC
v\'a\-s\'ar\-l\'a\-sa volt le\-het\-s\'e\-ges -- \'{\i}gy ezt egy $4\times 4\times 2$
rend\-sze\-r\H u r\'acs\-ba le\-he\-tett el\-ren\-dez\-ni. Az e\-gyes PC kom\-po\-nen\-sek \'a\-ra a
prog\-ram el\-kez\-d\'e\-se\-kor 350, a kom\-mu\-ni\-k\'a\-ci\-\'os k\'ar\-ty\'ak \'a\-ra to\-v\'ab\-bi
40 dol\-l\'ar volt. A PC kom\-po\-nen\-sek 100 MHZ-es SO\-YO SY-5EHM a\-lap\-la\-pot,
450 MHz-es AMD K6-I\-I pro\-cesszort, 128 Mbyte SDRAM-ot, 2.1 Gbyte-os
me\-rev\-le\-mezt, \'es 10 Mbit \'at\-vi\-te\-l\H u Et\-her\-net k\'ar\-ty\'at tar\-tal\-maz\-tak.
(Az al\-kat\-r\'e\-szek \'a\-ra\-it il\-le\-t\H o\-leg l\'asd \cite{pms/ref1}-t.)
\fancyhead[CO]{\hst{\thesection \quad A PMS fel\'ep\'{\i}t\'e\-se}}

A k\"onnyebb ke\-zel\-he\-t\H o\-s\'eg \'er\-de\-k\'e\-ben a PC-ket a\-la\-po\-san
\'at\-ren\-dez\-t\"uk; a t\'a\-pegy\-s\'e\-gek a szu\-per\-sz\'a\-m\'{\i}\-t\'o\-g\'ep em\-be\-res m\'e\-re\-t\H u
\'all\-v\'a\-ny\'a\-nak al\-j\'a\-ra ke\-r\"ul\-tek, a\-hon\-nan egy ha\-tal\-mas
ven\-til\-l\'a\-tor t\'a\-vo\-l\'{\i}t\-ja el a ter\-melt h\H ot. (A n\'o\-du\-sok \'al\-tal ter\-melt
h\H o a g\'ep te\-te\-j\'en l\'e\-v\H o n\'egy ven\-til\-l\'a\-to\-ron ke\-resz\-t\"ul t\'a\-vo\-zik.
A g\'ep h\H o\-ter\-me\-l\'e\-se o\-lyan nagy, hogy a sz\'a\-m\'{\i}\-t\'o\-g\'ep-te\-rem\-ben b\H o
\"ot fok\-kal is ma\-ga\-sabb le\-het a h\H o\-m\'er\-s\'ek\-let, mint a szom\-sz\'e\-dos
szo\-b\'a\-ban.)

Az a\-lap\-la\-pok \'{\i}gy min\-tegy m\'e\-ter\-nyi t\'a\-vol\-s\'ag\-ba ke\-r\"ul\-tek a
t\'a\-pegy\-s\'e\-gek\-t\H ol, a\-mi b\H o\-s\'e\-ges al\-kal\-mat biz\-to\-s\'{\i}\-tott for\-rasz\-t\'a\-si
k\'es\-zs\'e\-ge\-ink to\-v\'abb\-fej\-lesz\-t\'e\-s\'e\-re. Az a\-lap\-la\-pok 4, egy\-m\'as fe\-lett
el\-he\-lye\-zett t\'al\-c\'an kap\-tak he\-lyet, me\-lyek ol\-da\-li\-r\'any\-ba ki\-h\'uz\-ha\-t\'ok
-- a k\'e\-s\H obb el\-ke\-r\"ul\-he\-tet\-len sze\-re\-l\'e\-si mun\-k\'a\-la\-tok
meg\-k\"onny\'{\i}\-t\'e\-s\'e\-re. Az e\-gyes t\'al\-c\'a\-kon $2 \times 4$ n\'o\-dus
he\-lyez\-ke\-dik el. Az el\-ren\-de\-z\'es j\'ol l\'at\-ha\-t\'o a k\"o\-vet\-ke\-z\H o \'ab\-r\'an:
\bef[ht]
\bc
\epsfig{file=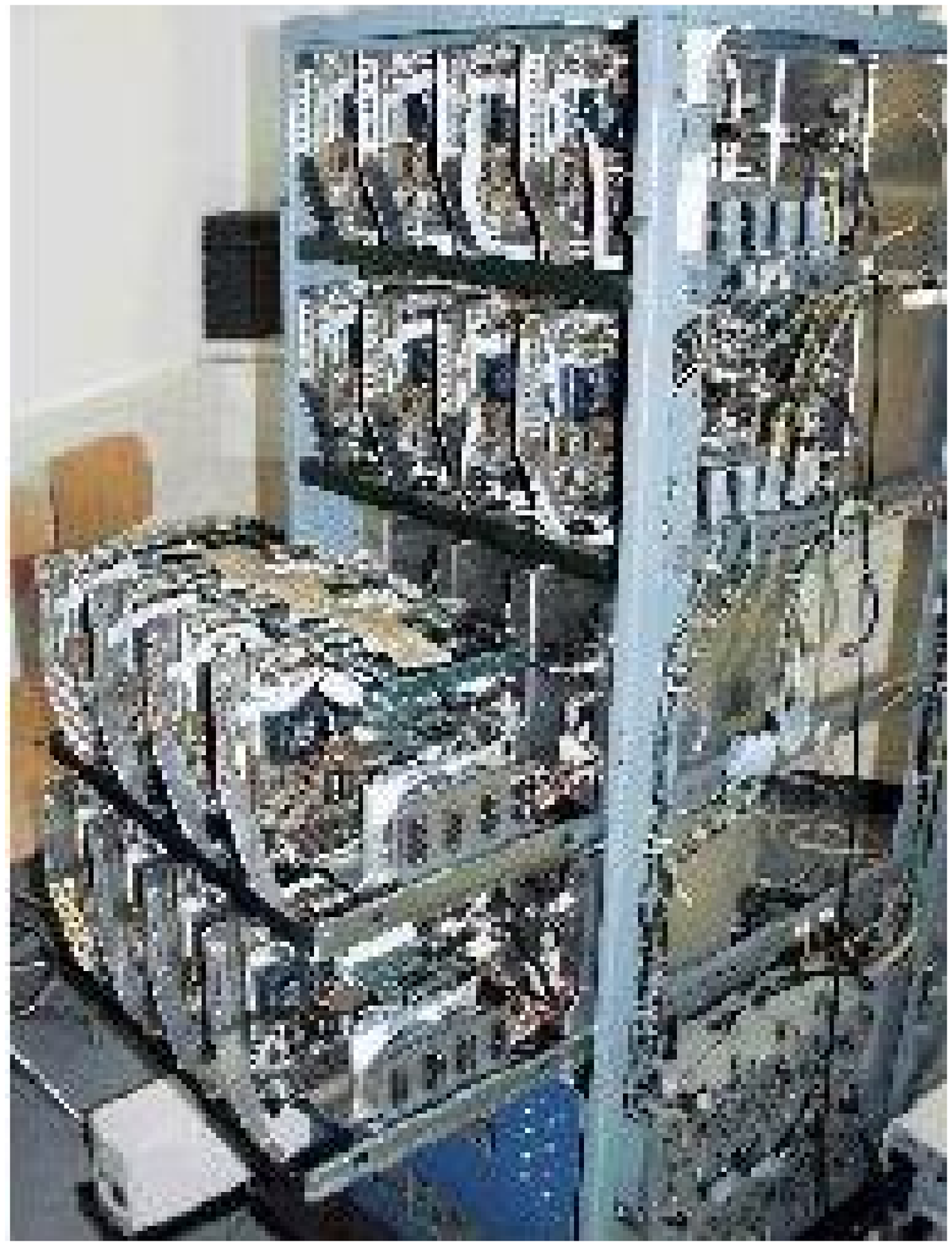,width=8cm}\quad
\epsfig{file=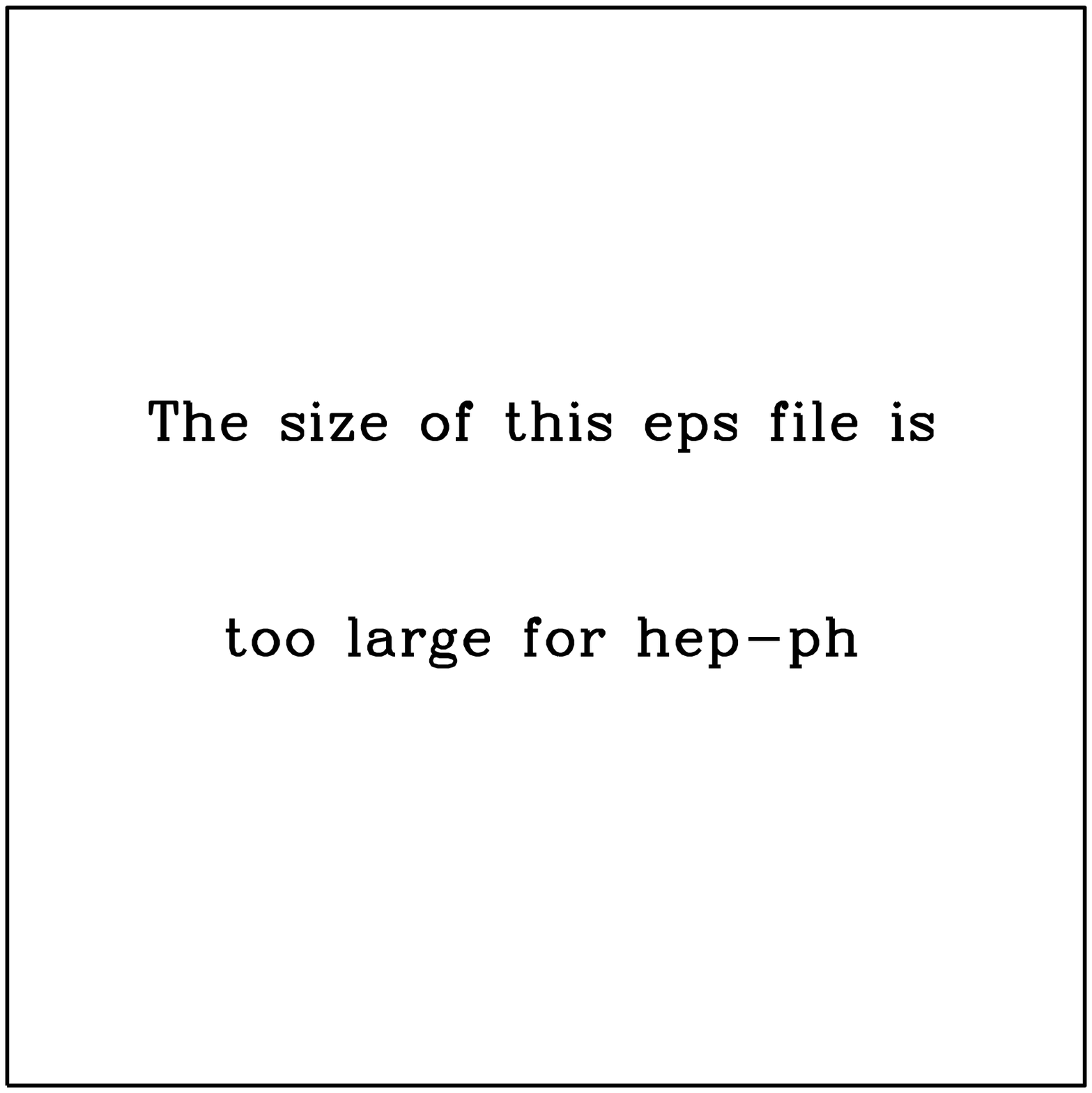,width=8cm}
\caption{A PMS szu\-per\-sz\'a\-m\'{\i}\-t\'o\-g\'ep \'es 8 egy t\'al\-c\'an el\-he\-lye\-zett
n\'o\-du\-sa \label{pms1}}
\ec
\enf

A szom\-sz\'e\-dos n\'o\-du\-sok k\"oz\-ti kom\-mu\-ni\-k\'a\-ci\-\'o se\-bes\-s\'e\-ge d\"on\-t\H o
fon\-tos\-s\'a\-g\'u. A k\"oz\-pon\-ti sz\'a\-m\'{\i}\-t\'o\-g\'ep\-pel va\-l\'o kom\-mu\-ni\-k\'a\-ci\-\'ot
meg\-te\-rem\-t\H o Et\-her\-net kap\-cso\-lat itt nem ki\-e\-l\'e\-g\'{\i}\-t\H o, egy\-r\'eszt mert
en\-nek ki\-\'e\-p\'{\i}\-t\'e\-s\'e\-hez t\'ul\-s\'a\-go\-san hossz\'u i\-d\H o sz\"uk\-s\'e\-ges,
m\'as\-r\'eszt mi\-vel a ma\-xi\-m\'a\-lis a\-da\-t\'at\-vi\-te\-li se\-bes\-s\'eg 1 Mbyte/sec.
A\-zok\-ban az e\-se\-tek\-ben, a\-mi\-kor az e\-g\'esz szu\-per\-sz\'a\-m\'{\i}\-t\'o\-g\'ep\-re e\-gyet\-len
r\'a\-csot he\-lye\-z\"unk, a k\"u\-l\"on\-b\"o\-z\H o n\'o\-du\-sok k\"oz\-ti kom\-mu\-ni\-k\'a\-ci\-\'o
a\-r\'any\-ta\-la\-nul sok i\-d\H ot e\-m\'eszt fel az e\-gyes n\'o\-du\-so\-kon v\'eg\-re\-haj\-tott
sz\'a\-mo\-l\'as i\-de\-j\'e\-hez k\'e\-pest. \'Igy az Et\-her\-net kap\-cso\-la\-tot hasz\-n\'a\-l\'o
szu\-per\-sz\'a\-m\'{\i}\-t\'o\-g\'e\-pek (pl.\ In\-di\-a\-na) tel\-je\-s\'{\i}\-m\'eny/\'ar h\'a\-nya\-do\-sa
t\'a\-vol\-r\'ol sem op\-ti\-m\'a\-lis.

A Myri\-net kap\-cso\-la\-tot hasz\-n\'a\-l\'o sz\'a\-m\'{\i}\-t\'o\-g\'e\-pek e\-se\-t\'e\-ben (pl.\
A\-li\-ce \cite{eicker}, Al\-tac\-lus\-ter) a g\'e\-pek k\"oz\-ti kom\-mu\-ni\-k\'a\-ci\-\'os
rend\-szer \'a\-ra nagy\-s\'ag\-ren\-di\-leg me\-ge\-gye\-zik a sz\'a\-m\'{\i}\-t\'o\-g\'e\-pek \'a\-r\'a\-val.
\'Igy a PMS \'e\-pi\-t\'e\-s\'e\-nek e\-gyik legk\-ri\-ti\-ku\-sabb pont\-ja a ha\-t\'e\-kony \'es
ol\-cs\'o kom\-mu\-ni\-k\'a\-ci\-\'os rend\-szer meg\-ter\-ve\-z\'e\-se volt, mely el\-s\H o\-sor\-ban
Hor\-v\'ath Vik\-tor \'er\-de\-me.

A kom\-mu\-ni\-k\'a\-ci\-\'os rend\-szer r\'esz\-le\-tes is\-mer\-te\-t\'e\-se t\'ul\-s\'a\-go\-san
messzi\-re ve\-zet\-ne; az \'er\-dek\-l\H o\-d\H o ol\-va\-s\'o a \cite{pmscikk} cikk\-ben
ta\-l\'al r\'esz\-le\-tes le\-\'{\i}\-r\'ast. Itt csak annyit em\-l\'{\i}\-te\-n\'ek meg, hogy
n\'o\-du\-son\-k\'ent k\'et k\'ar\-tya, az a\-da\-t\'at\-vi\-te\-li \'a\-ram\-k\"o\-r\"o\-ket
tar\-tal\-ma\-z\'o \emph{CPU Card}, \'es a szom\-sz\'e\-dos n\'o\-du\-sok\-kal va\-l\'o
kap\-cso\-la\-tot l\'et\-re\-ho\-z\'o sza\-lag\-k\'a\-be\-lek csat\-la\-ko\-z\'o\-it tar\-tal\-ma\-z\'o
\emph{Re\-lay Card} ke\-r\"ult be\-\"ul\-te\-t\'es\-re. A n\'o\-du\-so\-kat \"ossze\-k\"o\-t\H o
sza\-lag\-k\'a\-belt i\-gye\-kez\-t\"unk mi\-n\'el r\"o\-vi\-debb\-re ter\-vez\-ni; a je\-len\-le\-gi 2
Mbyte/sec kom\-mu\-ni\-k\'a\-ci\-\'os se\-bes\-s\'eg\-n\'el en\-nek je\-len\-t\H o\-s\'e\-ge ki\-csi,
a\-zon\-ban az e\-l\H o\-re\-l\'at\-ha\-t\'o fej\-lesz\-t\'e\-sek so\-r\'an ez d\"on\-t\H o
t\'e\-nye\-z\H o\-v\'e l\'ep\-het e\-l\H o.

Az Et\-her\-net kap\-cso\-lat\-n\'al a PMS kom\-mu\-ni\-k\'a\-ci\-\'os rend\-sze\-re nem csak a
fen\-ti kb.\ 2-es fak\-tor\-ral gyor\-sabb: a kap\-cso\-lat ki\-\'e\-p\'{\i}\-t\'e\-s\'e\-hez
sz\"uk\-s\'e\-ges i\-d\H o l\'e\-nye\-g\'e\-ben el\-ha\-nya\-gol\-ha\-t\'o, m\'as\-r\'eszt egy\-szer\-re
a\-k\'ar 16 p\'ar n\'o\-dus k\"o\-z\"ott is l\'e\-te\-s\'{\i}t\-he\-t\H o kom\-mu\-ni\-k\'a\-ci\-\'o -- ez
pe\-dig min\-tegy k\'et nagy\-s\'ag\-rend\-del na\-gyobb se\-bes\-s\'e\-get k\'e\-pes
biz\-to\-s\'{\i}\-ta\-ni. A n\'o\-du\-sok sz\'a\-m\'a\-nak e\-set\-le\-ges n\"o\-ve\-l\'e\-se e\-se\-t\'en a
kom\-mu\-ni\-k\'a\-ci\-\'o tel\-je\-s\'{\i}t\-m\'e\-nye to\-v\'abb ja\-vul, mint\-hogy az e\-gyes
n\'o\-du\-sok to\-v\'abb\-ra is csak hat szom\-sz\'e\-duk\-kal kom\-mu\-ni\-k\'al\-nak,
szi\-mul\-t\'an m\'o\-don.

\section{A PMS-en fu\-t\'o szoft\-ve\-rek}
\fancyhead[CO]{\hst{\thesection \quad A PMS-en fut\'o szoft\-ve\-rek}}
A sz\'a\-m\'{\i}\-t\'o\-g\'ep e\-gyes n\'o\-du\-sa\-in Li\-nux o\-pe\-r\'a\-ci\-\'os rend\-szer fut --
b\'ar 32 bi\-tes \emph{Ex\-ten\-ded DOS} is lett te\-le\-p\'{\i}t\-ve r\'a\-juk. A
kom\-mu\-ni\-k\'a\-ci\-\'os k\'ar\-tya ve\-z\'er\-l\'e\-s\'e\-hez va\-la\-mint a sz\'a\-m\'{\i}\-t\'o\-g\'e\-pen
fut\-ta\-tan\-d\'o prog\-ra\-mok me\-g\'{\i}\-r\'a\-sa C++ -- e\-se\-ten\-k\'ent Fort\-ran --
nyel\-ven t\"or\-t\'ent. A kom\-mu\-ni\-k\'a\-ci\-\'os k\'ar\-ty\'ak\-ra \'{\i}rt C nyel\-v\H u
p\'el\-dap\-rog\-ram a \cite{pmscikk} cikk f\"ug\-ge\-l\'e\-k\'e\-ben \'er\-he\-t\H o el.

A szu\-per\-sz\'a\-m\'{\i}\-t\'o\-g\'ep 32 n\'o\-du\-sa (s000, \ldots, s003, s010, \ldots
s133) Et\-her\-net kap\-cso\-lat\-ban \'all a ve\-z\'er\-l\H o\-g\'ep\-pel. A
szi\-mu\-l\'a\-ci\-\'ok\-ban hasz\-n\'alt pa\-ra\-m\'e\-te\-rek ki\-osz\-t\'a\-sa \'es a szi\-mu\-l\'a\-ci\-\'os
e\-red\-m\'e\-nyek \"ossze\-gy\H uj\-t\'e\-se je\-len\-ti az Et\-her\-net kap\-cso\-lat f\H o
al\-kal\-ma\-z\'a\-s\'at.

\section{A PMS tel\-je\-s\'{\i}t\-m\'e\-nye}
\fancyhead[CO]{\hst{\thesection \quad A PMS tel\-jes\'{\i}tm\'e\-nye}}
A b\H o m\'as\-f\'el \'e\-vig \'e\-p\'{\i}\-tett PMS g\'ep 2000.\ feb\-ru\-\'ar\-j\'a\-ban ju\-tott
el ar\-ra a pont\-ra, hogy a kom\-mu\-ni\-k\'a\-ci\-\'o mind a 32 n\'o\-dus k\"o\-z\"ott
meg\-b\'{\i}z\-ha\-t\'o\-an \'es gyor\-san m\H u\-k\"o\-d\"ott. Fi\-zi\-ka\-i sz\'a\-m\'{\i}\-t\'a\-sok\-ra
a\-zon\-ban sok\-kal ko\-r\'ab\-ban, a L\'agy\-m\'a\-nyos\-ra va\-l\'o k\"ol\-t\"o\-z\'es i\-de\-j\'en
kezd\-t\"uk hasz\-n\'al\-ni -- ek\-kor a k\"u\-l\"on\-b\"o\-z\H o n\'o\-du\-sok e\-gyen\-k\'ent
m\H u\-k\"od\-tek, te\-h\'at i\-ga\-z\'an nagy\-m\'e\-re\-t\H u r\'a\-csok szi\-mu\-l\'a\-l\'a\-s\'a\-ra eb\-ben
az i\-d\H o\-szak\-ban nem volt le\-he\-t\H o\-s\'eg.

A kom\-mu\-ni\-k\'a\-ci\-\'os rend\-szer me\-g\'e\-p\"u\-l\'e\-s\'e\-vel va\-l\'o\-di
szu\-per\-sz\'a\-m\'{\i}\-t\'o\-g\'ep\-p\'e v\'alt PMS tel\-je\-s\'{\i}t\-m\'e\-ny\'et r\'acs\-t\'e\-rel\-m\'e\-le\-ti
szi\-mu\-l\'a\-ci\-\'ok se\-g\'{\i}t\-s\'e\-g\'e\-vel m\'er\-he\-t\H o fel.
\begin{itemize}
\item Tisz\-ta SU(3) m\'er\-t\'e\-kel\-m\'e\-let a le\-gegy\-sze\-r\H ubb Wil\-son-f\'e\-le
ha\-t\'as\-sal. A link\-v\'al\-to\-z\'o\-kat fris\-s\'{\i}\-t\H o el\-j\'a\-r\'as o\-ver\-re\-la\-x\'a\-ci\-\'os
l\'e\-p\'e\-sek\-kel kom\-bi\-\'alt h\H o\-f\"ur\-d\H o al\-go\-rit\-must hasz\-n\'al. A $3\times3$-s
m\'at\-ri\-xok szor\-z\'a\-s\'a\-nak meggyor\-s\'{\i}\-t\'a\-s\'at egy \emph{assembly} nyel\-v\H u
prog\-ram tet\-te le\-he\-t\H o\-v\'e.
\item A PMS me\-g\'e\-p\'{\i}\-t\'e\-s\'et mo\-ti\-v\'a\-l\'o MSSM szi\-mu\-l\'a\-ci\-\'o\-ja. A
v\'al\-to\-z\'ok fris\-s\'{\i}\-t\'e\-s\'e\-re az SU(3) mo\-dell\-ben hasz\-n\'al\-ta\-kon t\'ul egy
\'uj m\'od\-szer, a ska\-l\'ar kvar\-kok ke\-ze\-l\'e\-s\'e\-re szol\-g\'a\-l\'o
mik\-ro\-ka\-no\-ni\-kus o\-ver\-re\-la\-x\'a\-ci\-\'o sz\"uk\-s\'e\-ges.
\end{itemize}
Dup\-la pon\-tos\-s\'a\-g\'u m\H u\-ve\-le\-tek e\-se\-t\'en a szu\-per\-sz\'a\-m\'{\i}\-t\'o\-g\'ep
tel\-je\-s\'{\i}t\-m\'e\-nye a \ref{sustain} \'ab\-r\'an l\'at\-ha\-t\'o m\'o\-don f\"ug\-g\"ott a
szi\-mu\-l\'alt r\'acs m\'e\-re\-t\'e\-t\H ol.

\bef[ht]
\bc
\epsfig{file=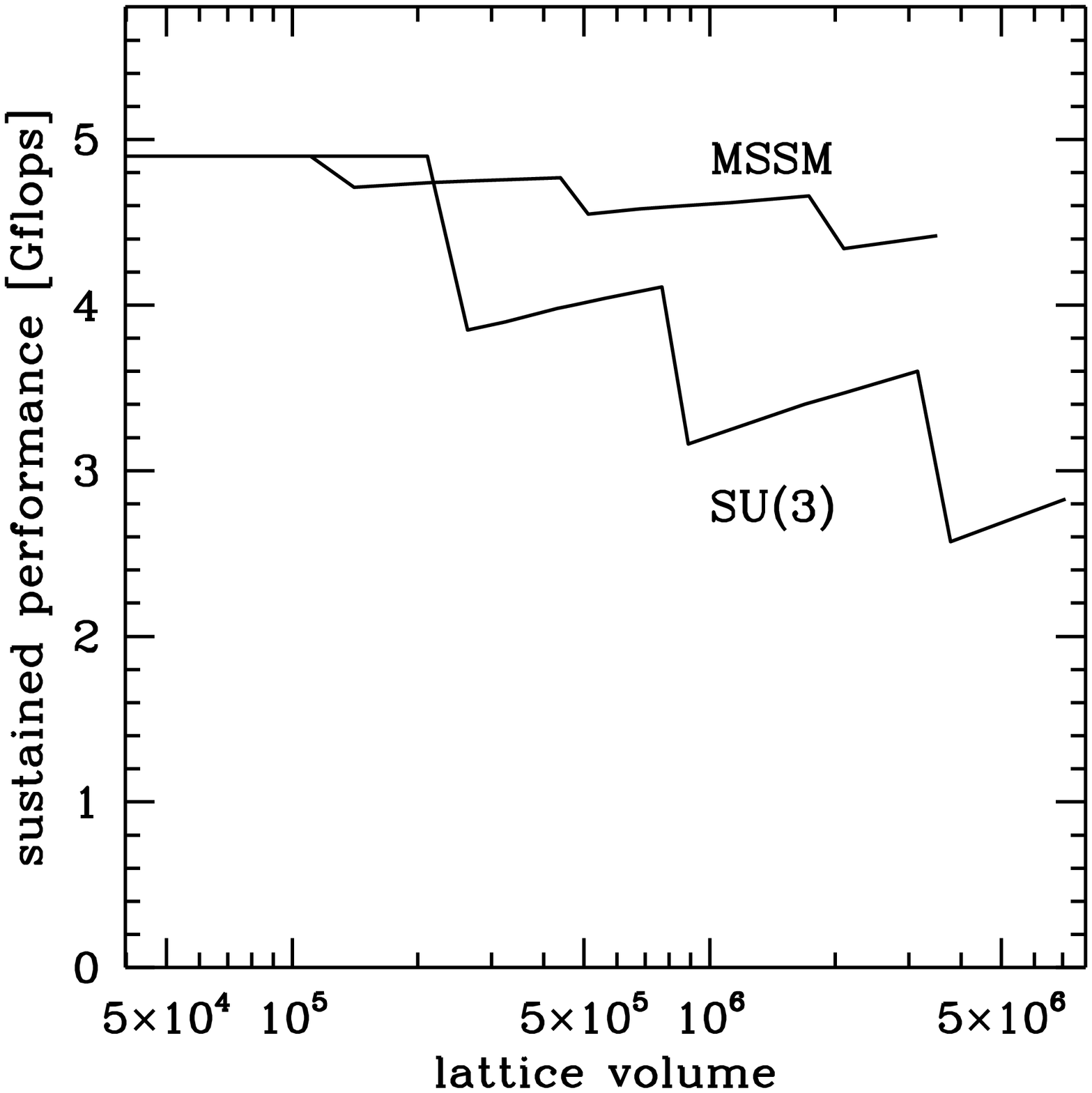,width=8cm}
\caption{A PMS tel\-je\-s\'{\i}t\-m\'e\-nye a szi\-mu\-l\'alt r\'acs\-t\'er\-fo\-gat
f\"ugg\-v\'e\-ny\'e\-ben \label{sustain}}
\ec
\enf

A fen\-ti e\-red\-m\'e\-nyek v\'a\-ra\-ko\-z\'a\-sunk\-nak meg\-fe\-le\-l\H o\-ek: az MSSM min\-tegy
k\'et\-szer annyi v\'al\-to\-z\'ot tar\-tal\-maz, mint az SU(3) mo\-dell; a
le\-be\-g\H o\-pon\-tos m\H u\-ve\-le\-tek sz\'a\-ma a\-zon\-ban kb.\ 1 nagy\-s\'ag\-rend\-del
na\-gyobb. \'Igy u\-gya\-nak\-ko\-ra r\'acs\-m\'e\-ret e\-se\-t\'en az e\-gyes n\'o\-du\-so\-kon a
szi\-mu\-l\'a\-ci\-\'o\-ra for\-d\'{\i}\-tott i\-d\H o me\-re\-de\-keb\-ben n\H o a r\'acs sz\'e\-l\'en
le\-v\H o a\-da\-tok to\-v\'ab\-b\'{\i}\-t\'a\-s\'a\-hoz k\'e\-pest az MSSM-ben, mint a tisz\-ta
SU(3) m\'er\-t\'e\-kel\-m\'e\-let\-ben. A hir\-te\-len le\-e\-s\'e\-sek az \'uj kom\-mu\-ni\-k\'a\-ci\-\'os
i\-r\'a\-nyok meg\-nyi\-t\'a\-s\'a\-n\'al mu\-tat\-koz\-nak -- pl.\ 4 \"ossze\-kap\-csolt g\'ep
he\-lyett 8-on oszlt\-juk sz\'et a tel\-jes r\'a\-csot stb.

A szi\-mu\-l\'a\-ci\-\'ok\-ban hasz\-n\'alt MSSM Lag\-ran\-ge-f\"ugg\-v\'eny e\-se\-t\'e\-ben az
egy n\'o\-dus\-ra e\-s\H o tel\-je\-s\'{\i}t\-m\'eny (dup\-la pon\-tos\-s\'ag e\-se\-t\'en)
r\'acs\-pon\-ton\-k\'ent, \'es \emph{sweep}en\-k\'ent kb.\ 3 ez\-red\-m\'a\-sod\-per\-cet
je\-len\-tett, te\-h\'at a leg\-ki\-sebb sz\'a\-molt r\'acs\-m\'e\-ret\-n\'el -- $4^3 * 2$ --
egy \emph{sweep}re min\-tegy f\'el m\'a\-sod\-per\-cet sz\'a\-mol\-ha\-tunk. Na\-gyobb
r\'a\-csok e\-se\-t\'e\-ben a m\'e\-ren\-d\H o mennyi\-s\'e\-gek (pl.\ Wil\-son-hur\-kok)
sz\'a\-m\'a\-nak n\"o\-ve\-ke\-d\'e\-se mi\-att a szi\-mu\-l\'a\-ci\-\'o\-hoz sz\"uk\-s\'e\-ges i\-d\H o a
r\'acs\-t\'er\-fo\-gat\-n\'al va\-la\-mi\-vel gyor\-sabb n\H o.

Az SU(3) m\'er\-t\'e\-kel\-m\'e\-let a\-lap\-j\'an k\"onnyeb\-ben \"ossze\-vet\-he\-t\H o a PMS
\'es m\'as szu\-per\-sz\'a\-m\'{\i}\-t\'o\-g\'e\-pek tel\-je\-s\'{\i}t\-m\'e\-ny\'et; l\'at\-ha\-t\'o\-an nem
be\-cs\"ul\-j\"uk t\'ul a PMS tel\-je\-s\'{\i}t\-m\'e\-ny\'et, ha azt 4 gi\-gaf\-lop\-ban
\'al\-la\-p\'{\i}t\-juk meg. Ez 3\$/me\-gaf\-lop \'ar/tel\-je\-s\'{\i}t\-m\'eny a\-r\'anyt je\-lent.

Ha egy\-sze\-res pon\-tos\-s\'a\-got k\"o\-ve\-te\-l\"unk meg, a tel\-je\-s\'{\i}t\-m\'eny
l\'e\-nye\-ge\-sen meg\-n\H o: az MMX prog\-ram\-cso\-mag hasz\-n\'a\-la\-t\'a\-val el\-vi\-leg
8-szo\-ros se\-bes\-s\'eg\-n\"o\-ve\-ke\-d\'es \'er\-he\-t\H o el; a ta\-pasz\-ta\-la\-tok azt
mu\-tat\-j\'ak, hogy en\-nek min\-tegy 80\%-a meg is va\-l\'o\-sul. \'Igy 27
gi\-gaf\-lop tel\-je\-s\'{\i}t\-m\'eny, a\-zaz 0.45\$/me\-gaf\-lop \'er\-he\-t\H o el. Ez\-zel a
tel\-je\-s\'{\i}t\-m\'ennyel a PMS a leg\-na\-gyobb tel\-je\-s\'{\i}t\-m\'e\-ny\H u ma\-gyar
sz\'a\-m\'{\i}\-t\'o\-g\'ep \cite{nepsz, chip}.
A k\"o\-vet\-ke\-z\H o \'ab\-r\'an a r\'acs\-t\'e\-rel\-m\'e\-le\-ti szi\-mu\-l\'a\-ci\-\'ok\-ban hasz\-n\'alt
szu\-per\-sz\'a\-m\'{\i}\-t\'o\-g\'e\-pek tel\-je\-s\'{\i}t\-m\'e\-nye l\'at\-ha\-t\'o. B\'ar l\'e\-tez\-nek
nagy\-s\'ag\-ren\-dek\-kel na\-gyobb tel\-je\-s\'{\i}t\-m\'e\-ny\H u g\'e\-pek
(CP-PACS \cite{iwasaki}, QCDSP \cite{chen}), a\-zon\-ban
\'ar/tel\-je\-s\'{\i}t\-m\'eny vi\-szony\-lat\-ban a PMS ki\-e\-mel\-ke\-d\H o\-en j\'o.

\bef[htb]
\bc
\epsfig{file=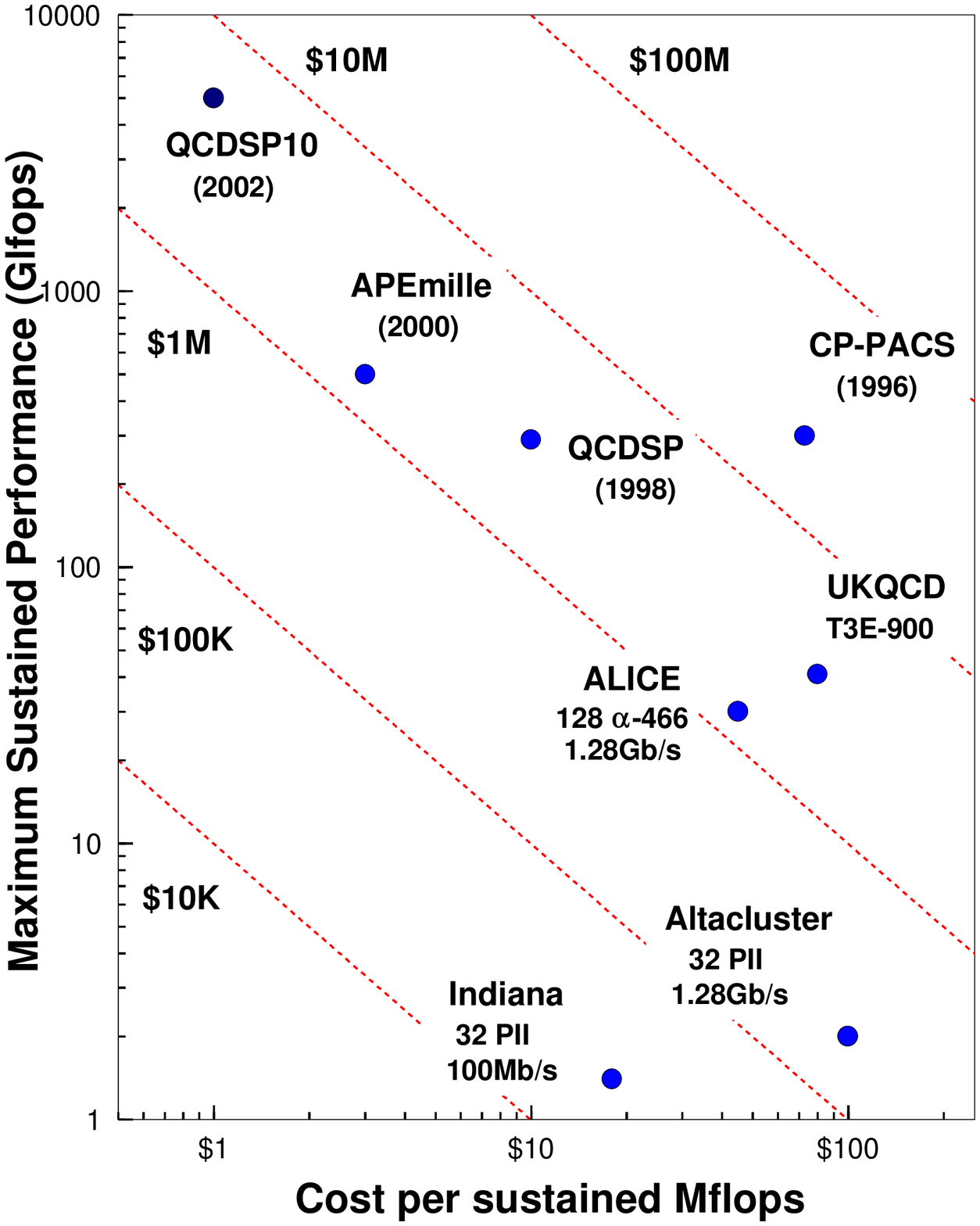,width=9cm,height=9cm} \\
\vspace{-9cm}
\epsfig{file=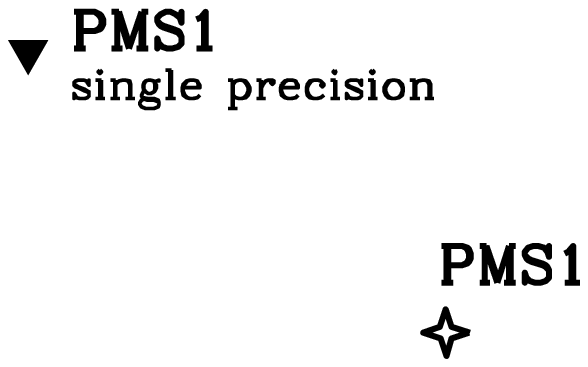,width=9cm,height=9cm}
\caption{A r\'acs\-t\'e\-rel\-m\'e\-let\-ben hasz\-n\'alt szu\-per\-sz\'a\-m\'{\i}\-t\'o\-g\'e\-pek \'ar
\'es tel\-je\-s\'{\i}t\-m\'eny sze\-rin\-ti \"ossze\-ve\-t\'e\-se \label{perform}}
\ec
\enf

(A \ref{perform} \'ab\-r\'an l\'at\-ha\-t\'o PMS1 fe\-li\-rat sej\-te\-ti, hogy egy PMS2
van sz\"u\-le\-t\H o\-ben. Az en\-nek a\-lap\-j\'at k\'e\-pe\-z\H o 64 PC m\'ar meg\-van, \'es a
2.135 szu\-per\-sz\'a\-m\'{\i}\-t\'o\-g\'ep-te\-rem\-ben meg is
te\-kint\-he\-t\H o. A PMS(1)-re
jel\-lem\-z\H o kom\-pakt el\-ren\-de\-z\'es \'es a 64 n\'o\-dus k\"oz\-ti kom\-mu\-ni\-k\'a\-ci\-\'o
a\-zon\-ban (m\'eg) hi\-\'any\-zik, \'{\i}gy e\-zek f\"ug\-get\-len n\'o\-du\-sok\-k\'ent
m\H u\-k\"od\-nek. A k\"o\-vet\-ke\-z\H o fe\-je\-ze\-tek\-ben be\-mu\-ta\-t\'as\-ra ke\-r\"u\-l\H o
e\-red\-m\'e\-nyek egy r\'e\-sze ter\-m\'e\-szet\-sze\-r\H u\-leg e\-zek\-r\H ol a g\'e\-pek\-r\H ol
sz\'ar\-ma\-zik.)

\chapter{Az MSSM r\'acsszi\-mu\-l\'a\-ci\-\'o\-ja \label{nonper}}
\fancyhead[CE]{\hst{\thechapter{}.\ fe\-je\-zet \quad Az MSSM
r\'acsszi\-mul\'a\-ci\'o\-ja}}
\section{A Lag\-ran\-ge-f\"ugg\-v\'eny}
Eb\-ben a fe\-je\-zet\-ben az MSSM n\'egy\-di\-men\-zi\-\'os r\'acsszi\-mu\-l\'a\-ci\-\'o\-j\'a\-nak
a\-lap\-ja\-it te\-kin\-tem \'at \cite{cikk2}. A szi\-mu\-l\'a\-ci\-\'ok\-ban hasz\-n\'alt
mo\-dell\-b\H ol -- a stan\-dard mo\-dell\-ben l\'a\-tot\-tak a\-lap\-j\'an k\'e\-zen\-fek\-v\H o
m\'o\-don -- a fer\-mi\-o\-no\-kat el\-s\H o l\'e\-p\'es\-ben el\-hagy\-juk; e\-zek k\'e\-s\H obb
egy per\-tur\-ba\-t\'{\i}v l\'e\-p\'es se\-g\'{\i}t\-s\'e\-g\'e\-vel ve\-he\-t\H ok fi\-gye\-lem\-be.

Ha\-son\-l\'o\-k\'ep\-pen az U(1) szek\-tor is per\-tur\-ba\-t\'{\i}v kor\-rek\-ci\-\'o\-k\'ent van
ke\-zel\-ve \cite{kaj97}; a kis Yu\-ka\-wa-csa\-to\-l\'as\-sal ren\-del\-ke\-z\H o
ska\-l\'ar\-r\'e\-szecs\-k\'ek el\-s\H o k\"o\-ze\-l\'{\i}\-t\'es\-be\-li el\-ha\-gy\'a\-sa szin\-t\'en
lo\-gi\-kus. Et\-t\H ol el\-te\-kint\-ve az MSSM tel\-jes bo\-zo\-ni\-kus szek\-to\-r\'at, te\-h\'at
az SU(3) il\-let\-ve SU(2) sze\-rint transz\-for\-m\'a\-l\'o\-d\'o m\'er\-t\'ek\-bo\-zo\-no\-kat,
a k\'et Higgs-dub\-let\-tet, va\-la\-mint a har\-ma\-dik ge\-ne\-r\'a\-ci\-\'os kvar\-kok
szu\-per\-szim\-met\-ri\-kus p\'ar\-ja\-it (stop, sbot\-tom) r\'acs\-ra tessz\"uk.

Az \'{\i}gy v\'eg\-re\-haj\-tan\-d\'o nu\-me\-ri\-kus szi\-mu\-l\'a\-ci\-\'ok k\'et je\-len\-t\H os
e\-l\H onnyel ren\-del\-kez\-nek a di\-men\-zi\-\'os re\-duk\-ci\-\'o\-val ka\-pott 3D
szi\-mu\-l\'a\-ci\-\'ok\-hoz k\'e\-pest.
\begin{itemize}
\item Ja\-v\'{\i}\-tat\-lan ha\-t\'as al\-kal\-ma\-z\'a\-sa e\-se\-t\'en a 4D szi\-mu\-l\'a\-ci\-\'ok
v\'e\-ges r\'a\-cs\'al\-lan\-d\'o\-b\'ol fa\-ka\-d\'o hi\-b\'a\-ja $\ordo{a^2}$ nagy\-s\'ag\-ren\-d\H u;
3D szi\-mu\-l\'a\-ci\-\'ok e\-se\-t\'en ez $\ordo{a}$.
\item Az ed\-di\-gi 3D mo\-del\-lek \cite{la98} e\-gyet\-len Higgs-dub\-let\-tet
tar\-tal\-maz\-tak, szi\-mu\-l\'a\-ci\-\'onk\-ban mind\-ket\-t\H o je\-len van. En\-nek
je\-len\-t\H o\-s\'e\-ge ab\-ban \'all, hogy az e\-gyes v\'a\-ku\-um v\'ar\-ha\-t\'o \'er\-t\'e\-kek
($v_1, v_2$) h\'a\-nya\-do\-s\'at le\-\'{\i}\-r\'o $\beta$ pa\-ra\-m\'e\-ter ($\tan \beta =
v_1 / v_2$) bu\-bo\-r\'ek\-fal\-be\-li v\'al\-to\-z\'a\-sa a f\'a\-zi\-s\'at\-me\-net so\-r\'an
ter\-melt ba\-ri\-o\-na\-szim\-met\-ri\-\'a\-val e\-gye\-ne\-sen a\-r\'a\-nyos \cite{moore}.
\end{itemize}

A n\'egy\-di\-men\-zi\-\'os szi\-mu\-l\'a\-ci\-\'ok Lag\-ran\-ge-f\"ugg\-v\'e\-ny\'et a
kon\-ti\-nu\-um-el\-m\'e\-let Lag\-ran\-ge-f\"ugg\-v\'e\-ny\'e\-b\H ol a r\'acs\-t\'e\-rel\-m\'e\-let\-ben
szo\-k\'a\-sos m\'o\-don kap\-ha\-t\'o meg \cite{cikk2}. A r\'acs\-meg\-fo\-gal\-ma\-z\'as\-ban
az e\-gyes r\'acs\-pon\-tok\-ra he\-lye\-zett lo\-k\'a\-lis j\'a\-ru\-l\'e\-ko\-kon k\'{\i}\-v\"ul
pla\-kett-ta\-gok \'es hop\-ping-ta\-gok van\-nak je\-len. A
r\'acs-Lag\-ran\-ge-f\"ugg\-v\'eny a\-lap\-j\'a\-ul szol\-g\'a\-l\'o
kon\-ti\-nu\-um\-t\'e\-rel\-m\'e\-let\-be\-li ki\-fe\-je\-z\'es
\beq
{\cal L}={\cal L}_g+{\cal L}_k+{\cal L}_V+{\cal L}_{sm}+{\cal L}_Y+
{\cal L}_w+{\cal L}_s, \label{racslag1}
\enq
mely\-ben a m\'er\-t\'ek-k\"ol\-cs\"on\-ha\-t\'ast le\-\'{\i}\-r\'o
\beq
{\cal L}_g={\tst \frac14} \cdot F^{(w)}_{\mu\nu}F^{(w)\mu\nu}+
{\tst \frac14} \cdot F^{(s)}_{\mu\nu}F^{(s)\mu\nu}
\enq
tag egy e\-r\H os \'es egy gyen\-ge k\"ol\-cs\"on\-ha\-t\'a\-s\'u r\'esz \"ossze\-ge; a
ki\-ne\-ti\-kus tag a k\'et Higgs-dub\-lett ($H_1,H_2$), a bal\-ke\-zes
stop-sbot\-tom dub\-lett ($Q$) \'es a jobb\-ke\-zes stop \'es sbot\-tom szing\-lett
ko\-va\-ri\-\'ans de\-ri\-v\'alt\-j\'a\-nak \"ossze\-ge:
\beqar
{\cal L}_k &=&
(\DD^{(w)}_\mu \ha)^\dagger (\DD^{(w) \mu} \ha)+
(\DD^{(w)}_\mu \hb)^\dagger (\DD^{(w) \mu} \hb)+
(\DD^{(ws)}_\mu Q)^\dagger  (\DD^{(ws) \mu} Q)+ \nonumber \\
&& \quad (\DD^{(s)}_\mu U^*)^\dagger (\DD^{(s) \mu} U^*)+
(\DD^{(s)}_\mu D^*)^\dagger (\DD^{(s) \mu} D^*).
\enqar
A Higgs-te\-rek\-b\H ol a\-d\'o\-d\'o po\-ten\-ci\-\'al
\beqar
{\cal L}_V &=& m_{12}^2 [\alpha_1|\ha|^2+ \alpha_2|\hb|^2-
(\ha^\dagger\tilde{\hb}+ \mbox{h.c.})]+ \nonumber \\*
&& \quad {\tst \frac{g_w^2}{8}} \cdot (|\ha|^2+ |\hb|^2- 2|\ha|^2|\hb|^2+
4|\ha^\dagger\hb|^2),
\enqar
mely\-ben k\'et di\-men\-zi\-\'ot\-la\-n\'{\i}\-tott t\"o\-meg\-tag sze\-re\-pel,
$\alpha_1=m_1^2/m_{12}^2$ \'es $\alpha_2=m_2^2/m_{12}^2.$
\fancyhead[CO]{\hst{\thesection \quad A Lag\-ran\-ge-f\"uggv\'eny}}

A skvark-t\"o\-me\-get tar\-tal\-ma\-z\'o Lag\-ran\-ge-s\H u\-r\H u\-s\'eg
\beq
{\cal L}_{sm}= m_Q^2 |Q|^2+m_U^2 |U|^2+m_D^2 |D|^2,
\enq
m\'{\i}g a do\-mi\-n\'ans Yu\-ka\-wa-csa\-to\-l\'a\-s\'u r\'eszt
\beq
{\cal L}_Y= h_t^2(|QU|^2+ |\hb|^2|U|^2+ |Q^\dagger\tilde{\hb}|^2)
\enq
ad\-ja. A skvark\-te\-rek n\'e\-gyes\-csa\-to\-l\'a\-s\'at
\beq
{\cal L}_w= {\tst \frac{g_w^2}{8}} \cdot [2 \{Q \}^4- |Q|^4+
4|\ha^\dagger Q|^2+ 4|\hb^\dagger Q|^2-
2|\ha|^2 |Q|^2- 2 |\hb|^2 |Q|^2]
\enq
\'es
\beqar
{\cal L}_s &=& {\tst \frac{g_s^2}{8}} \cdot \left[3 \{Q \}^4- |Q|^4+
2|U|^4+ 2|D|^4- 6|QU|^2 \right. \nonumber \\*
&& \quad - 6|QD|^2+ 6|U^\dagger D|^2+ 2|Q|^2|U|^2
\left. + 2|Q|^2|D|^2- 2 |U|^2|D|^2 \right],
\enqar
ad\-ja meg, mely\-ben
\beq
\{Q\}^4=Q^*_{i\alpha}Q^*_{j\beta}Q_{i\beta}Q_{j\alpha}.
\label{racslag2}
\enq

A fen\-ti Lag\-ran\-ge-f\"ugg\-v\'eny pa\-ra\-m\'e\-ter\-te\-re -- a h\'a\-rom\-di\-men\-zi\-\'os
e\-set\-hez ha\-son\-l\'o\-an -- sok\-di\-men\-zi\-\'os. Tel\-jes fel\-t\'er\-k\'e\-pe\-z\'e\-se \'{\i}gy
re\-m\'eny\-te\-len, a per\-tur\-ba\-t\'{\i}v \'es a h\'a\-rom\-di\-men\-zi\-\'os e\-red\-m\'e\-nyek
bir\-to\-k\'a\-ban a\-zon\-ban k\"onnyen v\'a\-laszt\-ha\-t\'o fi\-zi\-ka\-i\-lag \'er\-de\-kes
tar\-to\-m\'any. El\-s\H od\-le\-ges c\'e\-lunk az e\-r\H o\-sen k\"ol\-cs\"on\-ha\-t\'o
Lag\-ran\-ge-f\"ugg\-v\'eny r\'esz ha\-t\'a\-s\'a\-nak vizs\-g\'a\-la\-ta volt, a vizs\-g\'a\-lat
so\-r\'an en\-nek meg\-fe\-le\-l\H o\-en v\'a\-lasz\-tot\-tuk meg k\'et
pa\-ra\-m\'e\-ter-hal\-ma\-zun\-kat.

A gyen\-ge csa\-to\-l\'a\-si \'al\-lan\-d\'ot \'es a Yu\-ka\-wa-csa\-to\-l\'a\-so\-kat fi\-zi\-ka\-i
\'er\-t\'e\-k\"uk r\"og\-z\'{\i}\-te\-ti. A szu\-per\-szim\-met\-ri\-\'at s\'er\-t\H o l\'agy
t\"o\-meg\-ta\-gok\-ban a csu\-pasz pa\-ra\-m\'e\-te\-re\-ket $m_Q\!=\!m_D\!=\!250$ GeV,
sze\-rint v\'a\-lasz\-tot\-tuk meg. A jobb\-ke\-zes stop\-t\"o\-meg\-gel k\"oz\-vet\-len
kap\-cso\-lat\-ban \'al\-l\'o $m_U$ pa\-ra\-m\'e\-ter a per\-tur\-ba\-t\'{\i}v sz\'a\-mo\-l\'a\-sok
e\-red\-m\'e\-ny\'e\-nek t\"uk\-r\'e\-ben v\'al\-toz\-tat\-ha\-t\'o.

Az \'uj kon\-fi\-gu\-r\'a\-ci\-\'ok e\-l\H o\-\'al\-l\'{\i}\-t\'a\-sa o\-ver\-re\-la\-x\'a\-ci\-\'os \'es h\H o\-f\"ur\-d\H o
al\-go\-rit\-mus r\'e\-v\'en t\"or\-t\'e\-nik (k\"u\-l\"on--k\"u\-l\"on az e\-gyes me\-z\H ok\-re),
mely\-nek a\-lap\-ja\-i nagy\-r\'eszt me\-ge\-gyez\-nek az SU(2)--Higgs-mo\-dell
szi\-mu\-l\'a\-ci\-\'o\-j\'a\-n\'al hasz\-n\'al\-tak\-kal \cite{fod94}.

\section{A szi\-mu\-l\'a\-ci\-\'ok\-ban m\'ert mennyi\-s\'e\-gek \label{mssm_wilson}}
\fancyhead[CO]{\hst{\thesection \quad A szi\-mul\'a\-ci\'o\-ban m\'ert
mennyis\'e\-gek}}
A r\'acsszi\-mu\-l\'a\-ci\-\'ok so\-r\'an m\'er\-t\'e\-kin\-va\-ri\-\'ans mennyi\-s\'e\-gek m\'e\-r\'e\-se
b\'{\i}r fi\-zi\-ka\-i je\-len\-t\H o\-s\'eg\-gel. I\-lye\-nek
\begin{itemize}
\item
Az e\-gyes r\'acs\-pon\-to\-kon \"u\-l\"o, egy\-m\'as\-t\'ol f\"ug\-get\-len $\Phi^\dag(x)
\Phi(x)$ t\'{\i}\-pu\-s\'u ta\-gok \"ossze\-ge\-i,
\item
A k\'et szom\-sz\'e\-dos r\'acs\-pon\-ton \"u\-l\H o mennyi\-s\'e\-gek \'es a k\'et r\'acs\-pont
k\"o\-z\"ot\-ti \'el\-v\'al\-to\-z\'o meg\-fe\-le\-l\H o szor\-za\-ta, $\Phi^\dag(x+\hat \mu)
U(x,\mu) \Phi(x)$, \'es en\-nek \'al\-ta\-l\'a\-no\-s\'{\i}\-t\'a\-sa\-i,
\item
A Wil\-son-hur\-kok, te\-h\'at a r\'acs ten\-ge\-lye\-i\-vel p\'ar\-hu\-za\-mos \'e\-l\H u, $n
\times m$ ki\-ter\-je\-d\'e\-s\H u t\'eg\-la\-la\-pok, mely men\-t\'en az \'el\-v\'al\-to\-z\'o\-kat
szo\-roz\-zuk \"ossze,
\item
Pol\-ja\-kov-hur\-kok, te\-h\'at a pe\-ri\-o\-di\-kus ha\-t\'ar\-fel\-t\'e\-te\-lek\-kel el\-l\'a\-tott
r\'acs\-ban egy a\-dott ten\-gellyel mind\-v\'e\-gig p\'ar\-hu\-za\-mo\-san ha\-la\-d\'o,
\"on\-ma\-g\'a\-ba z\'a\-r\'o\-d\'o vo\-nal, stb.
\end{itemize}
E\-zen fe\-l\"ul m\'eg sz\'a\-mos m\'as mennyi\-s\'eg is de\-fi\-ni\-\'al\-ha\-t\'o, pl.\
tet\-sz\H o\-le\-ges z\'art hu\-rok men\-t\'en \"ossze\-szo\-roz\-hat\-juk az
\'el\-v\'al\-to\-z\'o\-kat. B\'ar bi\-zo\-nyos r\'acs\-t\'e\-rel\-m\'e\-le\-ti mun\-k\'ak\-ban fon\-tos
sze\-rep\-hez jut\-nak az i\-lyen ob\-jek\-tu\-mok, pl.\ a ja\-v\'{\i}\-tott ha\-t\'a\-sok
e\-se\-t\'e\-ben a pla\-kett\-v\'al\-to\-z\'on a\-la\-pu\-l\'o ha\-t\'as fi\-no\-m\'{\i}\-t\'a\-s\'a\-hoz 6
\'e\-l\H u hur\-ko\-kat, \'{\i}gy a csa\-vart hat\-sz\"og a\-la\-k\'u hur\-kot is fi\-gye\-lem\-be
kell ven\-n\"unk \cite{lus}, a to\-v\'ab\-bi\-ak\-ban is\-mer\-te\-t\'es\-re ke\-r\"u\-l\H o
szi\-mu\-l\'a\-ci\-\'ok\-ban e\-zek nem ju\-tot\-tak sze\-rep\-hez.

K\"u\-l\"on ki\-e\-me\-len\-d\H o a fen\-ti\-ek\-ben de\-fi\-ni\-\'alt mennyi\-s\'e\-gek a\-lap\-j\'an
\'er\-tel\-me\-zett kor\-re\-l\'a\-ci\-\'os f\"ugg\-v\'e\-nyek vizs\-g\'a\-la\-ta; a k\"oz\-pon\-ti
je\-len\-t\H o\-s\'e\-g\H u t\"o\-me\-ge\-ket (W-t\"o\-meg, Higgs-t\"o\-meg) e\-zen kor\-re\-l\'a\-ci\-\'os
f\"ugg\-v\'e\-nyek -- $\langle \Phi^\dag \Phi \rangle$, $\langle \Phi^\dag U
\Phi \rangle$ -- le\-csen\-g\'e\-s\'e\-b\H ol ha\-t\'a\-roz\-hat\-juk meg. (A le\-csen\-g\'es
majd\-nem ex\-po\-nen\-ci\-\'a\-lis; a pe\-ri\-o\-di\-kus ha\-t\'ar\-fel\-t\'e\-te\-lek e\-he\-lyett
gyak\-ran ch f\"ugg\-v\'e\-nyek il\-lesz\-t\'e\-s\'et in\-do\-kol\-j\'ak.)

A Wil\-son-hur\-kok, me\-lyek a \ref{p_bev} sza\-kasz\-ban be\-ve\-ze\-tett sta\-ti\-kus
kvark po\-ten\-ci\-\'al\-lal szo\-ros kap\-cso\-lat\-ban \'all\-nak, gya\-kor\-la\-ti\-lag
sz\"uk\-s\'e\-ges\-s\'e te\-szik a r\'a\-cson t\"or\-t\'e\-n\H o m\'er\-t\'ek\-r\"og\-z\'{\i}\-t\'est.
A\-mennyi\-ben az \"osszes a\-dott s\'{\i}\-k\'u Wil\-son-hur\-kot meg k\'{\i}\-v\'a\-nunk
m\'er\-ni, \'ugy egy $V = L_1 \times L_2 \times L_3 \times L_4$ m\'e\-re\-t\H u
r\'a\-cson mind a $V$ r\'acs\-pont le\-het a vizs\-g\'alt Wis\-lon-hu\-rok
kez\-d\H o\-pont\-ja; $l_1 \times l_2$ m\'e\-re\-t\H u hu\-rok e\-se\-t\'e\-ben $2 \times
(l_1 + l_2)$ da\-rab \'e\-lek k\"oz\-ti (SU(2), vagy SU(3)) szor\-z\'ast kell
v\'eg\-re\-haj\-ta\-ni; az \"osszes i\-lyen Wil\-son-hu\-rok\-n\'al v\'eg\-re\-haj\-tan\-d\'o
m\H u\-ve\-le\-tek sz\'a\-ma te\-h\'at
\beq
2 V \times \sum_1^{L_1} \sum_1^{L_2} (l_1 + l_2) \approx V (L_1^2 +
L_2^2) \label{muvszam}
\enq
A\-mennyi\-ben m\'er\-t\'ek\-r\"og\-z\'{\i}\-t\'est haj\-tunk v\'eg\-re, c\'el\-sze\-r\H u a ma\-xi\-m\'a\-lis
a\-xi\-\'al\-m\'er\-t\'e\-ket v\'a\-lasz\-ta\-ni. En\-nek so\-r\'an e\-l\H o\-sz\"or az e\-gyik i\-r\'any
men\-t\'en e\-gyen\-l\H o\-v\'e tessz\"uk az \"osszes le\-het\-s\'e\-ges \'el\-v\'al\-to\-z\'ot
1-gyel, a\-mi a szom\-sz\'e\-dos \'el\-v\'al\-to\-z\'ok \'es a r\'acs\-pon\-tok\-ban \"u\-l\H o
mennyi\-s\'e\-gek al\-kal\-mas meg\-v\'al\-toz\-ta\-t\'a\-s\'a\-val \'er\-he\-t\H o el. (Mi\-vel a
Pol\-ja\-kov-hu\-rok \'er\-t\'e\-ke m\'er\-t\'e\-kin\-va\-ri\-\'ans, e\-z\'ert az a\-dott i\-r\'any\-ban
(hur\-kon\-k\'ent) e\-gyet\-len ki\-v\'e\-tel\-lel az \"osszes \'el\-v\'al\-to\-z\'o \'er\-t\'e\-ke
1-gy\'e te\-he\-t\H o.) Ezt k\"o\-ve\-t\H o\-en a n\'egy\-di\-men\-zi\-\'os r\'acs
h\'a\-rom\-di\-men\-zi\-\'os fe\-l\"u\-le\-t\'en, me\-lyen az \'el\-v\'al\-to\-z\'o\-kat nem tud\-tuk
1-gyel e\-gyen\-l\H o\-v\'e ten\-ni, ha\-son\-l\'o m\'er\-t\'ek\-r\"og\-z\'{\i}\-t\'est hajt\-ha\-tunk
v\'eg\-re egy m\'a\-sik i\-r\'any\-ban.

J\'ol l\'at\-ha\-t\'o\-an a m\'er\-t\'ek\-r\"og\-z\'{\i}\-t\'es\-hez sz\"uk\-s\'e\-ges m\H u\-ve\-le\-tek
sz\'a\-ma a r\'acs\-pon\-tok sz\'a\-m\'a\-val a\-r\'a\-nyos; az e\-gyes r\'acs\-pon\-tok\-ban
n\'e\-h\'any (SU(2) vagy SU(3)) szor\-z\'ast kell v\'eg\-re\-haj\-ta\-ni. A
m\'er\-t\'ek\-r\"og\-z\'{\i}\-t\'es u\-t\'an a\-zon\-ban a Wil\-son-hur\-kok sz\'a\-m\'{\i}\-t\'a\-sa sok\-kal
egy\-sze\-r\H ubb lesz: a ki\-v\'a\-lasz\-tott i\-r\'any men\-ti \'e\-lek\-kel va\-l\'o
szor\-z\'ast v\'eg\-re sem kell haj\-ta\-ni; a\-dott kez\-d\H o\-pont\-b\'ol a
ki\-v\'a\-lasz\-tott i\-r\'any\-ra me\-r\H o\-le\-ges \"osszes le\-het\-s\'e\-ges $l_2$
ki\-ter\-je\-d\'es $L_2$ l\'e\-p\'es\-sel ki\-sz\'a\-m\'{\i}t\-ha\-t\'o, a\-mi az e\-g\'esz r\'acs\-ra $V
\times L_2$ da\-rab m\H u\-ve\-le\-tet je\-lent. Az \'{\i}gy ka\-pott sz\'a\-mok\-b\'ol
to\-v\'ab\-bi $V \times L_1$ m\H u\-ve\-let\-tel az \"osszes le\-het\-s\'e\-ges
Wil\-son-hu\-rok \'er\-t\'e\-ke me\-gad\-ha\-t\'o. Az e\-g\'esz el\-j\'a\-r\'as so\-r\'an $V \times
(\mathrm{const.} + L_1 + L_2)$ m\H u\-ve\-le\-tet haj\-tot\-tunk v\'eg\-re, mely $L$
egy hat\-v\'a\-ny\'a\-val ki\-sebb, mint (\ref{muvszam}). J\'ol l\'at\-szik az
is, hogy mi\-n\'el na\-gyobb r\'acs\-r\'ol van sz\'o, an\-n\'al je\-len\-t\H o\-sebb a
k\"u\-l\"onb\-s\'eg.

A szi\-mu\-l\'a\-ci\-\'ok\-ban m\'ert mennyi\-s\'e\-gek k\"o\-z\"ul az a\-l\'ab\-bi\-ak\-ban a Higgs-
\'es a stop\-t\'er n\'egy\-ze\-t\'e\-nek vizs\-g\'a\-la\-t\'a\-val fog\-lal\-ko\-zom. B\'ar a
szim\-met\-ri\-a\-s\'er\-t\H o f\'a\-zist a szim\-met\-ri\-kus\-t\'ol a\-zon tu\-laj\-don\-s\'a\-ga
a\-lap\-j\'an le\-gegy\-sze\-r\H ubb el\-k\"u\-l\"o\-n\'{\i}\-te\-ni, hogy ab\-ban az a\-dott t\'er
v\'a\-ku\-um v\'ar\-ha\-t\'o \'er\-t\'e\-ke nem z\'e\-rus, en\-nek Mon\-te Car\-lo
szi\-mu\-l\'a\-ci\-\'ok\-ban va\-l\'o m\'e\-r\'e\-se nem ma\-g\'a\-t\'ol \'er\-te\-t\H o\-d\H o. A
nul\-la--nem-nul\-la el\-k\"u\-l\"o\-n\'{\i}\-t\'es\-n\'el ke\-v\'es\-b\'e \'e\-les az el\-len\-t\'et az
a\-dott t\'er n\'egy\-ze\-t\'e\-nek v\'ar\-ha\-t\'o \'er\-t\'e\-k\'e\-nek vizs\-g\'a\-la\-ta\-kor
\cite{far}, \'am ez is b\H o\-s\'e\-ge\-sen e\-le\-gen\-d\H o ar\-ra, hogy a k\'et f\'a\-zist
el\-k\"u\-l\"o\-n\'{\i}t\-s\"uk, a\-mint az a \ref{simoutput} \'ab\-r\'an j\'ol l\'at\-ha\-t\'o.

\bef[ht]
\bc
\epsfig{file=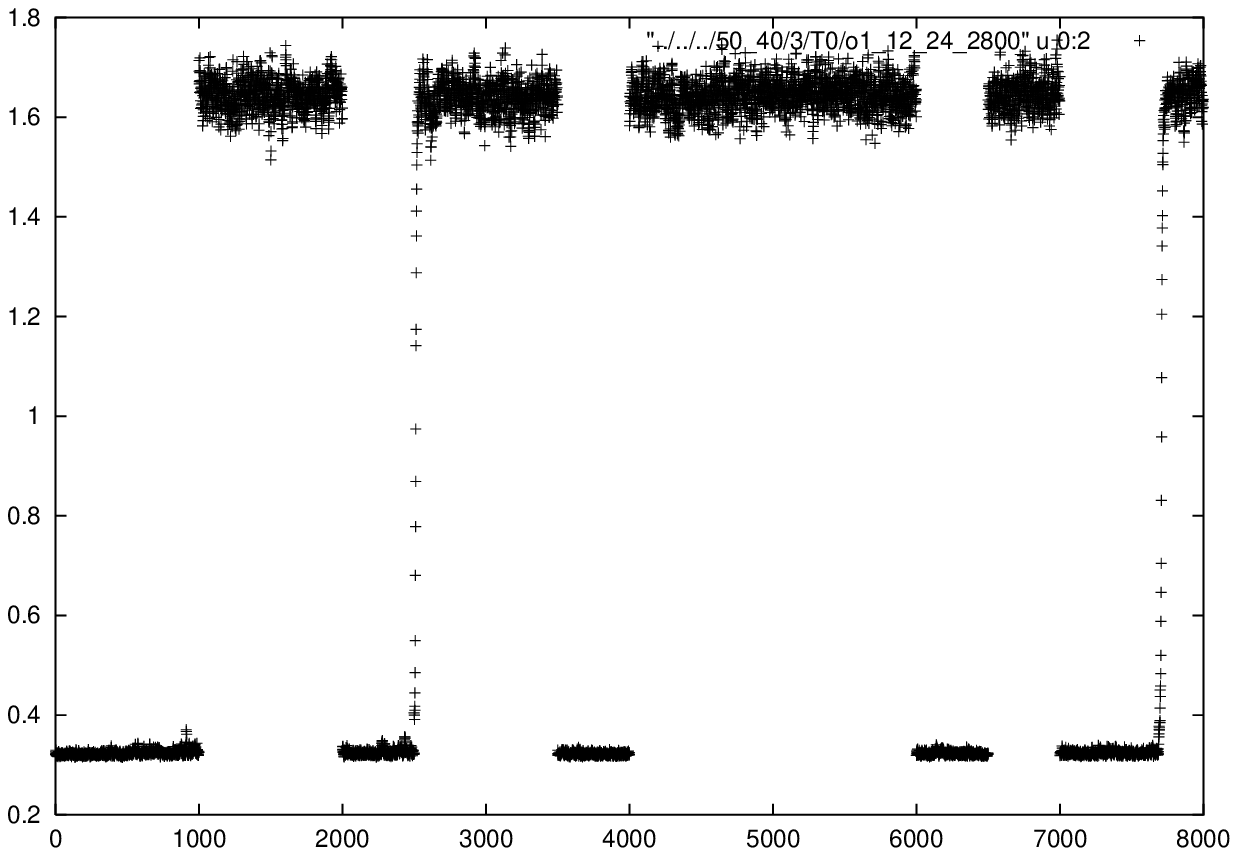,width=8cm}
\caption{A szim\-met\-ri\-kus \'es a szim\-met\-ri\-a\-s\'er\-t\H o f\'a\-zis
el\-k\"u\-l\"o\-n\'{\i}\-t\'e\-se a sz\'o\-ban\-for\-g\'o t\'er n\'egy\-ze\-te a\-lap\-j\'an is
le\-het\-s\'e\-ges \label{simoutput}}
\ec
\enf

Ha\-son\-l\'o\-k\'epp f\'a\-zi\-s\'at\-me\-net\-re u\-tal egy hisz\-te\-r\'e\-zis\-hu\-rok je\-len\-l\'e\-te,
\'es ez a f\'a\-zi\-s\'at\-me\-ne\-ti pont dur\-va meg\-ha\-t\'a\-ro\-z\'a\-s\'at is le\-he\-t\H o\-v\'e
te\-szi.

\bef[ht]
\bc
\epsfig{file=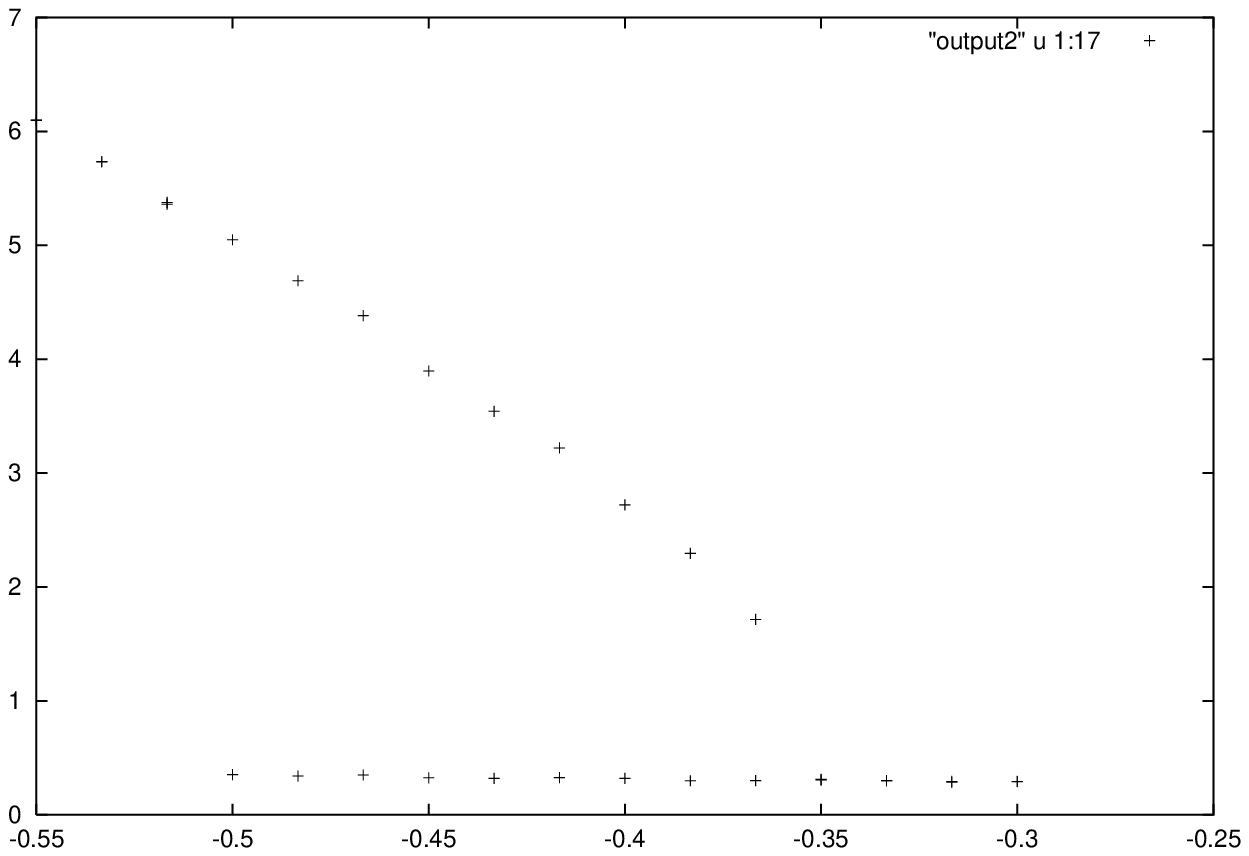,width=8cm}
\caption{A hisz\-te\-r\'e\-zis\-hu\-rok je\-len\-l\'e\-te a f\'a\-zi\-s\'at\-me\-ne\-ti pont
k\"o\-zel\-s\'e\-g\'e\-re u\-tal \label{hister}}
\ec
\enf

Az a\-l\'ab\-bi\-ak\-ban a\-zon\-ban nem en\-nek, ha\-nem a pon\-to\-sabb e\-red\-m\'enyt
a\-d\'o Le\-e--Yang-f\'e\-le m\'od\-szer\-nek a se\-g\'{\i}t\-s\'e\-g\'e\-vel ha\-t\'a\-roz\-zuk meg a
kri\-ti\-kus pon\-to\-kat.
Eh\-hez e\-l\H o\-sz\"or v\'e\-ges h\H o\-m\'er\-s\'ek\-le\-t\H u $(T \neq 0)$ szi\-mu\-l\'a\-ci\-\'o\-kat
kell v\'eg\-re\-haj\-ta\-ni, jel\-lem\-z\H o\-en $L_t = 2,3,4,5$ i\-d\H o\-i\-r\'a\-ny\'u
ki\-ter\-je\-d\'e\-s\H u r\'a\-cso\-kon. A r\'acs\-t\'e\-rel\-m\'e\-let\-ben szo\-k\'a\-sos m\'o\-don az
\'{\i}gy ka\-pott e\-red\-m\'e\-nyek sta\-tisz\-ti\-kus hi\-b\'a\-j\'an k\'{\i}\-v\"ul k\'et
szisz\-te\-ma\-ti\-kus hi\-ba\-for\-r\'as is van:
\begin{itemize}
\item
a v\'e\-ges r\'acs\-t\'er\-fo\-gat\-b\'ol a\-d\'o\-d\'o hi\-ba, me\-lyet v\'eg\-te\-len
r\'acs\-t\'er\-fo\-gat\-ra va\-l\'o ext\-ra\-po\-l\'a\-l\'as\-sal kor\-ri\-g\'al\-ha\-tunk, \'es
\item
a v\'e\-ges r\'a\-cs\'al\-lan\-d\'o\-b\'ol fa\-ka\-d\'o hi\-ba -- bo\-zo\-ni\-kus el\-m\'e\-let
e\-se\-t\'e\-ben ez $\ordo{a^2}$ ren\-d\H u -- mely kon\-ti\-nu\-um li\-mesz\-re va\-l\'o
ext\-ra\-po\-l\'a\-l\'as\-sal kor\-ri\-g\'al\-ha\-t\'o.
\end{itemize}
A v\'eg\-te\-len r\'acs\-t\'er\-fo\-gat\-ra va\-l\'o ext\-ra\-po\-l\'a\-l\'ast \'ugy hajt\-juk
v\'eg\-re, hogy a\-dott $L_t$ \'er\-t\'ek e\-se\-t\'en egy\-re na\-gyobb \'es na\-gyobb
t\'er\-be\-li ki\-ter\-je\-d\'e\-s\H u r\'a\-cso\-kon hajt\-juk v\'eg\-re a szi\-mu\-l\'a\-ci\-\'ot. A
kon\-ti\-nu\-um\-li\-meszt egy\-re fi\-no\-mabb r\'a\-csok hasz\-n\'a\-la\-t\'a\-val, te\-h\'at
u\-gya\-nak\-ko\-ra r\'acs\-t\'er\-fo\-gat mel\-lett egy\-re s\H u\-r\H ubb fel\-bon\-t\'as\-sal --
$L_t a = \mrm{const.}$ mel\-lett $L_t$ n\"o\-ve\-l\'e\-s\'e\-vel \'er\-het\-j\"uk el.
A k\"o\-vet\-ke\-z\H o t\'ab\-l\'a\-zat\-ban te\-h\'at a so\-rok ext\-ra\-po\-l\'a\-l\'a\-s\'a\-val a
v\'eg\-te\-len t\'er\-fo\-ga\-t\'u li\-mesz kap\-ha\-t\'o meg, az osz\-lo\-po\-k\'e\-val a
kon\-ti\-nu\-um-li\-mesz.

\begin{center}
\begin{tabular}{ccccc}
\hline
$2 * 4^3$ & $2 * 6^3$ & $2 * 8^3$ & $2 * 10^3$ & $2 * 12^3$ \\
$3 * 6^3$ & $3 * 9^3$ & $3 * 12^3$& $3 * 15^3$ & $3 * 18^3$ \\
$4 * 8^3$ & $4 * 12^3$& $4 * 16^3$& $4 * 20^3$ & $4 * 24^3$ \\
$5 * 10^3$& $5 * 15^3$& $5 * 20^3$& $5 * 25^3$ & $5 * 30^3$ \\
\hline
\end{tabular}
\end{center}

(A gya\-kor\-lat\-ban nem min\-dig a fen\-ti t\'ab\-l\'a\-zat sze\-rint szok\-tuk
meg\-v\'a\-lasz\-ta\-ni a szi\-mu\-l\'a\-ci\-\'o\-ban sze\-rep\-l\H o r\'a\-csok m\'e\-re\-te\-it; a
pe\-ri\-o\-di\-kus ha\-t\'ar\-fel\-t\'e\-te\-lek mi\-att rit\-k\'an hasz\-n\'a\-lunk o\-lyan r\'a\-csot,
mely\-nek t\'er\-be\-li ki\-ter\-je\-d\'e\-se p\'a\-rat\-lan.)

A fen\-ti\-ek gya\-kor\-la\-ti al\-kal\-ma\-z\'a\-sa a\-zon\-ban ko\-moly ne\-h\'e\-zs\'e\-ge\-ket is
fel\-vet. Ah\-hoz, hogy a kon\-ti\-nu\-um\-li\-mesz\-nek \'er\-tel\-me le\-gyen, u\-gya\-na\-zon
fi\-zi\-ka\-i pont\-ban kell v\'eg\-re\-haj\-ta\-ni a szi\-mu\-l\'a\-ci\-\'ot, te\-h\'at a
rend\-szer\-re jel\-lem\-z\H o fi\-zi\-ka\-i mennyi\-s\'e\-gek \'er\-t\'e\-k\'et -- $R_{HW} = M_H
/ M_W$ t\"o\-me\-ga\-r\'any, stop\-t\"o\-me\-gek, top\-t\"o\-meg stb. -- a k\"u\-l\"on\-b\"o\-z\H o
m\'e\-re\-t\H u r\'a\-csok\-n\'al u\-gya\-no\-lyan \'er\-t\'ek\-re kell be\-\'al\-l\'{\i}\-ta\-ni. \'Igy
p\'el\-d\'a\-ul az $2*4^3$ r\'acs, va\-la\-mint az u\-gya\-nak\-ko\-ra t\'er\-fo\-ga\-tot
fi\-no\-mabb fel\-bon\-t\'as\-sal le\-\'{\i}\-r\'o $3*6^3$ r\'acs e\-se\-t\'e\-ben e\-zen
pa\-ra\-m\'e\-te\-rek \'er\-t\'e\-ke a\-zo\-nos kell, hogy le\-gyen -- ez\-zel
biz\-to\-s\'{\i}t\-hat\-juk azt, hogy a re\-nor\-m\'a\-l\'a\-si cso\-port m\'od\-szer\-n\'el
meg\-szo\-kott m\'o\-don az \'al\-lan\-d\'o fi\-zi\-ka vo\-na\-l\'an (\emph{li\-ne of
cons\-tant physics}, LCP) mo\-zog\-has\-sunk. A fen\-ti mennyi\-s\'e\-gek a\-zon\-ban
a kor\-re\-l\'a\-ci\-\'os f\"ugg\-v\'e\-nyek le\-csen\-g\'e\-s\'e\-b\H ol, a szi\-mu\-l\'a\-ci\-\'ok u\-t\'an
ha\-t\'a\-roz\-ha\-t\'ok meg. Ho\-gyan le\-het a szi\-mu\-l\'a\-ci\-\'os pa\-ra\-m\'e\-te\-re\-ket \'ugy
han\-gol\-ni, hogy a t\"o\-meg\-re\-nor\-m\'a\-l\'o\-d\'ast meg\-fe\-le\-l\H o\-k\'epp ke\-zel\-ni
tud\-juk? Ha csak e\-gyet\-len t\"o\-meg\-pa\-ra\-m\'e\-ter\-r\H ol len\-ne sz\'o,
pr\'o\-b\'al\-ga\-t\'as\-sal is e\-l\'eg gyor\-san c\'elt \'er\-het\-n\'enk, eb\-ben a
bo\-nyo\-lul\-tabb e\-set\-ben a\-zon\-ban ez a me\-gol\-d\'as nem j\"on sz\'o\-ba.

E\-l\H o\-sz\"or te\-kint\-s\"uk az $m_U^2$ pa\-ra\-m\'e\-tert! A re\-nor\-m\'alt t\"o\-meg \'es a
csu\-pasz t\"o\-meg kap\-cso\-la\-ta
\beq
m_R^2 = m_0^2 + d a^{-2},
\enq
a\-hol $da^{-2}$ a le\-v\'a\-g\'a\-si j\'a\-ru\-l\'ek. A\-mennyi\-ben $m_R$ \'er\-t\'e\-k\'et
r\"og\-z\'{\i}t\-j\"uk, $m_0$ han\-go\-l\'a\-sa r\'e\-v\'en a $d$ kons\-tans \'er\-t\'e\-ke
meg\-ha\-t\'a\-roz\-ha\-t\'o. $m_R$ r\"og\-z\'{\i}\-t\'e\-s\'e\-re a z\'e\-rus h\H o\-m\'er\-s\'ek\-le\-t\H u
szi\-mu\-l\'a\-ci\-\'o f\'a\-zi\-s\'at\-me\-ne\-ti pont\-ja ny\'ujt le\-he\-t\H o\-s\'e\-get, itt $m_R =
0$. Eb\-b\H ol $d$ meg\-ha\-t\'a\-ro\-z\'a\-sa u\-t\'an az u\-gya\-na\-zon fi\-zi\-ka\-i pont\-hoz
tar\-to\-z\'o k\"u\-l\"on\-b\"o\-z\H o fi\-nom\-s\'a\-g\'u r\'a\-csok meg\-va\-l\'o\-s\'{\i}\-t\'a\-s\'a\-hoz
al\-kal\-mas $m_0$ pa\-ra\-m\'e\-tert min\-den e\-set\-ben k\"onnyen meg tud\-juk
ha\-t\'a\-roz\-ni -- a f\'a\-zi\-s\'at\-me\-ne\-ti pont\-t\'ol t\'a\-vol is.

A Higgs-szek\-tor ta\-nul\-m\'a\-nyo\-z\'a\-s\'at meg\-k\"onny\'{\i}\-ti a r\'acs\-ha\-t\'as egy
szim\-met\-ri\-\'a\-j\'a\-nak fe\-lis\-me\-r\'e\-se: a\-mennyi\-ben $m_{12}$-t meg\-szo\-roz\-zuk
$-1$-gyel, \'es ez\-zel e\-gy\"utt az e\-gyik Higgs-v\'al\-to\-z\'o he\-ly\'e\-be an\-nak
el\-len\-tett\-j\'et \'{\i}r\-juk (pl.\ $H_2 \rightarrow -H_2$), \'ugy a r\'acs\-ha\-t\'as
nem v\'al\-to\-zik. E\-zen szim\-met\-ri\-a a\-lap\-j\'an azt v\'ar\-juk, hogy $m_{12}$ nem
kap $a$-t\'ol f\"ug\-g\H o kor\-rek\-ci\-\'ot, hi\-szen b\'ar\-mi\-lyen e\-l\H o\-je\-let is
v\'a\-lasz\-ta\-n\'ank en\-nek, az a fen\-ti szim\-met\-ri\-\'at s\'er\-te\-n\'e. \\
Kap-e $a^2$-s kor\-rek\-ci\-\'ot $m_1^2$ \'es $m_2^2$? Mi\-vel $\alpha_1 = m_1^2
/ m_{12}^2$ szi\-mu\-l\'a\-ci\-\'ok\-ban hasz\-n\'a\-la\-tos \'er\-t\'e\-ke i\-gen nagy -- 40
k\"o\-r\"u\-li -- e\-z\'ert itt nem sz\'a\-m\'{\i}\-tunk je\-len\-t\H os re\-nor\-m\'a\-l\'a\-si
ef\-fek\-tu\-sok\-ra. Az $m_2^2$-tel kap\-cso\-la\-tos $\alpha_2$ pa\-ra\-m\'e\-ter pe\-dig
\'ep\-pen a han\-go\-lan\-d\'o mennyi\-s\'eg, mely\-nek se\-g\'{\i}t\-s\'e\-g\'e\-vel a
f\'a\-zi\-s\'at\-me\-ne\-ti pon\-tot r\"og\-z\'{\i}t\-j\"uk.

A fen\-ti gon\-do\-lat\-me\-net a\-lap\-j\'an v\'eg\-re\-haj\-tott v\'al\-toz\-ta\-t\'a\-sok\-kal
va\-la\-mi\-vel k\"o\-ze\-lebb ju\-tunk a kons\-tans fi\-zi\-ka vo\-na\-l\'a\-hoz. A\-zon\-ban hogy
a kons\-tans fi\-zi\-ka meg\-va\-l\'o\-sul\-has\-son, m\'as pa\-ra\-m\'e\-te\-rek \'a\-t\'al\-l\'{\i}\-t\'a\-sa
is sz\"uk\-s\'e\-ges, min\-de\-nek e\-l\H ott az $R_{HW}$ t\"o\-me\-ga\-r\'any han\-go\-l\'a\-sa.

A v\'e\-ges h\H o\-m\'er\-s\'ek\-le\-t\H u szi\-mu\-l\'a\-ci\-\'ok c\'el\-ja a f\'a\-zi\-s\'at\-me\-ne\-ti pont
meg\-ha\-t\'a\-ro\-z\'a\-sa. Ez a t\"ob\-bi pa\-ra\-m\'e\-ter \'al\-lan\-d\'o \'er\-t\'e\-ken tar\-t\'a\-sa
mel\-lett az $\alpha_2$ csa\-to\-l\'a\-si \'al\-lan\-d\'o han\-go\-l\'a\-s\'a\-val t\"or\-t\'e\-nik. A
Le\-e--Yang-z\'e\-rus\-he\-lyek meg\-ha\-t\'a\-ro\-z\'a\-s\'a\-val me\-g\'al\-la\-p\'{\i}t\-ha\-t\'o az
$\alpha_{2,c}$ kri\-ti\-kus pont. Az \'{\i}gy ka\-pott m\'e\-r\'e\-si pon\-tok\-b\'ol
v\'eg\-re\-hajt\-ha\-t\'o a v\'eg\-te\-len t\'er\-fo\-ga\-ti li\-mesz\-re t\"or\-t\'e\-n\H o
ext\-ra\-po\-l\'a\-l\'as; en\-nek so\-r\'an fel\-hasz\-n\'al\-juk a csa\-to\-l\'a\-si \'al\-lan\-d\'o
in\-verz t\'er\-fo\-gat ($1/V$) sze\-rin\-ti ha\-la\-d\'a\-s\'a\-nak sk\'a\-la\-t\"or\-v\'e\-ny\'et.

Az \'{\i}gy meg\-ka\-pott v\'eg\-te\-len t\'er\-fo\-ga\-t\'u $\alpha_2$ kri\-ti\-kus \'er\-t\'ek
mel\-lett z\'e\-rus h\H o\-m\'er\-s\'ek\-le\-t\H u szi\-mu\-l\'a\-ci\-\'o\-kat kell v\'eg\-re\-haj\-ta\-ni,
konk\-r\'e\-tan az $L_t =2$ e\-set\-ben ka\-pott v\'e\-ges h\H o\-m\'er\-s\'ek\-le\-t\H u
szi\-mu\-l\'a\-ci\-\'os e\-red\-m\'e\-nyek v\'eg\-te\-len t\'er\-fo\-ga\-t\'u li\-me\-sze\-k\'ent
ki\-sz\'a\-m\'{\i}\-tott $\alpha_{2,c}^\infty$ kri\-ti\-kus pa\-ra\-m\'e\-ter mel\-lett ez $8^3
* 16$ m\'e\-re\-t\H u r\'a\-cson t\"or\-t\'e\-nik.\footnote{Az e\-gyes
n\'o\-du\-sok\-ra te\-he\-t\H o leg\-na\-gyobb r\'acs m\'e\-re\-te $12^3 * 24$ k\"o\-r\"ul van;
az e\-g\'esz PMS1-re egy r\'a\-csot he\-lyez\-ve $24^3*48$ m\'eg
meg\-va\-l\'o\-s\'{\i}t\-ha\-t\'o.} Mi\-vel a ``h\H o\-m\'er\-s\'ek\-le\-tet'' a r\'acs leg\-ki\-sebb
m\'e\-re\-te szok\-ta jel\-le\-mez\-ni, itt a h\H o\-m\'er\-s\'ek\-let leg\-fel\-jebb ne\-gye\-de az
$L_t=2$ v\'e\-ges h\H o\-m\'er\-s\'ek\-le\-t\H u szi\-mu\-l\'a\-ci\-\'o\-ban meg\-va\-l\'o\-su\-l\'o\-nak.
Mi\-vel a h\H o\-m\'er\-s\'ek\-le\-ti in\-teg\-r\'a\-lok\-ban $T^2$-s szor\-z\'o sze\-re\-pel, az
e\-l\H o\-z\H o\-ek\-hez k\'e\-pest a h\H o\-m\'er\-s\'ek\-le\-ti tag egy $1/16$-os fak\-tor\-ral el
van nyom\-va, e\-z\'ert a ``z\'e\-rus h\H o\-m\'er\-s\'ek\-le\-t\H u szi\-mu\-l\'a\-ci\-\'o'' n\'ev
b\'ar nem pon\-tos, jo\-gos\-nak mond\-ha\-t\'o.

Mint\-hogy az $\alpha_2$ pa\-ra\-m\'e\-ter \'al\-lan\-d\'o \'er\-t\'e\-ken tar\-t\'a\-sa
mel\-lett cs\"ok\-kent a h\H o\-m\'er\-s\'ek\-le\-tet, a rend\-szer s\'er\-tett
f\'a\-zis\-ba ke\-r\"ult. A $W$ \'es Higgs-r\'e\-szecs\-ke t\"o\-me\-g\'et me\-ga\-d\'o
kor\-re\-l\'a\-ci\-\'os f\"ugg\-v\'eny (k\"o\-ze\-l\'{\i}\-t\H o\-leg) ex\-po\-nen\-ci\-\'a\-lis
le\-csen\-g\'e\-s\'e\-nek ka\-rak\-te\-risz\-ti\-kus hossza 2--4, \'{\i}gy $L_t = 16$ mel\-lett
e\-zen t\"o\-meg\-pa\-ra\-m\'e\-te\-rek meg\-ha\-t\'a\-ro\-z\'a\-sa kel\-l\H o\-en pon\-tos.
Meg\-ha\-t\'a\-roz\-ha\-t\'o to\-v\'ab\-b\'a a Higgs-t\'er ug\-r\'a\-sa, a k\"u\-l\"on\-b\"o\-z\H o
f\'a\-zi\-so\-kat el\-v\'a\-lasz\-t\'o bu\-bo\-r\'ek fa\-l\'a\-nak a\-lak\-ja, a $\beta$ pa\-ra\-m\'e\-ter
f\'a\-zis\-ha\-t\'a\-ron t\"or\-t\'e\-n\H o meg\-v\'al\-to\-z\'a\-sa stb.

Mi\-e\-l\H ott me\-gad\-n\'am az e\-r\H os csa\-to\-l\'a\-si \'al\-lan\-d\'o szi\-mu\-l\'a\-ci\-\'o\-ban
hasz\-n\'alt k\'et\-faj\-ta \'er\-t\'e\-ke mel\-lett \'{\i}gy ka\-pott \'er\-t\'e\-ke\-ket,
r\'esz\-le\-te\-seb\-ben ki\-fej\-tem a Le\-e--Yang-z\'e\-rus\-he\-lyek m\'od\-sze\-r\'et,
\'es me\-ga\-dom a v\'e\-ges h\H o\-m\'er\-s\'ek\-le\-t\H u szi\-mu\-l\'a\-ci\-\'o\-b\'ol e\-zen az \'u\-ton
nyer\-he\-t\H o $\alpha_2$ \'er\-t\'e\-ke\-ket.
\section{A Le\-e--Yang-z\'e\-rus\-he\-lyek}
\fancyhead[CO]{\hst{\thesection \quad A Le\-e--Yang-z\'e\-rus\-he\-lyek}}
A $Z$ \'al\-la\-po\-t\"osszeg Le\-e--Yang-z\'e\-rus\-he\-lye\-i\-nek
is\-me\-re\-t\'e\-ben \cite{kar, lyang, itz, lycikk} az el\-s\H o\-ren\-d\H u
f\'a\-zi\-s\'at\-me\-ne\-ti pon\-tok k\"onnyen meg\-ha\-t\'a\-roz\-ha\-t\'ok. Az f\'a\-zi\-s\'at\-me\-ne\-ti
pont k\"o\-ze\-l\'e\-ben
\beq
Z = Z_s + Z_b \propto \exp(-V f_s) + \exp(-V f_b),
\enq
mely ki\-fe\-je\-z\'es\-ben $s$, il\-let\-ve $b$ je\-l\"o\-li az e\-gyes f\'a\-zi\-so\-kat (pl.\
szim\-met\-ri\-kus \'es a szim\-met\-ri\-a\-s\'er\-t\H o), $f$ a
sza\-ba\-de\-ner\-gi\-a\-s\H u\-r\H u\-s\'e\-get, $V$ pe\-dig a rend\-szer t\'er\-fo\-ga\-t\'at. Mi\-vel a
f\'a\-zi\-s\'at\-me\-net so\-r\'an a sza\-ba\-de\-ner\-gi\-a\-s\H u\-r\H u\-s\'eg foly\-to\-no\-san v\'a\-lot\-zik,
\'{\i}gy el\-s\H o rend\-ben
\beq
f_b = f_s + \alpha(\kappa - \kappa_c),
\enq
a\-hol $\kappa$ a fen\-ti e\-set\-ben a Higgs-t\'er\-re vo\-nat\-ko\-z\'o hop\-ping
pa\-ra\-m\'e\-ter; $\kappa_c$ en\-nek a f\'a\-zi\-s\'at\-me\-ne\-ti pont\-ban fel\-vett
\'er\-t\'e\-ke. Ek\-kor az \'al\-la\-po\-t\"osszeg
\beq
Z \approx \exp \left[-V (f_s + f_b)/2 \right] \mathrm{ch} \left[-V
\alpha (\kappa - \kappa_c) /2 \right]
\enq
a\-la\-k\'u, mely komp\-lex $\kappa$ e\-se\-t\'en el\-t\H u\-nik, ha az
\beq
\mrm{Im}(\kappa) = 2 \pi (n - 1/2) / (V \alpha) \label{skala}
\enq
\"ossze\-f\"ug\-g\'es fen\-n\'all. A fen\-ti ki\-fe\-je\-z\'es\-ben $n$ e\-g\'esz, \'{\i}gy a
komp\-lex $\kappa$ s\'{\i}k\-ra el\-foly\-ta\-tott \'al\-la\-po\-t\"osszeg\-nek sok
Le\-e--Yang-z\'e\-rus\-he\-lye van, mely a $V \to \infty$ ha\-t\'a\-re\-set\-ben a
va\-l\'os ten\-gely\-hez tart. A\-mennyi\-ben nincs el\-s\H o\-ren\-d\H u f\'a\-zi\-s\'at\-me\-net,
\'ugy a Le\-e--Yang-z\'e\-rus\-he\-lyek a v\'eg\-te\-len t\'er\-fo\-ga\-t\'u li\-mesz\-ben nem
k\"o\-ve\-tik a $V \cdot \mrm{Im}(\kappa) = \mathrm{const.}$ sk\'a\-l\'a\-z\'ast.
Eh\-hez ter\-m\'e\-sze\-te\-sen sz\"uk\-s\'eg van ar\-ra, hogy u\-gya\-nannyi\-a\-dik --
lo\-gi\-kus m\'o\-don az el\-s\H o -- Le\-e--Yang-z\'e\-rus\-hely vi\-sel\-ke\-d\'e\-s\'et
vizs\-g\'al\-juk.

Az \'al\-la\-po\-t\"osszeg el\-foly\-ta\-t\'a\-sa u\-t\'an te\-h\'at a f\'a\-zi\-s\'at\-me\-ne\-ti pont
egy\-sze\-r\H u\-en meg\-ha\-t\'a\-roz\-ha\-t\'o. A komp\-lex s\'{\i}k\-ra va\-l\'o el\-foly\-ta\-t\'as
le\-he\-t\H o\-s\'e\-g\'et a Fer\-ren\-berg--Swend\-sen-f\'e\-le \'at\-s\'u\-lyo\-z\'a\-si m\'od\-szer
te\-szi le\-he\-t\H o\-v\'e.
\subsection{A Fer\-ren\-berg--Swend\-sen-f\'e\-le \'at\-s\'u\-lyo\-z\'a\-si m\'od\-szer}
A Fer\-ren\-berg--Swend\-sen-f\'e\-le m\'od\-szer egy\-sze\-r\H u kap\-cso\-la\-tot \'al\-l\'{\i}t
fel a rend\-pa\-ra\-m\'e\-ter k\'et egy\-m\'as\-hoz k\"o\-ze\-li pa\-ra\-m\'e\-ter\-hal\-maz e\-se\-t\'en
m\'er\-he\-t\H o e\-losz\-l\'as\-f\"ugg\-v\'e\-ny\'e\-ben \cite{ferr-swen}. Va\-la\-mely $O$
m\'er\-he\-t\H o mennyi\-s\'eg v\'ar\-ha\-t\'o \'er\-t\'e\-ke a
\beq
Z = \int dE N(E) e^{- \beta E}
\enq
\'al\-la\-po\-t\"osszeg\-b\H ol (mely\-ben $N(E)$ az $E$ e\-ner\-gi\-\'a\-j\'u \'al\-la\-po\-tok
sz\'a\-m\'at je\-l\"o\-li) a k\"o\-vet\-ke\-z\H o m\'o\-don ha\-t\'a\-roz\-ha\-t\'o meg. Je\-l\"ol\-je
$P_{\beta_0} (O, E)$ az $O$ o\-pe\-r\'a\-tor \'er\-t\'e\-ke \'es az e\-ner\-gi\-a sze\-rin\-ti
k\'et\-v\'al\-to\-z\'os e\-losz\-l\'as\-f\"ugg\-v\'enyt a $\beta = \beta_0$
pa\-ra\-m\'e\-te\-r\'er\-t\'ek mel\-lett. Ek\-kor nyil\-v\'an
\beq
P_{\beta_0}(O,E) \, dE \, dO = N(O,E)\, e^{-\beta_0 E} dE \, dO,
\enq
a\-hon\-nan $N(E)$ ki\-fe\-jez\-he\-t\H o:
\beq
N(E) = \int dO \, N(O,E) = \int dO \, P_{\beta_0} \, e^{+ \beta_0 E}.
\enq
In\-nen a\-zon\-ban az $O$ o\-pe\-r\'a\-tor v\'ar\-ha\-t\'o \'er\-t\'e\-ke nem csak a
szi\-mu\-l\'a\-ci\-\'o\-ban hasz\-n\'alt $\beta_0$ pa\-ra\-m\'e\-ter\-n\'el, ha\-nem egy at\-t\'ol
kis\-s\'e el\-t\'e\-r\H o $\beta$ pa\-ra\-m\'e\-ter\-n\'el is meg\-ha\-t\'a\-roz\-ha\-t\'o, hi\-szen
\beqar
\langle O \rangle_\beta &=
\frac{\dst \int dE\, dO\, O\, N(O,E)\, e^{-\beta E}}
{\dst \int dE\, dO\, N(O,E)\, e^{-\beta E}} = &
\frac{\dst \int dE\, dO\, O\, N(O,E)\, e^{-\beta_0 E} e^{-\beta (E-E_0)}}
{\dst \int dE\, dO\, N(O,E)\, e^{-\beta_0 E} e^{-\beta (E-E_0)}} =
\nonumber \\
&&
\frac{\dst \int dE\, dO\, O\, P_{\beta_0}(O,E)\, e^{-\beta (E-E_0)}}
{\dst \int dE\, dO\, P_{\beta_0}(O,E)\, e^{-\beta (E-E_0)}}.
\enqar
A fen\-ti\-ek\-ben nyil\-v\'an \"u\-gyel\-ni kell ar\-ra, hogy $\beta-\beta_0$ ne
le\-gyen t\'ul nagy, hi\-szen a Mon\-te Car\-lo szi\-mu\-l\'a\-ci\-\'ok ek\-kor i\-ga\-z\'an
meg\-b\'{\i}z\-ha\-t\'o\-ak.

\section{A szim\-met\-ri\-kus, a sz\'{\i}n\-s\'er\-t\H o \'es a Higgs-f\'a\-zis}
\fancyhead[CO]{\hst{\thesection \quad A szim\-metr\-kus, a
sz\'{\i}ns\'ert\H{o} \'es a Higgs-f\'a\-zis}}
Az MSSM szi\-mu\-l\'a\-ci\-\'ok so\-r\'an az \'at\-s\'u\-lyo\-z\'a\-si m\'od\-szert k\'et\-f\'e\-le
m\'o\-don is hasz\-n\'al\-tam. Az el\-s\H o le\-he\-t\H o\-s\'eg az, hogy egy szi\-mu\-l\'a\-ci\-\'o
e\-red\-m\'e\-nye\-k\'epp ka\-pott a\-da\-to\-kat oly m\'o\-don s\'u\-lyo\-zunk \'at, hogy
k\'et\-cs\'u\-cs\'u e\-losz\-l\'as\-g\"or\-b\'et kap\-junk; ez nyil\-v\'an a k\'et f\'a\-zis
e\-gy\"ut\-tes je\-len\-l\'e\-t\'e\-re u\-tal, te\-h\'at a f\'a\-zi\-s\'at\-me\-ne\-ti pont
k\"o\-zel\-s\'e\-g\'et jel\-zi, mint az a \ref{twopeax} \'ab\-r\'an l\'at\-ha\-t\'o.

\bef[ht]
\bc
\epsfig{file=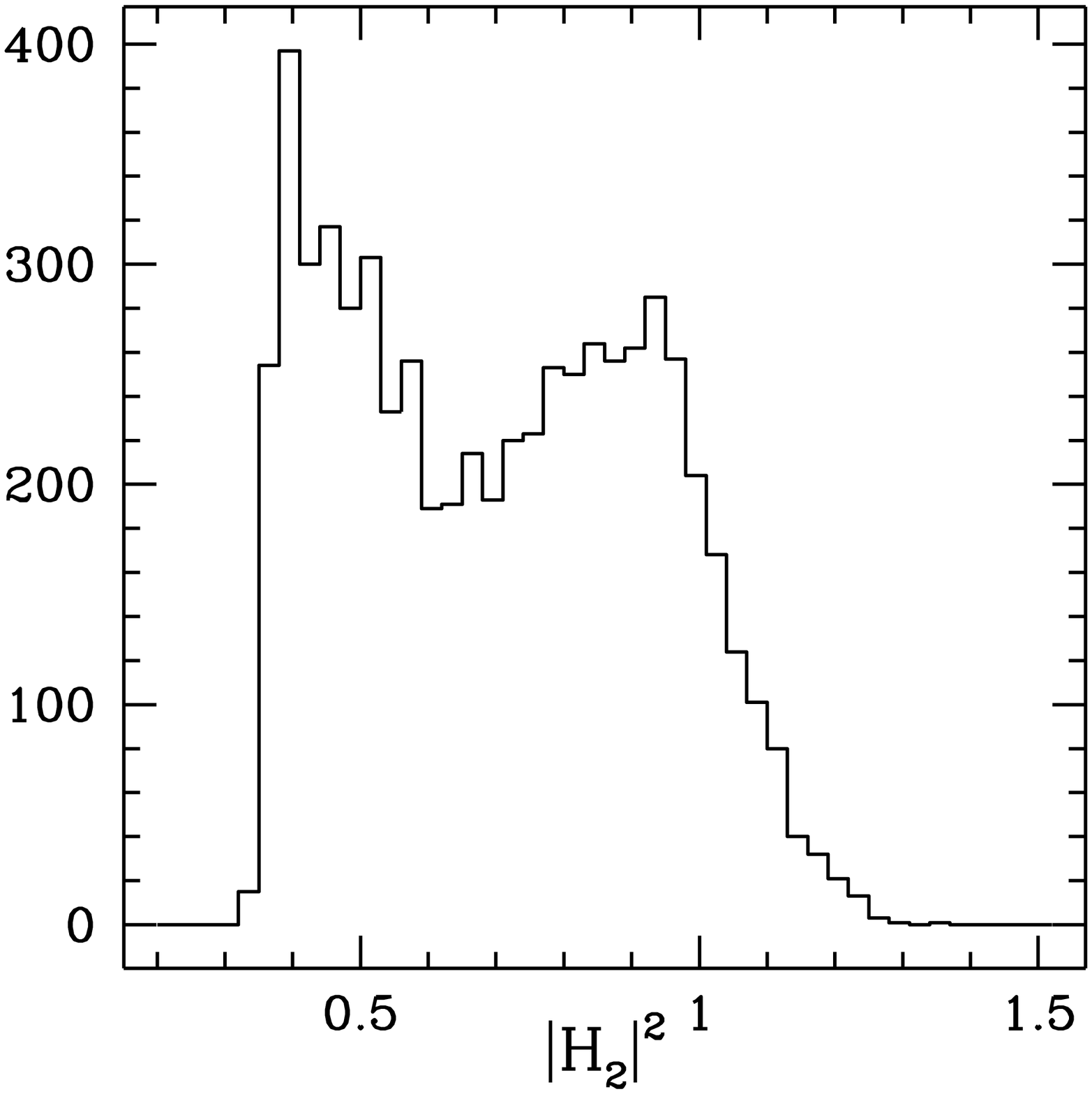,width=8cm}
\caption{A k\'et\-cs\'u\-cs\'u hisz\-tog\-ram a k\'et f\'a\-zis e\-gy\"ut\-tes
je\-len\-l\'e\-t\'e\-re u\-tal \label{twopeax}}
\ec
\enf

A m\'a\-sik le\-he\-t\H o\-s\'eg ter\-m\'e\-sze\-te\-sen a Le\-e--Yang-z\'e\-rus\-he\-lyek
meg\-ke\-re\-s\'e\-se, mely a SM vizs\-g\'a\-la\-t\'a\-hoz \'{\i}rt prog\-ram se\-g\'{\i}t\-s\'e\-g\'e\-vel
t\"or\-t\'e\-nik \cite{csik99}. Ez a\-dott be\-me\-n\H o\-a\-da\-tok\-b\'ol a szi\-mu\-l\'a\-ci\-\'os
pont k\"or\-nye\-ze\-t\'e\-ben vizs\-g\'al\-ta, van-e Le\-e--Yang-z\'e\-rus\-hely, oly
m\'o\-don, hogy (meg\-v\'a\-laszt\-ha\-t\'o sz\'a\-m\'u \'es m\'e\-re\-t\H u) ki\-csiny
l\'e\-p\'e\-sek\-kel csi\-ga\-vo\-nal\-ban k\"or\-be\-j\'ar\-ta a vizs\-g\'a\-lan\-d\'o pont
k\"or\-nye\-ze\-t\'et; a\-mennyi\-ben az \'al\-la\-po\-t\"osszeg va\-la\-mely l\'e\-p\'es so\-r\'an
az e\-l\H o\-re me\-ga\-dott kor\-l\'at\-n\'al k\"o\-ze\-lebb volt z\'e\-rus\-hoz, ott
Le\-e--Yang-z\'e\-rus\-he\-lyet \'al\-la\-p\'{\i}\-tott meg. Ezt gyak\-ran az \'{\i}gy ta\-l\'alt
z\'e\-rus\-pont\-ban v\'eg\-re\-haj\-tott \'u\-jabb szi\-mu\-l\'a\-ci\-\'ok se\-g\'{\i}t\-s\'e\-g\'e\-vel
el\-le\-n\H o\-riz\-tem il\-let\-ve pon\-to\-s\'{\i}\-tot\-tam.

Jel\-leg\-ze\-tes szi\-mu\-l\'a\-ci\-\'os el\-j\'a\-r\'as az, hogy m\'e\-lyen a szim\-met\-ri\-kus-
il\-let\-ve a szim\-met\-ri\-a\-s\'er\-t\H o tar\-to\-m\'any\-ban l\'et\-re\-ho\-zunk egy-egy
pre\-pa\-r\'alt kon\-fi\-gu\-r\'a\-ci\-\'ot, majd a f\'a\-zi\-s\'at\-me\-ne\-ti pont\-ban e\-zek\-b\H ol a
kon\-fi\-gu\-r\'a\-ci\-\'ok\-b\'ol in\-d\'{\i}t\-juk a szi\-mu\-l\'a\-ci\-\'ot. Jel\-lem\-z\H o m\'o\-don az
e\-gyes kon\-fi\-gu\-r\'a\-ci\-\'ok k\"o\-z\"ott van \'at\-j\'a\-r\'as, a\-zon\-ban az egy\-m\'as
u\-t\'a\-ni fris\-s\'{\i}\-t\'e\-sek kor\-re\-l\'a\-ci\-\'o\-ja nagy: a f\'a\-zi\-sok k\"o\-z\"ott rit\-ka az
\'at\-me\-net \'es \'at\-me\-net u\-t\'an a rend\-szer hossz\'u i\-de\-ig tar\-t\'oz\-ko\-dik az
\'uj f\'a\-zis\-ban (\ref{konfig} \'ab\-ra).

\bef[ht]
\bc
\epsfig{file=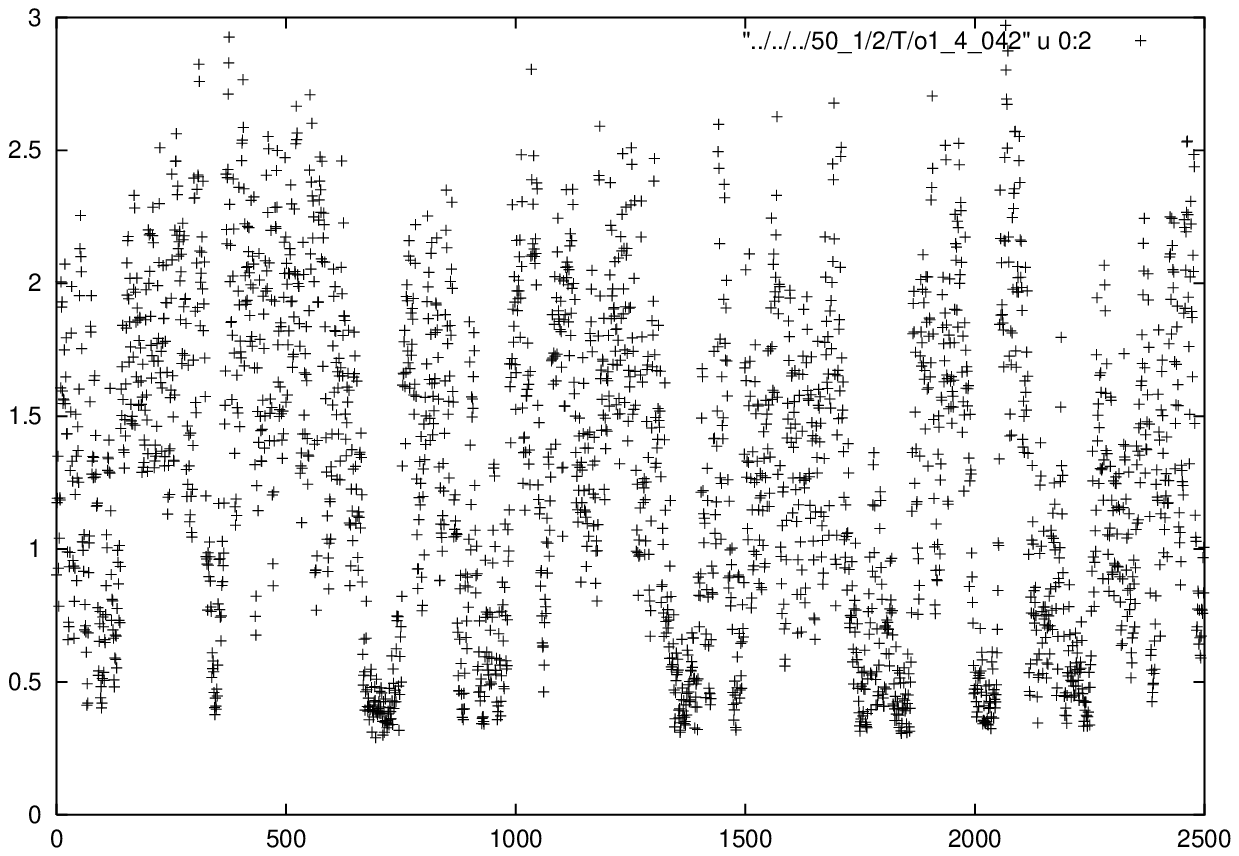,width=8cm}
\caption{A k\'et f\'a\-zis k\"oz\-ti \'at\-me\-net sok\-szor t\"obb e\-zer \'uj
kon\-fi\-gu\-r\'a\-ci\-\'ot i\-g\'e\-nyel \label{konfig}}
\ec
\enf

Ez jel\-leg\-ze\-tes k\'et\-cs\'u\-cs\'u struk\-t\'u\-r\'at mu\-tat a hisz\-tog\-ram\-mon (l\'asd
a \ref{twopeax} \'ab\-r\'at). \bigskip

Kis\-m\'e\-re\-t\H u r\'a\-cso\-kon vi\-szony\-lag r\"o\-vid szi\-mu\-l\'a\-ci\-\'ok se\-g\'{\i}t\-s\'e\-g\'e\-vel
is gyak\-ran meg\-ta\-l\'al\-ha\-t\'o a Le\-e--Yang-f\'e\-le z\'e\-rus\-hely; itt a
szi\-mu\-l\'a\-ci\-\'os pa\-ra\-m\'e\-te\-rek\-nek nem kell t\'ul k\"o\-zel es\-ni\-e a
Le\-e--Yang-z\'e\-rus\-hely\-hez. E\-zek\-b\H ol az a\-da\-tok\-b\'ol sok e\-set\-ben meg\-fe\-le\-l\H o
pon\-tos\-s\'ag\-gal tud\-tam ext\-ra\-po\-l\'al\-ni a na\-gyobb r\'acs\-t\'er\-fo\-ga\-tok\-ra, a\-hol
a ki\-sebb min\-t\'ak el\-le\-n\'e\-re pon\-to\-sab\-ban tud\-tuk meg\-ha\-t\'a\-roz\-ni a
Le\-e--Yang-f\'e\-le z\'e\-rus\-he\-lye\-ket. Az \'{\i}gy ka\-pott a\-da\-tok a\-lap\-j\'an a
v\'eg\-te\-len t\'er\-fo\-gat\-ra t\"or\-t\'e\-n\H o ext\-ra\-po\-l\'a\-l\'as kel\-l\H o pon\-tos\-s\'ag\-gal
le\-het\-s\'e\-ges -- ezt t\"obb e\-set\-ben v\'eg\-re\-haj\-tot\-tam. A v\'eg\-te\-len
t\'er\-fo\-ga\-t\'u li\-mesz\-ben a
\beq
\mrm{Im}\, \kappa_0(V) = \kappa_0^c + C V^{-\nu}
\enq
il\-lesz\-t\'es\-ben $\nu$-t c\'el\-sze\-r\H u pa\-ra\-m\'e\-ter\-nek tar\-ta\-ni (nem pe\-dig az
el\-s\H o\-ren\-d\H u f\'a\-zi\-s\'at\-me\-net e\-se\-t\'en fel\-vett \'er\-t\'e\-k\'e\-ben, 1-ben
r\"og\-z\'{\i}\-te\-ni), majd az a\-da\-tok\-ra il\-lesz\-tett g\"or\-b\'e\-b\H ol meg\-ha\-t\'a\-roz\-ni.
I\-lyen m\'o\-don ha\-t\'a\-roz\-ha\-t\'o meg a stan\-dard mo\-dell\-be\-li e\-lekt\-ro\-gyen\-ge
f\'a\-zi\-s\'at\-me\-net v\'eg\-pont\-ja is: a (\ref{skala}) sk\'a\-l\'a\-z\'ast mu\-ta\-t\'o
\'es az azt s\'er\-t\H o Higgs-t\"o\-meg tar\-to\-m\'any ha\-t\'a\-ra e\-l\'eg pon\-to\-san
me\-g\'al\-la\-p\'{\i}t\-ha\-t\'o \cite{lycikk}.

A szim\-met\-ri\-kus \'es a Higgs-f\'a\-zis k\"oz\-ti \'at\-me\-ne\-ti pon\-to\-kat
k\'et\-f\'e\-le $\alpha_2$ \'er\-t\'ek\-n\'el ha\-t\'a\-roz\-tuk meg, $L_t=2,3,4,5$-\"os
r\'a\-csok mel\-lett, k\"u\-l\"on\-b\"o\-z\H o $L_s$ t\'er\-be\-li r\'acs\-ki\-ter\-je\-d\'e\-sek mel\-lett.
Az e\-gyik $\alpha_2$ \'er\-t\'ek mel\-lett v\'eg\-re\-haj\-tott szi\-mu\-l\'a\-ci\-\'ok\-b\'ol
Le\-e--Yang-m\'od\-szer\-rel meg\-ha\-t\'a\-ro\-zott f\'a\-zi\-s\'at\-me\-ne\-ti pon\-to\-kat a
\ref{lytable} t\'ab\-l\'a\-zat fog\-lal\-ja \"ossze.

\begin{table}[htb]
\begin{center}
\footnotesize
\begin{tabular}{||c||c|c|c|c|c|c|c||}
\hline
\hline
$L_t=2$ & $L_s=4$   & $L_s=6$	& $L_s=8$   & $L_s=10$ &&&	\\
	& -1.0381(4)& -1.0247(6)& -1.0192(4)& -1.0178(2) &&&\\
\hline
$L_t=3$ & $L_s=6$   & $L_s=7$	& $L_s=8$   & $L_s=9$	 & $L_s=10$ &&\\
	& -0.9876(6)& -0.9807(4)& -0.9785(11)& -0.9768(3)& -0.9768(2) &&
\\
\hline
$L_t=4$ & $L_s=8$   & $L_s=10$	& $L_s=12$  & $L_s=14$	 & $L_s=16$    &
	  $L_s=18$  & $L_s=20$	\\
	& -0.9738(5)& -0.9718(1)& -0.9718(2)& -0.9710(1) & -0.97184(3) &
	  -0.97174(5)&-0.97130(4) \\
\hline
$L_t=5$ & $L_s=10$  & $L_s=11$	& $L_s=12$  & $L_s=14$	 & $L_s=16$ &&\\
	& -0.9759(1)& -0.9750(2)& -0.9765(1)& -0.9765(1) & -0.9765(1) &&\\
\hline
\hline
\end{tabular}
\caption{A ki\-seb\-bik $\alpha_s$ \'er\-t\'ek mel\-let\-ti szi\-mu\-l\'a\-ci\-\'ok
e\-red\-m\'e\-nye\-k\'ent ka\-pott v\'e\-ges t\'er\-fo\-ga\-t\'u Le\-e--Yang-z\'e\-rus\-he\-lyek}
\label{lytable}
\end{center}
\end{table}

Ha\-son\-l\'o\-k\'epp ki\-m\'er\-he\-t\H o\-ek a sz\'{\i}n\-s\'er\-t\H o f\'a\-zi\-s\'at\-me\-ne\-ti pon\-tok, me\-lyek
egy f\'a\-zis\-di\-ag\-ram fel\-v\'e\-te\-l\'e\-hez sz\"uk\-s\'e\-ge\-sek. Eh\-hez a k\"o\-vet\-ke\-z\H o
el\-j\'a\-r\'as a leg\-c\'el\-ra\-ve\-ze\-t\H obb: $L_t=3$ i\-d\H o\-be\-li r\'acs\-ki\-ter\-je\-d\'es
mel\-lett fel\-ve\-he\-t\H o egy ``$\alpha_2-m_U^2$ f\'a\-zis\-di\-ag\-ram'', a\-zaz
eb\-ben a s\'{\i}k\-ban meg\-ha\-t\'a\-roz\-zuk a f\'a\-zi\-s\'at\-me\-ne\-ti pon\-to\-kat \'es a
h\'ar\-mas\-pon\-tot. Az e\-red\-m\'e\-nye\-ket a \ref{alphamu} t\'ab\-l\'a\-zat tar\-tal\-maz\-za.

\begin{table}[htb]
\bc
\begin{tabular}{||c|c||}
\hline
$\alpha_2$ & $m_U^2$ (GeV\tsc{2}) \\
\hline
-0.8100    & -33115(17) \\
-1.0000    & -32756(45) \\
\hline
-1.0343(3) & 0 \\
-1.0011(9) & -6000 \\
-0.9869(6) & -10000 \\
-0.9717(5) & -20000 \\
-0.9540(2) & -25000 \\
-0.9364(5) & -30000 \\
\hline
-0.9364(5) & -32899 \\
\hline
\end{tabular}
\caption{Az $L_t =3$ szi\-mu\-l\'a\-ci\-\'ok a\-lap\-j\'an meg\-ha\-t\'a\-ro\-zott
f\'a\-zi\-s\'at\-me\-ne\-ti pon\-tok \'es a h\'ar\-mas\-pont (u\-tol\-s\'o sor) az
$\alpha_2$--$m_U^2$ s\'{\i}\-kon \label{alphamu}}
\ec
\end{table}

A h\'ar\-mas\-pont\-ban v\'eg\-re\-haj\-tott z\'e\-rus h\H o\-m\'er\-s\'ek\-le\-t\H u szi\-mu\-l\'a\-ci\-\'o\-ban
ka\-pott kor\-re\-l\'a\-ci\-\'os f\"ugg\-v\'e\-nyek le\-csen\-g\'e\-s\'e\-b\H ol meg\-ha\-t\'a\-roz\-ha\-t\'o a
r\'a\-csegy\-s\'e\-gek\-ben m\'ert W- \'es Higgs-t\"o\-meg. C\'el\-sze\-r\H u a h\'ar\-mas\-pont
kis k\"or\-nye\-ze\-t\'e\-ben is fel\-t\'er\-k\'e\-pez\-ni a t\"o\-me\-gek \'es az $\alpha_2$
pa\-ra\-m\'e\-ter kap\-cso\-la\-t\'at. A W-t\"o\-me\-get fi\-zi\-ka\-i \'er\-t\'e\-k\'en r\"og\-z\'{\i}t\-ve
me\-gad\-ha\-t\'o a r\'a\-cs\'al\-lan\-d\'o, me\-lyet a r\'acs $L_t$ ki\-ter\-je\-d\'e\-se
se\-g\'{\i}t\-s\'e\-g\'e\-vel kri\-ti\-kus h\H o\-m\'er\-s\'ek\-let\-t\'e kon\-ver\-t\'al\-ha\-tunk.

A f\'a\-zis\-di\-ag\-ram fel\-v\'e\-te\-l\'e\-hez m\'eg le\-ga\-l\'abb h\'a\-rom pont\-ra van sz\"uk\-s\'eg
(mind\-h\'a\-rom \'a\-gon egy pont\-ra). Eh\-hez e\-l\H o\-sz\"or c\'el\-sze\-r\H u meg\-fi\-gyel\-ni,
hogy a sz\'{\i}n\-s\'er\-t\H o f\'a\-zis ha\-t\'a\-r\'at jel\-lem\-z\H o g\"or\-be gya\-kor\-la\-ti\-lag csak
$m_U$-t\'ol f\"ugg, az $\alpha_2$ pa\-ra\-m\'e\-ter\-t\H ol va\-l\'o f\"ug\-g\'e\-se na\-gyon
gyen\-ge (a\-mint az a \ref{alphamu} t\'ab\-l\'a\-zat el\-s\H o k\'et so\-r\'a\-b\'ol j\'ol
le\-ol\-vas\-ha\-t\'o).

A $T_c-m_U^2$ s\'{\i}\-kon fel\-ve\-en\-d\H o f\'a\-zis\-g\"or\-be k\'et \'a\-g\'a\-nak
meg\-ha\-t\'a\-ro\-z\'a\-s\'a\-hoz te\-h\'at az $m_h/m_W$ Higgs-W t\"o\-me\-ga\-r\'any biz\-to\-s\'{\i}\-t\'a\-sa
mel\-lett kel\-le\-ne a h\H o\-m\'er\-s\'ek\-le\-tet v\'al\-toz\-tat\-nunk. Ez az $L_t=3$
he\-lyett $L_t =2, 4$ ki\-ter\-je\-d\'e\-s\H u r\'a\-cso\-kon v\'eg\-re\-haj\-tott szi\-mu\-l\'a\-ci\-\'ok
se\-g\'{\i}t\-s\'e\-g\'e\-vel le\-het\-s\'e\-ges. (Az e\-re\-de\-ti $10^3 \times 3$ r\'acs he\-lyett
$10^3 \times \{2,4\}$ r\'a\-csok sze\-re\-pel\-nek.) A sz\'{\i}n\-s\'er\-t\H o
f\'a\-zi\-s\'at\-me\-ne\-tet jel\-lem\-z\H o $m_U$ pa\-ra\-m\'e\-ter meg\-ha\-t\'a\-ro\-z\'a\-sa u\-t\'an
el\-le\-n\H o\-riz\-ni kell, nem v\'al\-to\-zott-e az $m_h$ pa\-ra\-m\'e\-ter t\'ul\-s\'a\-go\-san.
A\-mennyi\-ben i\-gen, az $L_t=3$ a\-da\-tok\-b\'ol v\'eg\-re\-haj\-tott z\'e\-rus
h\H o\-m\'er\-s\'ek\-le\-t\H u szi\-mu\-l\'a\-ci\-\'ok e\-red\-m\'e\-nye a\-lap\-j\'an pon\-to\-s\'{\i}t\-ha\-t\'o az
e\-red\-m\'eny.

A har\-ma\-dik g\"or\-be\-\'ag fel\-v\'e\-te\-l\'e\-hez egy \'u\-jabb $L_t=3$
szi\-mu\-l\'a\-ci\-\'o sz\"uk\-s\'e\-ges; eh\-hez a Higgs- \'es a szim\-met\-ri\-kus f\'a\-zis
k\"oz\-ti va\-la\-me\-lyik $\alpha'_2, m_U^2$ pon\-tot kell te\-kin\-te\-ni. E\-zek
bir\-to\-k\'a\-ban fel\-raj\-zol\-ha\-t\'o a f\'a\-zis\-di\-ag\-ram (\ref{phdiag} \'ab\-ra).

\bef[ht]
\bc
\epsfig{file=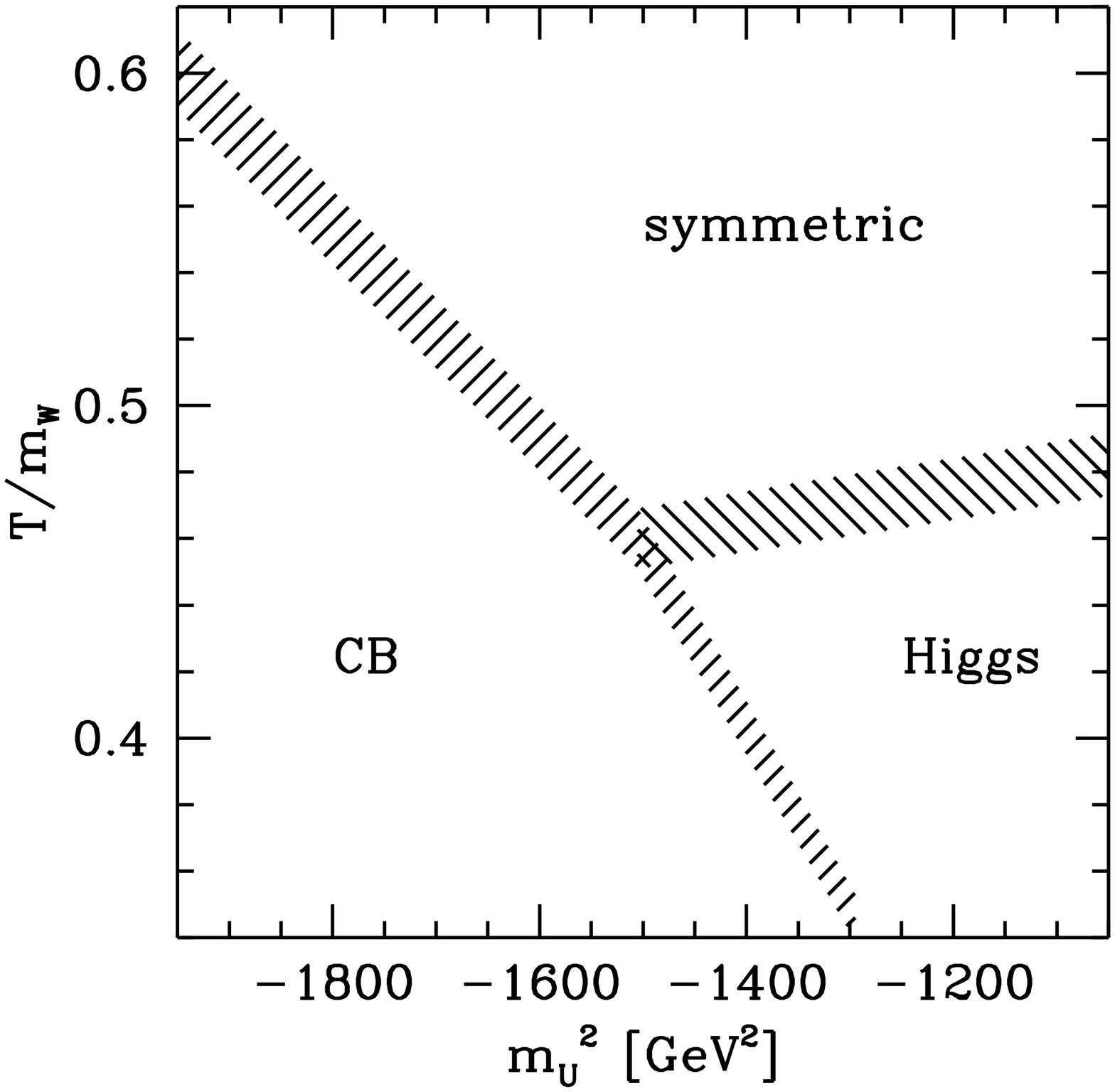,width=8cm}
\caption{A r\'acsszi\-mu\-l\'a\-ci\-\'os e\-re\-m\'e\-nyek a\-lap\-j\'an ka\-pott
f\'a\-zis\-di\-ag\-ram \label{phdiag}}
\ec
\enf

A f\'a\-zis\-di\-ag\-ra\-mon sze\-rep\-l\H o vo\-na\-lak gya\-kor\-la\-ti\-lag s\'a\-vok; en\-nek o\-ka a
t\"o\-meg\-meg\-ha\-t\'a\-ro\-z\'as hi\-b\'a\-ja.
Ezt a f\'a\-zis\-di\-ag\-ra\-mot a per\-tur\-ba\-t\'{\i}v meg\-k\"o\-ze\-l\'{\i}\-t\'es\-ben ka\-pot\-tak\-kal
\"ossze\-vet\-ve j\'o kva\-li\-ta\-t\'{\i}v e\-gye\-z\'es \'al\-la\-p\'{\i}t\-ha\-t\'o meg. A di\-men\-zi\-\'os
re\-duk\-ci\-\'on a\-la\-pu\-l\'o szi\-mu\-l\'a\-ci\-\'ok e\-red\-m\'e\-nye\-k\'ent ka\-pott
f\'a\-zis\-di\-ag\-ram \cite{la98} na\-gyon ha\-son\-l\'o \ref{phdiag}-hoz. A ko\-ra\-i
u\-ni\-ver\-zum a\-la\-ku\-l\'a\-s\'a\-ra e\-set\-le\-ges nagy ve\-sz\'e\-lye\-ket rej\-t\H o
sz\'{\i}n\-s\'er\-t\H o f\'a\-zis te\-h\'at a n\'egy\-di\-men\-zi\-\'os nem\-per\-tur\-ba\-t\'{\i}v
meg\-k\"o\-ze\-l\'{\i}\-t\'es\-ben is je\-len van. (A fen\-ti \'ab\-r\'a\-hoz hasz\-n\'alt
pa\-ra\-m\'e\-te\-rek mel\-lett nincs k\'et\-l\'ep\-cs\H os f\'a\-zi\-s\'at\-me\-net.)

A sz\'{\i}n\-s\'e\-r\H o f\'a\-zi\-s\'at\-me\-net a szi\-mu\-l\'a\-ci\-\'ok a\-lap\-j\'an sok\-kal e\-r\H o\-sebb,
mint a szim\-met\-ri\-kus f\'a\-zis \'es a Higgs-f\'a\-zis k\"oz\-ti f\'a\-zi\-s\'at\-me\-net.

A \ref{lytable} t\'ab\-l\'a\-zat a\-lap\-j\'an v\'eg\-re\-hajt\-va a v\'eg\-te\-len
t\'er\-fo\-ga\-t\'u ext\-ra\-po\-l\'a\-l\'ast $\alpha_2$-re, ott el\-v\'e\-gez\-he\-t\H o\-ek a
z\'e\-rus h\H o\-m\'er\-s\'ek\-le\-t\H u szi\-mu\-l\'a\-ci\-\'ok. A szi\-mu\-l\'a\-ci\-\'ok\-b\'ol nyer\-he\-t\H o
$W$ \'es Higgs-t\"o\-me\-ge\-ket a \ref{tomeg} t\'ab\-l\'a\-zat fog\-lal\-ja \"ossze. A
t\'ab\-l\'a\-zat m\'a\-so\-dik \'es har\-ma\-dik so\-r\'a\-ban sze\-rep\-l\H o t\"o\-me\-gek
r\'a\-csegy\-s\'e\-gek\-ben \'er\-ten\-d\H ok; az e\-zek a\-lap\-j\'an sz\'a\-molt $R_{HW}$
t\"o\-me\-ga\-r\'any a t\'ab\-l\'a\-zat u\-tol\-s\'o osz\-lo\-p\'a\-ban sze\-re\-pel. Ah\-hoz, hogy a
kons\-tans fi\-zi\-ka vo\-na\-l\'an mo\-zog\-junk, en\-nek a mennyi\-s\'eg\-nek kons\-tans\-nak
kel\-le\-ne len\-ni\-e.

\begin{table}[htb]
\bc
\begin{tabular}{|c|c|c|c|c|}\hline
$L_t$ & $\alpha_2$   &	 $m_H$	 &   $m_W$   &	 $R_{HW}$ \\ \hline\hline
2     & -0.9856(6)   & 0.308(7)  & 0.561(7)  & 0.55(1)	  \\ \hline
3     & -0.9542(3)   & 0.150(11) & 0.357(14) & 0.42(4)	  \\ \hline
4     & -0.9496(1)   & 0.114(11) & 0.274(19) & 0.42(4)	  \\ \hline
5     & -0.94545(5)  & 0.079(6)  & 0.228(11) & 0.36(3)	  \\ \hline\hline
2     & -1.0162(6)   & 0.375(5)  & 0.642(14) & 0.58(2)	  \\ \hline
3     & -0.9745(3)   & 0.162(15) & 0.399(12) & 0.41(4)	  \\ \hline
\end{tabular}
\caption{A z\'e\-rus\-h\H o\-m\'er\-s\'ek\-le\-t\H u szi\-mu\-l\'a\-ci\-\'ok\-ban m\'ert t\"o\-me\-gek
r\'a\-csegy\-s\'e\-gek\-ben, \'es a t\"o\-me\-gek a\-r\'a\-nya. \label{tomeg} Az e\-r\H os
csa\-to\-l\'a\-si \'al\-lan\-d\'o \'er\-t\'e\-ke\-i $\alpha_s=0.1$, $0.05$}
\ec
\end{table}

\section{A bu\-bo\-r\'ek\-fal vizs\-g\'a\-la\-ta}
\fancyhead[CO]{\hst{\thesection \quad A bu\-bor\'ek\-fal vizsg\'a\-la\-ta}}
A ba\-ri\-on\-kel\-t\'es az e\-lekt\-ro\-gyen\-ge f\'a\-zi\-s\'at\-me\-net so\-r\'an l\'et\-re\-j\"o\-v\H o,
k\"u\-l\"on\-b\"o\-z\H o f\'a\-zi\-so\-kat el\-v\'a\-lasz\-t\'o bu\-bo\-r\'ek\-fa\-lak\-ban le\-j\'at\-sz\'o\-d\'o
CP-s\'er\-t\H o fo\-lya\-ma\-tok r\'e\-v\'en t\"or\-t\'e\-nik \cite{brhlik, clin-gdm99,
coh-kap}. Az itt ke\-let\-ke\-z\H o bal\-ke\-zes kvar\-kok s\H u\-r\H u\-s\'e\-ge meg\-ha\-lad\-ja a
bal\-ke\-zes an\-tik\-var\-ko\-k\'et, u\-gya\-no\-lyan m\'er\-t\'ek\-ben a\-hogy a jobb\-ke\-zes
an\-tik\-var\-kok s\H u\-r\H u\-s\'e\-ge meg\-ha\-lad\-ja a jobb\-ke\-zes kvar\-ko\-k\'et. A z\'e\-rus
\"ossz\-ba\-ri\-on\-sz\'a\-m\'u \'al\-la\-pot\-ban a bal\-ke\-zes kvar\-kok t\'ul\-s\'u\-lya
ha\-t\'a\-s\'a\-ra a\-no\-m\'a\-lis szfa\-le\-ro\-n\'at\-me\-ne\-tek j\'at\-sz\'od\-nak le, mely
meg\-bont\-ja a ba\-ri\-on--an\-ti\-ba\-ri\-on a\-szim\-met\-ri\-\'at. A ke\-let\-ke\-z\H o
ba\-ri\-on\-t\"obb\-let egy r\'e\-sze a n\"o\-vek\-v\H o bu\-bo\-r\'e\-kok bel\-se\-j\'e\-be
dif\-fun\-d\'al, a\-hol a (\ref{vtpert}) fel\-t\'e\-tel tel\-je\-s\"u\-l\'e\-se e\-se\-t\'en a
szfa\-le\-ron \'at\-me\-ne\-tek be\-fagy\-nak.

A fen\-ti egy\-sze\-r\H u k\'ep kvan\-ti\-ta\-t\'{\i}v ke\-ze\-l\'e\-se meg\-le\-he\-t\H o\-sen
bo\-nyo\-lult \cite{clin00}. A prob\-l\'e\-ma szo\-k\'a\-sos ke\-ze\-l\'es\-ben klasszi\-kus
e\-r\H ok ha\-t\'a\-s\'a\-ra be\-k\"o\-vet\-ke\-z\H o dif\-f\'u\-zi\-\'ot t\'e\-te\-le\-z\"unk
fel \cite{gdm99,smit99}, mely k\"o\-ze\-l\'{\i}\-t\'es ab\-ban az e\-set\-ben
i\-ga\-zol\-ha\-t\'o, ha a bu\-bo\-r\'ek\-fal vas\-tag\-s\'a\-ga nagy az in\-verz
h\H o\-m\'er\-s\'ek\-let\-hez k\'e\-pest \cite{joy}.

A bu\-bo\-r\'ek\-fal vas\-tag\-s\'a\-ga te\-h\'at el\-s\H o\-ren\-d\H u fon\-tos\-s\'a\-g\'u mennyi\-s\'eg.
A per\-tur\-ba\-t\'{\i}v meg\-k\"o\-ze\-l\'{\i}\-t\'es \cite{mor, john} e\-red\-m\'e\-nye
\beq
l_w = (11.2 \pm 1.5) / T_c. \label{lw}
\enq
Az ed\-di\-gi h\'a\-rom\-di\-men\-zi\-\'os szi\-mu\-l\'a\-ci\-\'ok\-ban csak egy Higgs-dub\-let\-tet
hasz\-n\'al\-tak, \'{\i}gy a Higgs-te\-rek v\'a\-ku\-um v\'ar\-ha\-t\'o \'er\-t\'e\-ke\-i\-nek
h\'a\-nya\-do\-sa\-k\'ent de\-fi\-ni\-\'alt $\beta$ pa\-ra\-m\'e\-ter vizs\-g\'a\-la\-t\'a\-ra ez a
m\'od\-szer al\-kal\-mat\-lan. \'Igy a fen\-ti mennyi\-s\'eg n\'egy\-di\-men\-zi\-\'os
szi\-mu\-l\'a\-ci\-\'ok\-ban t\"or\-t\'e\-n\H o vizs\-g\'a\-la\-t\'a\-nak sz\"uk\-s\'e\-ges\-s\'e\-ge
nyil\-v\'an\-va\-l\'o.

A fal-pro\-fil meg\-ha\-t\'a\-ro\-z\'a\-sa (\ref{lw}) a\-lap\-j\'an o\-lyan r\'a\-csot
k\'{\i}\-v\'an, mely\-nek e\-gyik ($z$) i\-r\'any men\-ti ki\-ter\-je\-d\'e\-se i\-gen nagy
(a fen\-ti $l_w$ k\'et\-sze\-re\-s\'e\-n\'el l\'e\-nye\-ge\-sen na\-gyobb). A
szi\-mu\-l\'a\-ci\-\'o\-kat \'{\i}gy e\-l\H o\-sz\"or egy $2*12^2*192$ m\'e\-re\-t\H u r\'a\-cson
haj\-tot\-tuk v\'eg\-re; a szi\-mu\-l\'a\-ci\-\'o so\-r\'an a rend\-pa\-ra\-m\'e\-ter a k\'et
Higgs-t\'er $|H_1|^2$ \'es $|H_2|^2$ hossz\-n\'egy\-ze\-t\'e\-nek v\'ar\-ha\-t\'o
\'er\-t\'e\-ke volt. An\-nak biz\-to\-s\'{\i}\-t\'a\-s\'a\-ra, hogy a k\'et f\'a\-zis
meg\-ha\-t\'a\-ro\-zott a\-r\'any\-ban le\-gyen je\-len, e\-le\-gen\-d\H o a fris\-s\'{\i}\-t\'e\-si
al\-go\-rit\-mus r\'e\-sze\-k\'ent a Higgs-t\'er e\-g\'esz kon\-fi\-gu\-r\'a\-ci\-\'o\-ra vett
\'at\-la\-g\'er\-t\'e\-k\'et r\"og\-z\'{\i}\-te\-ni: a\-mennyi\-ben a szo\-k\'a\-sos fris\-s\'{\i}\-t\'e\-si
al\-go\-rit\-mus \'al\-tal ja\-va\-solt \'uj kon\-fi\-gu\-r\'a\-ci\-\'o nem tesz e\-le\-get en\-nek a
fel\-t\'e\-tel\-nek, \'ugy azt nem fo\-gad\-juk el, ha\-nem m\'eg \'u\-jab\-bat
ge\-ne\-r\'a\-lunk he\-lyet\-te. A f\'a\-zi\-s\'at\-me\-ne\-tek\-n\'el szo\-k\'a\-sos m\'o\-don a k\'et
f\'a\-zis\-ra jel\-lem\-z\H o rend\-pa\-ra\-m\'e\-ter \'er\-t\'ek k\"o\-z\"ott r\"og\-z\'{\i}t\-ve a
rend\-pa\-ra\-m\'e\-ter \'at\-la\-g\'er\-t\'e\-k\'et k\'et\-f\'a\-zi\-s\'u rend\-szer j\"on l\'et\-re.

Az e\-gyes f\'a\-zi\-sok k\"o\-z\"ot\-ti bu\-bo\-r\'ek\-fal a sza\-ba\-de\-ner\-gi\-a
mi\-ni\-mum\-fel\-t\'e\-te\-le mi\-att me\-r\H o\-le\-ges a hossz\'u i\-r\'any\-ra. A bu\-bo\-r\'ek\-fal
e\-l\'eg v\'e\-kony (a 192-s r\'acs\-ki\-ter\-je\-d\'es\-hez k\'e\-pest), a\-zon\-ban az e\-gyes
kon\-fi\-gu\-r\'a\-ci\-\'ok e\-se\-t\'e\-ben m\'as he\-lye\-ken he\-lyez\-ked\-het el, \'{\i}gy a
sta\-tisz\-ti\-kus \'at\-la\-go\-l\'as\-hoz a falp\-ro\-fil meg\-fe\-le\-l\H o el\-to\-l\'a\-s\'a\-ra is
sz\"uk\-s\'eg van. Ez egy\-faj\-ta kor\-re\-l\'a\-ci\-\'o ma\-xi\-ma\-li\-z\'a\-l\'a\-s\'a\-val
t\"or\-t\'e\-nik. Az e\-gyes m\'e\-r\'e\-sek e\-red\-m\'e\-nye\-k\'epp hosszi\-r\'any\-ban 192
a\-dat-cso\-port a\-d\'o\-dik; e\-ze\-ket az a\-dat\-hal\-ma\-zo\-kat -- a pe\-ri\-o\-di\-kus
ha\-t\'ar\-fel\-t\'e\-te\-lek fi\-gye\-lem\-be\-v\'e\-te\-l\'e\-vel -- egy\-m\'as\-hoz k\'e\-pest
el\-tol\-hat\-juk. Az (1) \'es a (2) a\-dat\-hal\-maz j\'ol kor\-re\-l\'alt, ha ki\-csi a
\beq
\sum_{i=1}^{192} \frac{(A_1(i) - A_2(i))^2}{\left(\sigma_1^2(i) +
\sigma_2^2(i)\right)^2}
\enq
\"osszeg, mely\-ben $A_1(i)$ \'es $A_2(i)$ az (1) il\-let\-ve a (2) min\-ta $i.$
met\-sze\-t\'e\-ben m\'ert \'er\-t\'ek, $\sigma_{1,2}(i)$ pe\-dig en\-nek hi\-b\'a\-ja.

A kor\-rek\-ci\-\'o u\-t\'an a falp\-ro\-fil\-ra a \ref{wall} \'ab\-ra a\-d\'o\-dik:

\bef[ht]
\bc
\epsfig{file=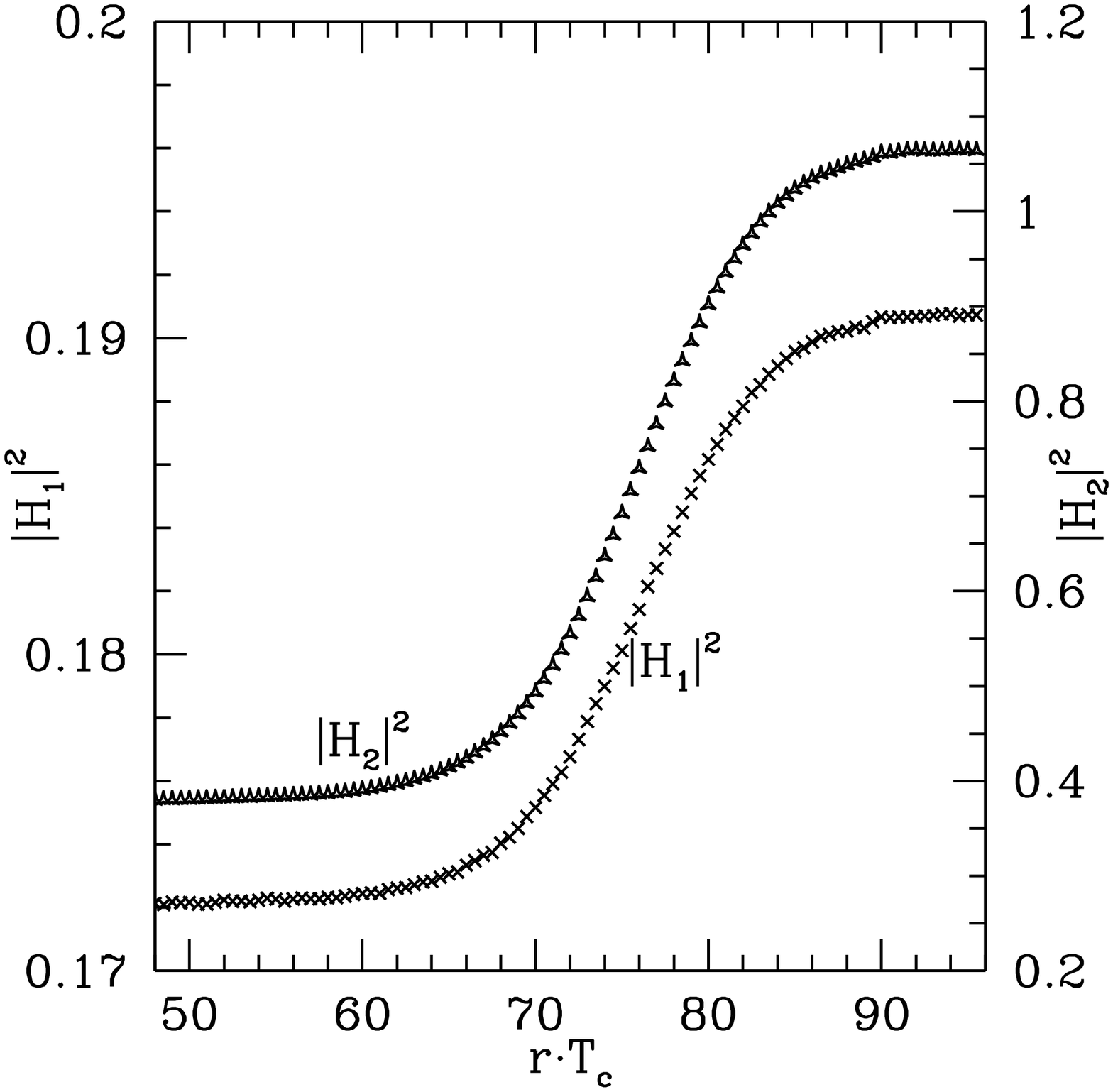,width=8cm}
\caption{A $2\cdot12^2\cdot192$ szi\-mu\-l\'a\-ci\-\'o\-b\'ol a\-d\'o\-d\'o
bu\-bo\-r\'ek\-fal-pro\-fil \label{wall}}
\ec
\enf

A fal\-vas\-tag\-s\'ag me\-g\'al\-la\-p\'{\i}\-t\'a\-sa c\'el\-j\'a\-b\'ol tan\-gens hi\-per\-bo\-li\-kusz
f\"ugg\-v\'eny il\-leszt\-he\-t\H o a \ref{wall} \'ab\-ra a\-da\-ta\-i\-ra; ez mind\-k\'et Higgs
e\-se\-t\'en i\-gen pon\-tos\-san fe\-di a m\'ert pon\-to\-kat. Az il\-lesz\-ten\-d\H o
f\"ugg\-v\'enyt
\beq
a_1 + a_2 * \mrm{th} \, \frac{x - x_0}{L_w/2}
\enq
a\-lak\-ba \'{\i}r\-va az $L_w$ fal\-vas\-tag\-s\'ag\-ra a leg\-jobb il\-lesz\-t\'es\-b\H ol
mind\-k\'et e\-set\-ben $L_w = (14.4 \pm 0.1) / T_c$ a\-d\'o\-dik. Ez az
e\-red\-m\'eny j\'ol e\-gye\-z\'es\-ben \'all a (\ref{lw}) egy\-hu\-rok-ren\-d\H u
per\-tur\-ba\-t\'{\i}v e\-red\-m\'ennyel. Mi\-vel a $B$ pa\-ra\-m\'e\-ter nem t\H u\-nik el,
jel\-leg\-ze\-tes \emph{roughening} t\'{\i}\-pu\-s\'u f\'a\-zi\-s\'at\-me\-net j\'at\-sz\'o\-dik le.

A k\'et Higgs-t\'er egy\-m\'as\-hoz na\-gyon ha\-son\-l\'o m\'o\-don v\'al\-to\-zik; a
ket\-t\H o k\"o\-z\"ott na\-gyon j\'o k\"o\-ze\-l\'{\i}\-t\'es\-sel li\-ne\-\'a\-ris kap\-cso\-lat \'all
fenn. $|H_1|^2$ f\"ugg\-v\'e\-ny\'e\-ben \'ab\-r\'a\-zol\-va $|H_2|^2$-t a
\ref{tanbeta} \'ab\-r\'at kap\-juk:

\bef[htb]
\bc
\epsfig{file=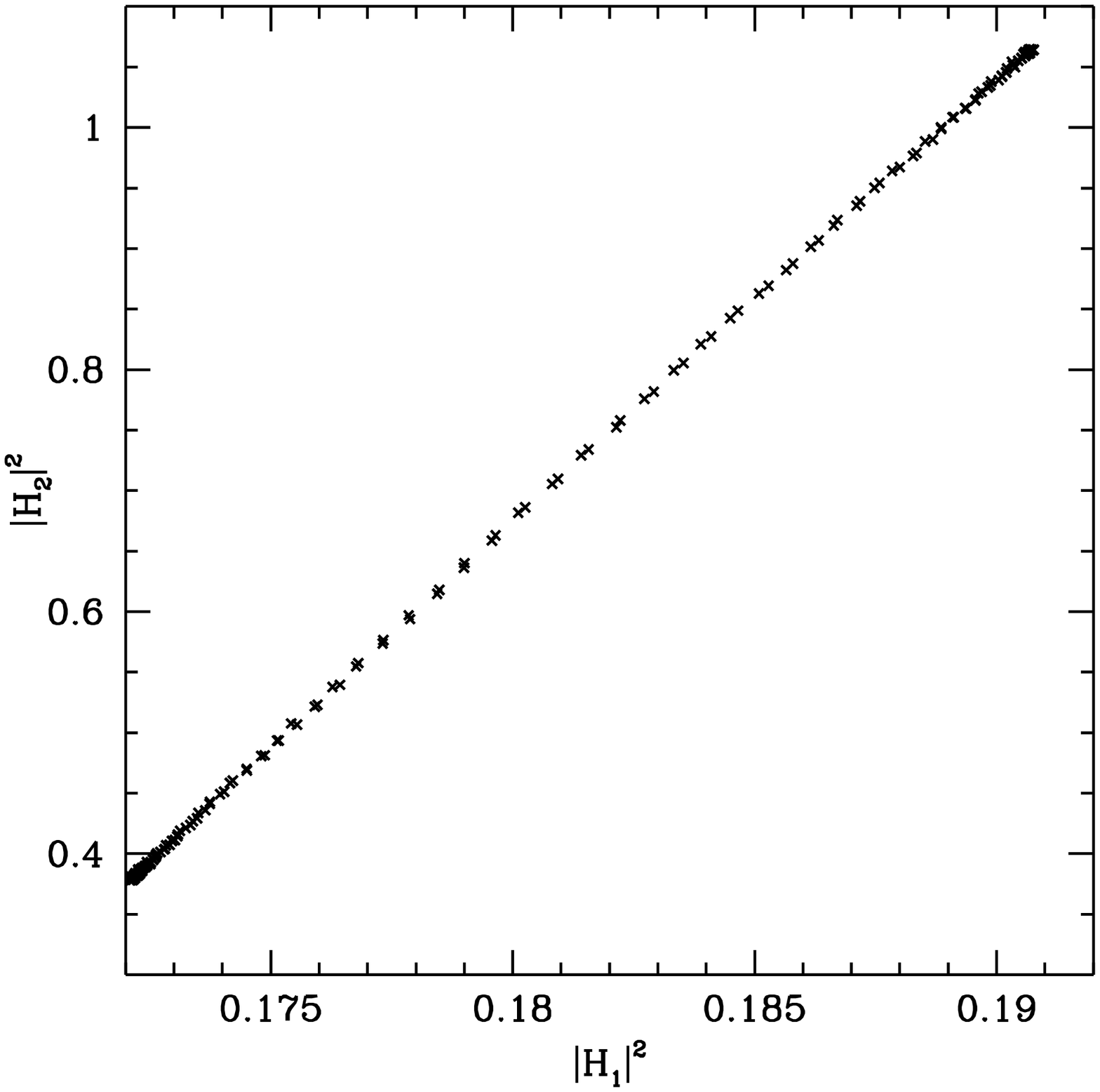,width=8cm}
\caption{A k\'et Higgs-t\'er kap\-cso\-la\-ta i\-gen j\'o k\"o\-ze\-l\'{\i}\-t\'es\-sel
li\-ne\-\'a\-ris \label{tanbeta}}
\ec
\enf

A li\-ne\-\'a\-ris kap\-cso\-lat\-t\'ol va\-l\'o el\-t\'e\-r\'es ki\-csi, \'am szig\-ni\-fi\-k\'ans; a
g\"or\-be al\-s\'o \'es fel\-s\H o v\'e\-g\'en il\-lesz\-tett e\-gye\-ne\-sek me\-re\-dek\-s\'e\-ge
kis\-s\'e el\-t\'e\-r\H o:
\beqar
\mrm{tg}^2\, \beta (\mrm{szimm}) &=& 38.00(27) \nonumber \\
\mrm{tg}^2\, \beta (\mrm{s\acute{e}rt}) &=& 35.43(27).
\enqar
E\-zek\-b\H ol a ba\-ri\-o\-ge\-n\'e\-zis szem\-pont\-j\'a\-b\'ol fon\-tos $\beta$ pa\-ra\-m\'e\-ter
k\'et f\'a\-zis k\"oz\-ti k\"u\-l\"onb\-s\'e\-ge meg\-ha\-t\'a\-roz\-ha\-t\'o:
\beq
\Delta \beta = 0.0061 \pm 0.0003.
\enq
U\-gya\-ne\-zen mennyi\-s\'eg per\-tur\-b\'a\-ci\-\'o\-sz\'a\-m\'{\i}\-t\'as\-sal meg\-ha\-t\'a\-ro\-zott
\'er\-t\'e\-ke \cite{mor, john}
\beq
\Delta \beta = 0.0046 \pm 0.0010,
\enq
a nem\-per\-tur\-ba\-t\'{\i}v e\-red\-m\'e\-nyek\-kel el\-fo\-gad\-ha\-t\'o e\-gye\-z\'es\-ben.

A szi\-mu\-l\'a\-ci\-\'ot $2 * \{L^2 = 8^2, 16^2\} * 192$-s r\'a\-cso\-kon is
v\'eg\-re\-hajt\-va meg\-ha\-t\'a\-roz\-ha\-t\'o a fal sz\'e\-les\-s\'e\-g\'e\-nek $L$-t\H ol va\-l\'o
f\"ug\-g\'e\-se. Er\-re az i\-ro\-da\-lom \cite{jas84} \'al\-tal j\'o\-solt
\beq
L_w = \left[A + B \cdot \log(aLT_c) \right] / T_c
\enq
f\"ugg\-v\'eny j\'ol il\-leszt\-he\-t\H o; $A=10.8 \pm 0.1$, $B = 2.1 \pm 0.1$
a\-d\'o\-dik \cite{cikk2}.

\section{A koz\-mo\-l\'o\-gi\-a\-i\-lag re\-le\-v\'ans pa\-ra\-m\'e\-ter\-tar\-to\-m\'any \label{kre}}
A r\'acsszi\-mu\-l\'a\-ci\-\'ok e\-gyik c\'el\-ja az len\-ne, hogy meg\-ha\-t\'a\-roz\-zuk a
pa\-ra\-m\'e\-ter\-t\'er a\-zon r\'e\-sz\'et, mely e\-le\-get tesz a ba\-ri\-o\-ge\-n\'e\-zis
(\ref{vtpert}) fel\-t\'e\-te\-l\'e\-nek.
\fancyhead[CO]{\hst{\thesection \quad A koz\-mol\'o\-gi\-a\-i\-lag re\-lev\'ans
pa\-ram\'e\-ter\-tar\-tom\'any}}
No\-ha az eh\-hez e\-len\-ged\-he\-tet\-le\-n\"ul sz\"uk\-s\'e\-ge kon\-ti\-nu\-um ha\-t\'a\-r\'at\-me\-net
j\'o k\"o\-ze\-l\'{\i}\-t\'es\-sel v\'eg\-re\-hajt\-ha\-t\'o, a pa\-ra\-m\'e\-ter\-t\'er r\'esz\-le\-tes
fel\-t\'er\-k\'e\-pe\-z\'e\-se t\'ul nagy g\'e\-pi\-d\H o-i\-g\'enyt t\'a\-masz\-ta\-na. E\-z\'ert
in\-di\-rekt m\'od\-szer\-hez kell fo\-lya\-mod\-nunk, mely ab\-b\'ol \'all, hogy k\'et
mennyi\-s\'eg, $v/T_c$ \'es $T_c/m_W$ e\-se\-t\'e\-ben meg\-ha\-t\'a\-roz\-zuk, mek\-ko\-ra
hi\-b\'a\-val be\-cs\"ul\-het\-j\"uk meg a kon\-ti\-nu\-um-li\-meszt, \'es becs\-l\'e\-s\"un\-ket
\"ossze\-vet\-j\"uk a per\-tur\-b\'a\-ci\-\'o\-sz\'a\-m\'{\i}\-t\'as e\-red\-m\'e\-ny\'e\-vel; ezt mu\-tat\-ja
a \ref{kozm1} \'ab\-ra.

\bef[ht]
\bc
\epsfig{file=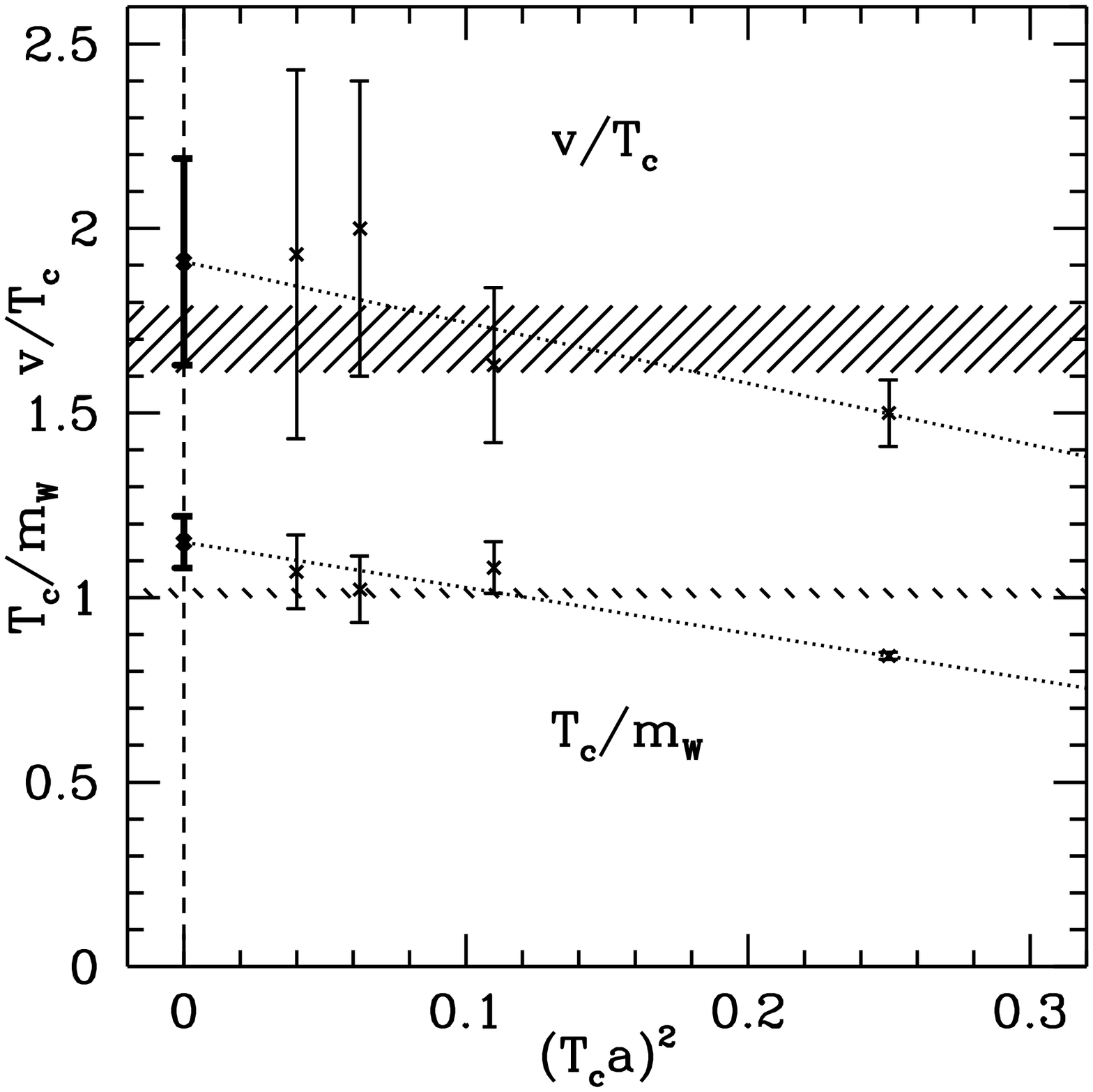,width=8cm}
\caption{A Higgs-t\'er nor\-ma\-li\-z\'alt ug\-r\'a\-sa \'es a nor\-ma\-li\-z\'alt
kri\-ti\-kus h\H o\-m\'er\-s\'ek\-let \label{kozm1}}
\ec
\enf

A pon\-tok a r\'acsszi\-mu\-l\'a\-ci\-\'ok\-ban m\'ert e\-red\-m\'e\-nyek; e\-zek\-nek hi\-b\'a\-i
leg\-na\-gyobb r\'eszt a kons\-tans fi\-zi\-ka vo\-na\-l\'a\-t\'ol va\-l\'o el\-t\'e\-r\'es\-b\H ol --
el\-s\H o\-sor\-ban $m_h$ nagy hi\-b\'a\-j\'a\-b\'ol -- fa\-kad\-nak. A $v/T_c$ pa\-ra\-m\'e\-ter
er\-re i\-gen \'er\-z\'e\-keny, $T_c/m_W$ l\'e\-nye\-ge\-sen ke\-v\'es\-b\'e. A 4 r\'acs\-pont\-ra
il\-lesz\-tett e\-gye\-nes $(T_c a)^2 = 0$-val va\-l\'o met\-sze\-te ad\-ja meg a
kon\-ti\-nu\-um-li\-meszt -- $v/T_c$ e\-se\-t\'e\-ben en\-nek hi\-b\'a\-ja t\'ul nagy a
koz\-mo\-l\'o\-gi\-a\-i k\"o\-vet\-kez\-te\-t\'e\-sek le\-vo\-n\'a\-s\'a\-hoz.

\'Igy a sa\-t\'{\i}\-ro\-zott per\-tur\-ba\-t\'{\i}v j\'os\-la\-tok\-kal ve\-tet\-t\"uk \"ossze az
e\-red\-m\'e\-nye\-ket; az eh\-hez hasz\-n\'alt per\-tur\-b\'a\-ci\-\'o\-sz\'a\-m\'{\i}\-t\'as egy
egy\-hu\-rok-ren\-d\H u el\-j\'a\-r\'as, mely nem \'e\-p\'{\i}t a ma\-gas h\H o\-m\'er\-s\'ek\-le\-t\H u
sor\-fej\-t\'es\-re, ha\-nem a r\'acsszi\-mu\-l\'a\-ci\-\'ok\-hoz i\-ga\-zo\-dik: a v\'e\-ges
re\-nor\-m\'a\-l\'a\-si ef\-fek\-tu\-so\-kat \'ugy ve\-szi fi\-gye\-lem\-be, hogy a
r\'acsszi\-mu\-l\'a\-ci\-\'ok\-ban m\'ert $T=0$ spekt\-rum\-mal mi\-n\'el t\"o\-k\'e\-le\-te\-sebb
e\-gye\-z\'est mu\-tas\-son \cite{jak}. \'Igy a k\'et meg\-k\"o\-ze\-l\'{\i}\-t\'es
k\"o\-z\"ott e\-l\'eg j\'o e\-gye\-z\'es va\-l\'o\-sul meg; a sk\'a\-l\'a\-z\'o tar\-to\-m\'a\-nyon
e\-set\-leg k\'{\i}\-v\"ul e\-s\H o $L_t=2$ a\-da\-tok ki\-ha\-gy\'a\-s\'a\-val ez to\-v\'abb
ja\-v\'{\i}t\-ha\-t\'o. A fen\-ti \'ab\-r\'a\-r\'ol l\'at\-szik, hogy a r\'a\-cse\-red\-m\'e\-nyek
kon\-ti\-nu\-um-li\-me\-sze mind\-k\'et e\-set\-ben na\-gyobb, mint a per\-tur\-ba\-t\'{\i}v \'u\-ton
ka\-pott e\-red\-m\'eny, \'{\i}gy a per\-tur\-ba\-t\'{\i}v e\-red\-m\'eny r\'acsszi\-mu\-l\'a\-ci\-\'os
e\-red\-m\'e\-nyek\-kel va\-l\'o \"ossze\-ve\-t\'e\-s\'e\-b\H ol egy kb.\ 14\%-os kor\-rek\-ci\-\'os
t\'e\-nye\-z\H ot kap.\label{14szaz}

A fen\-ti e\-gye\-z\'es a\-lap\-j\'an a per\-tur\-ba\-t\'{\i}v meg\-k\"o\-ze\-l\'{\i}\-t\'es
fel\-hasz\-n\'al\-ha\-t\'o ar\-ra, hogy a koz\-mo\-l\'o\-gi\-a\-i\-lag re\-le\-v\'ans
pa\-ra\-m\'e\-ter\-tar\-to\-m\'anyt fel\-t\'er\-k\'e\-pez\-z\"uk. (Az eh\-hez hasz\-n\'alt
per\-tur\-ba\-t\'{\i}v meg\-k\"o\-ze\-l\'{\i}\-t\'es a tel\-jes MSSM-re \'e\-p\"ul, te\-h\'at a
fer\-mi\-o\-no\-kat is fi\-gye\-lem\-be ve\-szi. A sz\'a\-mo\-l\'a\-sok\-ban $m_A=500$ GeV.) A
leg\-k\"onnyebb Higgs -- jobb\-ke\-zes (k\"onnyebb) stop s\'{\i}\-kon ezt k\'et\-faj\-ta
g\"or\-be fog\-ja jel\-le\-mez\-ni; az e\-gyi\-ket a $T=0$-hoz tar\-to\-z\'o ma\-xi\-m\'a\-lis
k\"onny\H u Higgs-t\"o\-meg ad\-ja -- e\-zek a g\"or\-b\'ek k\"o\-zel v\'{\i}z\-szin\-te\-sek --,
a  m\'a\-si\-kat a $v/T_c =1$ fel\-t\'e\-tel je\-l\"o\-li ki. A koz\-mo\-l\'o\-gi\-a\-i\-lag
\'er\-de\-kes tar\-to\-m\'any a k\'et g\"or\-be a\-latt he\-lyez\-ke\-dik el.

\bef[ht]
\bc
\epsfig{file=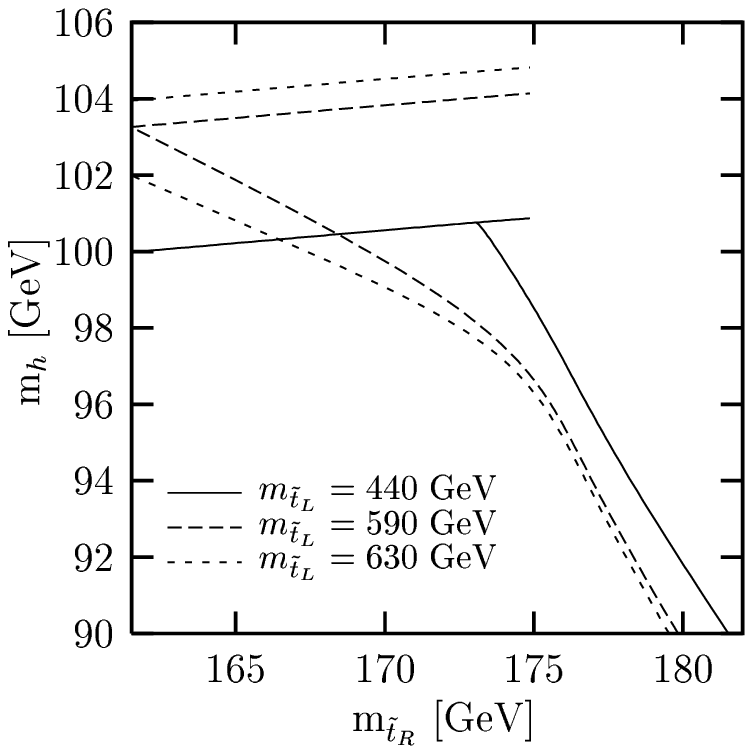,width=8cm}
\caption{A koz\-mo\-l\'o\-gi\-a\-i\-lag re\-le\-v\'ans $v/T_c > 1$ tar\-to\-m\'any
meg\-ha\-t\'a\-ro\-z\'a\-sa \label{kozm2}}
\ec
\enf

Az $m_Q$ pa\-ra\-m\'e\-ter v\'al\-toz\-ta\-t\'a\-s\'a\-val a bal\-ke\-zes Higgs t\"o\-me\-ge
v\'al\-toz\-tat\-ha\-t\'o; ha $m_Q$ n\H o, $m_{\tilde t_L}$ cs\"ok\-ken; u\-gya\-nek\-kor a
leg\-k\"onnyebb Higgs t\"o\-me\-g\'e\-re a\-d\'o\-d\'o ma\-xi\-mum\-kor\-l\'at fel\-jebb to\-l\'o\-dik.
Ha a Higgs-t\"o\-meg \'er\-t\'e\-ke n\H o, a f\'a\-zi\-s\'at\-me\-net gyen\-g\"u\-l\'e\-se az
$-m_U^2$ pa\-ra\-m\'e\-ter n\"o\-ve\-l\'e\-s\'e\-vel ke\-r\"ul\-he\-t\H o el,
m\'as sz\'o\-val a jobb\-ke\-zes stop-t\'er t\"o\-me\-g\'e\-nek cs\"ok\-ken\-t\'e\-s\'e\-vel.
E\-sze\-rint $m_Q$ n\"o\-ve\-l\'e\-se ha\-t\'a\-s\'a\-ra a $v/T_c$ g\"or\-be bal\-ra to\-l\'o\-dik.

Mi az a ma\-xi\-m\'a\-lis Higgs-t\"o\-meg, mely be\-le\-f\'er m\'eg a koz\-mo\-l\'o\-gi\-a\-i\-lag
re\-le\-v\'ans tar\-to\-m\'any\-ba? A k\'et kor\-l\'at \'al\-tal meg\-ha\-t\'a\-ro\-zott
met\-sz\'es\-pont, mely $m_{\tilde t_L} = 440$ GeV e\-se\-t\'en ki\-csit 100 GeV
f\"o\-l\"ott van, $m_{\tilde t_L} = 590$ GeV e\-se\-t\'en pe\-dig 103 GeV
k\"o\-r\"ul. A har\-ma\-dik g\"or\-b\'e\-r\H ol le\-ol\-vas\-ha\-t\'o, hogy b\'ar\-hogy is
v\'a\-laszt\-juk meg $m_Q$ \'er\-t\'e\-k\'et, a\-lig\-ha me\-he\-t\"unk 103 GeV f\"o\-l\'e.
R\'esz\-le\-te\-sebb sz\'a\-m\'{\i}\-t\'a\-sok a\-lap\-j\'an $m_{\tilde t_L} = 560$ GeV
mel\-lett le\-het $m_h$ a leg\-na\-gyobb; $m_h \approx 103$ GeV.

A fen\-ti el\-j\'a\-r\'as az e\-l\H obb em\-l\'{\i}\-tett 14\%-os kor\-rek\-ci\-\'ot fi\-gye\-lem\-be
ve\-szi; a \ref{kozm1} di\-ag\-ram \'al\-tal a r\'acs- \'es per\-tur\-ba\-t\'{\i}v
k\"o\-ze\-l\'{\i}\-t\'es k\"o\-z\"ott l\'e\-te\-s\'{\i}\-tett kap\-cso\-lat a\-zon\-ban je\-len\-t\'e\-keny
hi\-ba\-for\-r\'as, mely az egy- \'es k\'et\-hu\-rok rend k\"oz\-ti el\-t\'e\-r\'est sem
ve\-szi fi\-gye\-lem\-be \cite{cikk2}. \'Igy a fen\-ti e\-red\-m\'eny
kor\-rek\-teb\-ben
\beq
m_h = 103 \pm 4\ \mrm{GeV}. \label{mha}
\enq

A k\"u\-l\"on e c\'el\-ra ki\-fej\-lesz\-tett egy\-hu\-rok szin\-t\H u per\-tur\-ba\-t\'{\i}v
m\'od\-szer\-rel meg\-fe\-jelt n\'egy\-di\-men\-zi\-\'os r\'acsszi\-mu\-l\'a\-ci\-\'ok\-ra a\-la\-po\-zott
meg\-k\"o\-ze\-l\'{\i}\-t\'es te\-h\'at a ko\-r\'ab\-bi, 3-di\-men\-zi\-\'os e\-red\-m\'e\-nyek\-kel
\"ossze\-e\-gyez\-tet\-he\-t\H o j\'os\-la\-tot ad. A j\'os\-lat sze\-rint a je\-len\-le\-gi
szu\-per\-szim\-met\-ri\-kus Higgs t\"o\-meg\-kor\-l\'a\-tok mel\-lett l\'e\-te\-zik a
pa\-ra\-m\'e\-ter\-t\'er\-nek o\-lyan tar\-to\-m\'a\-nya, a\-hol a ba\-ri\-o\-ge\-n\'e\-zis fel\-t\'e\-te\-le\-it
ki\-e\-l\'e\-g\'{\i}\-t\H o e\-r\H os el\-s\H o\-ren\-d\H u f\'a\-zi\-s\'at\-me\-net j\'at\-sz\'o\-dik le.

A kor\-l\'at a\-zon\-ban na\-gyon a\-la\-csony; a nagy r\'e\-szecs\-ke\-gyor\-s\'{\i}\-t\'ok\-ban
ha\-ma\-ro\-san ki\-de\-r\"ul, l\'e\-te\-zik-e az (\ref{mha}) fe\-t\'e\-tel\-nek e\-le\-get te\-v\H o
MSSM Higgs-r\'e\-szecs\-ke. K\"onnyen ki\-de\-r\"ul\-het, hogy nem -- a\-mi an\-nak
je\-le, hogy a fen\-ti MSSM mo\-dellt to\-v\'abb kell fi\-no\-m\'{\i}\-ta\-nunk; ez a
szu\-per\-szim\-met\-ri\-a ny\'uj\-tot\-ta b\H o\-s\'e\-ges ke\-re\-ten be\-l\"ul le\-het\-s\'e\-ges lesz.
A\-mennyi\-ben a szu\-per\-szim\-met\-ri\-kus ba\-ri\-o\-ge\-n\'e\-zis el\-m\'e\-let \'al\-tal j\'o\-solt
Higgs fel\-s\H o t\"o\-meg\-kor\-l\'at a\-latt a r\'e\-szecs\-ke\-gyor\-s\'{\i}\-t\'ok\-ban
meg\-ta\-l\'al\-j\'ak a leg\-k\"o\-nyebb Higgs-r\'e\-szecs\-k\'et, az a ba\-ri\-o\-ge\-n\'e\-zis
mo\-del\-lek nagy si\-ke\-re lesz, mely a koz\-mo\-l\'o\-gi\-a \'es a r\'e\-szecs\-ke\-fi\-zi\-ka
szin\-t\'e\-zi\-se\-k\'ent l\'et\-re\-j\"ott r\'e\-szecs\-ke--aszt\-ro\-fi\-zi\-ka
\'e\-let\-k\'e\-pes\-s\'e\-g\'e\-nek \'u\-jabb \'e\-kes bi\-zo\-ny\'{\i}\-t\'e\-k\'a\-ul szol\-g\'al\-hat.

A\-zon\-ban meg\-v\'a\-la\-szo\-lat\-lan k\'er\-d\'es b\H o\-ven ma\-rad ak\-kor is, ha a
ba\-ri\-o\-ge\-n\'e\-zis prob\-l\'e\-m\'a\-j\'at a szu\-per\-szim\-met\-ri\-kus mo\-del\-lek
tisz\-t\'az\-z\'ak. A stan\-dard mo\-dell\-hez k\'e\-pest sok \'uj pa\-ra\-m\'e\-ter
sze\-re\-pel az MSSM-ben -- a bo\-nyo\-lul\-tabb szu\-per\-szim\-met\-ri\-kus
el\-m\'e\-le\-tek\-ben pe\-dig m\'eg t\"obb. \emph{A pri\-o\-ri} fe\-let\-t\'ebb
va\-l\'o\-sz\'{\i}\-n\H ut\-len\-nek t\H u\-nik, hogy e\-zek a pa\-ra\-m\'e\-te\-rek \'epp o\-lyan
\'er\-t\'e\-k\H u\-ek, hogy a ba\-ri\-o\-ge\-n\'e\-zis fel\-t\'e\-te\-le\-i\-nek e\-le\-get te\-gye\-nek.
V\'e\-let\-len egy\-be\-e\-s\'es, vagy m\'e\-lyebb fi\-zi\-ka\-i ok h\'u\-z\'o\-dik a\-m\"o\-g\"ott,
hogy (ha) ez a mo\-dell k\'e\-pes sz\'a\-mot ad\-ni a vi\-l\'a\-ge\-gye\-tem
ba\-ri\-on--an\-ti\-ba\-ri\-on szim\-met\-ri\-\'a\-j\'a\-r\'ol? Nyil\-v\'an fi\-zi\-ka\-i \'er\-vek\-kel
k\'{\i}\-v\'an\-juk a\-l\'a\-t\'a\-masz\-ta\-ni az egy\-be\-e\-s\'est -- \'es b\'ar eb\-b\H ol a
prog\-ram\-b\'ol ma c\'el\-j\'an k\'{\i}\-v\"ul i\-gen ke\-v\'es l\'at\-szik, a\-lig\-ha
k\'et\-s\'e\-ges, hogy a mik\-ro\-vi\-l\'ag m\'e\-lyebb me\-g\'er\-t\'e\-s\'e\-nek \'ut\-j\'an fon\-tos
m\'er\-f\"old\-k\H o lesz.

\chapter*{\"Ossze\-fog\-lal\'as}
\fancyhead[CE]{\hst{\"Ossze\-fog\-lal\'as}}
\addcontentsline{toc}{chapter}{\"Ossze\-fog\-lal\'as}
Az e\-lekt\-ro\-gyen\-ge f\'a\-zi\-s\'at\-me\-net vizs\-g\'a\-la\-t\'a\-nak e\-gyik f\H o
mo\-ti\-v\'a\-ci\-\'o\-j\'at az u\-ni\-ver\-zum\-ban meg\-fi\-gyel\-he\-t\H o ba\-ri\-on--an\-ti\-ba\-ri\-on
a\-szim\-met\-ri\-a vizs\-g\'a\-la\-t\'a\-nak le\-he\-t\H o\-s\'e\-ge ad\-ja. Kva\-li\-ta\-t\'{\i}v szin\-ten a
ba\-ri\-o\-ge\-n\'e\-zis\-hez sz\"uk\-s\'e\-ges fel\-t\'e\-te\-lek a stan\-dard mo\-dell\-ben is
meg\-van\-nak, \'{\i}gy elv\-ben le\-he\-t\H o\-s\'eg van a k\'er\-d\'es k\'{\i}\-s\'er\-le\-ti\-leg
a\-l\'a\-t\'a\-masz\-tott el\-m\'e\-le\-ti mo\-del\-len a\-la\-pu\-l\'o meg\-v\'a\-la\-szo\-l\'a\-s\'a\-ra.

A kvan\-ti\-ta\-t\'{\i}v vizs\-g\'a\-lat per\-tur\-ba\-t\'{\i}v \'es nem\-per\-tur\-ba\-t\'{\i}v
m\'od\-sze\-rek\-kel le\-het\-s\'e\-ges. A ma\-gas h\H o\-m\'er\-s\'ek\-le\-t\H u szim\-met\-ri\-kus
f\'a\-zis\-ban a per\-tur\-ba\-t\'{\i}v meg\-k\"o\-ze\-l\'{\i}\-t\'es \emph{a pri\-o\-ri} nem
meg\-b\'{\i}z\-ha\-t\'o, a\-zon\-ban van re\-m\'eny ar\-ra, hogy bi\-zo\-nyos pa\-ra\-m\'e\-te\-rek
(Higgs-t\"o\-meg) mel\-lett a per\-tur\-ba\-t\'{\i}v \'es nem\-per\-tur\-ba\-t\'{\i}v e\-red\-m\'e\-nyek
\"ossze\-e\-gyez\-tet\-he\-t\H o\-ek le\-gye\-nek. Eh\-hez a k\'et meg\-k\"o\-ze\-l\'{\i}\-t\'es a\-la\-pos
\"ossze\-ha\-son\-l\'{\i}\-t\'a\-s\'a\-ra van sz\"uk\-s\'eg, mely a nem\-per\-tur\-ba\-t\'{\i}v m\'od\-szer
ke\-re\-t\'e\-ben de\-fi\-ni\-\'alt r\'a\-cs\'al\-lan\-d\'o a\-lap\-j\'a\-ul szol\-g\'a\-l\'o szta\-ti\-kus
kvark po\-ten\-ci\-\'al per\-tur\-ba\-t\'{\i}v ki\-sz\'a\-m\'{\i}\-t\'a\-sa r\'e\-v\'en le\-het\-s\'e\-ges. A
po\-ten\-ci\-\'alt a n\'egy\-di\-men\-zi\-\'os szi\-mu\-l\'a\-ci\-\'ok a\-lap\-j\'a\-ul szol\-g\'a\-l\'o
SU(2)--Higgs-mo\-dell\-ben kell meg\-ha\-t\'a\-roz\-ni.

A dol\-goz\-ta\-ban Feyn\-man-m\'er\-t\'ek\-ben ki\-sz\'a\-m\'{\i}\-tot\-tam az egy\-hu\-rok-ren\-d\H u
im\-pul\-zus\-t\'er\-be\-li po\-ten\-ci\-\'alt. A csa\-to\-l\'a\-si \'al\-lan\-d\'ok k\"oz\-ti
kap\-cso\-lat\-hoz ezt (nu\-me\-ri\-ku\-san) Fo\-u\-ri\-er-transz\-for\-m\'al\-ni, majd
dif\-fe\-ren\-ci\-\'al\-ni kell. A
po\-ten\-ci\-\'al\-b\'ol t\"obb\-f\'e\-le m\'o\-don is de\-fi\-ni\-\'al\-ha\-t\'o a csa\-to\-l\'a\-si
\'al\-lan\-d\'o; mi\-vel c\'e\-lom a r\'acsszi\-mu\-l\'a\-ci\-\'os e\-red\-m\'e\-nyek\-kel va\-l\'o
\"ossze\-ve\-t\'es, c\'el\-sze\-r\H u mi\-n\'el job\-ban ra\-gasz\-kod\-ni az ott al\-kal\-ma\-zott
m\'od\-szer\-hez. Az \'{\i}gy de\-fi\-ni\-\'alt kap\-cso\-lat r\'e\-v\'en e\-li\-mi\-n\'al\-ha\-\'o a
per\-tur\-ba\-t\'{\i}v \'es nem\-per\-tur\-ba\-t\'{\i}v e\-red\-m\'e\-nyek \"ossze\-ve\-t\'e\-se\-kor az
el\-t\'e\-r\H o r\'a\-cs\'al\-lan\-d\'o-de\-fi\-n\'{\i}\-ci\-\'ok mi\-att fel\-l\'e\-p\H o hi\-ba\-for\-r\'ast. A
f\'a\-zi\-s\'at\-me\-net\-re jel\-lem\-z\H o ter\-mo\-di\-na\-mi\-ka\-i mennyi\-s\'e\-gek vizs\-g\'a\-la\-ta azt
mu\-tat\-ja, hogy a\-la\-csony Higgs-t\"o\-me\-gek e\-se\-t\'en a per\-tur\-ba\-t\'{\i}v
e\-red\-m\'e\-nyek e\-l\'eg j\'ol e\-gyez\-nek a n\'egy\-di\-men\-zi\-\'os (\'es a di\-men\-zi\-\'os
re\-duk\-ci\-\'o\-val ka\-pott h\'a\-rom\-di\-men\-zi\-\'os) e\-red\-m\'e\-nyek\-kel. A Higgs-t\"o\-me\-get
n\"o\-vel\-ve a per\-tur\-b\'a\-ci\-\'o\-sz\'a\-m\'{\i}\-t\'as el\-rom\-lik: a nem\-per\-tur\-ba\-t\'{\i}v
m\'od\-sze\-rek\-kel meg\-j\'o\-solt f\'a\-zi\-s\'at\-me\-ne\-ti v\'eg\-pont per\-tur\-ba\-t\'{\i}v
meg\-k\"o\-ze\-l\'{\i}\-t\'es\-ben nem is l\'e\-te\-zik.

A csa\-to\-l\'a\-si \'al\-lan\-d\'ok k\"oz\-ti kap\-cso\-lat a f\'a\-zi\-s\'at\-me\-ne\-ti v\'eg\-pont
pon\-to\-sabb meg\-ha\-t\'a\-ro\-z\'a\-s\'at is le\-he\-t\H o\-v\'e te\-szi; a tel\-jes stan\-dard
mo\-dell\-re a\-dott j\'os\-lat $72.1 \pm 1.4$ GeV. Ez l\'e\-nye\-ge\-sen ki\-sebb, mint
a Higgs-r\'e\-szecs\-ke t\"o\-me\-g\'e\-nek k\'{\i}\-s\'er\-le\-ti kor\-l\'at\-ja, \'{\i}gy a stan\-dard
mo\-dell nem ad\-hat sz\'a\-mot a ba\-ri\-o\-ge\-n\'e\-zis\-r\H ol. A fen\-ti vizs\-g\'a\-lat
po\-zi\-t\'{\i}v e\-red\-m\'e\-nye, hogy a per\-tur\-b\'a\-ci\-\'o\-sz\'a\-m\'{\i}\-t\'as a
nem\-per\-tur\-ba\-t\'{\i}\-ve meg\-j\'o\-solt f\'a\-zi\-s\'at\-me\-ne\-ti pont\-t\'ol t\'a\-vol
m\H u\-k\"o\-d\H o\-k\'e\-pes -- ez a bo\-nyo\-lul\-tabb el\-m\'e\-le\-tek\-ben vizs\-g\'alt
e\-lekt\-ro\-gyen\-ge f\'a\-zi\-s\'at\-me\-net so\-r\'an hasz\-nos t\'am\-pont.

A stan\-dard mo\-dell legp\-rag\-ma\-ti\-ku\-sabb ki\-ter\-jesz\-t\'e\-s\'e\-ben, az MSSM-ben a
sz\'a\-mos sza\-bad pa\-ra\-m\'e\-ter le\-he\-t\H o\-s\'e\-get ny\'ujt a ba\-ri\-o\-ge\-n\'e\-zis
ma\-gya\-r\'a\-za\-t\'a\-ra. A dol\-go\-zat\-ban e\-l\H o\-sz\"or egy egy\-sze\-r\H u per\-tur\-ba\-t\'{\i}v
mo\-dellt vizs\-g\'a\-lok, mely j\'ol mu\-tat n\'e\-h\'any \'al\-ta\-l\'a\-nos ten\-den\-ci\-\'at:
a ba\-ri\-o\-ge\-n\'e\-zis\-hez sz\"uk\-s\'e\-ges $\langle \phi \rangle / T_c > 1$
fel\-t\'e\-tel meg\-va\-l\'o\-su\-l\'a\-s\'a\-ra j\'o le\-he\-t\H o\-s\'eg ny\'{\i}\-lik, ha a top kvark
t\"o\-me\-ge na\-gyobb, mint jobb\-ke\-zes szu\-per\-szim\-met\-ri\-kus p\'ar\-j\'a\-\'e.
A\-mennyi\-ben a ket\-t\H o t\"o\-meg\-n\'egy\-zet k\"u\-l\"onb\-s\'e\-g\'e\-re jel\-lem\-z\H o $m_U^2$
pa\-ra\-m\'e\-ter ab\-szo\-l\'ut \'er\-t\'e\-k\'et e\-l\'eg nagy\-nak v\'a\-laszt\-juk,
sz\'{\i}n\-s\'er\-t\H o f\'a\-zi\-s\'at\-me\-net is le\-j\'at\-sz\'od\-hat. A per\-tur\-ba\-t\'{\i}v
vizs\-g\'a\-la\-tok \'{\i}gy h\'a\-rom f\'a\-zis je\-len\-l\'e\-t\'e\-re u\-tal\-nak, \'es nem
z\'ar\-j\'ak ki a ba\-ri\-o\-ge\-n\'e\-zis MSSM-re \'e\-p\"u\-l\H o ma\-gya\-r\'a\-za\-t\'at.

A stan\-dard mo\-dell\-hez ha\-son\-l\'o\-an itt is sz\"uk\-s\'eg van nem\-per\-tur\-ba\-t\'{\i}v
vizs\-g\'a\-la\-tok\-ra. A n\'egy\-di\-men\-zi\-\'os szi\-mu\-l\'a\-ci\-\'ok g\'e\-pi\-d\H o-i\-g\'e\-nye
l\'e\-nye\-ge\-sen na\-gyobb, mint a di\-men\-zi\-\'os re\-duk\-ci\-\'on a\-la\-pu\-l\'o m\'od\-sze\-r\'e,
\'{\i}gy en\-nek ki\-vi\-te\-le\-z\'e\-s\'e\-hez e\-len\-ged\-he\-tet\-len egy ol\-cs\'o
szu\-per\-sz\'a\-m\'{\i}\-t\'o\-g\'ep. E c\'el\-b\'ol \'e\-p\'{\i}\-tet\-t\"uk meg 1998.\ nya\-ra \'es
2000.\ feb\-ru\-\'ar\-ja k\"o\-z\"ott a PC e\-le\-mek\-b\H ol \'al\-l\'o PMS-t, mely
tel\-je\-s\'{\i}t\-m\'eny/\'ar vi\-szony\-ban a r\'acs\-t\'e\-rel\-m\'e\-let\-ben hasz\-n\'alt
szu\-per\-sz\'a\-m\'{\i}\-t\'o\-g\'e\-pek so\-r\'a\-ban az el\-s\H o.

A r\'acsszi\-mu\-l\'a\-ci\-\'ok c\'el\-ja a pa\-ra\-m\'e\-ter\-tar\-to\-m\'any va\-la\-mely -- a
ko\-r\'ab\-bi mun\-k\'ak \'al\-tal fa\-vo\-ri\-z\'alt -- cs\"ucs\-k\'e\-nek fel\-t\'er\-k\'e\-pe\-z\'e\-se.
A v\'e\-ges h\H o\-m\'er\-s\'ek\-le\-t\H u szi\-mu\-l\'a\-ci\-\'ok\-ban meg\-ha\-t\'a\-ro\-zott
f\'a\-zi\-s\'at\-me\-ne\-ti pon\-tok\-ban v\'eg\-re\-haj\-tott z\'e\-rus\-h\H o\-m\'er\-s\'ek\-le\-t\H u
szi\-mu\-l\'a\-ci\-\'ok\-ban meg\-ha\-t\'a\-roz\-ha\-t\'ok a t\"o\-me\-gek, a\-mi \'al\-tal a h\'a\-rom
f\'a\-zis (sz\'{\i}n\-s\'er\-t\H o-, Higgs-, szim\-met\-ri\-kus) ki\-mu\-ta\-t\'a\-sa le\-het\-s\'e\-ges.
A f\'a\-zis\-di\-ag\-ram fel\-v\'e\-te\-le u\-t\'an a\-zon\-ban a koz\-mo\-l\'o\-gi\-a\-i\-lag re\-le\-v\'ans
pa\-ra\-m\'e\-ter\-tar\-to\-m\'any (a stop-t\"o\-meg -- leg\-ki\-sebb Higgs-t\"o\-meg s\'{\i}k
meg\-fe\-le\-l\H o r\'e\-sz\'e\-nek ki\-je\-l\"o\-l\'e\-se) prob\-l\'e\-m\'as, mi\-vel a v\'e\-ges
r\'a\-cs\'al\-lan\-d\'o\-j\'u szi\-mu\-l\'a\-ci\-\'os e\-red\-m\'e\-nyek kon\-ti\-nu\-um-li\-me\-sz\'e\-nek
k\'ep\-z\'e\-se ko\-moly prob\-l\'e\-m\'a\-kat t\'a\-maszt. Ezt a r\'a\-cse\-red\-m\'e\-nyek\-hez
j\'ol il\-lesz\-ke\-d\H o, egy\-hu\-rok-szin\-t\H u per\-tur\-b\'a\-ci\-\'o\-sz\'a\-m\'{\i}\-t\'as
se\-g\'{\i}t\-s\'e\-g\'e\-vel le\-het vizs\-g\'al\-ni. En\-nek e\-red\-m\'e\-nye
azt mu\-tat\-ta, hogy $m_h \leq 103 \pm 4$ GeV sz\"uk\-s\'e\-ges az e\-r\H os
el\-s\H o\-ren\-d\H u e\-lekt\-ro\-gyen\-ge f\'a\-zi\-s\'at\-me\-net meg\-va\-l\'o\-su\-l\'a\-s\'a\-hoz. A ka\-pott
\'er\-t\'ek a ko\-r\'ab\-bi e\-red\-m\'e\-nyek\-kel \"ossz\-hang\-ban azt mu\-tat\-ja, hogy a
Higgs-t\"o\-meg meg\-ha\-t\'a\-ro\-z\'a\-s\'a\-ra i\-r\'a\-nyu\-l\'o k\'{\i}\-s\'er\-le\-ti
e\-r\H o\-fe\-sz\'{\i}\-t\'e\-sek a k\"o\-zel\-j\"o\-v\H o\-ben el\-d\"on\-tik az MSSM-en a\-la\-pu\-l\'o
ba\-ri\-o\-ge\-n\'e\-zis mo\-dell \'e\-let\-k\'e\-pes\-s\'e\-g\'et.

A r\'acsszi\-mu\-l\'a\-ci\-\'ok a ba\-ri\-o\-ge\-n\'e\-zis-mo\-del\-lek\-ben je\-len\-l\'e\-v\H o
bu\-bo\-r\'ek\-fal vas\-tag\-s\'a\-g\'a\-nak meg\-ha\-t\'a\-ro\-z\'a\-s\'a\-ra is al\-kal\-ma\-sak. Az
e\-red\-m\'e\-nyek a per\-tur\-ba\-t\'{\i}v meg\-k\"o\-ze\-l\'{\i}\-t\'es\-sel itt is
\"ossze\-e\-gyez\-tet\-he\-t\H o\-ek.

A n\'egy\-di\-men\-zi\-\'os r\'acsszi\-mu\-l\'a\-ci\-\'os e\-red\-m\'e\-nyek to\-v\'abb\-fej\-lesz\-t\'e\-se
fo\-lya\-mat\-ban van. A\-zon\-ban k\"onnyen el\-k\'ep\-zel\-he\-t\H o, hogy a k\'{\i}\-s\'er\-le\-ti
e\-red\-m\'e\-nyek r\"o\-vid \'u\-ton ki\-z\'ar\-j\'ak az MSSM-be\-li e\-lekt\-ro\-gyen\-ge
f\'a\-zi\-s\'at\-me\-net le\-he\-t\H o\-s\'e\-g\'et -- ek\-kor a szu\-per\-szim\-met\-ri\-a ke\-re\-t\'en
be\-l\"ul \'u\-jabb mo\-del\-lek vizs\-g\'a\-la\-ta k\"o\-vet\-kez\-het. A je\-len dol\-go\-zat\-ban
be\-mu\-ta\-tott m\'od\-sze\-rek e bo\-nyo\-lul\-tabb mo\-del\-lek vizs\-g\'a\-la\-t\'a\-ban is j\'o
ki\-in\-du\-l\'a\-sul szol\-g\'al\-hat\-nak.

\chapter*{K\"osz\"onetnyilv\'an\'{\i}t\'as}
\fancyhead[CE,CO]{}
\addcontentsline{toc}{chapter}{K\"osz\"onetnyilv\'an\'{\i}t\'as}
K\"o\-sz\"o\-ne\-te\-met sze\-ret\-n\'em ki\-fe\-jez\-ni t\'e\-ma\-ve\-ze\-t\H om\-nek, Fo\-dor
Zol\-t\'an\-nak, t\"ob\-b\'e\-ves f\'a\-rad\-s\'a\-gos mun\-k\'a\-j\'a\-\'ert, a sz\'a\-mo\-l\'a\-sok
so\-r\'an fel\-me\-r\"ult ne\-h\'e\-zs\'e\-gek tisz\-t\'a\-z\'a\-s\'a\-\'ert \'es az \'er\-te\-ke\-z\'es
k\'e\-zi\-ra\-t\'a\-nak a\-la\-pos \'at\-n\'e\-z\'e\-s\'e\-\'ert. K\"o\-sz\"o\-n\"om Csi\-kor Fe\-renc
se\-g\'{\i}t\-s\'e\-g\'et, a\-ki p\'ot-t\'e\-ma\-ve\-ze\-t\H om volt az el\-m\'ult \'e\-vek so\-r\'an,
\'es mind ku\-ta\-t\'o\-mun\-k\'am, mind a dok\-to\-ri \'er\-te\-ke\-z\'es me\-g\'{\i}\-r\'a\-sa so\-r\'an
nagy se\-g\'{\i}t\-s\'e\-get ny\'uj\-tott. K\"o\-sz\"o\-n\"om Katz S\'an\-dor\-nak a
r\'acsszi\-mu\-l\'a\-ci\-\'o\-val kap\-cso\-la\-tos esz\-me\-cse\-r\'e\-ket, He\-ge\-d\"us P\'al\-nak a
szta\-ti\-kus po\-ten\-ci\-\'al\-lal kap\-cso\-la\-tos sz\'a\-mo\-l\'a\-sok\-kal kap\-cso\-la\-tos
\'esz\-re\-v\'e\-te\-le\-it, \'es Ja\-ko\-v\'ac An\-tal\-nak az MSSM per\-tur\-ba\-t\'{\i}v
vizs\-g\'a\-la\-t\'a\-val kap\-cso\-la\-tos \'e\-p\'{\i}\-t\H o meg\-jegy\-z\'e\-se\-it.
K\"o\-sz\"o\-n\"om $\Gamma.$ Ko\-ut\-so\-um\-bas\-nak a szta\-ti\-kus kvark po\-ten\-ci\-\'al\-lal
kap\-cso\-la\-tos sz\'a\-mo\-l\'a\-sok\-ban ny\'uj\-tott se\-g\'{\i}t\-s\'e\-g\'et.

A dok\-to\-ri prog\-ram ke\-re\-t\'e\-ben Pe\-nis\-co\-l\'a\-ban \'es les Ho\-u\-ches-ban ny\'a\-ri
is\-ko\-l\'an ve\-het\-tem r\'eszt, Z\'ag\-r\'ab\-ban \'es Bu\-da\-pes\-ten e\-l\H o\-ad\-hat\-tam a
Tri\-ang\-le Sympo\-si\-u\-mon. K\"o\-sz\"o\-n\"om a Dok\-to\-ri Is\-ko\-la, \'es k\"u\-l\"on P\'o\-csik
Gy\"orgy t\'a\-mo\-ga\-t\'a\-s\'at.

\newpage
\appendix
\chapter{A $K'$ in\-teg\-r\'al ki\-sz\'a\-m\'{\i}\-t\'a\-sa \label{bgraf}}
\fancyhead[CE]{\hst{\thechapter{}.\ f\"ug\-gel\'ek\quad A $K'$
in\-tegr\'al}}
Eb\-ben a f\"ug\-ge\-l\'ek\-ben a (\ref{k'}) k\'ep\-let\-ben sze\-rep\-l\H o
\beq
K' (M,k) = \mu_0^{4-D} \int {{d^D q} \over {(2 \pi)^D}}{1 \over
{q^2-M^2+i \epsilon}} {1 \over {(k-q)^2-M^2+i \epsilon}} {1 \over
{q_0+i \epsilon }} {1 \over {-q_0+i \epsilon}}
\enq
in\-teg\-r\'alt sz\'a\-m\'{\i}\-tom ki. Ha ezt a (\ref{k}) $K$ in\-teg\-r\'al\-n\'al l\'a\-tott
m\'o\-don pr\'o\-b\'al\-juk v\'eg\-re\-haj\-ta\-ni, ke\-zel\-he\-tet\-len di\-ver\-gen\-ci\-\'ak\-kal
ta\-l\'al\-juk szem\-ben ma\-gun\-kat. E\-z\'ert a ne\-h\'ez kvark- \'es an\-tik\-vark
pro\-pa\-g\'a\-to\-rok\-ban a (\ref{propag}) ki\-fe\-je\-z\'es \'al\-ta\-l\'a\-no\-sabb a\-lak\-j\'at
hasz\-n\'al\-juk: $(qv+i \epsilon )^{-1}$-t, il\-let\-ve $(-qv'+i \epsilon
)^{-1}$-t. C\'el\-sze\-r\H u lesz a $K$ in\-teg\-r\'alt is eb\-be az a\-lak\-ba \'{\i}r\-nunk:
\beq
K (M,k) = \mu_0^{4-D} \int {{d^D q} \over {(2 \pi)^D}}{1 \over {q^2-M^2+i
\epsilon}} {1 \over {(k-q)^2-M^2+i \epsilon}} {1 \over {qv+i \epsilon }} {1
\over {qv'+i \epsilon}},
\enq
\beq
K' (M,k) = \mu_0^{4-D} \int {{d^D q} \over {(2 \pi)^D}}{1 \over {q^2-M^2+i
\epsilon}} {1 \over {(k-q)^2-M^2+i \epsilon}} {1 \over {qv+i \epsilon }} {1
\over {-qv'+i \epsilon}}.
\enq
A \ref{ksec} sza\-kasz\-ban l\'a\-tott ket\-t\H os Feyn\-man-pa\-ra\-m\'e\-te\-re\-z\'es\-sel
\begin{eqnarray}
K^{(')} &=& 6 \mu_0^{4-D} \int_0^1 d\alpha \int_0^\infty d\beta
\int_0^\infty d \gamma \int {{d^D q} \over {(2 \pi)^D}} \nonumber \\
&& {1 \over {[q^2 - 2(1- \alpha ) kq + (1- \alpha) k^2 - M^2 + \beta
q v \pm \gamma q v' + (\beta \pm \gamma) i \epsilon + i \epsilon
]^4}} \nonumber \\
&=& 6 \mu_0^{4-D} \int_0^1 d\alpha \int_0^\infty d\beta \int_0^\infty
d \gamma \int {{d^D q} \over {(2 \pi)^D}} \nonumber \\
&& [q^2 + (\beta v \pm \gamma v' - 2 (1- \alpha ) k) q + (\beta \pm
\gamma) i \epsilon + (1- \alpha) k^2 - M^2 + i \epsilon]^{-4},
\end{eqnarray}
a\-hol a fel\-s\H o e\-l\H o\-jel a $K$, az al\-s\'o a $K'$ in\-teg\-r\'al\-ra vo\-nat\-ko\-zik.
Az in\-teg\-r\'a\-l\'a\-si v\'al\-to\-z\'ot
\begin{eqnarray}
q \rightarrow q' = q+ {1 \over 2} (\beta v \pm \gamma v' -2 (1- \alpha)k)
\nonumber
\end{eqnarray}
sze\-rint el\-tol\-va a sz\"og\-le\-tes z\'a\-r\'o\-jel bel\-se\-j\'e\-ben le\-v\H o tag\-ra
\beq
q'^2 + \alpha (1- \alpha) k^2 - M^2 - {1 \over 4} (\beta v \pm \gamma
v')^2 - (1- \alpha) (\beta v \pm \gamma v') k + (\beta \pm \gamma) i
\epsilon + i \epsilon
\enq
a\-d\'o\-dik. Mint\-hogy $vk=v'k=0$, az \"o\-t\"o\-dik tag el\-t\H u\-nik, $v^2=v'^2=1$
mi\-att pe\-dig a $b=vv'$, $\gamma'=b \gamma$ je\-l\"o\-l\'e\-sek be\-ve\-ze\-t\'e\-s\'e\-vel
\beq
(\beta v \pm \gamma v')^2 = (\beta \pm vv' \gamma)^2 - {{(vv')^2 -
1} \over {(vv')^2}} (vv' \gamma)^2 = (\beta \pm \gamma')^2 - {{b^2 -
1} \over b^2} \gamma'^2,
\enq
Ve\-ze\-t\H o rend\-ben $v=v'= (1, {\bf 0})$, \'{\i}gy
\begin{eqnarray}
K^{(')} &=& 6 \mu_0^{4-D} \int_0^1 d\alpha \int_0^\infty d\beta
\int_0^\infty {1 \over b} d \gamma' \int {{d^D q} \over {(2 \pi)^D}}
\nonumber \\
&& \left[ q^2 + \alpha (1- \alpha) k^2 - M^2 - {1 \over 4} (\beta \pm
\gamma' )^2 + (\beta \pm \gamma') i \epsilon + {{b^2 - 1} \over
{4b^2}} \gamma'^2 + i \epsilon \right]^{-4}.
\end{eqnarray}
Wick-for\-ga\-t\'ast \'es az e\-g\'esz t\'er\-re t\"or\-t\'e\-n\H o in\-teg\-r\'a\-l\'ast
v\'eg\-re\-hajt\-va
\begin{eqnarray}
K^{(')} &=& {i \over {(4 \pi)^{D/2}}} \Gamma (4- {D \over 2})
\mu_0^{4-D} \int_0^1 d\alpha \int_0^\infty d\beta \int_0^\infty {1
\over b} d \gamma' \nonumber \\
&& \left[ \alpha (1- \alpha) {\bf k}^2 + M^2 + {1 \over 4} (\beta \pm
\gamma' )^2 - {{b^2 - 1} \over {4b^2}} \gamma'^2 + (\beta \pm
\gamma') i \epsilon \right]^{{D \over 2}-4},
\end{eqnarray}
a\-hol ki\-hasz\-n\'al\-tuk, hogy a ki\-cse\-r\'elt im\-pul\-zus t\'er\-sze\-r\H u.
\fancyhead[CO]{\hst{\thechapter{}.\ f\"ug\-gel\'ek\quad A $K'$
in\-tegr\'al}}

K\"o\-vet\-ke\-z\H o l\'e\-p\'es\-k\'ent $\beta$ \'es $\gamma'$ he\-lyett o\-lyan \'uj
in\-teg\-r\'a\-l\'a\-si v\'al\-to\-z\'o\-kat ve\-ze\-tek be, me\-lyek se\-g\'{\i}t\-s\'e\-g\'e\-vel $K$
\'es $K'$ u\-gya\-no\-lyan a\-la\-k\'u lesz.
\begin{eqnarray}
K: & \beta, \gamma' \rightarrow \xi, \psi: & \beta=(1-\psi)\xi \hskip
0.5truecm \gamma'=\psi \xi, \\
K': & \beta, \gamma' \rightarrow \xi, \psi: & \beta=(1+\psi)\xi \hskip
0.5truecm \gamma'=\psi \xi.
\end{eqnarray}
Mind\-k\'et e\-set\-ben a transz\-for\-m\'a\-ci\-\'o Ja\-co\-bi-de\-ter\-mi\-n\'an\-sa $|\xi|$. \\
Az \'uj v\'al\-to\-z\'ok se\-g\'{\i}t\-s\'e\-g\'e\-vel fe\-l\'{\i}rt in\-teg\-r\'a\-l\'a\-si tar\-to\-m\'a\-nyok
$K$ e\-se\-t\'e\-ben \\
$\xi=\beta+\gamma' \in [0, \infty)$, \'es $\beta / \gamma' = (1- \psi)
/ \psi$ mi\-att $\psi \in [0,1]$, \'{\i}gy
\begin{eqnarray}
K & = & {i \over {(4 \pi)^{D/2}}} \Gamma (4- {D \over 2}) \mu_0^{4-D}
\int_0^1 d\alpha \int_0^1 d\psi \int_0^\infty d\xi {\xi \over b}
\nonumber \\
&& \left[ \alpha (1- \alpha) {\bf k}^2 + M^2 + {1 \over 4} \xi^2 -
{{b^2 - 1} \over {4b^2}} \xi^2 \psi^2 + \xi i \epsilon \right]^{{D
\over 2}-4}.
\end{eqnarray}
A $K'$-re vo\-nat\-ko\-z\'o in\-teg\-r\'a\-l\'a\-si tar\-to\-m\'any meg\-ha\-t\'a\-ro\-z\'a\-s\'a\-hoz
c\'el\-sze\-r\H u a tar\-to\-m\'anyt k\'et r\'esz\-re v\'ag\-ni:
az el\-s\H o\-ben $(D_1)$ $\beta>\gamma'$, m\'{\i}g a m\'a\-so\-dik\-ban $(D_2)$
$\beta \leq \gamma'$ \'all fenn. \\
Ek\-kor $D_1$-re

$\xi = \beta-\gamma' \in (0, \infty)$, \qquad $\infty > \beta /
\gamma' = (1+ \psi)/\psi \ge 1 \rightarrow \psi \in [0,\infty)$, \\
$D_2$-re pe\-dig

$\xi = \beta-\gamma' \in [0, -\infty)$, \qquad $ \infty > \gamma' /
\beta = \psi / (1+\psi) \ge 1 \rightarrow \psi \in (-\infty,-1]$ \\
a\-d\'o\-dik. \'Igy
\begin{eqnarray}
\!\!\!\!\!K' \!& = & {i \over {(4 \pi)^{D/2}}} \Gamma (4- {D \over 2})
\mu_0^{4-D} \int_0^1 d\alpha \nonumber \\*
&& \left( \int_0^\infty d\psi \int_0^\infty d\xi {\xi \over b} \left[
\alpha (1- \alpha) {\bf k}^2 + M^2 + {1 \over 4} \xi^2 - {{b^2 - 1}
\over {4b^2}} \xi^2 \psi^2 + \xi i \epsilon \right]^{{D \over 2}-4} +
\right. \nonumber \\*
&& \left. \int_{-\infty}^{-1} d\psi \int_{0}^{-\infty} d\xi {{-\xi}
\over b} \left[ \alpha (1- \alpha) {\bf k}^2 + M^2 + {1 \over 4}
\xi^2 - {{b^2 - 1} \over {4b^2}} \xi^2 \psi^2 + \xi i \epsilon
\right]^{{D \over 2}-4} \right) \nonumber \\
&=& {i \over {(4 \pi)^{D/2}}} \Gamma (4- {D \over 2}) \mu_0^{4-D}
\int_0^1 d\alpha \nonumber \\*
&& \left( \int_{-\infty}^{-1} + \int_0^\infty \right) d\psi
\int_0^\infty d\xi {\xi \over b} \left[ \alpha (1- \alpha) {\bf k}^2 +
M^2 + {1 \over 4}  \xi^2 - {{b^2 - 1} \over {4b^2}} \xi^2 \psi^2 +
\xi i \epsilon \right]^{{D \over 2}-4}.
\end{eqnarray}
$K$-ban a $\psi$ sze\-rin\-ti in\-teg\-r\'a\-l\'as ha\-t\'a\-ra\-it $[0,1]$-r\H ol $[-1,
0]$-ra v\'a\-toz\-tat\-va
\begin{eqnarray}
K+K' & = & {i \over {(4 \pi)^{D/2}}} \Gamma (4- {D \over 2})
\mu_0^{4-D} \int_0^1 d\alpha \int_{-\infty}^\infty d\psi
\int_0^\infty d\xi \nonumber \\
&& {\xi \over b} \left[ \alpha (1- \alpha) {\bf k}^2 + M^2 + \left(
{1 \over 4} - {{b^2 - 1} \over {4b^2}} \psi^2 + {{i \epsilon} \over
\xi} \right) \xi^2 \right]^{{D \over 2}-4}.
\end{eqnarray}
Ek\-kor a $\xi^2$ e\-l\H ot\-ti z\'a\-r\'o\-jel\-ben $i \epsilon / \xi$,
he\-lyett $i \epsilon$ \'{\i}r\-ha\-t\'o, mi\-vel $\xi$ nem\-ne\-ga\-t\'{\i}v.

A $\xi$ sze\-rin\-ti in\-teg\-r\'a\-l\'as ek\-kor m\'ar v\'eg\-re\-hajt\-ha\-t\'o;
\beq
\int_0^\infty {1 \over {2b}} {{d \xi^2} \over {(A \xi^2 + C)^{4- {D\over
2}}}} = {1 \over {2bA}} {{\Gamma(1) \Gamma(3- {D\over 2})} \over {\Gamma(4-
{D \over 2})}} C^{{D \over 2} -3}
\enq
mi\-att e\-red\-m\'e\-ny\"ul
\begin{eqnarray}
K+K' &=& {i \over {(4 \pi)^{D/2}}} {{\Gamma (3- {D \over 2})} \over 2b}
\mu_0^{4-D} \int_0^1 d\alpha (\alpha (1- \alpha) {\bf k}^2 + M^2)^{{D
\over 2}-3} \nonumber \\*
&& \int_{-\infty}^\infty d\psi \left( {1 \over 4} - {{b^2 - 1} \over
{4b^2}} \psi^2 + i \epsilon \right)^{-1}.
\end{eqnarray}
a\-d\'o\-dik.

A $\psi$ sze\-rin\-ti in\-teg\-r\'al a $b^2 \rightarrow 1$ e\-set\-ben
szin\-gu\-l\'a\-ris. A szin\-gu\-la\-ri\-t\'as a\-zon\-ban le\-v\'a\-laszt\-ha\-t\'o a $b^2 \!
\searrow \! 1$ ha\-t\'a\-r\'at\-me\-net so\-r\'an. (Ezt a l\'e\-p\'est nem le\-het\-ne
meg\-ten\-ni, ha a na\-iv $(q_0 + i \epsilon)^{-1}$ pro\-pa\-g\'a\-tort
hasz\-n\'al\-n\'ank, u\-gya\-nis a $b^2=1$ r\"og\-z\'{\i}\-t\'es el\-fe\-di a szin\-gu\-la\-ri\-t\'as
szer\-ke\-ze\-t\'et.)

A $\psi$ sze\-rin\-ti in\-teg\-r\'a\-l\'as e\-red\-m\'e\-nye
\beq
\int_{-\infty}^\infty d\psi \left( {1 \over 4} - {{b^2 - 1} \over
{4b^2}} \psi^2 + i \epsilon \right)^{-1} = {{4b} \over
{\sqrt{b^2-1}}} \int_{-\infty }^{\infty} (1- {\tilde \psi}^2 + i
\epsilon)^{-1} d \tilde \psi = {{4b} \over {\sqrt{b^2-1}}} i \pi.
\enq
E\-zek u\-t\'an az $\alpha$ sze\-rin\-ti in\-teg\-r\'a\-l\'as a (\ref{alphaint1})-ben
\'es (\ref{alphaint2})-ben l\'a\-tot\-tak\-kal tel\-je\-sen a\-zo\-nos m\'o\-don
v\'eg\-re\-hajt\-ha\-t\'o. Az $M \neq 0$ e\-set\-ben
\beq
K + K' = {{-1} \over {8 \pi}} {1 \over {\sqrt{b^2-1}}} {1 \over
{\sqrt{{\bf k }^4 + 4M^2{\bf k}^2}}} \ln \left( {{\bf k}^2 + \sqrt
{{\bf k}^4 + 4M^2{\bf k }^2}} \over {{\bf k}^2 - \sqrt{{\bf k}^4 +
4M^2{\bf k}^2}} \right)^2.
\label{K+K'1}
\enq
a\-d\'o\-dik, m\'{\i}g $M=0$ e\-se\-t\'en
\beq
K + K' = {{-1} \over {4 \pi}} {1 \over {\sqrt{b^2-1}}} {1 \over {\bf
k}^2} \left[ {1 \over \epsilon_I} + \left( \gamma - \ln {{4 \pi
\mu_0^2} \over {\bf k}^2} \right) + \ordo{\epsilon} \right]. \label{K+
K'2}
\enq
Mind\-k\'et e\-red\-m\'eny szin\-gu\-l\'a\-ris, a\-zon\-ban az $M\neq 0$ e\-set\-ben a
szin\-gu\-la\-ri\-t\'as f\"ug\-get\-len a t\'e\-ri\-d\H o di\-men\-zi\-\'o\-sz\'a\-m\'a\-t\'ol.

\chapter{A QCD szta\-ti\-kus kvark po\-ten\-ci\-\'al\-ja \label{qcd}}
\fancyhead[CE]{\hst{\thechapter{}.\ f\"ug\-gel\'ek \quad A QCD
szta\-ti\-kus kvark po\-ten\-ci\'al\-ja}}
Eb\-ben a f\"ug\-ge\-l\'ek\-ben r\'esz\-le\-te\-sen ki\-sz\'a\-m\'{\i}\-tom a
kvan\-tum\-sz\'{\i}n\-di\-na\-mi\-ka\-i (tisz\-ta SU(3) m\'er\-t\'e\-kel\-m\'e\-let\-be\-li) sta\-ti\-kus
kvark po\-ten\-ci\-\'alt. Az e\-red\-m\'eny \'es a m\'od\-sze\-rek az i\-ro\-da\-lom\-b\'ol
is\-mer\-tek, a\-zon\-ban a ne\-he\-zebb sz\'a\-m\'{\i}\-t\'a\-sok\-ban hasz\-n\'alt m\'od\-sze\-rek
el\-le\-n\H or\-z\'e\-se v\'e\-gett c\'el\-sze\-r\H u ezt az egy\-sze\-r\H ubb e\-se\-tet
v\'e\-gi\-ga\-sz\'a\-mol\-ni. A sz\'a\-mo\-l\'as je\-len\-t\H o\-sen r\"o\-vi\-d\"ul, ha a
Feyn\-man-f\'e\-le m\'er\-t\'ek\-r\"og\-z\'{\i}\-t\'est \'{\i}r\-juk e\-l\H o.

A QCD e\-se\-t\'e\-ben a \ref{graph_lq} \'ab\-r\'an l\'at\-ha\-t\'o gr\'a\-fok k\"o\-z\"ul
csak az $L$, $M$, $N$ je\-l\H u\-ek ad\-nak j\'a\-ru\-l\'e\-kot. A tad\-po\-le gr\'a\-fok\-ban
az a\-l\'abb de\-fi\-ni\-\'a\-lan\-d\'o $J$ hu\-ro\-kin\-teg\-r\'al l\'ep fel, mely a 0
t\"o\-me\-g\H u e\-set\-ben el\-t\H u\-nik. Az e\-l\H ob\-bi h\'a\-rom gr\'af j\'a\-ru\-l\'e\-ka
\cite{muta} \medskip

{\underline {L gr\'af}}
\beqar
&& \! \! \! \! \! {1 \over 2} g^4 C^{acd} C^{bcd} T^a T^b {1 \over
{(k^2-M_W^2+i \epsilon)^2 }} \delta^{\mu 0} \delta^{\nu 0} \cdot
\nonumber \\*
&& \! \! \! \! \! \left[ g_{\mu \nu} (5k^2 I + 2k_\tau I_\tau + 2
I_2) + (4D-6) I_{\mu \nu} + (2D-3) (k_\mu I_\nu + k_\nu I_\mu) +
(D-6) k_\mu k_\nu I \right]\!\!,
\label{glres}
\enqar

{\underline {M gr\'af}}
\beq
g^4 (D-1) C^{lac}C^{lbc} T^a T^b {{\delta^{\mu 0} \delta^{\nu 0}}
\over {(k^2-M_W^2+i \epsilon)^2}} g_{\mu \nu} J, \label{gmres}
\enq

{\underline {N gr\'af}}
\beq
g^4 C^{acd} C^{bcd} T^a T^b \delta^{\mu 0} \delta^{\nu 0} {1 \over
{(k^2- M_W^2+i \epsilon)^2}} \left[ I_{\mu \nu} + k_\nu I_\mu \right].
\label{gnres}
\enq

Itt az \'al\-ta\-l\'a\-no\-sabb
\beqar
I (m,M,k) & = & \mu_0^{4-D} \int {{d^D q} \over {(2 \pi)^D}}{1 \over
{q^2-m^2 +i \epsilon}} {1 \over {(q+k)^2-M^2+i \epsilon}} \\
I_{\mu} (m,M,k) &=& \mu_0^{4-D} \int {{d^D q} \over {(2 \pi)^D}}
q_{\mu} {1 \over {q^2-m^2+i \epsilon}} {1 \over {(q+k)^2-M^2+i
\epsilon}} \\
I_{\mu \nu} (m,M,k) &=& \mu_0^{4-D} \int {{d^D q} \over {(2 \pi)^D}}
q_{\mu} q_{\nu} {1 \over {q^2-m^2+i \epsilon}} {1 \over {(q+k)^2-M^2+
i \epsilon}} \\
I_2 (m,M,k) &=& \mu_0^{4-D} \int {{d^D q} \over {(2 \pi)^D}} q^2 {1
\over {q^2-m^2+i \epsilon}} {1 \over {(q+k)^2-M^2+i \epsilon}} \\
J (m) &=& \mu_0^{4-D} \int {{d^D q} \over {(2 \pi)^D}}{1 \over
{q^2-m^2+ i \epsilon}}
\enqar
k\'ep\-let\-ben a t\"o\-me\-gek he\-ly\'e\-be 0 \'{\i}\-ran\-d\'o.

A fen\-ti in\-teg\-r\'a\-lok ki\-sz\'a\-m\'{\i}\-t\'a\-sa a QCD e\-se\-t\'e\-ben l\'e\-nye\-ge\-sen
egy\-sze\-r\H ubb, mint a t\"o\-me\-ges el\-m\'e\-le\-tek\-re. El\-s\H o l\'e\-p\'es\-ben
a
\beqar
I (m,M,k) & = & \mu_0^{4-D} \int_0^1 d \alpha \nonumber \\*
&& \int {{d^D q'} \over {(2 \pi)^D}} {1 \over {(q'^2 + \alpha
(1-\alpha) k^2 - \alpha m^2 - (1- \alpha )M^2 + i \epsilon)^2}}, \\
I_{\mu} (m,M,k) & = & \mu_0^{4-D} \int_0^1 d \alpha \nonumber \\*
&& \int {{d^D q'} \over {(2 \pi)^D}} {{q_{\mu}'+(\alpha-1)k_{\mu}}
\over {(q'^2 + \alpha (1-\alpha) k^2 - \alpha m^2 - (1- \alpha ) M^2 + i
\epsilon)^2}}, \\
I_{\mu \nu} (m,M,k) & = & \mu_0^{4-D} \int_0^1 d \alpha \nonumber \\*
&& \int {{d^D q'} \over {(2 \pi)^D}} {{(q_{\mu}'+(\alpha-1)k_{\mu})
(q_{\nu}' +(\alpha-1) k_{\nu})} \over {(q'^2 + \alpha(1-\alpha) k^2 -
\alpha m^2 - (1 - \alpha) M^2+i \epsilon)^2}}, \\
I_2 (m,M,k) & = & \mu_0^{4-D} \int_0^1 d \alpha \nonumber \\*
&& \int {{d^D q'} \over {(2 \pi)^D}} {{(q'+(\alpha -1)k)^2} \over
{(q'^2 + \alpha (1-\alpha) k^2 - \alpha m^2 - (1- \alpha )M^2 + i
\epsilon)^2}} \\
J (m) & = & \mu_0^{4-D} \int {{d^D q} \over {(2 \pi)^D}}{1 \over {q^2-
m^2 +i \epsilon}}
\enqar
\"ossze\-f\"ug\-g\'e\-sek a\-lap\-j\'an be\-ve\-zet\-he\-t\H ok az $M$ \'es $N$ in\-teg\-r\'a\-lok,
\begin{eqnarray}
M_1 & = & \mu_0^{4-D} \int_0^1 d\alpha \left[ \alpha m^2 + (1 -
\alpha) M^2 - \alpha (1-\alpha) k^2 \right]^{{D-4}\over 2}, \\
M_2 & = & \mu_0^{4-D} \int_0^1 d\alpha (1 - \alpha ) \left[ \alpha
m^2 + (1 - \alpha) M^2 - \alpha (1- \alpha) k^2 \right]^{{D-4}\over
2}, \\
M_3 & = & \mu_0^{4-D} \int_0^1 d\alpha (1 - \alpha)^2 \left[ \alpha
m^2 + (1 - \alpha) M^2 - \alpha (1 - \alpha) k^2 \right]^{{D-4}\over
2}, \\
N_1 & = & \mu_0^{4-D} \int_0^1 d\alpha \left[ \alpha m^2 + (1 -
\alpha) M^2 - \alpha (1-\alpha) k^2 \right]^{{D-2}\over 2}.
\end{eqnarray}
me\-lye\-ket 0 t\"o\-me\-g\H u e\-set\-ben k\"onnyen ki\-\'er\-t\'e\-kel\-he\-t\"unk. Az e\-red\-m\'eny
\beqar
M_1 & = & \left( {\mu_0^2 \over {-k^2}} \right)^\epsilon {{\Gamma (1-
\epsilon) \Gamma (1 - \epsilon)} \over {\Gamma (2-2 \epsilon)}} =
\left( 1 + \epsilon \ln {\mu_0^2 \over {-k^2}} \right) (1 + 2
\epsilon), \label{m1qcdres} \\
M_2 & = & \left( {\mu_0^2 \over {-k^2}} \right)^ \epsilon {{\Gamma (1-
\epsilon) \Gamma (2 - \epsilon)} \over {\Gamma (3-2 \epsilon)}} = {1
\over 2} \left(1 + \epsilon \ln {\mu_0^2 \over {-k^2}} \right) (1 + 2
\epsilon), \label{m2qcdres} \\
M_3 & = & \left( {\mu_0^2 \over {-k^2}} \right)^ \epsilon {{\Gamma (1-
\epsilon) \Gamma (3 - \epsilon)} \over {\Gamma (4-2 \epsilon)}} = {1
\over 3} \left(1 + \epsilon \ln {\mu_0^2 \over {-k^2}} \right) \left(
1 + {13 \over 6} \epsilon \right), \label{m3qcdres} \\
N_1 & = & -k^2 \left( {\mu_0^2 \over {-k^2}} \right)^ \epsilon
{{\Gamma (2- \epsilon) \Gamma (2 - \epsilon)} \over {\Gamma (4-2
\epsilon)}} = - {k^2 \over 6} \left( 1 + \epsilon \ln {\mu_0^2 \over
{-k^2}} \right) \left(1 + {5 \over 3} \epsilon \right). \hskip
1truecm \label{n1qcdres}
\end{eqnarray}
\fancyhead[CO]{\hst{\thechapter{}.\ f\"ug\-gel\'ek \quad A QCD
szta\-ti\-kus kvark po\-ten\-ci\'al\-ja}}
E\-z\'al\-tal a k\"o\-vet\-ke\-z\H ok a\-d\'od\-nak:
\beqar
I^{QCD} & = & {i \over {(4 \pi)^2}} \left( {1 \over \epsilon} -
\gamma + \ln (4 \pi) + \ln {\mu_0^2 \over {-k^2}} + 2 \right), \\
I_\mu^{QCD} &=& {{-i k_\mu} \over {2 (4 \pi)^2}} \left( {1 \over
\epsilon} - \gamma + \ln (4 \pi) + \ln {\mu_0^2 \over {-k^2}} + 2
\right), \\
I_2^{QCD} & = & 0, \\
J^{QCD} & = & 0, \\
I_{\mu \nu}^{QCD} & = & {i \over {(4 \pi)^2}} \left[ {1 \over 3}
k_\mu k_\nu \left( {1 \over \epsilon} - \gamma + \ln(4 \pi) + \ln
{\mu_0^2 \over {-k^2}} + {13 \over 6} \right) - \right. \nonumber \\*
&& \quad \left. {k^2 \over 12} g_{\mu \nu} \left( {1 \over \epsilon} -
\gamma + \ln (4 \pi) + \ln {\mu_0^2 \over {-k^2}} + {8 \over 3}
\right) \right].
\end{eqnarray}

A fen\-ti e\-red\-m\'e\-nyek di\-ver\-gen\-sek, \'{\i}gy az $\overline {\rm MS}$
e\-l\H o\-\'{\i}\-r\'as sze\-rint re\-nor\-m\'al\-va \H o\-ket
\begin{eqnarray}
I^R & = & {i \over {(4 \pi)^2}} \left( \ln {\mu_0^2 \over {-k^2}} + 2
\right), \\*
I_\mu^R & = & {{-i k_\mu} \over {2 (4 \pi)^2}} \left( \ln {\mu_0^2 \over
{-k^2}} + 2 \right), \\
I_2^R & = & 0, \\
J^R & = & 0, \\
I_{\mu \nu}^R & = & {i \over {(4 \pi)^2}} \left[ {1 \over 3} k_\mu
k_\nu \left( \ln {\mu_0^2 \over {-k^2}} + {13 \over 6} \right) - {k^2
\over 12} g_{\mu \nu} \left( \ln {\mu_0^2 \over {-k^2}} + {8 \over 3}
\right) \right].
\end{eqnarray}
a\-d\'o\-dik. Sz\"uk\-s\'eg lesz m\'eg a fen\-ti di\-ver\-gens in\-teg\-r\'a\-lok
di\-men\-zi\-\'o\-sz\'am\-mal szor\-zott ki\-fe\-je\-z\'e\-s\'e\-nek re\-nor\-m\'alt a\-lak\-j\'a\-ra is:
\begin{eqnarray}
D^n I^R & = & {{i 4^n} \over {(4 \pi)^2}} \left( \ln {\mu_0^2 \over
{-k^2}} + 2 - {n \over 2} \right), \\
D^n I_\mu^R & = & {{-i 4^n k_\mu} \over {2 (4 \pi)^2}} \left( \ln
{\mu_0^2 \over {-k^2}} + 2 - {n \over 2} \right), \\
D^n I_2^R & = & 0, \\
D^n J^R & = & 0, \\
D^n I_{\mu \nu}^R & = & {{i 4^n} \over {(4 \pi)^2}} \left[ {1 \over
3} k_\mu k_\nu \left( \ln {\mu_0^2 \over {-k^2}} + {13 \over 6} - {n
\over 2} \right) - {k^2 \over 12} g_{\mu \nu} \left( \ln {\mu_0^2
\over {-k^2}} + {8 \over 3} - {n\over 2} \right) \right].
\end{eqnarray}
Az $M$ gr\'af 0 j\'a\-ru\-l\'e\-kot ad, az $L$ \'es az $N$ gr\'a\-fok j\'a\-ru\-l\'e\-ka
pe\-dig
\begin{eqnarray}
G_{L+N}^{QCD} & = & {1 \over 2} g^4 C(G) T^a T^a {1 \over {({\bf
k}^2)^2}} \delta^{\mu 0} \delta^{\nu 0} \cdot \Bigl( \bigl[ g_{\mu
\nu} (5k^2 I + 2k_\tau I_\tau + 2 I_2) +
\nonumber \\*
&& \qquad (4D-6) I_{\mu \nu} + (2D-3) (k_\mu I_\nu + k_\nu I_\mu) +
(D-6) k_\mu k_\nu I - 2 (I_{\mu \nu} + k_\nu I_\mu) \bigr] \Bigr)
\nonumber \\
& = & {1 \over 2} g^4 C(G) T^a T^a {1 \over {({\bf k}^2)^2}}
\delta^{\mu 0} \delta^{\nu 0} \cdot \\*
&& \left[ g_{\mu \nu} (5k^2 I + 2k_\tau I_\tau + 2 I_2) + 4D I_{\mu
\nu} - 8I_{\mu \nu} + 4 k_\nu D I_\mu - 8 k_\nu I_\mu + k_\mu k_\nu D
I - 6 k_\mu k_\nu I \right] \nonumber
\end{eqnarray}
Be\-\'{\i}r\-va a meg\-fe\-le\-l\H o re\-nor\-m\'alt mennyi\-s\'e\-ge\-ket
\begin{eqnarray}
R_{L+N}^{QCD} &=& {i \over {2(4 \pi)^2}} g^4 C(G) T^a T^a {1 \over
{({\bf k} ^2)^2}} \delta^{\mu 0} \delta^{\nu 0} \cdot \nonumber \\*
&& \left[ g_{\mu \nu} \left( 5k^2 \left( \ln {\mu_0^2 \over {-k^2}} +
2 \right) - 2k_\tau {k_\tau \over 2} \left( \ln {\mu_0^2 \over
{-k^2}} + 2 \right) \right) \right. \nonumber \\*
&& \quad + 16 \left({1 \over 3} k_\mu k_\nu \left( \ln {\mu_0^2 \over
{-k^2}} + {5 \over 3} \right) - {k^2 \over 12} g_{\mu \nu} \left( \ln
{\mu_0^2 \over {-k^2}} + {13 \over 6} \right) \right) \nonumber \\
&& \quad - 8 \left({1 \over 3} k_\mu k_\nu \left( \ln {\mu_0^2 \over
{-k^2}} + {13 \over 6} \right) - {k^2 \over 12} g_{\mu \nu} \left(
\ln {\mu_0^2 \over {-k^2}} + {8 \over 3} \right) \right) \nonumber \\
&& \quad - 16 k_\nu {k_\mu \over 2} \left( \ln {\mu_0^2 \over {-k^2}} + {3
\over 2} \right) + 8 k_\nu {k_\mu \over 2} \left( \ln {\mu_0^2 \over
{-k^2}} + 2 \right) \nonumber \\
&& \quad \left. + 4 k_\mu k_\nu \left( \ln {\mu_0^2 \over {-k^2}} + {3
\over 2} \right) - 6 k_\mu k_\nu \left( \ln {\mu_0^2 \over {-k^2}} +
2 \right) \right] \nonumber \\
& = & {-i \over {2(4 \pi)^2}} g^4 C(G) T^a T^a {1 \over {({\bf
k}^2)^2}} \delta^{\mu 0} \delta^{\nu 0} \cdot
\left( {10 \over 3} \ln {\mu_0^2 \over {{\bf k}^2}} +{62 \over 9}
\right) ( k^2 g_{\mu \nu}- k_\mu k_\nu) \nonumber \\
& = & {i \over {(4 \pi)^2}} g^4 C(G) C(R) {1 \over {{\bf k}^2}}
\left( {5 \over 3} \ln {\mu_0^2 \over {{\bf k}^2}} + {31 \over 9}
\right).
\label{qcdrenprop}
\end{eqnarray}
A fen\-ti\-ek\-ben ki\-hasz\-n\'al\-tuk, hogy a ki\-cse\-r\'elt n\'e\-gye\-sim\-pul\-zus
t\'er\-sze\-r\H u.
\bigskip \\
A fag\-r\'af \'es a k\'et\-bo\-zon-cse\-r\'es gr\'a\-fok j\'a\-ru\-l\'e\-k\'at hoz\-z\'a\-ad\-va, az
egy\-hu\-rok-ren\-d\H u sta\-ti\-kus po\-ten\-ci\-\'al\-ra a
\begin{eqnarray}
R^{QCD} & = & ig^2 C(R) {1 \over { {\bf k}^2 }} + {{i} \over {8 \pi^2
{\bf k}^2}} g^4 C(R) C(G) \ln \left( {\mu_0^2 \over {\bf k}^2}
\right) \nonumber \\
&& + {i \over {(4 \pi)^2}} g^4 C(G) C(R) {1 \over {{\bf k}^2}} \left(
{5 \over 3} \ln {\mu_0^2 \over {{\bf k}^2}} + {31 \over 9} \right)
\nonumber \\
& = & {{ig^2 C(R)} \over { {\bf k}^2 }} \left[ 1 + {{g^2 C(G)} \over
{16 \pi^ 2}} \left( {11 \over 3} \ln {\mu_0^2 \over {{\bf k}^2}} +
{31 \over 9} \right) \right]
\end{eqnarray}
ki\-fe\-je\-z\'es a\-d\'o\-dik, a ko\-r\'ab\-bi e\-red\-m\'e\-nyek\-kel \cite{suss, fis, mark,
app, schr} \"ossz\-hang\-ban.

\vfill


\begin{thebibliography}{199}
\addcontentsline{toc}{chapter}{Irodalomjegyz\'ek}
%
\bibitem{sakharov} A.~D.~Sakharov, JETP Letters \tb{91B} (1967), 24
%
\bibitem{thoft} G.~'t Hooft, Phys.\ Rev.\ Lett.\ \tb{37} (1976), 8, \\
G.~'t Hooft, Phys.\ Rev.\ \tb{D14} (1976), 3432
%
\bibitem{rub1} V.~A.~Kuzmin, V.~A.~Rubakov, M.~E.~Shaposhnikov,
Phys.\ Lett.\ \tb{B155} (1985), 36
%
\bibitem{arn} P.~Arnold and O.~Espinosa, Phys.\ Rev.\ \tb{D47}
3546 (1993), \\
\qquad Erratum Phys.\ Rev.\ \tb{D50} 6662 (1994).
%
\bibitem{fod94/6} W.~Buchm\"uller et al Ann.\ Phys.\ (NY) \tb{234}
(1994), 260
%
\bibitem{fod-heb} Z.~Fodor, A.~Hebecker, Nucl.\ Phys.\ \tb{B432} (1994),
127
%
\bibitem{fod94} Fodor et al., Nucl.\ Phys.\ \tb{B439} (1994), 147
%
\bibitem{kaj} K.~Kajantie et al, Nucl.\ Phys.\ \tb{B407} (1993), 356,
\\
K.~Kajantie et al, Nucl.\ Phys.\ \tb{B466} (1996), 189
%
\bibitem{phil} O.~Philipsen et al, Nucl.\ Phys.\ \tb{B469} (1996), 445
%
\bibitem{suss} L.~Susskind, \emph{Coarse Grained Quantum
Chromodynamics} in R.~Balian and C.~H.~Llewellyn Smith (eds.),
\emph{Weak and Electromagnetic Interactions at High Energy} (North
Holland, Amsterdam, 1977).
%
\bibitem{cikk1} F.~Csikor, Z.~Fodor, P.~Heged\"us, A.~Pir\'oth Phys.\
Rev.\ \tb{D60} (1999) 114511
%
\bibitem{la} M.~Laine, J.\ High Energy Physics \tb{06} (1999), 020
%
\bibitem{velt} G.~Passarino, M.~Veltman, Nucl.\ Phys.\ \tb{B160}
(1979), 151
%
\bibitem{car97} M.~Carena, M.~Quir\'os, C.~E.~M.~Wagner, Nucl.\ Phys.\
\tb{B524} (1998), 3-22
%
\bibitem{bod} D.~B\"odeker et al, Nucl.\ Phys.\ \tb{B497} (1997), 387
%
\bibitem{pmscikk} F.~Csikor, Z.~Fodor, P.~Heged\"us, V.~K.~Horv\'ath,
S.~D.~Katz, A.~Pir\'oth, hep-lat/9912059
%
\bibitem{brhlik} M.~Brhlik, G.~J.~Good, G.~L.~Kane, hep-ph/9911243, \\
M.~Brhlik, hep-ph/0004042
%
\bibitem{clin-gdm99} J.~M.~Cline, G.~D.~Moore, G.~Servant, Phys.\
Rev.\ \tb{D60} (1999) 105035
%
\bibitem{jak2} Jakov\'ac Antal, jegyzet
%
\bibitem{los} M.~Losada, Nucl.\ Phys.\ \tb{B537} (1999), 3, \\
M.~Losada, hep-ph/9905441
%
\bibitem{la98} M.~Laine, K.~Rummukainen, Phys.\ Rev.\ Lett.\ \tb{80}
(1998), 5259, \\
M.~Laine, K.~Rummukainen, Nucl.\ Phys.\ \tb{B535} (1998), 423
%
\bibitem{fis} W.~Fischler, Nucl.\ Phys.\ \tb{B129} (1977), 157
%
\bibitem{physspeak} C.~C.~Gaither, A.~E.~Cavazos-Gaither,
\emph{Physically Speaking}, Inst.\ of Physics Publishing (Bristol and
Philadelphia) 1997
%
\bibitem{coh} A.~G.~Cohen, A.~de~Rujula, S.~L.~Glashow, Astrophys.\
J.\ \tb{495} (1998), 539
%
\bibitem{ol} K.~A.~Olive, Nucl.\ Phys.\ (Proc.\ Suppl.) \tb{70} (1999),
521
%
\bibitem{nanopoulos} D.~V.~Nanopoulos, \emph{Cosmological Implications
of Grand Unified Theories} in \emph{Proceedings of the International
School of Physics ``Enrico Fermi'', Course LXXXI (Theory of
Fundamental Interactions)}, North-Holland Publishing Company, 1982
%
\bibitem{fuk-yan} M.~Fukugita, T.~Yanagida, Phys.\ Lett.\ \tb{B174}
(1986), 45
%
\bibitem{clin00} J.~M.~Cline, PRAMANA \tb{54} vol.\ 4 (2000), 1
(hep-ph/0003029)
%
\bibitem{rub2} V.~A.~Rubakov, M.~E.~Shaposhnikov, Usp.\ Fiz.\ Nauk.\
\tb{166} (1996), 493 (hep-ph/9603298)
%
\bibitem{turok} N.~Turok, Les Houches lectures, 1999
%
\bibitem{asy97} P.~Arnold, D.~Son, L.~G.~Yaffe, Phys.\ Rev.\ \tb{D55}
(1997), 6264
%
\bibitem{bod99} D.~B\"odeker, G.~D.~Moore, K.~Rummukainen,
hep-lat/9909054
%
\fancyhead[CO,CE]{\hst{Irodalomjegyz\'ek}}
%
\bibitem{gdm99} G.~D.~Moore, hep-lat/9907009
%
\bibitem{smit99} J.~Smit, Nucl.\ Phys.\ (Proc.\ Suppl.) \tb{63}
(1998), 89
%
\bibitem{shap87} M.~E.~Shaposhnikov, Nuclear Physics \tb{B287} (1987),
757
%
\bibitem{fod94/5} M.~E.~Carrington, Phys.\ Rev.\ \tb{D47} (1993), 2933
%
\bibitem{kaj2} K.~Kajantie et al, Phys.\ Rev.\ Lett.\ \tb{77} (1996),
2887
%
\bibitem{kar} F.~Karsch et al,, Nucl.\ Phys.\ B (Proc.\ Suppl.)
\tb{54} (1997), 623
%
\bibitem{gur} M.~G\"urtler, E.-M.~Ilgenfritz, A.~Schiller, Phys.\
Rev.\ \tb{D56} (1997), 3888
%
\bibitem{far} K.~Farakos et al, Nucl.\ Phys.\ \tb{B425}
(1994), 67-109, \\
K.~Farakos et al, Nucl.\ Phys.\ \tb{B442} (1995), 317
%
\bibitem{jak} A.~Jakov\'ac, A.~Patk\'os, Phys.\ Lett.\ \tb{B334}
(1994), 54\\
A.~Jakov\'ac, A.~Patk\'os, Nucl.\ Phys.\ \tb{B494} (1997), 54
%
\bibitem{kog} J.~Kogut, Rev.\ Mod.\ Phys.\ \tb{55 3} (1983) 776
%
\bibitem{montvay} I.~Montvay, G.~M\"unster \emph{Quantum Fields on a
	      Lattice} {(Cambridge University Press)}
%
\bibitem{hursz} F.~Knechtli, R.~Sommer, Phys.\ Lett.\
\tb{B440} (1998), 345
%
\bibitem{wilson} K.~G.~Wilson, Phys.\ Rev.\ \tb{D10} (1974), 2445
%
\bibitem{hol} M.~B\"ohm, H.~Spiesberger, W.~Hollik, Fortschr.\
Phys.\ \tb{34} (1986), 687
%
\bibitem{muta} T.~Muta, \emph{Foundations of Quantum Chromodynamics},
World Scientific, 1988
%
\bibitem{jeg} F.~Jegerlehner, \emph{University of Colorado}
lectures, 1990
%
\bibitem{mark} M.~Peter, Phys.\ Rev.\ Lett.\ \tb{78} (1997), 602, \\
M.~Peter, Nucl.\ Phys.\ \tb{B501} (1997), 471
%
\bibitem{app} T.~Appelquist, M.~Dine, Phys.\ Lett.\ \tb{69B} (1977),
231
%
\bibitem{schr} Y.~Schr\"oder, Phys.\ Lett.\ \tb{B447} (1999), 321
%
\bibitem{som94} R.~Sommer, Nucl.\ Phys.\ \tb{B411} (1994), 839
%
\bibitem{numrec} H.~Press, \emph{Numerical Recipes in Fortran}, \
Cambridge University Press, 1992
%
\bibitem{csik99} F.~Csikor, Z.~Fodor, J.~Heitger, Phys.\ Rev.\ Lett.\
\tb{82} (1999), 21
%
\bibitem{arkhi} P.~Strathern, \emph{Arkhim\'ed\'esz}, Elektra Kiad\'o,
2000
%
\bibitem{goldstein} H.~Goldstein, \emph{Classical Mechanics}, Narosa
Publishing House, 1996
%
\bibitem{fey5} R.~P.~Feynman, R.~B.~Leighton, M.~Sands, \emph{Mai
fizika} 5.\ ko2tet, M\H{u}szaki K\"onyv\-kiad\'o, 1970
%
\bibitem{born} Max Born, \emph{Atomic Physics} Dover Publications
Inc., 1969
%
\bibitem{niedermayer} F.~Niedermayer, Nucl.\ Phys.\ \tb{B} (Proc.\ Suppl.)
\tb{73} (1999), 105
%
\bibitem{gin} P.~Ginsparg, Nucl.\ Phys.\ \tb{B170} (1980), 388
%
\bibitem{app2} T.~Appelquist, R.~Pisarki, Phys.\ Rev.\ \tb{D23}
(1981), 2305
%
\bibitem{csik96} F.~Csikor et al, Nucl.\ Phys.\ \tb{B474} (1996), 421
%
\bibitem{csik98} F.~Csikor, Z.~Fodor, J.~Heitger, Phys.\ Lett.\
\tb{B441} (1999), 354
%
\bibitem{cikk15} J.~Hein, J.~Heitger, Phys.\ Lett.\ \tb{B385} (1996),
242
%
\bibitem{efron} B.~Efron, SIAM Review \tb{21} (1979), 460, \\
R.~Gupta et al, Phys Rev \tb{D36} (1987), 2813
%
\bibitem{anizo} F.~Csikor, Z.~Fodor, J.~Heitger, Phys.\ Rev.\
\tb{D58} (1998), 094504
%
\bibitem{buch-fod-heb} W.~Buchm\"uller, Z.~Fodor, A.~Hebecker, Nucl.\
Phys.\ \tb{B447} (1995), 317
%
\bibitem{buchm} W.~Buchm\"uller, O.~Philipsen, Nucl.\ Phys.\ \tb{B443}
(1995), 47
%
\bibitem{bentley} www.snowflakebentley.com
%
\bibitem{penrose} R.~Penrose, \emph{A cs\'asz\'ar \'uj elm\'eje},
Akad\'emiai kiad\'o, Budapest 1993
%
\bibitem{weinberg} S.~Weinberg, \emph{The Quantum Theory of Fields},
Cambridge University Press, 1996
%
\bibitem{cerncour} Cern Courier, \cite{physspeak} alapj\'an
%
\bibitem{gu} G.~F.~Giudice, Phys.\ Rev.\ \tb{D45} (1992), 3177
%
\bibitem{esp} J.~R.~Espinosa et al, Phys.\ Lett.\ \tb{B307} (1993), 106
%
\bibitem{brig} A.~Brignole et al, Phys.\ Lett.\ \tb{B324} (1994), 181
%
\bibitem{esp2} J.~R.~Espinosa, Nucl.\ Phys.\ \tb{B475} (1996), 273
%
\bibitem{carl} B.~de~Carlos, J.~R.~Espinosa, Nucl.\ Phys.\ \tb{B503}
(1997), 24
%
\bibitem{clin} J.~M.~Cline, G.~D.~Moore, Phys.\ Rev.\ Lett.\ \tb{81}
(1998), 315
%
\bibitem{car96} M.~Carena et al, Phys.\ Lett.\ \tb{B380} (1996), 81
%
\bibitem{car98} M.~Carena et al, Nucl.\ Phys.\ \tb{B524} (1998), 3
%
\bibitem{fun} K.~Funakubo et al, Prog.\ Theor.\ Phys.\ \tb{99} (1998),
1045, \\
K.~Funakubo et al, Prog.\ Theor.\ Phys.\ \tb{102} (1999), 389
%
\bibitem{la99} M.~Laine, K.~Rummukainen, Nucl.\ Phys.\ \tb{B545}
(1999), 141, \\
M.~Laine, K.~Rummukainen, hep-lat/9908045
%
\bibitem{boyd} C.~G.~Boyd, D.~E.~Brahm, S.~D.~H.~Hsu, Phys.\ Rev.\
\tb{D48} (1993), 4952
%
\bibitem{kus96} A.~Kusenko, P.~Langacker, G.~Segre, Phys.\ Rev.\
\tb{D54} (1996), 5824
%
\bibitem{pms/ref1} http://www.pricewatch.com
%
\bibitem{eicker} N.~Eicker et al, hep-lat/9909146
%
\bibitem{nepsz} N\'epszabads\'ag, 2000.\ feb.\ 15
%
\bibitem{chip} Chip Magazin, 2000.\ m\'ajus 5.
%
\bibitem{iwasaki} Y.~Iwasaki, Nucl.\ Phys.\ (Proc.\ Suppl.) \tb{60A}
(1998), 246, \\
S.~Aoki et al, hep-lat/9903001, \\
http://www.rccp.tsukuba.ac.jp
%
\bibitem{chen} D.~Chen et al, Nucl.\ Phys.\ \tb{B} (Proc.\ Suppl.)
\tb{73} (1999), 808, \\
http://phys.columbia.edu/cqft
%
\bibitem{cikk2} F.~Csikor, Z.~Fodor, P.~Heged\"us, A.~Jakov\'ac,
S.~Katz, A.~Pir\'oth, hep-ph/0001087
%
\bibitem{kaj97} K.~Kajantie et al, Nucl.\ Phys.\ \tb{B493} (1997), 413
%
\bibitem{moore} G.~D.~Moore, Nucl.\ Phys.\ \tb{B523} (1998), 568
%
\bibitem{lus} M.~L\"uscher, P.~Weisz, Commun.\ Math.\ Phys.\ \tb{97}
(1985), 59, \\
P.~Weisz, R.~Wohlert, Nucl.\ Phys.\ \tb{B236} (1984), 397
%
\bibitem{lyang} C.~N.~Yang, T.~D.~Lee, Phys.\ Rev.\ \tb{87} (1952),
404
%
\bibitem{itz} C.~Itzykson, R.~B.~Pearson, J.~B.~Zuber, Nucl.\ Phys.\
\tb{B220}[FS8] (1983), 415
%
\bibitem{lycikk} Y.~Aoki et al, Phys.\ Rev.\ \tb{D60} (1999) 013001
%
\bibitem{ferr-swen} A.~M.~Ferrenberg, R.~Swendsen, Phys.\ Rev.\ Lett.\
\tb{61} (1988), 2058, \\
A.~M.~Ferrenberg, R.~Swendsen, Phys.\ Rev.\ Lett.\ \tb{63} (1989),
1195
%
\bibitem{coh-kap} A.~G.~Cohen, D.~B.~Kaplan, A.~E.~Nelson, Nucl.\
Phys.\ \tb{B349} (1991), 727
%
\bibitem{joy} M.~Joyce, T.~Prokopec, N.~Turok, Phys.\ Rev.\ Lett.\
\tb{75} (1995), 1695, \\
\qquad Erratum Phys.\ Rev.\ Lett.\ \tb{75} (1995), 3375, \\
M.~Joyce, T.~Prokopec, N.~Turok, Phys.\ Rev.\ \tb{D53} (1996), 2958
%
\bibitem{mor} J.~M.~Moreno et al, Nucl.\ Phys.\ \tb{B526} (1998), 489
%
\bibitem{john} P.~John, Phys.\ Lett.\ \tb{B452} (1999), 221
%
\bibitem{jas84} D.~Jasnow, Rep.\ Prog.\ Phys.\ \tb{47} (1984), 1059
%
\end{thebibliography}
\end{document}